\pdfoutput=1

\newcommand{\thesisTitleFrontmatter}{ENABLING FAST, ACCURATE, AND EFFICIENT\\ REAL-TIME GENOME ANALYSIS\\ VIA NEW ALGORITHMS AND TECHNIQUES}

\newcommand{\thesisTitlePlain}{Enabling Fast, Accurate, and Efficient Real-Time Genome Analysis via New Algorithms and Techniques}
\newcommand{\thesisDissNumber}{30696}
\newcommand{\thesisAuthor}{Can Firtina}
\newcommand{\thesisUni}{\protect{ETH Zurich}}

\newcommand{\thesisYear}{2024}
\newcommand{\thesisVersion}[0]{1.0}

\documentclass[12pt,oneside,a4paper]{ethzthesis}
\usepackage{listings}
\usepackage{fancyhdr}
\usepackage{datetime}
\usepackage{cite}
\usepackage{amsmath}
\usepackage{algorithmic}
\usepackage{graphicx}
\usepackage{textcomp}
\usepackage[table]{xcolor}
\usepackage{tikz}
\usepackage[utf8]{inputenc}
\usepackage[T1]{fontenc}
\usepackage{booktabs} %
\usepackage{setspace}
\usepackage[italic]{mathastext}
\usepackage{array}
\usepackage{titlesec}
\usepackage[normalem]{ulem}
\usepackage{multirow}
\usepackage{multicol}
\usepackage{color}
\usepackage[font={small,bf}]{caption}
\usepackage{float}
\usepackage[font={small,bf}]{subcaption}
\usepackage[linesnumbered,ruled]{algorithm2e}
\usepackage{makecell}
\usepackage{pifont}
\usepackage[]{microtype}
\usepackage{pdfpages}

\usepackage{todonotes}
\usepackage{ifthen}

\usepackage{balance}
\usepackage[framemethod=tikz]{mdframed}
\usepackage{xargs}
\usepackage[most]{tcolorbox}

\usepackage[resetlabels]{multibib}

\definecolor{bestresult}{HTML}{b3e2cd} %

\definecolor{blue}{rgb}{0.0, 0.0, 0.0}
\definecolor{red}{rgb}{0.0, 0.0, 0.0}
\definecolor{blue}{rgb}{0.0, 0.0, 0.0}
\definecolor{green}{rgb}{0.0, 0.0, 0.0}
\definecolor{yellow}{rgb}{0.0, 0.0, 0.0}

\definecolor{denim}{rgb}{0.08, 0.38, 0.74}
\definecolor{darkolivegreen}{rgb}{0.33, 0.42, 0.18}
\definecolor{dgreen}{rgb}{0.00, 0.75, 0.00}
\definecolor{darkpink}{rgb}{0.88, 0.28, 0.54}
\definecolor{forestgreen}{rgb}{0.0, 0.27, 0.13}
\definecolor{amber}{rgb}{1.0, 0.49, 0.0}
\definecolor{lightyellow}{rgb}{0.980, 0.956, 0.623}
\definecolor{lightblue}{rgb}{0.980, 0.956, 0.623}
\definecolor{darkamber}{rgb}{0.5, 0.19, 0.0}
\definecolor{dkgreen}{rgb}{0,0.6,0}
\definecolor{gray}{rgb}{0.5,0.5,0.5}
\definecolor{mauve}{rgb}{0.58,0,0.82}
\definecolor{lightmauve}{rgb}{0.68,0.4,0.92}
\definecolor{chocolate}{rgb}{0.48, 0.25, 0.0}
\definecolor{dollarbill}{rgb}{0.52,0.73,0.4}
\definecolor{dkdkgreen}{rgb}{0,0.45,0}
\definecolor{gfored}{rgb}{0.580, 0.050, 0.211}
\definecolor{darkwarmgray}{rgb}{0.15, 0.050, 0.05}
\definecolor{ups-truck}{rgb}{0.53, 0.28, 0.21}

\definecolor{worseresult}{HTML}{f4c7c3} %

\newcommand\rev[1]{{\color{black}{#1}}}
\newcommand\revb[1]{{\color{black}{#1}}}
\newcommand\revc[1]{{\color{black}{#1}}}
\newcommand\revd[1]{{\color{black}{#1}}}

\newboolean{showtodos}
\setboolean{showtodos}{false} %
\newcommand{\mytodo}[1]{%
  \ifthenelse{\boolean{showtodos}}%
    {\todo[size=\scriptsize]{#1}}{}%
}

\titlespacing*{\section}{0pt}{0ex}{0ex}
\titlespacing*{\subsection}{0pt}{0ex}{0ex}
\titlespacing*{\subsubsection}{0pt}{0ex}{0ex}
\titlespacing\section{1pt}{5pt plus 0.5pt minus 4pt}{5pt plus 0.5pt minus 4pt}
\titlespacing\subsection{1pt}{5pt plus 0.5pt minus 3pt}{5pt plus 0.5pt minus 3pt}
\titlespacing\subsubsection{1pt}{5pt plus 0.5pt minus 2pt}{5pt plus 0.5pt minus 2pt}

\newcommand{\circlednumber}[1]{\raisebox{0.1pt}{
  \protect\tikz[baseline=(myanchor.base)]{
  \protect\node[circle,fill=.,inner sep=0.8pt] (myanchor) {\color{-.}\scriptsize \textbf{#1}};}%
}}

\expandafter\def\expandafter\UrlBreaks\expandafter{\UrlBreaks
  \do\a\do\b\do\c\do\d\do\e\do\f\do\g\do\h\do\i\do\j
  \do\k\do\l\do\m\do\n\do\o\do\p\do\q\do\r\do\s\do\t
  \do\u\do\v\do\w\do\x\do\y\do\z\do\A\do\B\do\C\do\D
  \do\E\do\F\do\G\do\H\do\I\do\J\do\K\do\L\do\M\do\N
  \do\O\do\P\do\Q\do\R\do\S\do\T\do\U\do\V\do\W\do\X
  \do\Y\do\Z}

\newcommand{\head}[1]{{\noindent\textbf{#1.}\xspace}} %

\newcommand{\circled}[1]{{\tikz[baseline=(char.base)]{\node[shape=circle,inner sep=1.3pt,fill=black, text=white] (char) {\small \textbf{#1}};}}}

\newcommand{\citegenedit}{jinek_programmable_2012, doudna_promise_2020, pickar-oliver_next_2019, selvakumar_crisprcas9_2022,puchta_updates_2022,luo_crispr-cas9_2022,li_computational_2022,katti_crispr_2022,huang_high-throughput_2022,braun_tutorial_2022,hwang_current_2021,brakebusch_crispr_2021,bao_tools_2021,zhang_silico_2020,yan_benchmarking_2020,sledzinski_computational_2020,manghwar_crisprcas_2020,liu_computational_2020,lin_computational_2020,jeong_current_2020,hanna_design_2020,clement_technologies_2020,chaudhari_evaluation_2020,bodapati_benchmark_2020,bradford_benchmark_2019,yan_benchmarking_2018,sentmanat_survey_2018,chuai_silico_2017}
\newcommand{\citeneonetal}{miller_26-hour_2015,saunders_rapid_2012,soden_effectiveness_2014,priest_molecular_2014,willig_whole-genome_2015,mellis_diagnostic_2022,best_promises_2018,sabbagh_current_2020,monaghan_use_2020,wang_large-scale_2020,guadagnolo_prenatal_2021,emms_next_2022,chandler_lessons_2022,wang_diagnostic_2021,li_prenatal_2020,mone_evolving_2022,poljak_prenatal_2023,de_koning_prenatal_2022,kilby_role_2021,vora_prenatal_2023,tolusso_beyond_2021,bowling_genome_2022,rinaldi_next-generation_2020,lei_whole-exome_2021,zalusky_3-hour_2024}
\newcommand{\citeagriculture}{the_arabidopsis_genome_initiative_analysis_2000,zhu_applications_2020,choi_nanopore_2020,stevens_sequence_2016,campos_high_2021,gao_genome_2021,van_dijk_machine_2021,sun_twenty_2022,kim_application_2020,thudi_genomic_2021,michael_building_2020,shen_omics-based_2022,shahroodi_demeter_2022}
\newcommand{\citepersonalized}{alkan_personalized_2009,lightbody_review_2019,morganti_next_2019,branco_bioinformatics_2021,quazi_artificial_2022,aronson_building_2015,f_lochel_comparative_2020,papadopoulou_application_2023,tafazoli_applying_2021,gambardella_personalized_2020,leary_development_2010,hamburg_margaret_a_path_2010,van_der_lee_technologies_2020,moon_precision_2022,mohan_profiling_2020,chung_rapid_2020,bielinski_preemptive_2014,ho_enabling_2020,hussen_emerging_2022,russell_pharmacogenomics_2021,verma_nanopore_2024}
\newcommand{\citeevolution}{kanehisa_toward_2019,qing_whole_2022,wittkopp_cis-regulatory_2012,romero_comparative_2012,wang_population_2020,hill_molecular_2021,vaishnav_evolution_2022,zhang_haplotype-resolved_2021,fay_evaluating_2008,kanzi_next_2020,wray_evolution_2003,wu_one_2021,signor_evolution_2018,whitehead_variation_2006,coolon_tempo_2014}
\newcommand{\citecancer}{lawrence_mutational_2013,vogelstein_cancer_2013,ramskold_full-length_2012,baslan_unravelling_2017,shapiro_single-cell_2013,sakamoto_new_2020,jia_high-throughput_2022,lawson_tumour_2018,liu_mrna-based_2023,van_de_sande_applications_2023,chakravarty_clinical_2021,cortes-ciriano_computational_2022,deveson_evaluating_2021,xiao_toward_2021,bolton_cancer_2020,szustakowski_advancing_2021,navin_future_2011,hong_rna_2020,lei_applications_2021,han_single-cell_2022,federici_variants_2020,zhang_singlecell_2021,ren_understanding_2018,tian_cicero_2020,malone_molecular_2020,tang_single-cell_2019,ellsworth_single-cell_2017,zhong_application_2021,stadler_therapeutic_2021,tan_targeted_2022,degasperi_substitution_2022,xu_single-cell_2022,horak_comprehensive_2021,zhang_single-cell_2016,bruno_next_2020,de_luca_fgfr_2020,waarts_targeting_2022,lim_advancing_2020,colomer_when_2020,saadatpour_single-cell_2015,dizman_sequencing_2020,buzdin_rna_2020,xiao_tumor_2021,nandwani_lncrnas_2021,marchetti_error-corrected_2023,chen_next-generation_2021,navin_first_2015}
\newcommand{\citeoutbreak}{dunn_squigglefilter_2021,robinson_genomics_2013,fournier_clinical_2014,koser_routine_2012,eloit_diagnosis_2014,gardy_jennifer_l_whole-genome_2011,taylor_angela_j_characterization_2015,quainoo_scott_whole-genome_2017,goldberg_brittany_making_2015,gilchrist_carol_a_whole-genome_2015,arias_rapid_2016,besser_interpretation_2019,li_application_2021,deng_integrated_2021,bertelli_rapid_2013,kwong_whole_2015,deurenberg_application_2017,tang_infection_2017,croucher_application_2015,comin_investigation_2020}
\newcommand{\citemicrobiome}{johnson_evaluation_2019,land_insights_2015,galloway-pena_tools_2020,song_progress_2018,chiu_clinical_2019,lotstedt_spatial_2023,strong_microbial_2014,galazzo_how_2020,malla_exploring_2019,jovel_characterization_2016,lepage_metagenomic_2013,bhagwat_exploring_2021,gao_introduction_2021,comeau_andre_m_microbiome_2017,cummings_clinical_2016,cox_sequencing_2013,bharti_current_2021,bashiardes_use_2016,laudadio_quantitative_2018,wensel_next-generation_2022,mitchell_next-generation_2019,ojala_current_2023,ranjan_analysis_2016,kashyap_microbiome_2017,fricker_what_2019}
\newcommand{\citeforensics}{bruijns_massively_2018,ballard_massive_2020,smith_evolution_2024,ogden_nanopore_2021,yang_application_2014,tytgat_nanopore_2020,diepenbroek_pushing_2021,michaleas_parallel_2022,daniel_snapshot_2015,borsting_next_2015,liu_review_2018,kocher_inter-laboratory_2018,de_knijff_next_2019,gandotra_validation_2020,haas_forensic_2021,foley_global_2023,ariza_next-generation_2022,meiklejohn_current_2021,alvarez-cubero_next_2017}

\newcommand{\citesbs}{bentley_accurate_2008,margulies_genome_2005,shendure_accurate_2005,harris_single-molecule_2008,turcatti_new_2008,wu_termination_2007,fuller_rapid_2007,mckernan_reagents_2008,fuller_method_2011}
\newcommand{\citesbsrealtime}{lindner_hilive_2017,loka_reliable_2019,tausch_livekrakenreal-time_2018,stranneheim_rapid_2014,zhang_optimized_2022,tausch_patholivereal-time_2022,loka_prilive_2018}

\newcommand{\citebasecallsbs}{kao_naivebayescall_2010,cacho_base-calling_2018,erlich_alta-cyclic_2008,wang_adaptive_2017,rougemont_probabilistic_2008,shen_particlecall_2012,ji_bm-bc_2012,das_base_2013,kircher_improved_2009,massingham_all_2012,ye_blindcall_2014,renaud_freeibis_2013,das_onlinecall_2012,menges_totalrecaller_2011,bravo_model-based_2010,kao_bayescall_2009}

\newcommand{\citesmrt}{eid_real-time_2009}
\newcommand{\citesanger}{sanger_dna_1977}

\newcommand{\citenanopore}{menestrina_ionic_1986,cherf_automated_2012,manrao_reading_2012,laszlo_decoding_2014,deamer_three_2016,kasianowicz_characterization_1996,meller_rapid_2000,stoddart_single-nucleotide_2009,laszlo_detection_2013,schreiber_error_2013,butler_single-molecule_2008,derrington_nanopore_2010,song_structure_1996,walker_pore-forming_1994,wescoe_nanopores_2014,lieberman_processive_2010,bezrukov_dynamics_1996,akeson_microsecond_1999,stoddart_nucleobase_2010,ashkenasy_recognizing_2005,stoddart_multiple_2010,bezrukov_current_1993,zhang_single-molecule_2024}
\newcommand{\citebasecallnanodnn}{cavlak_targetcall_2024,xu_fast-bonito_2021,peresini_nanopore_2021,boza_deepnano_2017,boza_deepnano-blitz_2020,oxford_nanopore_technologies_dorado_2024,oxford_nanopore_technologies_guppy_2017,lv_end--end_2020,singh_rubicon_2024,zhang_nanopore_2020,xu_lokatt_2023,zeng_causalcall_2020,teng_chiron_2018,konishi_halcyon_2021,yeh_msrcall_2022,noordijk_baseless_2023,huang_sacall_2022,miculinic_mincall_2019}
\newcommand{\citebasecallnanohmm}{loman_complete_2015,david_nanocall_2017,timp_dna_2012,schreiber_analysis_2015}
\newcommand{\citesignalanalysis}{bao_squigglenet_2021,loose_real-time_2016,zhang_real-time_2021,kovaka_targeted_2021,senanayake_deepselectnet_2023,sam_kovaka_uncalled4_2024,lindegger_rawalign_2024,firtina_rawhash_2023,firtina_rawhash2_2024,firtina_rawsamble_2024,shih_efficient_2023,sadasivan_rapid_2023,dunn_squigglefilter_2021,shivakumar_sigmoni_2024,sadasivan_accelerated_2024,gamaarachchi_gpu_2020,samarasinghe_energy_2021}
\newcommand{\citesignalanalysismapped}{loose_real-time_2016,zhang_real-time_2021,kovaka_targeted_2021,lindegger_rawalign_2024,firtina_rawhash_2023,firtina_rawhash2_2024,shih_efficient_2023,dunn_squigglefilter_2021,shivakumar_sigmoni_2024,sadasivan_accelerated_2024,gamaarachchi_gpu_2020,samarasinghe_energy_2021}
\newcommand{\citebasecalledreal}{payne_readfish_2021,edwards_real-time_2019,ulrich_readbouncer_2022,mikalsen_coriolis_2023,samarakoon_genopo_2020,ahmed_pan-genomic_2021,ahmed_spumoni_2023,weilguny_dynamic_2023}
\newcommand{\citerealtimeall}{\citesignalanalysis,\citebasecalledreal}

\newcommand{\citemethbasecalled}{simpson_detecting_2017,yao_effective_2024,liu_nanomod_2019,bai_deepbam_2024,stanojevic_rockfish_2024,zhang_application_2021,bonet_deepmp_2022,yuen_systematic_2021,mcintyre_single-molecule_2019,rand_mapping_2017,ni_deepsignal_2019,liu_dna_2021,liu_detection_2019,sigurpalsdottir_comparison_2024,ahsan_signal_2024,ni_rna_2024}
\newcommand{\citemethraw}{laszlo_detection_2013,schreiber_error_2013,wescoe_nanopores_2014,stoiber_novo_2017,liu_nanomod_2019}

\newcommand{\citepolyaraw}{krause_tailfindr_2019,workman_nanopore_2019}

\newcommand{\citestrraw}{giesselmann_analysis_2019,de_roeck_nanosatellite_2019,sitarcik_warpstr_2023}

\newcommand{\citemaphashtable}{chakraborty_s-conlsh_2021,li_minimap_2016,li_minimap2_2018,chakraborty_conlsh_2020,lin_zoom_2008,david_shrimp2_2011,sahlin_effective_2021,ren_lra_2021,jiang_seqmap_2008,li_soap_2008,smith_using_2008,ma_patternhunter_2002,altschul_basic_1990,altschul_gapped_1997,kent_blatblast-like_2002,ning_ssaha_2001,egidi_better_2013,rizk_gassst_2010,wu_fast_2010,liu_rhat_2016,alkan_personalized_2009,hach_mrsfast_2010,rumble_shrimp_2009,weese_razersfast_2009,schneeberger_simultaneous_2009,homer_bfast_2009,ondov_efficient_2008,wu_gmap_2005,slater_automated_2005,schwartz_humanmouse_2003,sahlin_strobealign_2022,firtina_blend_2023,firtina_rawhash_2023,firtina_rawhash2_2024,li_mapping_2008,sovic_fast_2016,shaw_fast_2023,shiryev_indexing_2024,xi_bsmap_2009,kim_improving_2014,liu_desalt_2019,lee_mosaik_2014,li_wham_2012,schatz_cloudburst_2009,clement_gnumap_2010,eaves_mom_2009,campagna_pass_2009,chen_perm_2009,malhis_slidermaximum_2009,sedlazeck_nextgenmap_2013,au_detection_2010,bryant_supersplatspliced_2010,wang_duhi_2023,mu_fast_2012,faust_yaha_2012,hu_osa_2012,liao_subread_2013,gontarz_srmapper_2013,hach_mrsfast-ultra_2014,karami_designing_2024,ekim_efficient_2023}
\newcommand{\citemapsuffix}{marcais_mummer4_2018,chaisson_mapping_2012,li_soap2_2009,langmead_aligning_2010,ferragina_opportunistic_2000,shivakumar_sigmoni_2024,langmead_ultrafast_2009,vasimuddin_efficient_2019,ahmed_pan-genomic_2021,ahmed_spumoni_2023,lam_compressed_2008,wheeler_block-sorting_1994,boucher_r-indexing_2021,langmead_fast_2012,marco-sola_gem_2012,kim_hisat_2015,kim_graph-based_2019,chen_bs_2010,boucher_prefix-free_2019,hong_pfp-fm_2024,hoffmann_fast_2009,burkhardt_q-gram_1999,mccreight_space-economical_1976,grossi_compressed_2000,gagie_optimal-time_2018,gagie_fully_2020,kempa_dynamic_2022,meek_-_2003,de_bona_optimal_2008,trapnell_tophat_2009,harris_brat-bw_2012,tennakoon_batmis_2012,siragusa_fast_2013,tarraga_acceleration_2014,lin_kart_2017,lin_dart_2018,haghshenas_lordfast_2019,jung_bwa-meme_2022,ji_compressive_2024,lippert_space-efficient_2005,mun_matching_2020,rossi_moni_2022,rossi_finding_2022,subramaniyan_accelerated_2021,sadakane_new_2003,boucher_r-indexing_2024,guo_efficient_2024,li_aligning_2013,kielbasa_adaptive_2011,bertram_move-r_2024,kovaka_targeted_2021,li_bwt_2024}

\newcommand{\citemappairalign}{marco-sola_fast_2021,lindegger_scrooge_2023,marco-sola_optimal_2023,baeza-yates_new_1992,myers_fast_1999,needleman_general_1970,smith_identification_1981,senol_cali_genasm_2020,papamichail_improved_2009,suzuki_introducing_2018,waterman_biological_1976,wu_onp_1990,gotoh_improved_1982,wu_fast_1992,wagner_string--string_1974,sankoff_matching_1972,groot_koerkamp_exact_2024,sellers_theory_1974,ukkonen_algorithms_1985}

\newcommand{\citemapsketch}{zheng_improved_2020,li_minimap2_2018,jain_weighted_2020,ondov_mash_2016,roberts_reducing_2004,berlin_assembling_2015,baker_dashing_2019,sahlin_effective_2021,edgar_syncmers_2021,chin_human_2019,broder_resemblance_1997,sahlin_strobealign_2022,baker_dashing_2023,joudaki_aligning_2023,schleimer_winnowing_2003,karami_designing_2024,joudaki_fast_2021,irber_lightweight_2022,orenstein_designing_2017,dutta_parameterized_2022,manber_finding_1994,marcais_asymptotically_2018,frith_minimally_2021,frith_how_2023,xin_context-aware_2020}

\newcommand{\citemaplowcolhash}{simpson_abyss_2009,mohamadi_nthash_2016,kazemi_nthash2_2022,gontarz_srmapper_2013,gonnet_analysis_1990,cohen_recursive_1997,pibiri_locality-preserving_2023,karp_efficient_1987,lemire_recursive_2010,wong_aahash_2023,steinegger_clustering_2018,farach_perfect_1996,pibiri_weighted_2023,pibiri_sparse_2022,karcioglu_improving_2021}
\newcommand{\citemapmaskedhash}{chakraborty_s-conlsh_2021,chakraborty_conlsh_2020,petrucci_iterative_2020,mallik_ales_2021,lin_zoom_2008,david_shrimp2_2011,ma_patternhunter_2002,greenberg_lexichash_2023,girotto_fast_2017,girotto_efficient_2018,girotto_fsh_2018}
\newcommand{\citemaplsh}{chakraborty_s-conlsh_2021,ondov_mash_2016,berlin_assembling_2015,chakraborty_conlsh_2020,charikar_similarity_2002,manku_detecting_2007,sinha_fruit-fly_2021,chen_using_2020,sharma_improving_2018,firtina_blend_2023,joudaki_fast_2021,ryali_bio-inspired_2020,marcais_locality-sensitive_2019}

\newcommand{\citemaphash}{chakraborty_s-conlsh_2021,ondov_mash_2016,berlin_assembling_2015,chakraborty_conlsh_2020,charikar_similarity_2002,manku_detecting_2007,petrucci_iterative_2020,mallik_ales_2021,lin_zoom_2008,david_shrimp2_2011,sinha_fruit-fly_2021,chen_using_2020,sharma_improving_2018,ma_patternhunter_2002,firtina_blend_2023,simpson_abyss_2009,greenberg_lexichash_2023,mohamadi_nthash_2016,kazemi_nthash2_2022,gontarz_srmapper_2013,joudaki_fast_2021,gonnet_analysis_1990,cohen_recursive_1997,pibiri_locality-preserving_2023,karp_efficient_1987,lemire_recursive_2010,wong_aahash_2023,steinegger_clustering_2018,farach_perfect_1996,pibiri_weighted_2023,pibiri_sparse_2022,karcioglu_improving_2021,girotto_fast_2017,girotto_efficient_2018,girotto_fsh_2018,ryali_bio-inspired_2020,marcais_locality-sensitive_2019}

\newcommand{\citemapfilter}{xin_accelerating_2013,alser_gatekeeper_2017,xin_shifted_2015,kim_grim-filter_2018,alser_shouji_2019,alser_sneakysnake_2020,alser_magnet_2017,laguna_seed-and-vote_2020,xin_optimal_2016,liao_subread_2013}

\newcommand{\citemapchain}{li_minimap2_2018,abouelhoda_chaining_2005,abouelhoda_multiple_2003,jain_co-linear_2022,chandra_sequence_2023,rizzo_chaining_2023,otto_fast_2011,rajput_co-linear_2024,eppstein_sparse_1992,eppstein_sparse_1992-1,myers_chaining_1995,uricaru_novel_2011,jain_algorithms_2022,chandra_gap-sensitive_2023,makinen_chaining_2020}

\newcommand{\citemaplift}{kim_airlift_2024,chen_improved_2024,sadedin_bazam_2019,kim_fastremap_2022,shumate_liftoff_2021,zhao_crossmap_2014,talenti_nf-lo_2021,mun_leviosam_2021,gao_segment_liftover_2018}

\newcommand{\citeassembly}{kececioglu_combinatorial_1995,cheng_haplotype-resolved_2021,ekim_minimizer-space_2021,li_minimap_2016,nurk_hicanu_2020,fleischmann_whole-genome_1995,myers_whole-genome_2000,iqbal_novo_2012,chin_phased_2016,xiao_mecat_2017,chen_efficient_2021,li_genome_2024,jarvis_semi-automated_2022,kolmogorov_assembly_2019,shafin_nanopore_2020,di_genova_efficient_2021,bankevich_multiplex_2022,cheng_haplotype-resolved_2022,rautiainen_telomere--telomere_2023,ruan_fast_2020,cheng_scalable_2024,eche_bos_2023,vaser_time-_2021,chen_accurate_2021,pevzner_eulerian_2001,lin_assembly_2016,bonfield_new_1995,peltola_seqaid_1984,kamath_hinge_2017,butler_allpaths_2008,koren_canu_2017,myers_fragment_2005}

\newcommand{\citevariantcallers}{alkan_genome_2011,sedlazeck_accurate_2018,poplin_scaling_2018,weckx_novosnp_2005,kwok_comparative_1994,nickerson_polyphred_1997,marth_general_1999,poplin_universal_2018,conrad_origins_2010,mills_mapping_2011,cooper_systematic_2008,eichler_widening_2006,cameron_comprehensive_2019,han_functional_2020,mandiracioglu_ecole_2024,dou_accurate_2020,smolka_detection_2024,lin_svision_2022,popic_cue_2023,narzisi_genome-wide_2018,zheng_symphonizing_2022,zhang_improved_2012,layer_lumpy_2014,karaoglanoglu_valor2_2020,jiang_long-read-based_2020,ahsan_nanocaller_2021,minoche_clinsv_2021,baird_rapid_2008,zarate_parliament2_2020,odonnell_mumco_2020,xu_smcounter2_2019,heller_svim_2019,pedersen_cyvcf2_2017,li_fermikit_2015,eisfeldt_tiddit_2017,zheng_svsearcher_2023,medvedev_detecting_2010,garrison_haplotype-based_2012}

\newcommand{\citemetagenomics}{lapierre_metalign_2020,wood_improved_2019,segata_metagenomic_2012,alneberg_binning_2014,milanese_microbial_2019,blanco-miguez_extending_2023,zhao_keeping_2020,ounit_clark_2015,marcelino_ccmetagen_2020,song_centrifuger_2024,koslicki_david_metapalette_2016,piro_dudes_2016,wu_maxbin_2016,piro_ganon_2020,shen_kmcp_2023,kang_metabat_2015,kang_metabat_2019,lu_bracken_2017,kim_centrifuge_2016,breitwieser_krakenuniq_2018,jahshan_vital_2024}

\newcommand{\citeerrorcorrection}{firtina_hercules_2018,xiao_mecat_2017,stanojevic_telomere--telomere_2024,salmela_lordec_2014,kang_hybrid-hybrid_2023,holley_ratatosk_2021,morisse_hybrid_2018,wang_fmlrc_2018,zhu_lcat_2023,salmela_accurate_2017,bao_halc_2017,haghshenas_colormap_2016,goodwin_oxford_2015,hackl_proovread_2014,hu_lscplus_2016,salmela_correcting_2011,schroder_shrec_2009,salmela_correction_2010,koren_hybrid_2012,au_improving_2012}

\newcommand{\citeassemblypolishing}{loman_complete_2015,vaser_fast_2017,walker_pilon_2014,firtina_apollo_2020,chin_nonhybrid_2013,mastoras_highly_2024, hu_nextpolish2_2024,darian_constructing_2024,wick_polypolish_2022,huang_neuralpolish_2021,huang_blockpolish_2022,zimin_genome_2020,hu_nextpolish_2020,aury_hapo-g_2021,kundu_hypo_2019,warren_ntedit_2019}

\newcommand{\citehwbaseccust}{peresini_nanopore_2021,hammad_scalable_2021,wu_fpga_2022,wu_fpga-accelerated_2020,wu_fpga-based_2018,ramachandra_ont-x_2021}
\newcommand{\citehwbasecgpu}{cavlak_targetcall_2024,xu_fast-bonito_2021,boza_deepnano_2017,oxford_nanopore_technologies_bonito_2021,oxford_nanopore_technologies_dorado_2024,oxford_nanopore_technologies_guppy_2017,lv_end--end_2020,singh_rubicon_2024,zhang_nanopore_2020,xu_lokatt_2023,zeng_causalcall_2020,teng_chiron_2018,konishi_halcyon_2021,yeh_msrcall_2022,noordijk_baseless_2023,huang_sacall_2022,sneddon_language-informed_2022,miculinic_mincall_2019}
\newcommand{\citehwbasecpim}{lou_brawl_2018,lou_helix_2020,mao_genpip_2022,shahroodi_swordfish_2023}
\newcommand{\citehwbasecall}{cavlak_targetcall_2024,xu_fast-bonito_2021,lou_brawl_2018,peresini_nanopore_2021,boza_deepnano_2017,hammad_scalable_2021,wu_fpga_2022,wu_fpga-accelerated_2020,ulrich_readbouncer_2022,boza_deepnano-blitz_2020,oxford_nanopore_technologies_bonito_2021,oxford_nanopore_technologies_dorado_2024,lou_helix_2020,wu_fpga-based_2018,mao_genpip_2022,oxford_nanopore_technologies_guppy_2017,lv_end--end_2020,ramachandra_ont-x_2021,singh_rubicon_2024,shahroodi_swordfish_2023,zhang_nanopore_2020,xu_lokatt_2023,zeng_causalcall_2020,teng_chiron_2018,konishi_halcyon_2021,yeh_msrcall_2022,noordijk_baseless_2023,huang_sacall_2022,sneddon_language-informed_2022,grzesik_serverless_2021}

\newcommand{\citehwassdist}{ellis_dibella_2019,kundeti_pakman_2023,georganas_hipmer_2015,ghosh_pakman_2019,ghosh_pakman_2021,georganas_extreme_2018,mahadik_scalable_2017,guidi_parallel_2021,goswami_distributed_2020,pan_fast_2020,goswami_lazer_2016,meng_swap-assembler_2014,georganas_parallel_2014,qiu_parallelizing_2017,sinha_dsim_2022,poje_first_2024,kalyanaraman_assembling_2006,guidi_distributed-memory_2023}
\newcommand{\citehwasscust}{varma_accelerating_2014,varma_high_2014,varma_hardware_2017,meng_hardware_2014,varma_fpga-based_2016,poirier_dna_2015,poirier_dna_2018,kuo_parallel_2013,varma_fassem_2013,hu_real-time_2016,galanos_fpga-based_2021}
\newcommand{\citehwassgpu}{goswami_gpu-accelerated_2018,jain_gams_2016,garg_ggake_2013,jain_gagm_2013,mahmood_gpu-euler_2011,meng_hardware_2014,qiu_parallelizing_2019,lu_gpu-accelerated_2013,ahmed_gpu_2020}
\newcommand{\citehwasspim}{angizi_panda_2024,zhou_ultra_2021,angizi_pim-assembler_2020,sinha_dsim_2022}
\newcommand{\citehwassall}{ellis_dibella_2019,kundeti_pakman_2023,georganas_hipmer_2015,ghosh_pakman_2019,ghosh_pakman_2021,georganas_extreme_2018,mahadik_scalable_2017,guidi_parallel_2021,goswami_gpu-accelerated_2018,angizi_panda_2024,goswami_distributed_2020,zhou_ultra_2021,pan_fast_2020,jain_gams_2016,goswami_lazer_2016,garg_ggake_2013,jain_gagm_2013,mahmood_gpu-euler_2011,varma_accelerating_2014,varma_high_2014,varma_hardware_2017,meng_hardware_2014,meng_swap-assembler_2014,varma_fpga-based_2016,poirier_dna_2015,poirier_dna_2018,kuo_parallel_2013,qiu_parallelizing_2019,georganas_parallel_2014,qiu_parallelizing_2017,angizi_pim-assembler_2020,sinha_dsim_2022,poje_first_2024,varma_fassem_2013,hu_real-time_2016,galanos_fpga-based_2021,lu_gpu-accelerated_2013,kalyanaraman_assembling_2006,ahmed_gpu_2020,guidi_distributed-memory_2023}

\newcommand{\citehwmetacust}{saavedra_mining_2020,zhang_genomix_2023,cervi_metagenomic_2022,soto_jacc-fpga_2023}
\newcommand{\citehwmetagpu}{kobus_metacache-gpu_2021,jia_metabing_2011,wang_gpmeta_2023,kobus_accelerating_2017,su_parallel-meta_2012,su_gpu-meta-storms_2013,yano_clast_2014}
\newcommand{\citehwmetaisp}{mansouri_ghiasi_megis_2024,wu_abakus_2024}
\newcommand{\citehwmetapim}{shahroodi_demeter_2022,khalifa_clapim_2023,shahroodi_krakenonmem_2022,wu_sieve_2021,jahshan_dash-cam_2023,hanhan_edam_2022,zou_biohd_2022,harary_gcoc_2024,merlin_diper_2024,garzon_hamming_2022}
\newcommand{\citehwmetaall}{shahroodi_demeter_2022,mansouri_ghiasi_megis_2024,khalifa_clapim_2023,kobus_metacache-gpu_2021,shahroodi_krakenonmem_2022,jia_metabing_2011,wang_gpmeta_2023,kobus_accelerating_2017,su_parallel-meta_2012,su_gpu-meta-storms_2013,yano_clast_2014,saavedra_mining_2020,zhang_genomix_2023,cervi_metagenomic_2022,wu_sieve_2021,jahshan_dash-cam_2023,hanhan_edam_2022,zou_biohd_2022,soto_jacc-fpga_2023,harary_gcoc_2024,wu_abakus_2024,merlin_diper_2024,garzon_hamming_2022,besta_communication-efficient_2020}

\newcommand{\citehwrawcust}{shih_efficient_2023,dunn_squigglefilter_2021,samarasinghe_energy_2021}
\newcommand{\citehwrawgpu}{sadasivan_accelerated_2024,gamaarachchi_gpu_2020}
\newcommand{\citehwrawall}{shih_efficient_2023,dunn_squigglefilter_2021,sadasivan_accelerated_2024,gamaarachchi_gpu_2020,samarasinghe_energy_2021}

\newcommand{\citehwaligncust}{banerjee_asap_2019,fei_fpgasw_2018,waidyasooriya_hardware-acceleration_2015,rucci_swifold_2018,haghi_fpga_2021,fujiki_genax_2018,chen_accelerating_2014,fujiki_seedex_2020,chen_novel_2015,li_pipebsw_2021,firtina_aphmm_2024,lindegger_scrooge_2023,madhavan_race_2014,senol_cali_genasm_2020,doblas_gmx_2023,liao_adaptively_2018,teng_adapting_2023,gu_gendp_2023,haghi_wfa-fpga_2023,haghi_wfasic_2023,walia_talco_2024}
\newcommand{\citehwaligngpu}{aguado-puig_accelerating_2022,aguado-puig_wfa-gpu_2022,houtgast_hardware_2018,zeni_logan_2020,ahmed_gasal2_2019,de_oliveira_sandes_cudalign_2016,liu_gswabe_2015,liu_cudasw_2013,liu_cudasw_2009,liu_cudasw_2010,wilton_arioc_2015,goenka_segalign_2020,nishimura_accelerating_2017,lindegger_algorithmic_2022,lindegger_scrooge_2023,park_saloba_2022,park_agatha_2024,kong_cuk-band_2024}
\newcommand{\citehwalignpim}{zokaee_aligner_2018,kaplan_resistive_2017,kaplan_rassa_2018,chowdhury_dna_2020,houtgast_hardware_2018,gupta_rapid_2019,angizi_exploring_2020,huangfu_radar_2018,diab_framework_2024,diab_high-throughput_2022,akbari_customized_2018,kaplan_bioseal_2020,xu_rapidx_2023,lanius_multi-function_2022,zhang_aligner-d_2023}
\newcommand{\citehwalignall}{zokaee_aligner_2018,kaplan_resistive_2017,kaplan_rassa_2018,chowdhury_dna_2020,aguado-puig_accelerating_2022,aguado-puig_wfa-gpu_2022,houtgast_hardware_2018,zeni_logan_2020,ahmed_gasal2_2019,de_oliveira_sandes_cudalign_2016,liu_gswabe_2015,liu_cudasw_2013,liu_cudasw_2009,liu_cudasw_2010,wilton_arioc_2015,banerjee_asap_2019,fei_fpgasw_2018,waidyasooriya_hardware-acceleration_2015,rucci_swifold_2018,gupta_rapid_2019,angizi_exploring_2020,huangfu_radar_2018,goenka_segalign_2020,haghi_fpga_2021,fujiki_genax_2018,nishimura_accelerating_2017,chen_accelerating_2014,fujiki_seedex_2020,chen_novel_2015,li_pipebsw_2021,diab_framework_2024,lindegger_algorithmic_2022,firtina_aphmm_2024,diab_high-throughput_2022,lindegger_scrooge_2023,madhavan_race_2014,senol_cali_genasm_2020,akbari_customized_2018,kaplan_bioseal_2020,doblas_gmx_2023,liao_adaptively_2018,xu_rapidx_2023,teng_adapting_2023,gu_gendp_2023,haghi_wfa-fpga_2023,haghi_wfasic_2023,walia_talco_2024,lanius_multi-function_2022,park_saloba_2022,park_agatha_2024,kong_cuk-band_2024,roodi_memalign_2019,zhang_aligner-d_2023}

\newcommand{\citehwchaincust}{guo_hardware_2019,liyanage_efficient_2023,gu_gendp_2023,liyanage_accelerating_2024}
\newcommand{\citehwchaingpu}{guo_hardware_2019,sadasivan_accelerating_2023}
\newcommand{\citehwchainpim}{chen_parc_2020}
\newcommand{\citehwchainall}{chen_parc_2020,guo_hardware_2019,sadasivan_accelerating_2023,liyanage_efficient_2023,gu_gendp_2023,liyanage_accelerating_2024}

\newcommand{\citehwfiltcust}{alser_gatekeeper_2017,alser_shouji_2019,alser_sneakysnake_2020,singh_fpga-based_2021,alser_magnet_2017,turakhia_darwin_2018}
\newcommand{\citehwfiltgpu}{bingol_gatekeeper-gpu_2021}
\newcommand{\citehwfiltisp}{mansouri_ghiasi_genstore_2022}
\newcommand{\citehwfiltpim}{kim_grim-filter_2018,kaplan_rassa_2018,turakhia_darwin_2018,laguna_seed-and-vote_2020,khalifa_filtpim_2021,nag_gencache_2019,angizi_pim-aligner_2020,shahroodi_rattlesnakejake_2023,shahroodi_sievemem_2023,hameed_alpha_2022}
\newcommand{\citehwfiltall}{alser_gatekeeper_2017,kim_grim-filter_2018,alser_shouji_2019,alser_sneakysnake_2020,singh_fpga-based_2021,kaplan_rassa_2018,alser_magnet_2017,turakhia_darwin_2018,laguna_seed-and-vote_2020,mansouri_ghiasi_genstore_2022,khalifa_filtpim_2021,bingol_gatekeeper-gpu_2021,nag_gencache_2019,angizi_pim-aligner_2020,shahroodi_rattlesnakejake_2023,shahroodi_sievemem_2023,hameed_alpha_2022}

\newcommand{\citehwrmapgpu}{houtgast_efficient_2017,cheng_bitmapper2_2018}
\newcommand{\citehwrmappim}{mao_genpip_2022,khatamifard_genvom_2021,huangfu_beacon_2022,elisseev_scalable_2022,lanius_fully_2024}
\newcommand{\citehwrmapall}{chen_high-throughput_2021,chen_hybrid_2013,houtgast_efficient_2017,goyal_ultra-fast_2017,mao_genpip_2022,khatamifard_genvom_2021,cheng_bitmapper2_2018,chen_when_2016,huangfu_beacon_2022,elisseev_scalable_2022,preuber_short-read_2012,liyanage_efficient_2023,lanius_fully_2024,pavon_quetzal_2024}

\newcommand{\citehwstgcust}{senol_cali_segram_2022,zhang_harp_2024,zeng_high-performance_2024}
\newcommand{\citehwstgall}{senol_cali_segram_2022,zhang_harp_2024,zeng_high-performance_2024}

\newcommand{\citehwseedcust}{han_bless_2024}
\newcommand{\citehwseedgpu}{nisa_distributed-memory_2021,cheng_rapidgkc_2024}
\newcommand{\citehwseedpim}{zokaee_finder_2019,huangfu_nest_2020,jahshan_majork_2024,jahshan_majork_2024,huang_casa_2023,zhang_pim-quantifier_2021,huangfu_medal_2019}
\newcommand{\citehwseeddist}{kundeti_pakman_2023,pan_kmerind_2016,pan_optimizing_2018,nisa_distributed-memory_2021,gao_bloomfish_2017,mocheng_topkmer_2022}
\newcommand{\citehwseedall}{zokaee_finder_2019,huangfu_nest_2020,han_bless_2024,jahshan_majork_2024,kundeti_pakman_2023,pan_kmerind_2016,pan_optimizing_2018,nisa_distributed-memory_2021,gao_bloomfish_2017,mocheng_topkmer_2022,huang_casa_2023,zhang_pim-quantifier_2021,huangfu_medal_2019,cheng_rapidgkc_2024}

\newcommand{\citehwvccust}{wu_fpga_2019,wu_high-throughput_2021,wu_975-mw_2021,wertenbroek_acceleration_2019,huang_hardware_2017,banerjee_accelerating_2017,ham_genesis_2020,lo_algorithm-hardware_2020,xu_fpga_2023,chen_banded_2019}
\newcommand{\citehwvcgpu}{li_improved_2021,das_efficient_2023}
\newcommand{\citehwvcpim}{wu_repair_2022,abecassis_gapim_2023}
\newcommand{\citehwvcall}{wu_fpga_2019,li_improved_2021,wu_high-throughput_2021,wu_975-mw_2021,wertenbroek_acceleration_2019,huang_hardware_2017,banerjee_accelerating_2017,wu_repair_2022,ham_genesis_2020,lo_algorithm-hardware_2020,xu_fpga_2023,das_efficient_2023,chen_banded_2019,roodi_memalign_2019,abecassis_gapim_2023}

\newcommand{\citehwcustall}{alser_gatekeeper_2017,alser_shouji_2019,alser_sneakysnake_2020,singh_fpga-based_2021,chen_high-throughput_2021,peresini_nanopore_2021,hammad_scalable_2021,wu_fpga_2022,wu_fpga-accelerated_2020,alser_magnet_2017,chen_hybrid_2013,turakhia_darwin_2018,goyal_ultra-fast_2017,banerjee_asap_2019,fei_fpgasw_2018,waidyasooriya_hardware-acceleration_2015,rucci_swifold_2018,wu_fpga_2019,shih_efficient_2023,wu_fpga-based_2018,dunn_squigglefilter_2021,ramachandra_ont-x_2021,guo_hardware_2019,haghi_fpga_2021,fujiki_genax_2018,chen_when_2016,chen_accelerating_2014,fujiki_seedex_2020,chen_novel_2015,li_pipebsw_2021,firtina_aphmm_2024,lindegger_scrooge_2023,wu_high-throughput_2021,wu_975-mw_2021,wertenbroek_acceleration_2019,huang_hardware_2017,banerjee_accelerating_2017,madhavan_race_2014,samarasinghe_energy_2021,senol_cali_genasm_2020,saavedra_mining_2020,zhang_genomix_2023,cervi_metagenomic_2022,soto_jacc-fpga_2023,preuber_short-read_2012,doblas_gmx_2023,liyanage_efficient_2023,liao_adaptively_2018,teng_adapting_2023,gu_gendp_2023,haghi_wfa-fpga_2023,haghi_wfasic_2023,walia_talco_2024,pavon_quetzal_2024,ham_genesis_2020,lo_algorithm-hardware_2020,xu_fpga_2023,chen_banded_2019,liyanage_accelerating_2024,varma_accelerating_2014,varma_high_2014,varma_hardware_2017,meng_hardware_2014,varma_fpga-based_2016,poirier_dna_2015,poirier_dna_2018,kuo_parallel_2013,varma_fassem_2013,hu_real-time_2016,galanos_fpga-based_2021}
\newcommand{\citehwispall}{mansouri_ghiasi_genstore_2022,mansouri_ghiasi_megis_2024,wu_abakus_2024}
\newcommand{\citehwgpuall}{cavlak_targetcall_2024,xu_fast-bonito_2021,boza_deepnano_2017,aguado-puig_accelerating_2022,aguado-puig_wfa-gpu_2022,houtgast_hardware_2018,houtgast_efficient_2017,zeni_logan_2020,ahmed_gasal2_2019,de_oliveira_sandes_cudalign_2016,liu_gswabe_2015,liu_cudasw_2013,liu_cudasw_2009,liu_cudasw_2010,wilton_arioc_2015,oxford_nanopore_technologies_bonito_2021,oxford_nanopore_technologies_dorado_2024,oxford_nanopore_technologies_guppy_2017,lv_end--end_2020,bingol_gatekeeper-gpu_2021,guo_hardware_2019,goenka_segalign_2020,cheng_bitmapper2_2018,nishimura_accelerating_2017,lindegger_algorithmic_2022,lindegger_scrooge_2023,singh_rubicon_2024,sadasivan_accelerating_2023,li_improved_2021,zhang_nanopore_2020,xu_lokatt_2023,zeng_causalcall_2020,teng_chiron_2018,konishi_halcyon_2021,yeh_msrcall_2022,noordijk_baseless_2023,huang_sacall_2022,sadasivan_accelerated_2024,gamaarachchi_gpu_2020,kobus_metacache-gpu_2021,sneddon_language-informed_2022,jia_metabing_2011,wang_gpmeta_2023,kobus_accelerating_2017,su_parallel-meta_2012,su_gpu-meta-storms_2013,yano_clast_2014,park_saloba_2022,park_agatha_2024,kong_cuk-band_2024,das_efficient_2023,miculinic_mincall_2019,goswami_gpu-accelerated_2018,jain_gams_2016,garg_ggake_2013,jain_gagm_2013,mahmood_gpu-euler_2011,meng_hardware_2014,qiu_parallelizing_2019,lu_gpu-accelerated_2013,ahmed_gpu_2020}
\newcommand{\citehwpimall}{kim_grim-filter_2018,shahroodi_demeter_2022,lou_brawl_2018,zokaee_aligner_2018,kaplan_resistive_2017,kaplan_rassa_2018,chowdhury_dna_2020,turakhia_darwin_2018,houtgast_hardware_2018,lou_helix_2020,mao_genpip_2022,gupta_rapid_2019,chen_parc_2020,khatamifard_genvom_2021,angizi_exploring_2020,laguna_seed-and-vote_2020,khalifa_filtpim_2021,huangfu_radar_2018,nag_gencache_2019,diab_framework_2024,diab_high-throughput_2022,angizi_pim-aligner_2020,shahroodi_swordfish_2023,huangfu_beacon_2022,khalifa_clapim_2023,shahroodi_rattlesnakejake_2023,shahroodi_sievemem_2023,akbari_customized_2018,elisseev_scalable_2022,kaplan_bioseal_2020,shahroodi_krakenonmem_2022,wu_sieve_2021,jahshan_dash-cam_2023,hanhan_edam_2022,zou_biohd_2022,harary_gcoc_2024,lanius_fully_2024,xu_rapidx_2023,lanius_multi-function_2022,wu_repair_2022,abecassis_gapim_2023,merlin_diper_2024,garzon_hamming_2022,angizi_panda_2024,zhou_ultra_2021,angizi_pim-assembler_2020,sinha_dsim_2022,zhang_aligner-d_2023,hameed_alpha_2022}
\newcommand{\citehwdistall}{ellis_dibella_2019,kundeti_pakman_2023,georganas_hipmer_2015,ghosh_pakman_2019,ghosh_pakman_2021,besta_communication-efficient_2020,georganas_extreme_2018,mahadik_scalable_2017,guidi_parallel_2021,goswami_distributed_2020,pan_fast_2020,goswami_lazer_2016,meng_swap-assembler_2014,georganas_parallel_2014,qiu_parallelizing_2017,sinha_dsim_2022,poje_first_2024,kalyanaraman_assembling_2006,guidi_distributed-memory_2023}

\newcommand{\citehwall}{alser_gatekeeper_2017,kim_grim-filter_2018,alser_shouji_2019,alser_sneakysnake_2020,singh_fpga-based_2021,shahroodi_demeter_2022,chen_high-throughput_2021,cavlak_targetcall_2024,xu_fast-bonito_2021,lou_brawl_2018,zokaee_aligner_2018,kaplan_resistive_2017,kaplan_rassa_2018,zokaee_finder_2019,chowdhury_dna_2020,peresini_nanopore_2021,boza_deepnano_2017,hammad_scalable_2021,wu_fpga_2022,wu_fpga-accelerated_2020,alser_magnet_2017,chen_hybrid_2013,turakhia_darwin_2018,aguado-puig_accelerating_2022,aguado-puig_wfa-gpu_2022,houtgast_hardware_2018,houtgast_efficient_2017,zeni_logan_2020,ahmed_gasal2_2019,de_oliveira_sandes_cudalign_2016,liu_gswabe_2015,liu_cudasw_2013,liu_cudasw_2009,liu_cudasw_2010,wilton_arioc_2015,goyal_ultra-fast_2017,banerjee_asap_2019,fei_fpgasw_2018,waidyasooriya_hardware-acceleration_2015,rucci_swifold_2018,wu_fpga_2019,ulrich_readbouncer_2022,boza_deepnano-blitz_2020,shih_efficient_2023,oxford_nanopore_technologies_bonito_2021,oxford_nanopore_technologies_dorado_2024,lou_helix_2020,wu_fpga-based_2018,mao_genpip_2022,gupta_rapid_2019,chen_parc_2020,khatamifard_genvom_2021,angizi_exploring_2020,laguna_seed-and-vote_2020,mansouri_ghiasi_genstore_2022,khalifa_filtpim_2021,huangfu_radar_2018,dunn_squigglefilter_2021,oxford_nanopore_technologies_guppy_2017,lv_end--end_2020,ramachandra_ont-x_2021,bingol_gatekeeper-gpu_2021,guo_hardware_2019,goenka_segalign_2020,nag_gencache_2019,haghi_fpga_2021,senol_cali_segram_2022,fujiki_genax_2018,cheng_bitmapper2_2018,nishimura_accelerating_2017,chen_when_2016,chen_accelerating_2014,fujiki_seedex_2020,chen_novel_2015,li_pipebsw_2021,diab_framework_2024,lindegger_algorithmic_2022,firtina_aphmm_2024,diab_high-throughput_2022,lindegger_scrooge_2023,singh_rubicon_2024,sadasivan_accelerating_2023,li_improved_2021,wu_high-throughput_2021,wu_975-mw_2021,wertenbroek_acceleration_2019,huang_hardware_2017,banerjee_accelerating_2017,ellis_dibella_2019,angizi_pim-aligner_2020,madhavan_race_2014,mutlu_accelerating_2023,mansouri_ghiasi_megis_2024,shahroodi_swordfish_2023,zhang_nanopore_2020,xu_lokatt_2023,zeng_causalcall_2020,teng_chiron_2018,konishi_halcyon_2021,yeh_msrcall_2022,noordijk_baseless_2023,huang_sacall_2022,sadasivan_accelerated_2024,gamaarachchi_gpu_2020,samarasinghe_energy_2021,senol_cali_genasm_2020,huangfu_beacon_2022,huangfu_nest_2020,khalifa_clapim_2023,shahroodi_rattlesnakejake_2023,shahroodi_sievemem_2023,akbari_customized_2018,elisseev_scalable_2022,kaplan_bioseal_2020,kobus_metacache-gpu_2021,shahroodi_krakenonmem_2022,sneddon_language-informed_2022,grzesik_serverless_2021,jia_metabing_2011,wang_gpmeta_2023,kobus_accelerating_2017,su_parallel-meta_2012,su_gpu-meta-storms_2013,yano_clast_2014,saavedra_mining_2020,zhang_genomix_2023,cervi_metagenomic_2022,wu_sieve_2021,jahshan_dash-cam_2023,hanhan_edam_2022,zou_biohd_2022,soto_jacc-fpga_2023,harary_gcoc_2024,preuber_short-read_2012,doblas_gmx_2023,liyanage_efficient_2023,lanius_fully_2024,liao_adaptively_2018,xu_rapidx_2023,zhang_harp_2024,zeng_high-performance_2024,teng_adapting_2023,gu_gendp_2023,haghi_wfa-fpga_2023,haghi_wfasic_2023,walia_talco_2024,lanius_multi-function_2022,han_bless_2024,wu_abakus_2024,pavon_quetzal_2024,wu_repair_2022,park_saloba_2022,park_agatha_2024,kong_cuk-band_2024,ham_genesis_2020,lo_algorithm-hardware_2020,xu_fpga_2023,das_efficient_2023,chen_banded_2019,roodi_memalign_2019,liyanage_accelerating_2024,miculinic_mincall_2019,flynn_very_1966,jahshan_majork_2024,abecassis_gapim_2023,merlin_diper_2024,garzon_hamming_2022,kundeti_pakman_2023,georganas_hipmer_2015,ghosh_pakman_2019,ghosh_pakman_2021,besta_communication-efficient_2020,georganas_extreme_2018,pan_kmerind_2016,pan_optimizing_2018,mahadik_scalable_2017,guidi_parallel_2021,goswami_gpu-accelerated_2018,angizi_panda_2024,goswami_distributed_2020,zhou_ultra_2021,nisa_distributed-memory_2021,gao_bloomfish_2017,pan_fast_2020,mocheng_topkmer_2022,jain_gams_2016,goswami_lazer_2016,garg_ggake_2013,jain_gagm_2013,mahmood_gpu-euler_2011,varma_accelerating_2014,varma_high_2014,varma_hardware_2017,meng_hardware_2014,meng_swap-assembler_2014,varma_fpga-based_2016,poirier_dna_2015,poirier_dna_2018,kuo_parallel_2013,qiu_parallelizing_2019,georganas_parallel_2014,qiu_parallelizing_2017,angizi_pim-assembler_2020,sinha_dsim_2022,huang_casa_2023,zhang_pim-quantifier_2021,huangfu_medal_2019,zhang_aligner-d_2023,poje_first_2024,hameed_alpha_2022,cheng_rapidgkc_2024,varma_fassem_2013,hu_real-time_2016,galanos_fpga-based_2021,lu_gpu-accelerated_2013,kalyanaraman_assembling_2006,ahmed_gpu_2020,guidi_distributed-memory_2023}

\newcommand{\citesafaripim}{singh_fpga-based_2021,lee_fast_2015,mutlu_accelerating_2023,boroumand_google_2018,gomez-luna_experimental_2022,gomez-luna_machine_2022,singh_accelerating_2021,boroumand_polynesia_2022,giannoula_sparsep_2022,giannoula_towards_2022,denzler_casper_2021,orosa_codic_2021,mutlu_intelligent_2021,boroumand_google_2021,gomez-luna_benchmarking_2022,gomez-luna_benchmarking_2021,mutlu_main_2015,ferreira_pluto_2021,mutlu_modern_2023,boroumand_mitigating_2021,hajinazar_simdram_2021,oliveira_damov_2021,giannoula_syncron_2021,ghose_processing--memory_2019,vijaykumar_locality_2018,kim_evanesco_2020,song_improving_2020,muralidhara_reducing_2011,hajinazar_virtual_2020,seshadri_evicted-address_2012,luo_warm_2015,cai_error_2013,cai_error_2012,cai_flash_2012,cai_program_2013,cai_threshold_2013,cai_data_2015,cai_vulnerabilities_2017,li_utility-based_2017,yoon_row_2012,haj-yahya_techniques_2020,kim_thread_2010,boroumand_conda_2019,luo_clr-dram_2020,pekhimenko_linearly_2012,pekhimenko_linearly_2013,singh_nero_2020,singh_napel_2019,fernandez_natsa_2020,seshadri_-dram_2020,rezaei_nom_2020,kim_solar-dram_2018,mutlu_rowhammer_2019,wang_reducing_2018,mutlu_recent_2018,mutlu_enabling_2019,mutlu_processing_2019,ghose_workload_2019,kanellopoulos_smash_2019,ausavarungnirun_mask_2018,ausavarungnirun_mosaic_2018,jog_owl_2013,jog_exploiting_2016,ghose_processing--memory_2019-1,ghose_enabling_2018,cai_errors_2018,hsieh_accelerating_2016,hassan_softmc_2017,kim_ramulator_2015,ebrahimi_prefetch-aware_2011,liu_concurrent_2017,kim_case_2012,lee_architecting_2009,yoon_efficient_2014,hashemi_continuous_2016,lee_tiered-latency_2013,liu_raidr_2012,kim_flipping_2014,meza_revisiting_2015,mutlu_research_2014,mutlu_memory_2013,chang_understanding_2016,boroumand_lazypim_2016,hsieh_pointer_2016,pattnaik_scheduling_2016,srinath_feedback_2007,mutlu_techniques_2005,lee_simultaneous_2016,ahn_scalable_2015,hsieh_transparent_2016,ahn_pim-enabled_2015,seshadri_dirty-block_2014,seshadri_rowclone_2013,seshadri_fast_2015,joao_flexible_2009,seshadri_buddy-ram_2016,seshadri_simple_2017,seshadri_processing_2016,chang_low-cost_2016,seshadri_ambit_2017,kultursay_evaluating_2013,lee_phase-change_2010,lee_phase_2010,meza_case_2013,zhao_firm_2014,ren_thynvm_2015,kim_dram_2018,subramanian_mise_2013,subramanian_application_2015,mutlu_stall-time_2007,mutlu_parallelism-aware_2008,ausavarungnirun_exploiting_2015,cai_error_2017,tavakkol_flin_2018,ipek_self-optimizing_2008,singh_sibyl_2022,oliveira_accelerating_2022}

\makeatletter
\newcommand\requiredelimiter[2][########]{%
  \ifdefined#2%
    \def\@temp{\def#2#1}%
    \expandafter\@temp\expandafter{#2}%
  \else
    \@latex@error{\noexpand#2undefined}\@ehc
  \fi
}
\@onlypreamble\requiredelimiter
\makeatother

\newcites{S}{Survey Sources}

\usepackage{fancyhdr}
\usepackage{xcolor}
\newcommand{\versionnum}{3.0}
 \fancypagestyle{firststyle}
 {
   \fancyhead[C]{\textcolor{blue}{\emph{Version \versionnum~---~\today, \xxivtime \ UTC}}}%
   \fancyfoot[C]{\thepage}
 }

\newcommand{\setchapterbasednumbering}{
    \setcounter{secnumdepth}{3}
    \renewcommand{\theequation}{\thechapter.\arabic{equation}}
    \renewcommand{\thetable}{\thechapter.\arabic{table}}
    \renewcommand{\thefigure}{\thechapter.\arabic{figure}}
    \renewcommand{\thesection}{\thechapter.\arabic{section}}
    \renewcommand{\thesubsection}{\thesection.\arabic{subsection}}
}

\newcommand{\setsuppbasednumbering}{
    \setcounter{secnumdepth}{3}
    \setcounter{section}{0}
    \setcounter{equation}{0}
    \setcounter{figure}{0}
    \setcounter{table}{0}
    \renewcommand{\theequation}{S\thechapter.\arabic{equation}}
    \renewcommand{\thetable}{S\thechapter.\arabic{table}}
    \renewcommand{\thefigure}{S\thechapter.\arabic{figure}}
    \renewcommand{\thesection}{S\thechapter}
    \renewcommand{\thesubsection}{\thesection.\arabic{subsection}}
    \renewcommand{\thesubsubsection}{\thesubsection.\arabic{subsubsection}}
}

\begin{document}
\frenchspacing
\raggedbottom
\selectlanguage{english}
\pagenumbering{roman}
\pagestyle{plain}

\newcommand\blend{BLEND\xspace}
\newcommand\blendltitle{BLEND: A Fast, Memory-Efficient, and Accurate Mechanism to Find Fuzzy Seed Matches in Genome Analysis\xspace}
\newcommand{\blendrelease}{\href{https://github.com/CMU-SAFARI/BLEND}{https://github.com/CMU-SAFARI/BLEND}\xspace}

\newcommand\blendovmaxseed{$27.3\times$\xspace}

\newcommand\blendavgovpM{$19.3\times$\xspace}
\newcommand\blendavgovpHM{$40.3\times$\xspace}
\newcommand\blendovpM{$83.9\times$\xspace}
\newcommand\blendmovpM{$2.4\times$\xspace}

\newcommand\blendavgovmM{$3.8\times$\xspace}
\newcommand\blendavgovmHM{$7.2\times$\xspace}
\newcommand\blendovmM{$14.1\times$\xspace}
\newcommand\blendmovmM{$0.9\times$\xspace}

\newcommand\blendavgovpMH{$808.2\times$\xspace}
\newcommand\blendavgovpHMH{$1580.0\times$\xspace}
\newcommand\blendovpMH{$4367.8\times$\xspace}
\newcommand\blendmovpMH{$28.4\times$\xspace}

\newcommand\blendavgovmMH{$127.8\times$\xspace}
\newcommand\blendavgovmHMH{$214.0\times$\xspace}
\newcommand\blendovmMH{$234.7\times$\xspace}
\newcommand\blendmovmMH{$36.0\times$\xspace}

\newcommand\blendavgrmpM{$1.7\times$\xspace}
\newcommand\blendavgrmpHM{$2.7\times$\xspace}
\newcommand\blendrmpM{$4.1\times$\xspace}
\newcommand\blendmrmpM{$0.8\times$\xspace}

\newcommand\blendavgrmmM{$1.0\times$\xspace}
\newcommand\blendavgrmmHM{$0.9\times$\xspace}
\newcommand\blendrmmM{$1.1\times$\xspace}
\newcommand\blendmrmmM{$0.5\times$\xspace}

\newcommand\blendavgrmpL{$6.8\times$\xspace}
\newcommand\blendavgrmpHL{$6.1\times$\xspace}
\newcommand\blendrmpL{$18.6\times$\xspace}
\newcommand\blendmrmpL{$1.2\times$\xspace}

\newcommand\blendavgrmmL{$0.6\times$\xspace}
\newcommand\blendavgrmmHL{$0.6\times$\xspace}
\newcommand\blendrmmL{$1.0\times$\xspace}
\newcommand\blendmrmmL{$0.3\times$\xspace}

\newcommand\blendavgrmpW{$4.3\times$\xspace}
\newcommand\blendavgrmpHW{$5.9\times$\xspace}
\newcommand\blendrmpW{$9.9\times$\xspace}
\newcommand\blendmrmpW{$1.1\times$\xspace}

\newcommand\blendavgrmmW{$1.5\times$\xspace}
\newcommand\blendavgrmmHW{$1.0\times$\xspace}
\newcommand\blendrmmW{$4.1\times$\xspace}
\newcommand\blendmrmmW{$0.9\times$\xspace}

\newcommand\blendavgrmpS{$13.3\times$\xspace}
\newcommand\blendavgrmpHS{$19.7\times$\xspace}
\newcommand\blendrmpS{$29.8\times$\xspace}
\newcommand\blendmrmpS{$1.4\times$\xspace}

\newcommand\blendavgrmmS{$1.6\times$\xspace}
\newcommand\blendavgrmmHS{$1.7\times$\xspace}
\newcommand\blendrmmS{$4.2\times$\xspace}
\newcommand\blendmrmmS{$0.2\times$\xspace}

\newcommand\blendavgrmpSt{$0.5\times$\xspace}
\newcommand\blendrmpSt{$0.5\times$\xspace}
\newcommand\blendmrmpSt{$0.5\times$\xspace}

\newcommand\blendavgrmmSt{$1.5\times$\xspace}
\newcommand\blendrmmSt{$1.5\times$\xspace}
\newcommand\blendmrmmSt{$1.5\times$\xspace}

\newcommand\rh{RawHash\xspace}
\newcommand\rhcap{RawHash\xspace}
\newcommand\rhltitle{RawHash: Enabling Fast and Accurate Real-Time Analysis\\ of Raw Nanopore Signals for Large Genomes\xspace}

\newcommand{\rhrelease}{\href{https://github.com/CMU-SAFARI/RawHash}{https://github.com/CMU-SAFARI/RawHash}\xspace}

\newcommand{\ont}{{ONT}\xspace}
\newcommand{\pb}{{PacBio\xspace}}
\newcommand\unc{UNCALLED\xspace}
\newcommand\sig{Sigmap\xspace}
\newcommand\ru{Read Until\xspace}

\newcommand\rhavgthrU{$25.8\times$\xspace}
\newcommand\rhmaxthrU{$82.6\times$\xspace}
\newcommand\rhminthrU{$0.1\times$\xspace}

\newcommand\rhavgthrS{$3.4\times$\xspace}
\newcommand\rhmaxthrS{$6.1\times$\xspace}
\newcommand\rhminthrS{$1.3\times$\xspace}

\newcommand\rhavgmtU{$32.1\times$\xspace}
\newcommand\rhmaxmtU{$178.0\times$\xspace}
\newcommand\rhminmtU{$0.2\times$\xspace}

\newcommand\rhavgmtS{$2.1\times$\xspace}
\newcommand\rhmaxmtS{$3.4\times$\xspace}
\newcommand\rhminmtS{$1.1\times$\xspace}

\newcommand\rht{RawHash2\xspace}
\newcommand\rhtcap{RawHash2\xspace}

\newcommand{\rhmin}{{RawHash2-Minimizer}\xspace}

\newcommand\rhtltitle{RawHash2: Mapping Raw Nanopore Signals\\ Using Hash-Based Seeding and Adaptive Quantization\xspace}

\newcommand{\rhtrelease}{\href{https://github.com/CMU-SAFARI/RawHash}{https://github.com/CMU-SAFARI/RawHash}\xspace}

\newcommand\rhtavgthrU{$26.5\times$\xspace}
\newcommand\rhtmaxthrU{$60.3\times$\xspace}
\newcommand\rhtminthrU{$1.1\times$\xspace}

\newcommand\rhtavgthrS{$19.2\times$\xspace}
\newcommand\rhtmaxthrS{$48.4\times$\xspace}
\newcommand\rhtminthrS{$0.9\times$\xspace}

\newcommand\rhtavgthrR{$4.0\times$\xspace}
\newcommand\rhtmaxthrR{$9.9\times$\xspace}
\newcommand\rhtminthrR{$0.8\times$\xspace}

\newcommand\rhtavgthrRM{$0.4\times$\xspace}
\newcommand\rhtmaxthrRM{$0.6\times$\xspace}
\newcommand\rhtminthrRM{$0.2\times$\xspace}

\newcommand\rhtavgAccRHMin{$4.38\%$\xspace}
\newcommand\rhtavgAccRH{$10.57\%$\xspace}
\newcommand\rhtavgAccUNCALLED{$13.17\%$\xspace}
\newcommand\rhtavgAccSigmap{$18.04\%$\xspace}

\newcommand\rhtmaxAccRHMin{$12.84\%$\xspace}
\newcommand\rhtmaxAccRH{$20.25\%$\xspace}
\newcommand\rhtmaxAccUNCALLED{$44.02\%$\xspace}
\newcommand\rhtmaxAccSigmap{$43.30\%$\xspace}

\newcommand\rhtavgseqU{$1.9\times$\xspace}
\newcommand\rhtmaxseqU{$3.8\times$\xspace}
\newcommand\rhtminseqU{$0.4\times$\xspace}

\newcommand\rhtavgseqS{$1.5\times$\xspace}
\newcommand\rhtmaxseqS{$2.2\times$\xspace}
\newcommand\rhtminseqS{$1.0\times$\xspace}

\newcommand\rhtavgseqR{$1.9\times$\xspace}
\newcommand\rhtmaxseqR{$3.4\times$\xspace}
\newcommand\rhtminseqR{$1.1\times$\xspace}

\newcommand\rs{Rawsamble\xspace}
\newcommand\rscap{Rawsamble\xspace}
\newcommand\rsltitle{Rawsamble: Overlapping and Assembling Raw Nanopore Signals\\using a Hash-based Seeding Mechanism\xspace}
\newcommand\rsasm{Rawasm\xspace}

\newcommand{\rsrelease}{\href{https://github.com/CMU-SAFARI/RawHash}{https://github.com/CMU-SAFARI/RawHash}\xspace}

\newcommand\mm{minimap2\xspace}
\newcommand\mmc{Minimap2\xspace}

\newcommand\rsavgthr{$2{,}432{,}561$\xspace}

\newcommand\rsavgcpummt{$0.12\times$\xspace}
\newcommand\rsmaxcpummt{$0.58\times$\xspace}
\newcommand\rsavgelmmt{$0.16\times$\xspace}
\newcommand\rsmaxelmmt{$0.33\times$\xspace}
\newcommand\rsavgpeakmmt{$1.11\times$\xspace}
\newcommand\rsmaxpeakmmt{$3.98\times$\xspace}
\newcommand\rsavgcpufcpu{$16.45\times$\xspace}
\newcommand\rsmaxcpufcpu{$97.79\times$\xspace}
\newcommand\rsavgelfcpu{$16.36\times$\xspace}
\newcommand\rsmaxelfcpu{$41.59\times$\xspace}
\newcommand\rsavgpeakfcpu{$11.73\times$\xspace}
\newcommand\rsmaxpeakfcpu{$41.99\times$\xspace}
\newcommand\rsavgcpuhcpu{$36.93\times$\xspace}
\newcommand\rsmaxcpuhcpu{$165.36\times$\xspace}
\newcommand\rsavgelhcpu{$59.70\times$\xspace}
\newcommand\rsmaxelhcpu{$149.18\times$\xspace}
\newcommand\rsavgpeakhcpu{$11.23\times$\xspace}
\newcommand\rsmaxpeakhcpu{$54.19\times$\xspace}
\newcommand\rsavgelhgpu{$1.99\times$\xspace}
\newcommand\rsmaxelhgpu{$33.67\times$\xspace}
\newcommand\rsavgpeakhgpu{$1.53\times$\xspace}
\newcommand\rsmaxpeakhgpu{$3.98\times$\xspace}
\newcommand\rsavgelsgpu{$7.40\times$\xspace}
\newcommand\rsmaxelsgpu{$515.67\times$\xspace}
\newcommand\rsavgpeaksgpu{$1.70\times$\xspace}
\newcommand\rsmaxpeaksgpu{$3.98\times$\xspace}

\newcommand\rsavgshared{$36.57\%$\xspace}
\newcommand\rsavgru{$16.25\%$\xspace} %
\newcommand\rsavgmmu{$47.17\%$\xspace} %

\newcommand\rsavgnc{$4.14\times$\xspace}
\newcommand\rsavgcl{$6.19\times$\xspace}
\newcommand\rsavgmcl{$3.22\times$\xspace}

\bstctlcite{IEEEexample:BSTcontrol}
\bstctlcite[@auxoutS]{IEEEexample:BSTcontrol}
\setbiblabelwidth{1000} %

\begin{titlepage}
    \large
    \begin{center}
        \begingroup
        \MakeUppercase{Diss. ETH No. \thesisDissNumber{}}
        \endgroup
    
        \hfill

        \vfill

        \begingroup
            \textit{\textbf{\thesisTitleFrontmatter}}
        \endgroup

        \vfill

        \begingroup
            A thesis submitted to attain the degree of\\
            \vspace{1em}
            \MakeUppercase{Doctor of Sciences}\\
            (Dr. sc. \thesisUni)\\
        \endgroup

        \vfill

        \begingroup
            presented by\\
            \vfill
            \textit{\MakeUppercase{\thesisAuthor}}\\
            \vspace{1em}
            \textit{M.Sc. from Bilkent University}\\
            \vspace{1em}
            born on \textit{19.04.1992}\\
        \endgroup

        \vfill

        \begingroup
            accepted on the recommendation of\\
            \vspace{0.5em}
            Prof.\ Dr.\ Onur Mutlu, examiner\\
            \vspace{0.5em}
            Prof. Dr. Reetuparna Das, co-examiner \\
            \vspace{0.5em}
            Dr. Hasindu Gamaarachchi, co-examiner\\
            \vspace{0.5em}
            Prof. Dr. Benjamin Langmead, co-examiner \\
            \vspace{0.5em}
            Prof. Dr. Heng Li, co-examiner
        \endgroup

        \vfill

        \thesisYear%

        \vfill
    \end{center}
\end{titlepage}

\thispagestyle{empty}

\hfill

\vfill

\noindent\thesisAuthor: \textit{\thesisTitlePlain,}
\textcopyright\ \thesisYear

\cleardoublepage

\thispagestyle{empty}

\vspace*{3cm}

\begin{center}
    \textit{``...Bak dedin buna\\Bak dedin ona\\Bak dedin göğe\\Bak dedin bana\\Bak şu ne?\\Bak baba ne?\\Aldım soruyu,\\Bu ilk saf soruyu,\\Dedim;\\Bu budur\\Bu budur\\Bu da bir diğeri.''}\\
    \hspace{3cm} \textit{Turan Fırtına}\\
    
    \vspace{1cm}
    
    My very first questions in life were met with my parents' patient and loving answers, which sparked a curiosity leading to many more—some of which shaped this thesis.

    With every question that followed, I have been fortunate to have a life partner\\who shares in my wonder and journey with her love and patience.

    This thesis is dedicated to them—my parents, Emine and Turan, and my wife, Çiçek.
\end{center}

\medskip

\clearpage
\chapter*{Acknowledgments}
\addcontentsline{toc}{chapter}{Acknowledgments}

This thesis is the culmination of six years of growth and is the product of support from many individuals and institutions that contributed to its completion. The following acknowledgements reflect my gratitude for their influence.

First, I express my sincere thanks to my advisor, Onur Mutlu, for enabling me to reach the potential that I had not imagined possible. Under his guidance, I have grown both as a researcher and as an individual. As a researcher, I am grateful to him for reshaping my mindset to see the fact that we, as researchers, are mainly problem solvers, and we should be passionate about solving important and impactful research problems in our field to change the world positively. Along with this mindset shift, his constructive feedback and patience have played a key role in making this dissertation possible. Under his guidance, I have grown as an individual, as I now believe it pays off to be patient and caring for the people you trust. I thank him for his trust in me, even when it was not convenient for him.

I wish to thank my PhD committee members, Reetu Das, Hasindu Gamaarachchi, Ben Langmead, and Heng Li, for their time and support in reviewing my thesis and providing valuable feedback. In addition, I am grateful to the anonymous peer reviewers who provided useful comments that improved my work in several journal and conference submissions.

I also acknowledge the funding agencies and industry sponsors who made my PhD work possible: BioPIM, SNSF, Intel, Google, Huawei, Microsoft, VMware, and SRC.

Constructive feedback is essential for growth as a researcher. The SAFARI Research Group provided an ideal environment for receiving such feedback. I thank all the members of the SAFARI Research Group for the stimulating and scholarly intellectual environment they created and for their friendship. I am especially grateful to Jeremie Kim for being an exceptionally good and kind friend, my first flatmate, and a reliable mentor during the early stages of my PhD. Our long, nice, and sometimes weird discussions, especially while drinking Turkish coffee and watching \textit{Brooklyn Nine-Nine}, helped me survive depressing times during the COVID-19 pandemic. I am grateful to Abdullah Giray Yağlıkçı and Geraldo Francisco de Oliveira Junior for their great and long-standing friendship. We started our PhD journeys around the same time, and we are now finishing at nearly the same time. This unique journey with them gave me the emotional, psychological, and professional support I needed throughout my PhD journey. In addition, I thank Can Alkan, Damla Şenol Çalı, Nika Mansouri Ghiasi, Zülal Bingöl, Ataberk Olgun, Joël Lindegger, F. Nisa Bostancı, Ismail E. Yüksel, Minesh Patel, Hasan Hassan, Mohammad Sadrosadati, Konstantina Koliogeorgi, Klea Zambaku, Juan Gómez Luna, Gagandeep Singh, Sreenivas Subramoney, Harun Mustafa, Andre Kahles, Maximilian Mordig, Sayan Goswami, Furkan Eriş, Taha Michael Shahroodi, Jisung Park, Talu Güloğlu, Ezgi Ebren, Haiyu Mao, Rakesh Nadig, Konstantinos Kanellopoulos, Rahul Bera, Stefano Mercogliano, Nastaran Hajinazar, Roknoddin Azizibarzoki, Haocong Luo, Yahya C. Tuğrul, Oğuzhan Canpolat, İrem Boybat Kara, Kaan Kara, Lukas Breitwieser, Kaveh Razavi, Yan Zhu, M. Banu Çavlak, Julien Eudine, Arvid Gollwitzer, Marie-Louise Dugua, Melina Soysal, Ulysse McConnell, Christina Giannoula, Lois Orosa, Giulia Guidi, Qingcai Jiang, Batuhan Ceylan, Merve Gürel, and many others for their support and friendship. I specifically thank Tracy Ewen and Tulasi Blake for their administrative and personal support throughout my PhD journey.

During my time in Zürich, I met many wonderful people and received support from long-standing friendships. I thank Serdar Yonar, İdil Kanpolat, Arif C. Güngör, and Emre Aksan for their initial and ongoing support, especially during my adjustment to life in Zürich. I am grateful to Volkan and Yuliia Çalışkan for their continuous support before and after my move to Zürich. I also thank Berna Kabadayı, Kadir Çetinkaya, Betül Taşkoparan Yağlıkçı, İrem Aylakdurmaz, Zeynep Aydın, and Tobias Bischof for their friendship, not only to me but also essentially to Çiçek. I am grateful to Uğur Utku Atmaca, Onur Tan, Ayhun Tekat, Sevgi G. Kafalı Tekat, İlker Burak Kurt, Feyza G. Özbay Kurt, Cem Sevim, and Mert Albaba for their unwavering support and friendship.

I consider myself extremely fortunate to have been raised by two wonderful parents, Emine and Turan, whose unconditional love and support have given me the courage to take bold steps in life. I am eternally grateful to them. Finally, for my wife, Çiçek, through everything we experienced together, thank you for being my life partner. The PhD journey is not always easy or predictable, and many events may divert one from the well-worn path. I have never been afraid to take these paths, knowing that you are by my side.

\clearpage
\chapter*{Abstract}
\addcontentsline{toc}{chapter}{Abstract}

The advent of high-throughput sequencing (HTS) technologies has revolutionized genome analysis by enabling the rapid and cost-effective sequencing of large genomes. Despite these advancements, the increasing complexity and volume of genomic data present significant challenges related to accuracy, scalability, and computational efficiency. These challenges are mainly due to various forms of unwanted and unhandled variations in sequencing data, collectively referred to as noise. Addressing these challenges requires a deep understanding of \rev{different types of noise in genomic data} and the development of techniques to mitigate \rev{the impact of noise} on genome analysis.

In this dissertation, we aim to understand the \rev{types of noise} that affect the genome analysis pipeline and address challenges \revb{posed by such noise} by developing \revb{new computational techniques} to tolerate or reduce \rev{noise} for faster, more accurate, and scalable analysis of \rev{different types of sequencing data (e.g., raw electrical signals from nanopore sequencing).}

First, we introduce \blend, a noise-tolerant hashing mechanism that quickly identifies both \rev{exactly matching} and highly similar sequences with arbitrary differences using a single lookup of their hash values.
Second, to enable scalable and accurate analysis of noisy raw nanopore signals, we propose \rh, a novel mechanism that effectively reduces noise in raw nanopore signals and enables accurate, real-time analysis by proposing the \emph{first} hash-based similarity search \rev{technique} for raw nanopore signals.
Third, we extend the capabilities of \rh with \rht, an improved mechanism that 1)~provides a better understanding of noise in raw nanopore signals to reduce it more effectively and 2)~improves the robustness of mapping decisions.
Fourth, we explore the broader implications and new applications of raw nanopore signal analysis by introducing \rs, the first mechanism for all-vs-all overlapping of raw signals using hash-based search. \rs enables the construction of \emph{de novo} assemblies directly from raw signals without basecalling, which opens \rev{up} new directions \rev{and uses} for raw nanopore signal analysis.

This dissertation builds a comprehensive understanding of how \rev{noise in different types of genomic data} affects the genome analysis pipeline and provides novel solutions to mitigate the impact of noise. Our findings demonstrate that by effectively tolerating and reducing noise \revb{using new computational techniques}, we can 1)~significantly improve the performance, accuracy, and scalability of genome analysis \rev{and 2)~expand the scope of raw signal analysis by enabling new applications and directions}. We hope and believe that the methods and insights presented in this dissertation will contribute to the \rev{invention and} development of more robust, efficient, \rev{and capable} genomic analysis tools, especially in the field of raw signal analysis.

\clearpage
\chapter*{Zusammenfassung}
\addcontentsline{toc}{chapter}{Zusammenfassung}

Der Aufstieg der Hochdurchsatz-Sequenzierungstechnologien (HTS) hat die Genomanalyse revolutioniert, indem er die schnelle und kosteneffiziente Sequenzierung grosser Genome ermoglicht. Trotz dieser Fortschritte stellen die zunehmende Komplexitat und das grosse Volumen genomischer Daten erhebliche Herausforderungen in Bezug auf Genauigkeit, Skalierbarkeit und rechnerische Effizienz dar. Diese Herausforderungen sind hauptsaechlich auf verschiedene Formen unerwuenschter und unbehandelt bleibender Variationen in Sequenzierungsdaten zurueckzufuhren, die zusammenfassend als Rauschen bezeichnet werden. Die Bewaltigung dieser Herausforderungen erfordert ein tiefes Verstondnis der unterschiedlichen Arten von Rauschen in genomischen Daten sowie die Entwicklung von Techniken, um den Einfluss von Rauschen auf die Genomanalyse zu mindern.

In dieser Dissertation zielen wir darauf ab, die Arten von Rauschen, die die Genomanalyse-Pipeline beeinflussen, zu verstehen und die durch solches Rauschen hervorgerufenen Herausforderungen anzugehen, indem wir neue rechnerische Techniken entwickeln, um Rauschen zu tolerieren oder zu reduzieren. Dies ermoeglicht eine schnellere, genauere und skalierbarere Analyse verschiedener Arten von Sequenzierungsdaten (z. B. rohe elektrische Signale aus der Nanoporen-Sequenzierung).
Erstens, stellen wir \blend vor, einen rausch-toleranten Hashing-Mechanismus, der sowohl exakt uebereinstimmende als auch hochaehnliche Sequenzen mit beliebigen Unterschieden mit einem einzigen Nachschlageverfahren ihrer Hash-Werte schnell identifiziert.
Zweitens, um eine skalierbare und genaue Analyse verrauschter roher Nanoporen-Signale zu ermoeglichen, schlagen wir \rh vor, einen neuartigen Mechanismus, der Rauschen in rohen Nanoporen-Signalen effektiv reduziert und eine genaue, Echtzeit-Analyse ermoeglicht, indem er die \emph{erste} hash-basierte Suchmethode zur Bestimmung der Aehnlichkeit von rohen Nanoporen-Signalen vorschlaegt.
Drittens, erweitern wir die Faehigkeiten von \rh mit \rht, einem verbesserten Mechanismus, der 1) ein tieferes Verstondnis des Rauschens in rohen Nanoporen-Signalen liefert, um es effektiver zu reduzieren, und 2) die Robustheit der Mapping-Entscheidungen verbessert.
Viertens, ergrunden wir die weiterreichenden Implikationen und neuen Anwendungsmoglichkeiten der rohen Nanoporen-Signalanalyse, indem wir \rs vorstellen, den ersten Mechanismus zur All-vs-All-Ueberlappung von rohen Signalen mittels hash-basierter Suche. \rs ermoeglicht den Aufbau von \emph{de novo}-Assemblierungen direkt aus rohen Signalen ohne Basecalling, was neue Richtungen und Anwendungsmoglichkeiten in der rohen Nanoporen-Signalanalyse eroeffnet.

Diese Dissertation baut ein umfassendes Verstondnis dafuer auf, wie Rauschen in verschiedenen Arten von genomischen Daten die Genomanalyse-Pipeline beeinflusst, und bietet neuartige Losungen, um den Einfluss von Rauschen zu mindern. Unsere Ergebnisse zeigen, dass durch die effektive Toleranz und Reduktion von Rauschen mit neuen rechnerischen Techniken 1) die Leistung, Genauigkeit und Skalierbarkeit der Genomanalyse erheblich verbessert werden kann und 2) der Anwendungsbereich der rohen Signalanalyse erweitert wird, indem neue Anwendungen und Richtungen ermoeglicht werden. Wir hoffen und glauben, dass die in dieser Dissertation vorgestellten Methoden und Einsichten zur Entwicklung robusterer, effizienterer und leistungsfaehigerer genomischer Analysetools beitragen werden, insbesondere im Bereich der rohen Signalanalyse.

\pagestyle{headings}
\cleardoublepage
\tableofcontents
\newpage
\listoffigures
\newpage
\listoftables

\cleardoublepage
\pagenumbering{arabic}%
\setstretch{1.3}
\setchapterbasednumbering
\chapter{Introduction}

Genome analysis plays a crucial role in various fields such as personalized medicine\revb{~\cite{\citepersonalized}, genome editing~\cite{\citegenedit}, evolutionary biology~\cite{\citeevolution}, cancer research~\cite{\citecancer}, prenatal and neonatal screening~\cite{\citeneonetal}, outbreak tracing~\cite{\citeoutbreak}, microbiome studies~\cite{\citemicrobiome}, agriculture~\cite{\citeagriculture}, and forensics~\cite{\citeforensics}.} The advent of high-throughput sequencing (HTS) technologies, such as \revb{sequencing-by-synthesis (SBS)~\cite{\citesbs}, Single Molecule Real-Time (SMRT)~\cite{\citesmrt}, and nanopore sequencing~\cite{\citenanopore},} has revolutionized genome analysis, enabling faster and more cost-effective sequencing of genomes by generating large \rev{amounts} of genomic data at relatively low cost~\cite{shendure_dna_2017} \revb{compared to Sanger sequencing~\cite{\citesanger}}. However, the analysis of genomic data is challenging due to a variety of reasons: 1)~HTS technologies can only sequence relatively short fragments of genomes, called \emph{reads}, whose locations in their corresponding genome are unknown~\cite{fleischmann_whole-genome_1995, myers_whole-genome_2000, alkan_limitations_2011, nagarajan_sequence_2013, heather_sequence_2016}, 2)~these reads can contain sequencing errors~\cite{alkan_limitations_2011, shendure_dna_2017, senol_cali_nanopore_2019, firtina_hercules_2018, firtina_apollo_2020}, leading to differences from their original sequences, 3)~the sequenced genome may not (and usually does not) exactly match recorded genomes in a reference database, known as \emph{reference genomes}, due to variations between individuals within and across species~\cite{senol_cali_nanopore_2019, alser_technology_2021}. Despite significant improvements in computational tools since the 1980s~\cite{alser_technology_2021} to overcome such challenges, the rapid growth in genomic data~\cite{stephens_big_2015} has led to ever larger computational overheads in the genome analysis pipeline, posing large challenges for efficient, accurate and timely analysis of genomes~\cite{alser_accelerating_2020, alser_molecules_2022, mutlu_accelerating_2023}.
    
\begin{figure}[tbh]
  \centering
  \includegraphics[width=0.9\linewidth]{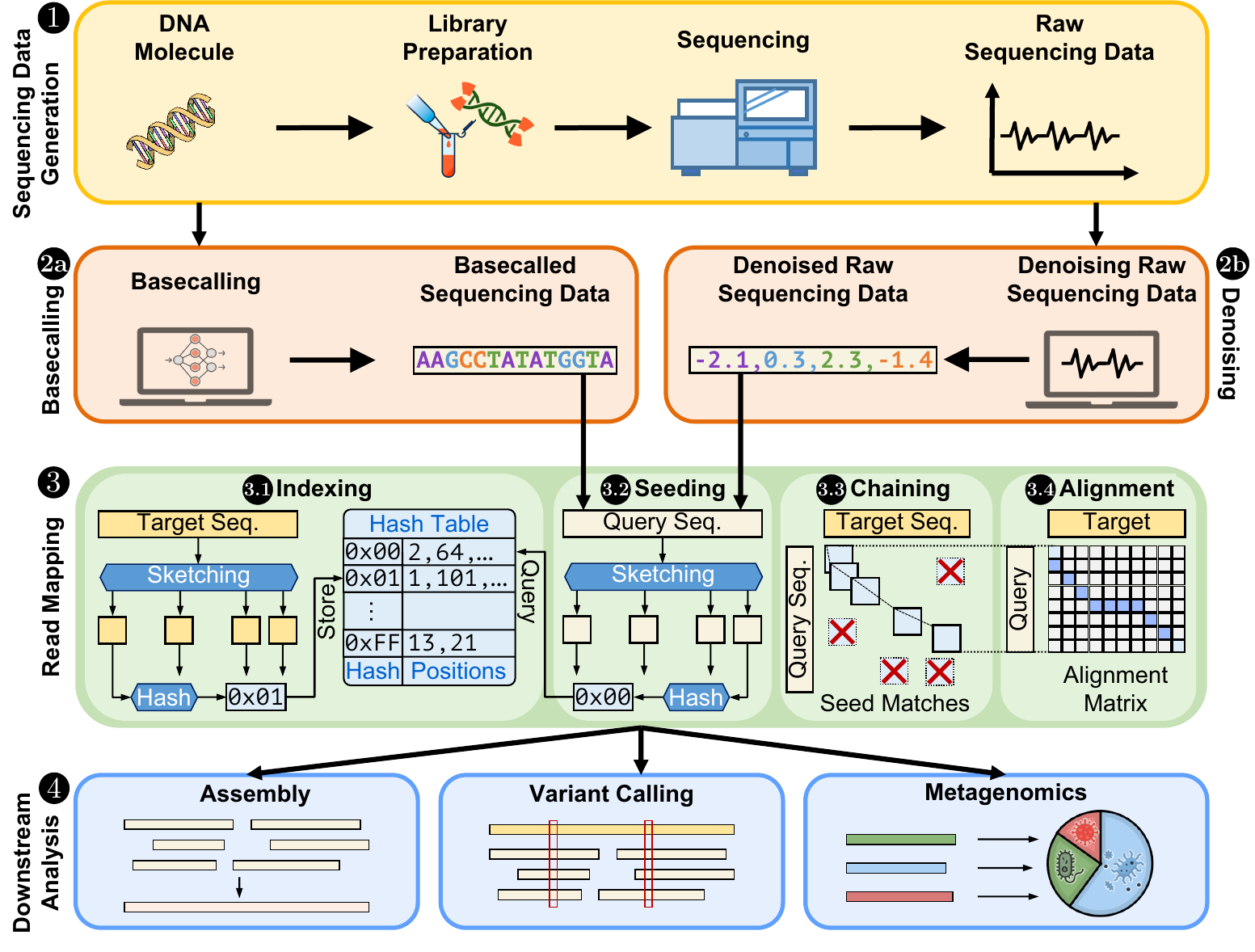}
  \caption{Key steps in the genome analysis pipeline.}
  \label{int:fig:pipeline}
\end{figure}

\rev{Figure~\ref{int:fig:pipeline} shows multiple key steps of a genome analysis pipeline, each of which affects the accuracy, speed, and energy consumption of genome analysis.}

First, a biological molecule (e.g., a DNA molecule~\cite{watson_molecular_1953}) is sequenced using an HTS technology to generate the raw sequencing data (e.g., measured raw electrical signals in nanopore sequencing)\revb{\circlednumber{1}}. \revc{Among HTS technologies, nanopore sequencing uniquely allows for significant reductions in genome analysis time and cost by enabling early termination of sequencing, either for individual reads or the entire run, based on real-time data analysis. This capability is known as \emph{adaptive sampling}~\cite{loose_real-time_2016}.} Such an analysis needs to be quick and, ideally, should be made as soon as the raw signal data is generated by the sequencer. \revc{Analyzing raw nanopore signals during sequencing at a computational speed that matches or exceeds the sequencer's data generation speed (i.e., throughput) is known as \emph{real-time analysis} and is essential for applications such as adaptive sampling.} \revc{Real-time analysis is mainly useful for 1)~reducing the genome analysis latency by overlapping the sequencing with analysis, especially needed for urgent cases that require rapid analysis (e.g., neonatal screening~\cite{zalusky_3-hour_2024}), and 2)~reducing the sequencing time and cost by reducing unnecessary sequencing~\cite{firtina_rawhash2_2024}.} \revc{However, real-time analysis introduces unique challenges, particularly in meeting the stringent throughput and latency requirements demanded by nanopore sequencing. The analysis speed must keep pace with data generation to avoid bottlenecks, which is often difficult due to the large data volumes and the need for real-time decision-making during sequencing~\cite{firtina_rawhash_2023}.} \revc{Section~\ref{bg:sec:hts} provides detailed background on sequencing technologies.}

Second, raw sequencing data is typically translated into sequences of nucleotide characters or \emph{bases} (e.g., A, C, G, and Ts in DNA) via \emph{basecalling}\revb{\circlednumber{2a}}. \revb{Basecalling tools~\cite{\citebasecallnanodnn, \citebasecallnanohmm} mainly} rely on compute-intensive approaches that process large chunks of \emph{noisy} and error-prone raw data to accurately infer the actual nucleotide sequences~\cite{purnell_nucleotide_2008, timp_dna_2012, alser_molecules_2022}. Alternatively \revb{\circlednumber{2b}}, \emph{raw sequencing data} can directly be analyzed \revb{\emph{without} basecalling~\cite{\citesignalanalysis}.} \revc{While direct analysis of raw data can avoid the computational overhead of basecalling, it introduces challenges due to increased noise, requiring specialized techniques for accurate denoising.} \revb{These denoising techniques usually include time series analysis~\revb{\cite{ruxton_unequal_2006, david_nanocall_2017, zhang_real-time_2021}} and quantization~\revb{\cite{firtina_rawhash_2023, firtina_rawhash2_2024, shivakumar_sigmoni_2024}}}.

Third, read mapping\circlednumber{3} aims to find similarities and differences between genomic sequence pairs (e.g., between sequenced \emph{query} reads and \emph{target} reference genomes of one or more species). To facilitate practical similarity identification, \revb{read mapping includes several steps.} \revc{These steps include} constructing (\revc{i.e., } indexing\revc{\circlednumber{3.1}}) and using (\revc{i.e., } seeding\revc{\circlednumber{3.2}}) a database~\cite{\citemaphashtable,\citemapsuffix} for efficient similarity search by utilizing various sketching (i.e., sampling)~\cite{\citemapsketch} and hashing~\cite{\citemaphash} methods. Filtering~\cite{\citemapfilter} and co-linear chaining (i.e., sparse dynamic programming)~\cite{\citemapchain} \revc{steps\revc{\circlednumber{3.3}} in read mapping aim to reduce the workload of the next computationally costly steps in read mapping by quickly identifying highly dissimilar or similar regions between query and target sequences. The alignment~\cite{\citemappairalign} step\revc{\circlednumber{3.4}} identifies the exact differences and similarities between query and target sequences, which overall demand considerable processing power and memory due to the large scale of genomic sequences~\cite{alser_technology_2021, kim_airlift_2024, kim_fastremap_2022}.} \revc{Section~\ref{bg:sec:mapping} provides detailed background on the keys steps in read mapping.}

Fourth, the output generated in the read mapping step can be used in the subsequent steps of genome analysis \revb{(i.e., \emph{downstream analysis}\circlednumber{4}), such as \emph{de novo} genome assembly (\revc{i.e., construction of a genome from scratch})~\cite{\citeassembly},} variant calling (\revc{i.e., identifying genetic variations in an individual's genome compared to a reference genome})~\cite{\citevariantcallers}, and metagenomics (\revc{i.e., identifying and profiling organisms present in an environment})~\cite{\citemetagenomics}.
Downstream analysis often requires additional computationally intensive steps~\cite{alser_molecules_2022}, including set intersection~\cite{liu_cmash_2022}, graph processing~\cite{li_minimap_2016, firtina_aphmm_2024}, and the use of deep neural networks (DNNs)~\cite{poplin_universal_2018}. These additional steps further contribute to the overall computational overhead and energy consumption of the genome analysis pipeline~\cite{alser_molecules_2022}. \revc{Section~\ref{bg:subsec:downstream} provides detailed background on key steps in the downstream genome analysis pipeline after read mapping.}

Many algorithmic, software, and hardware techniques aim to address the computational challenges in the genome analysis pipeline. These works improve the performance and accuracy of the computational tools by 1)~reducing overall computational and space complexity~\cite{marco-sola_fast_2021, marco-sola_optimal_2023,firtina_rawhash_2023,poplin_universal_2018, poplin_scaling_2018}, 2)~eliminating useless work~\cite{cavlak_targetcall_2024,singh_rubicon_2024,firtina_rawhash_2023,singh_rubicon_2024,shivakumar_sigmoni_2024,li_minimap2_2018,firtina_aphmm_2024,\citemapfilter}, 3)~optimizing data structures and memory access patterns~\cite{hach_mrsfast_2010, pan_kmerind_2016, guidi_bella_2021, ellis_dibella_2019,kanellopoulos_smash_2019,senol_cali_segram_2022}, 4)~exploiting parallelism and distributed computing in multi-core, many-core, and SIMD architectures~\cite{xin_shifted_2015, firtina_blend_2023, \citehwassdist, \citehwassgpu,\citehwaligngpu,\citehwchaingpu,\citehwfiltgpu,\citehwrmapgpu,\citehwseeddist,\citehwseedgpu,\citehwvcgpu,\citehwmetagpu,\citehwrawgpu}, and \revb{5) designing specialized} hardware accelerators for many steps in genome analysis~\cite{\citehwall}. \revc{We provide a detailed description of these approaches in Sections~\ref{bg:sec:accelerating} and~\ref{chap:related}}.

\section{Problem Discussion}

Unfortunately, the increasing complexity of genomic data analysis introduces significant challenges in terms of accuracy, scalability, and computational efficiency. These challenges are mainly due to the heuristic techniques that need to analyze \rev{large volumes} of sequencing data in the presence of various forms of \emph{unwanted} and \emph{unhandled} variations. In this dissertation, we refer to these variations collectively as noise. \rev{Noise} can originate from sequencing errors, biological variations, or technical limitations in both the sequencing and analysis processes, which substantially impacts the heuristic decisions and sensitivity of genome analysis.

\head{Noise in Seed Matching} 
A critical step in genome analysis is seed matching~\cite{sahlin_survey_2023}\mytodo{Cite more}, where short sequence segments, called \emph{seeds}, are identified between biological sequences (e.g., DNA reads) to find similarities. Traditional hashing methods used in seed matching~\cite{\citemaplowcolhash} require \emph{exact} matches of these seeds. The requirement for exact matching \rev{necessitates} the use of limited parameter choices, which either increases the number of computationally costly steps or reduces the sensitivity in detecting relevant genomic regions.

Although identifying \revb{exactly matching} seeds is crucial for finding similarities, seeds may still have arbitrary differences—what we define as \emph{noise}—due to sequencing errors or biological variations. Conventional methods either 1)~fail to detect these near-matches when the differences occur at arbitrary positions or 2)~resort to computationally expensive steps to identify these differences. \revc{The lack of a mechanism that can efficiently \emph{tolerate noise} during a single hash value lookup limits the performance, accuracy, and efficiency of genome analysis, as we demonstrate in Chapter~\ref{chap:blend}.}
 
\head{Noise in Raw Nanopore Signals} 
Nanopore sequencing offers unique opportunities for genome analysis, particularly with its capability to stop the sequencing of a read or the entire sequencing run based on real-time analysis. However, the raw electrical signals generated by nanopore sequencers are inherently noisy~\cite{gamaarachchi_simulation_2024,wen_guide_2021,munro_icarust_2024,mordig_simreaduntil_2024,rang_squiggle_2018}. Traditional approaches to handling this noise rely on computationally expensive basecalling that usually utilize GPUs~\cite{\citehwbasecgpu}. \rev{Although \revb{various} works that analyze raw nanopore signals without basecalling~\cite{\citesignalanalysis} can provide better scalability and efficiency, many of these techniques \revc{fall short in speed and accuracy when} analyzing larger genomes~\cite{firtina_rawhash_2023}, such as human genomes. This is because these works handle noise in raw nanopore signals ineffectively \revb{(as shown in Chapters~\ref{chap:rh} and~\ref{chap:rht})}.} \emph{Understanding and reducing noise effectively} in raw nanopore signals provides the opportunity for enabling scalable and accurate analysis of raw nanopore signals without basecalling.

\head{Limited Scope of Applications for Raw Nanopore Signal Analysis} 
\revc{Raw nanopore signals can be analyzed without the intermediate basecalling step; however, the scope of new applications, such as constructing genomes from scratch using raw nanopore signals, remains limited. This is mainly due to the lack of techniques that can enable new uses and directions in raw nanopore signal analysis, particularly in the absence of a reference genome (as Chapter~\ref{chap:rs} shows).} The full potential of raw nanopore signal analysis remains largely untapped \rev{due to limited techniques designed for analyzing raw nanopore signals.}

\head{Impact on Genome Analysis Pipeline} 
The sources of noise discussed above lead to significant computational overhead and limit the applicability of current genome analysis methods \revb{(as Chapters~\ref{chap:blend},~\ref{chap:rh},~\ref{chap:rht} demonstrate)}. There is a pressing need for the development of new techniques that can effectively tolerate or reduce noise throughout the genome analysis pipeline. Such techniques can enable more accurate, scalable, and real-time genome analysis, opening up new directions and applications in the field \revb{(as we show in Chapters~\ref{chap:rs} and~\ref{chap:conc})}.

\section{Goal}

In this dissertation, our goal is to 1)~\emph{understand} how \rev{certain types of noise} in sequencing data and its analysis impact the performance, accuracy, and scalability of the analysis and 2)~build mechanisms to tolerate or reduce \rev{this noise} for faster, scalable, more accurate, and real-time analysis of genomic sequencing data.

\section{Thesis Statement}

Our approach is encompassed by the following thesis statement:

\begin{center}
\parbox{12.5cm}{\textit{The imperfections \revb{(i.e., noise)} in DNA sequencing data and its analysis can be mitigated by\\\rev{1)}~building a better understanding of \rev{the types of noise}, and\\\rev{2)}~developing new algorithms and techniques that can tolerate and reduce \rev{noise,}\\\rev{thereby providing} accurate, scalable, and real-time analysis of sequencing data and \rev{enabling} new applications in genome analysis.}}
\end{center}

\section{Our Approach}

\revc{To identify and address the challenges introduced by imperfections in genomic sequencing data and its analysis, we pursue three main strategies:
1)~we integrate a noise-tolerant hashing mechanism in seeding to quickly identify highly similar seeds with arbitrary differences between them;
2)~we build a detailed understanding of the variations in raw nanopore signals and reduce noise caused by these variations to enable hash-based search for raw nanopore signals; and
3)~we enable new applications and directions by providing new overlap detection and assembly mechanisms for raw nanopore signals.}

\subsection{Tolerating Arbitrary Noise with Hash-based Fuzzy Seeding}

Generating the hash values of seeds enables quickly identifying similarities between genomic sequences by matching seeds with a single lookup of their hash values. However, these hash values can be used only for finding \revb{exactly matching} seeds, as conventional hashing methods~\cite{\citemaplowcolhash} assign distinct hash values for different seeds, including highly similar seeds. Due to the limited parameter choices when finding only \revb{exactly matching} seeds, the selected seeds either 1)~increase the use of the computationally costly steps after seeding or 2)~provide limited sensitivity.

To address these challenges, we introduce \emph{\blend}, the first mechanism that can identify \emph{both} \revb{exactly matching} and highly similar seeds with arbitrary differences \rev{(i.e., noise)} using a single lookup of their hash values, called \emph{fuzzy seeds}. \revc{To achieve this, \blend 1)~utilizes SimHash~\cite{charikar_similarity_2002, manku_detecting_2007}, which can generate identical hash values for a similar list of values (i.e., vectors), and 2)~introduces mechanisms that treat seeds as vectors, allowing SimHash to efficiently identify fuzzy seed matches.}

We show the benefits of \blend when \rev{it is} used in read overlapping and read mapping. For read overlapping, \blend is faster by \blendmovpM-\blendovpM (on average \blendavgovpM), has a lower memory footprint by \blendmovmM-\blendovmM (on average \blendavgovmM), and finds higher quality overlaps leading to accurate \emph{de novo} assemblies than the state-of-the-art tool, minimap2. For read mapping, \blend is faster by \blendmrmpM-\blendrmpM (on average \blendavgrmpM) than minimap2.

\subsection{Enabling Hash-based Search of Raw Nanopore Signals by Reducing Noise}

\rev{A unique and important feature of nanopore sequencing, adaptive sampling~\cite{loose_real-time_2016},} can eject \revc{single DNA or RNA molecules (i.e., strands)} from sequencers without fully sequencing them, which provides opportunities to computationally reduce the sequencing time and cost. However, due to the inherent noise in raw electrical signals, existing works analyzing these raw signals~\cite{\citerealtimeall} either 1)~require powerful computational resources such as GPUs to eliminate this noise by converting the signals to DNA bases~\cite{\citebasecalledreal} or 2)~lack scalability for large genomes due to their mechanisms when analyzing noisy raw signals without converting them bases~\cite{kovaka_targeted_2021, zhang_real-time_2021}.

To understand and reduce noise in raw nanopore signals and to enable efficient analysis of these signals, we propose \rh, the first mechanism that can accurately and efficiently perform real-time analysis of nanopore raw signals for large genomes using hash-based similarity search. To enable this, \rh provides accurate hash-based similarity search \rev{via effective noise reduction} with quantization of the raw signals such that signals corresponding to the same \rev{genomic} content can have the same quantized value and, subsequently, the same hash value.

We evaluate \rh on three applications: 1)~read mapping, 2)~relative abundance estimation, and 3)~contamination analysis. When compared to the state-of-the-art techniques, \unc~\cite{kovaka_targeted_2021} and \sig~\cite{zhang_real-time_2021}, \rh provides 1)~\rhavgthrU and \rhavgthrS better average throughput and 2)~significantly better accuracy for large genomes, respectively. Our evaluations show that \rh is the only tool that can provide high accuracy and high throughput for analyzing large genomes in real-time compared to these prior state-of-the-art works~\cite{kovaka_targeted_2021,zhang_real-time_2021}.

\subsection{Better Noise Reduction \rev{for} and Robust Real-Time Mapping \rev{of} Raw Nanopore Signals}

Real-time analysis of raw signals is essential to \rev{utilizing} the unique features that nanopore sequencing provides, enabling the early stopping of the sequencing of a read or the entire sequencing run based on the analysis. The state-of-the-art mechanism \rev{we introduce}, \rh, offers the first hash-based efficient similarity identification between raw signals and a reference genome by effectively reducing noise in raw signals to enable quick matching between hash values.

\rev{To build a better understanding \revc{of noise} and \rev{better} mechanisms for reducing noise in raw nanopore signals, we introduce \rht, which provides major improvements over \rh in several ways. First, to provide more sensitive noise reduction in raw signals, \rht provides \revc{a new} adaptive noise reduction technique based on the characteristics of raw nanopore signals rather than a fixed method \revc{(as proposed in \rh)} for reducing noise. Second, to perform similarity search with fewer and more \revc{useful} seeds (\revc{in order to improve both} performance and accuracy), \rht 1)~provides a sampling mechanism, known as \emph{minimizer sketching}, to reduce the volume of raw signals used in analysis, 2)~provides a filter to reduce ambiguous matching seeds before chaining, 3)~improves the sensitivity of the chaining algorithm, and 4)~identifies the best mapping for a read based on a weighted decision mechanism that takes several metrics into account instead of only one \revc{(as in \rh)}.}

We evaluate \rht in a similar evaluation setup we use for \rh. \rht provides better accuracy (on average by \rhtavgAccRH and up to \rhtmaxAccRH) and better throughput (on average by \rhtavgthrR and up to \rhtmaxthrR) than \rh, while providing \revc{even more} substantial \revc{performance and accuracy} improvements compared to \revb{two} other existing works, \unc~\cite{kovaka_targeted_2021} \revc{(e.g., \rhtavgthrU better throughput on average)} and \sig~\cite{zhang_real-time_2021} \revc{(e.g., \rhtavgthrS better throughput on average)}.

\subsection{Enabling New Applications by Overlapping and Assembling Raw Nanopore Signals}

\rev{Existing works that directly analyze raw signals without basecalling cannot interpret raw signals directly \revc{if a reference genome is unknown due to several major challenges. These challenges mainly stem from a lack of accurate mechanisms to handle increased noise and data volume in pairwise raw signal comparison}.
Our goal is to enable the \revc{direct} analysis of raw signals \revb{\emph{without}} a reference genome. To this end, we propose \rs, the \textit{first} mechanism that can 1)~\rev{identify regions of similarity between all raw signal pairs, known as \textit{all-vs-all overlapping}, using a hash-based search mechanism and 2)~use these to construct genomes from scratch, called \textit{de novo} assembly. To enable overlapping and assembly of signals, \rs provides 1)~an aggressive filtering mechanism to minimize the increased noise caused by overlapping two raw signals, 2)~adjusts the chaining and outputting strategies to enable \revb{the generation of a large number of useful overlapping read pairs.}}}

Our extensive evaluations across multiple genomes of varying sizes show that \rs provides a significant speedup (on average by \rsavgelfcpu~\rev{and up to $41.59\times$}) and reduces peak memory usage (on average by \rsavgpeakfcpu~\rev{and up to $41.99\times$}) compared to \rev{a conventional genome assembly} pipeline \rev{using} the state-of-the-art \rev{tools} for basecalling (Dorado~\cite{oxford_nanopore_technologies_dorado_2024}) and overlapping (minimap2~\cite{li_minimap2_2018}) on a CPU. We find that \rsavgshared of overlapping pairs generated by \rs~\rev{are} identical to those generated by minimap2. \revc{Using the overlaps from \rs, we construct the first \emph{de novo} assemblies \emph{ever} constructed directly from raw signals without basecalling. We show that we can construct contiguous assembly segments (i.e., unitigs~\cite{kececioglu_combinatorial_1995}) up to 2.7 million bases in length (half the genome length of \textit{E. coli}).} We identify \revc{other} \rev{previously unexplored} directions that can be enabled by finding overlaps and constructing \textit{de novo} assemblies.

\section{Contributions}

This dissertation makes the following contributions:

\begin{enumerate}
\item \rev{We provide a detailed review and overview of the state-of-the-art in raw signal analysis, covering a comprehensive spectrum of existing methodologies and their limitations. We systematically analyze various methods for managing noise, computational efficiency, and accuracy in the analysis of raw nanopore signals. This thorough review serves as a foundational framework for our subsequent contributions and sets the stage for the novel techniques proposed in this dissertation. \revc{Our review also includes acceleration methods in genome analysis and key aspects of downstream analysis.} Chapters~\ref{chap:bg} and~\ref{chap:related} provide this comprehensive review.}

\item \revc{We develop a detailed understanding of how noise in genomic sequencing data impacts computational mechanisms, hindering improvements in speed, accuracy, efficiency, and applicability in genome analysis.} Based on our understanding of the sources and types of noise, we provide mechanisms to better \rev{handle noise} in sequencing data, which enables new directions with faster, more accurate, and scalable techniques. Chapters~\ref{chap:blend}-\ref{chap:rs} describe how we build this understanding and the \revc{resulting} mechanisms.

\item We introduce \blend, to enable quick identification of highly similar genomic sequences. \blend is a novel mechanism that can quickly and efficiently find highly-similar seed matches between sequences with a single lookup of their hash values by designing a noise-tolerant hashing mechanism for fuzzy seed matching. Chapter~\ref{chap:blend} describes \blend and its evaluations in detail. \rev{\blend is open source~\cite{noauthor_blend_2023}. A paper describing \blend was published in the Nucleic Acids Research (NAR) Genomics and Bioinformatics journal in~\cite{firtina_blend_2023} and presented in the highlights track of the RECOMB 2023 conference\revb{~~\cite{firtina_blend_talk_2023}}.}

\item We introduce \rh, the first mechanism that, \revb{without basecalling}, can accurately and quickly map raw nanopore signals to large genomes (e.g., a human genome) \revb{using 1)~a} novel quantization (i.e., bucketing) technique for reducing noise in raw signals and \revb{2)~a} hash-based search mechanism. \rh provides a fast solution that can match the throughput of a nanopore sequencer to perform analysis during sequencing \rev{(i.e., real-time analysis)}. Chapter~\ref{chap:rh} describes \rh and its evaluations in detail. \rev{\rh is open source~\cite{noauthor_rawhash_2024}. \rh was presented at the ISMB/ECCB 2023 conference\revb{~~\cite{firtina_rawhash_talk_2023}}, and a paper describing \rh was published in the Bioinformatics journal~\cite{firtina_rawhash_2023} as part of the conference proceedings.}

\item We propose \emph{Sequence Until}, the first mechanism that can stop the entire sequencing run for relative abundance estimation when an accurate estimation can be made without sequencing the entire sample. \emph{Sequence Until} provides opportunities for mechanisms that can perform real-time analysis, such as \rh, to substantially reduce the sequencing time and costs without sacrificing accuracy. Chapter~\ref{chap:rh} describes \emph{Sequence Until} and its evaluations in detail. \revc{\emph{Sequence Until} is open source~\cite{noauthor_rawhash_2024}.} \rev{\emph{Sequence Until} was presented at the ISMB/ECCB 2023 conference\revb{~\cite{firtina_rawhash_talk_2023}}, and a paper describing \emph{Sequence Until} was published in the Bioinformatics journal~\cite{firtina_rawhash_2023} as part of the conference proceedings.}

\item We introduce \rht, \rev{an improved mechanism} for 1)~\revb{better} reducing noise in raw nanopore signals with an adaptive noise reduction technique designed based on the characteristics of raw nanopore signals and 2)~improving the seed matching and mapping mechanisms to achieve faster and more accurate real-time mapping of raw nanopore signals to reference genomes. Chapter~\ref{chap:rht} describes \rht and its evaluations in detail. \rev{\rht is open source~\cite{noauthor_rawhash_2024}. A paper describing \rht was published in the Bioinformatics journal~\cite{firtina_rawhash2_2024}.}

\item We propose \rs, the first mechanism that can find all-vs-all overlapping of raw signals to enable new use cases for raw signal analysis, such as assembling raw signals without basecalling. Using the overlaps that \rs generates, we construct the first \emph{de novo} assemblies from raw signals without basecalling. Chapter~\ref{chap:rs} describes \rs and its evaluations in detail. \rev{\rs is open source~\cite{noauthor_rawhash_2024}. \rs was presented in the HitSeq track of the ISMB 2024 conference\revb{~~\cite{firtina_rawsamble_talk_2024}}.} \revc{A preprint describing \rs is available~\cite{firtina_rawsamble_2024}.}

\item We describe the remaining challenges in overcoming the problems to enable new applications in raw signal and real-time analysis and discuss new directions enabled by the contributions described in this thesis. Chapter~\ref{chap:conc} discusses these challenges and directions, including the challenges for constructing \emph{de novo} assemblies in real-time.

\end{enumerate}

\section{Outline}

This dissertation is organized into~\ref{chap:conc} chapters. Chapter~\ref{chap:bg} gives relevant background information about sequencing technologies, the challenges, and steps to analyze the noisy sequencing data. Chapter~\ref{chap:related} discusses related works that address the relevant problems in tolerating noise when mapping basecalled sequences and the works that aim to perform real-time analysis with or without basecalling the sequencing data. Chapter~\ref{chap:blend} introduces \blend and its evaluations. Chapter~\ref{chap:rh} introduces \rh and \emph{Sequence Until} and their respective evaluations. Chapter~\ref{chap:rht} introduces \rht and its evaluations. Chapter~\ref{chap:rs} proposes \rs and provides its evaluations. Chapter~\ref{chap:conc} provides a summary of this dissertation as well as future research directions and concluding remarks.

\chapter{Background}
\label{chap:bg}

This chapter provides an overview of the background material necessary to understand our discussions, analyses, and contributions. Section~\ref{bg:sec:hts} describes the main sequencing technologies, real-time analysis, and the challenges in analyzing sequencing data. Section~\ref{bg:sec:mapping} provides the main steps taken to provide a practical sequence analysis. Section~\ref{bg:subsec:downstream} describes the subsequent steps that can utilize the output from read mapping to perform genome analysis. \revc{Section~\ref{bg:sec:accelerating} describes the software and hardware approaches for accelerating genome analysis.} Section~\ref{bg:sec:summary} provides a summary and suggests further reading for interested readers.

\section{High-Throughput Sequencing Technologies}
\label{bg:sec:hts}

\begin{figure}[tbh]
  \centering
  \includegraphics[width=0.9\columnwidth]{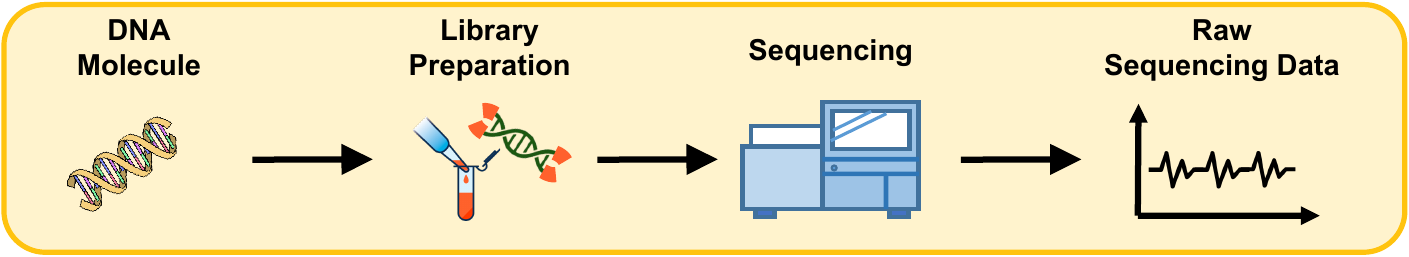}
  \caption{Main steps in sequencing data generation.}
  \label{rht:fig:sequencing}
\end{figure}

High-throughput sequencing (HTS) technologies produce raw sequencing data, the content of which depends on the type of sequencing technology. Figure~\ref{rht:fig:sequencing} shows the flow of main steps for generating the sequencing data from HTS technologies. There are three main types of HTS technologies: sequencing by synthesis (SBS)~\cite{\citesbs}, Single Molecule Real-Time (SMRT)~\cite{\citesmrt}, and nanopore sequencing~\cite{\citenanopore}.

\subsection{Sequencing by Synthesis (SBS)}
\label{bg:subset:sbs}

Sequencing by synthesis (SBS)~\cite{\citesbs} is a widely adopted high-throughput sequencing technology that enables accurate and cost-effective sequencing of nucleic acid molecules. The SBS process involves several steps to determine the nucleotide sequences~\cite{turcatti_new_2008}.

\revc{First, the nucleic acid molecule (e.g., DNA) is fragmented into smaller pieces, called \emph{reads}, each around a few hundred nucleotides long (e.g., approximately 200 bases).} To enable the attachment of these reads to a specialized glass slide known as a \emph{flow cell} for sequencing, certain short nucleic acid molecules known as \emph{adapters} are added to molecules' ends.
Second, to increase the redundancy for improved reliability during sequencing, clusters of identical DNA fragments on the flow cell are created with amplification.
Third, to sequence DNA fragments, a complementary DNA strand from a read is synthesized base by base. To do so, an enzyme called \emph{DNA polymerase} adds a single fluorescently labeled nucleotide to the growing DNA strand in each sequencing \emph{cycle}. To allow for the identification of each individual nucleotide, these labeled nucleotides emit a specific light at a certain frequency (i.e., a different light color for each nucleotide) when incorporated into the strand.
Fourth, after each nucleotide is added, the flow cell is imaged to capture the fluorescent signal emitted by the incorporated nucleotide. This nucleotide incorporation and imaging cycle is repeated for each base in the DNA fragment. The sequence of the entire DNA fragment is determined base by base as the fluorescent signals are recorded across multiple cycles.

SBS generates a series of images as \emph{raw data}, where each image corresponds to a single cycle of synthesis, and the color intensity at a specific position in the image represents a particular nucleotide. Translating the raw data into nucleotide sequences is called \emph{basecalling}. Basecalling tools~\cite{\citebasecallsbs} developed for SBS aim to accurately associate these colors with their corresponding bases while correcting sequencing errors~\cite{cacho_comparison_2016}. \revc{Although SBS generates highly accurate raw sequencing data, it also introduces certain limitations, primarily due to the cycling procedure. Specifically, SBS generates \emph{short reads}, typically a few hundred bases long, which can complicate downstream analysis, especially when assembling large genomes or resolving complex genomic regions~\cite{alkan_limitations_2011, firtina_genomic_2016}.}

\subsection{Single Molecule Real-Time (SMRT) Sequencing}
\label{bg:subsec:smrt}

\revd{Single-molecule real-time (SMRT) sequencing is a high-throughput sequencing technology without sequencing each nucleotide cycle by cycle~\cite{\citesmrt}. By avoiding the cycling procedure in SBS, longer reads up to a few thousand bases can be generated with SMRT sequencing~\cite{logsdon_long-read_2020}, which is crucial for resolving complex genomic regions~\cite{logsdon_long-read_2020}. SMRT sequencing mainly works in several steps~\cite{eid_real-time_2009}.
First, similar to SBS, the nucleic acid molecule (e.g., DNA) is prepared by fragmenting it into smaller pieces (i.e., reads).
Second, these fragmented molecules are circularized by adding special adapters to their ends (i.e., hairpin adapters). These adapters include hairpin loops that allow the DNA to form a closed circular molecule, which is essential to continuously sequence the same molecule multiple times in SMRT.
Third, similar to SBS, to initiate the synthesis of a complementary strand of the circularized DNA template, a single DNA polymerase enzyme is attached to a hairpin adapter.
Fourth, SMRT sequencing introduces the circularized fragments into specialized wells, known as Zero-Mode Waveguide (ZMW) wells~\cite{levene_zero-mode_2003}. These wells are critical for SMRT sequencing as they create a nano-scale environment that is small enough for the circularized molecule to be inserted and extract accurate information when sequencing each nucleotide~\cite{levene_zero-mode_2003}.
Fifth, unlike SBS, which involves discrete sequencing cycles, SMRT sequencing operates continuously in real-time. As fluorescently labeled nucleotides are introduced in the nano-scale ZMW wells, the DNA polymerase continuously incorporates these nucleotides into the growing DNA strand. Each incorporation event emits light at a certain frequency as each of the four nucleotides (A, C, G, T) is labeled with a distinct fluorescent dye. The emitted light provides a high signal-to-noise ratio with a reduced background noise compared to SBS due to the small volume of ZMW wells~\cite{iizuka_zero-mode_2022}. Thus, the propagation of the emitted light can be blocked to create a high signal-to-noise ratio~\cite{tsai_chapter_2016} and enable producing a continuous stream of image data, unlike the cycle-by-cycle imaging.

SMRT sequencing can be performed in different modes depending on the application. One common mode is Circular Consensus Sequencing (CCS)~\cite{eid_real-time_2009, travers_flexible_2010}, where the polymerase repeatedly sequences the same DNA fragment as it loops around the circular template. This approach generates multiple reads of the same fragment, which are combined to produce a highly accurate consensus sequence. Alternatively, in the Continuous Long Read (CLR) mode~\cite{sharon_single-molecule_2013}, the polymerase reads the DNA fragment continuously without repeated sequencing, which is useful for generating longer reads but may have lower accuracy due to the absence of the error correction provided by CCS.

SMRT sequencing generates a continuous series of images in a movie format while capturing the fluorescent signals in real-time. Although these images can be quickly converted into corresponding nucleotide sequences, the noise inherent in SMRT sequencing requires additional steps to correct sequencing errors~\cite{\citeerrorcorrection}. These steps include sequence alignment~\cite{li_minimap2_2018}, consensus assembly construction~\cite{wenger_accurate_2019}, and assembly polishing~\cite{\citeassemblypolishing}, which are crucial for ensuring the high accuracy of the final sequences, particularly in applications that benefit from the long-read capability of SMRT sequencing.}

\subsection{Nanopore Sequencing}
\label{bg:subsec:nanopore}

\begin{figure}[tbh]
  \centering
  \includegraphics[width=0.9\columnwidth]{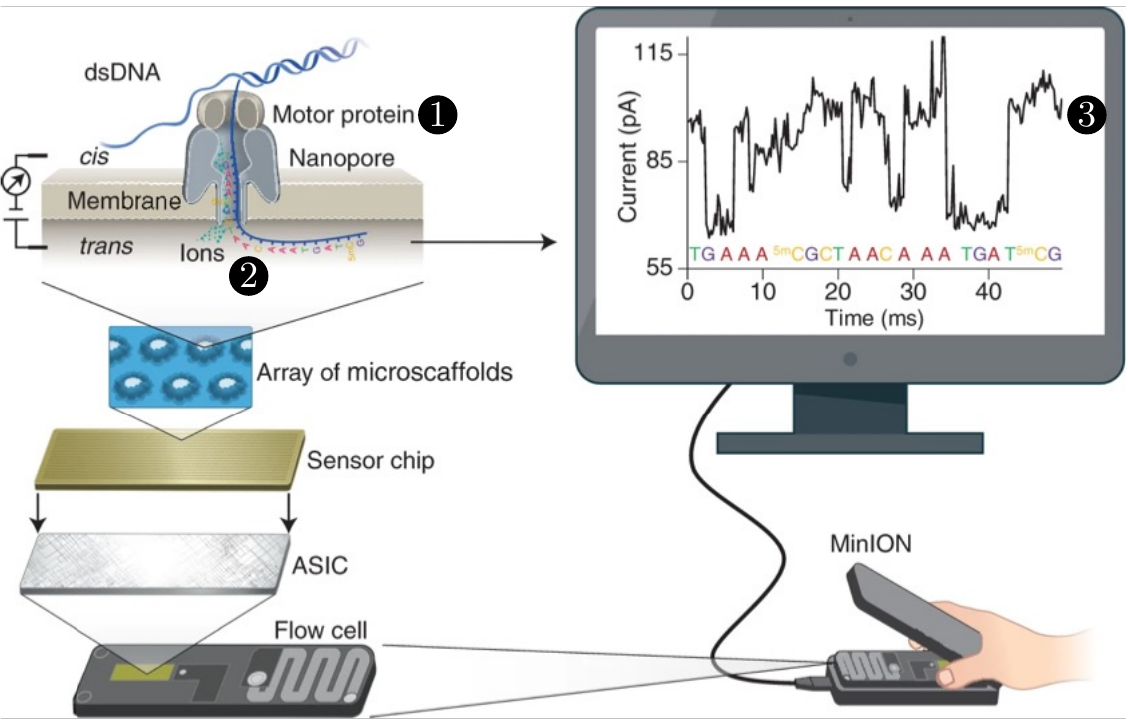}
  \caption{Structure of a nanopore sequencer and its sequencing steps. Image taken from~\cite{wang_nanopore_2021}.}
  \label{rht:fig:nanopore-seq}
\end{figure}

\revb{Nanopore sequencing is a high-throughput sequencing technology that enables the sequencing of nucleic acid molecules \revc{(e.g., double-stranded DNA, \emph{dsDNA})} as they pass through tiny nanoscale pores called \emph{nanopores}~\cite{\citenanopore}, as shown in Figure~\ref{rht:fig:nanopore-seq}.} These pores are embedded in a \revc{\emph{membrane}} \revd{that work as an isolator between two sides containing charged ions (i.e., \emph{cis} and \emph{trans} sides). By isolating with a membrane layer, the only \emph{communication} channel between the two sides is the open channel inside a nanopore. To enable an \emph{ionic current flow} inside this open channel in a nanopore, a voltage is applied in a specific direction. By doing so, the net current flow inside the open channel in a nanopore can be measured as an electrical signal. The idea in nanopore sequencing is to \emph{disrupt this net current flow} with the presence of ionized nucleic acid molecules inside a nanopore and report these disruptions as electrical signals. These \emph{informative} electrical signals can be used to identify each nucleotide passing through a nanopore.} Due to the nature of its sequencing method, nanopore sequencing enables sequencing \emph{ultra long} reads up to a few million bases~\cite{pugh_current_2023} as nanopores can enable the passing of such large molecules without practical limitations in terms of read length that SBS and SMRT sequencing technologies introduce. \revb{Figure~\ref{rht:fig:nanopore-seq} shows the three main steps in nanopore sequencing~\cite{wang_nanopore_2021}:} \revc{1)~guiding the movement of a nucleic acid molecule through the nanopore, 2)~disrupting the ionic current flow, and 3)~generating electrical signals from these disruptions.}

\revd{First, to facilitate accurate and controlled passing of nucleic acid molecules through the pores, nucleic acid molecules floating in the \emph{cis} side of the solution need to 1)~move towards and attach to a nanopore and 2)~pass inside the nanopore at a controlled speed towards the \emph{trans} side. To enable the moving of nucleic acid molecules, the voltage is applied in a specific direction such that the upper part of the membrane (i.e., \emph{cis} side) is negatively charged, and the lower part (i.e., \emph{trans} side) is positively charged. Since DNA and RNA molecules are negatively charged, they are naturally driven through the nanopore from the \emph{cis} side to the \emph{trans} side by this electric potential. To \emph{dock} nucleic acid molecules on nanopores, these molecules utilize a particular enzyme called a \emph{motor protein}\revc{\circlednumber{1}}. This motor protein 1)~attaches to one end of the molecule, 2)~guides the molecule towards a nanopore, 3)~separates two strands of a molecule (for dsDNA), and 4)~controls the movement and speed of a single-stranded molecule through the nanopore.} Controlling the \emph{translocation speed} is crucial, as it improves the accuracy and resolution of the sequencing process, which was a significant challenge in the early development of nanopore sequencing in the late 1980s~\cite{deamer_three_2016}.

\revd{Second, as the nucleic acid molecule passes through the nanopore, it partially disrupts the net ionic current flow inside the nanopore\revc{\circlednumber{2}}. These disruptions are caused by the presence of a specific sequence of $k$-many nucleotides inside a nanopore at a time, called \emph{k-mers}. As the molecule translocates (i.e., moves) through the pore, each k-mer causes characteristic changes in the ionic current, allowing the identification of the $k$-many nucleotide sequences via sensing regions (i.e., narrow heads inside nanopores, also called reader heads) located near these k-mers~\cite{wang_nanopore_2021,deamer_three_2016}.} The \revc{number ($k$)} of nucleotides disrupting the ionic current measurements is usually specific to the design of the nanopore (i.e., chemistry). This number $k$ is usually between 6 and 9~\cite{samarakoon_leveraging_2024, deamer_three_2016}, determined based on the characteristics, the number of sensing regions, and length of nanopores~\cite{wang_nanopore_2021,deamer_three_2016}. A recent sequencing by expansion (SBX) technique~\cite{kokoris_sequencing_2025} provides a solution to sequence a single base inside a nanopore at a time by \emph{expanding} the span that a single nucleotide covers with a special polymerase. However, several sources of variance, or \emph{noise}~\cite{smeets_noise_2008}, can affect the accuracy of these \revc{measurements} in nanopore sequencing. These sources include stochastic fluctuations in the ionic current~\cite{deamer_three_2016}, variations in the speed of the molecule as it translocates through the pore~\cite{bhattacharya_molecular_2012}, and environmental factors such as temperature~\cite{kawano_controlling_2009}.

Third, from each nanopore, raw electrical signals are generated at a certain frequency (around 5,000 signals per second~\cite{zhang_single-molecule_2024})\revc{\circlednumber{3}}. A high-throughput nanopore sequencer usually allows many reads to be sequenced simultaneously across multiple pores in the sequencer's flow cell~\cite{wang_nanopore_2021}. These electrical signals contain the characteristics and content of the nucleic acid molecules passing through the pores.

\begin{figure}[tbh]
  \centering
  \includegraphics[width=0.9\columnwidth]{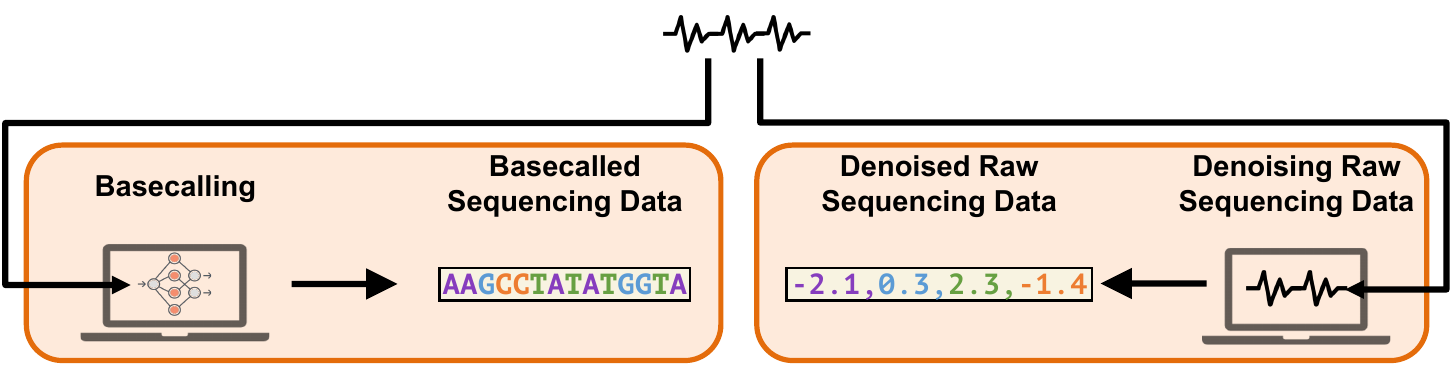}
  \caption{Main steps for processing raw sequencing data.}
  \label{rht:fig:processing}
\end{figure}

Figure~\ref{rht:fig:processing} shows the two subsequent steps that use \emph{raw nanopore signals}: 1)~\emph{basecalling}, which translates these raw signals into nucleotide sequences (A, C, G, T), and 2)~\emph{raw signal analysis}, where the raw nanopore signals are analyzed without basecalling.

\subsubsection{Basecalling}
\label{bg:subsubsec:basecalling}

Many basecallers have been proposed specifically for nanopore sequencing~\cite{\citehwbasecall} \revc{, as opposed to} basecallers designed for other sequencing technologies~\cite{\citebasecallsbs}\mytodo{Cite more for SMRT}, due to two main reasons. First, the abundance of information and noise in raw signals makes the basecalling task more challenging compared to the relatively simpler and more direct signal-to-base conversion in SBS and SMRT sequencing methods~\cite{alser_molecules_2022, singh_rubicon_2024}. Second, nanopore sequencing provides unique opportunities that can be used to reduce the time and cost of sequence analysis~\cite{alser_molecules_2022, firtina_rawhash_2023}, \revc{which we explain in Section~\ref{bg:subsubsec:adaptive}.}

To translate the raw nanopore signals to sequences of nucleotide characters, modern basecalling techniques utilize deep learning models, \revc{i.e., deep neural networks (DNNs)}~\cite{\citebasecallnanodnn}, to improve the precision of identifying a nucleotide base from raw signals compared to traditional non-deep learning-based basecallers~\cite{\citebasecallnanohmm}. Deep learning models can accurately basecall raw signals because their architectures have advanced to effectively model and recognize spatial characteristics in the raw signal data~\cite{pages-gallego_comprehensive_2023}.

Deep learning models used in basecalling generally consist of convolutional neural networks (CNNs)~\cite{miculinic_mincall_2019}\mytodo{Cite more} and recurrent neural networks (RNNs)~\cite{zhang_nanopore_2020}\mytodo{Cite more}, or a combination thereof~\cite{oxford_nanopore_technologies_bonito_2021,oxford_nanopore_technologies_dorado_2024}\mytodo{Cite more}. These models can capture the spatial and temporal dependencies in the signal data in two ways. First, CNNs process the raw signal input to extract features indicative of the underlying nucleotide sequence. \revc{This feature extraction step is crucial, as it converts raw electrical signals into a structured format that deep learning models can use to predict the nucleotide sequence.}
Second, many basecalling models employ RNNs, specifically a bidirectional LSTM network~\cite{oxford_nanopore_technologies_bonito_2021, oxford_nanopore_technologies_dorado_2024}, to predict the sequence of nucleotides. The bidirectional approach allows the model to use the temporal context from signals to improve accuracy in noisy conditions such as variance due to translocation speed. This is particularly important in nanopore sequencing, where the signal may be affected by various sources of noise, such as stochastic fluctuations in the ionic current and variations in the speed of the molecule as it translocates through the pore.

\revb{To assign bases to a window of signals where the window length can be variable due to the varying translocation speed, several basecalling models use \emph{decoder} mechanisms, usually in the form of a Connectionist Temporal Classification (CTC) layer~\cite{graves_connectionist_2006}, Conditional Random Field (CRF) decoders~\cite{lafferty_conditional_2001}, or a combination of both~\cite{pages-gallego_comprehensive_2023, oxford_nanopore_technologies_bonito_2021}. This flexibility in the window length is essential for identifying the variable lengths of signal segments that correspond to a particular nucleotide. \revc{These variabilities arise due to fluctuations in the ionic current and changes in the translocation speed of the molecule through the nanopore, as discussed in Section~\ref{bg:subsec:nanopore}.}}

\revc{The complexity of deep neural networks for basecalling~\cite{singh_rubicon_2024} often necessitates powerful computational resources, such as GPUs~\cite{\citehwbasecgpu} or custom hardware designs~\cite{\citehwbaseccust}, to perform the required computations quickly in a massively parallel manner.} Such a large amount of computation is usually costly in terms of power and latency, mainly caused by data movements~\cite{boroumand_google_2018, boroumand_google_2021, shahroodi_swordfish_2023} and a large number of operations~\cite{singh_rubicon_2024}, which leads to challenges in using them in computationally (or power-wise) resource-constrained environments such as for in-field analysis with mobile devices~\cite{shih_efficient_2023}.

\subsubsection{Raw Signal Analysis}
\label{bg:subsubsec:rawsignal}

\begin{figure}[tbh]
  \centering
  \includegraphics[width=\columnwidth]{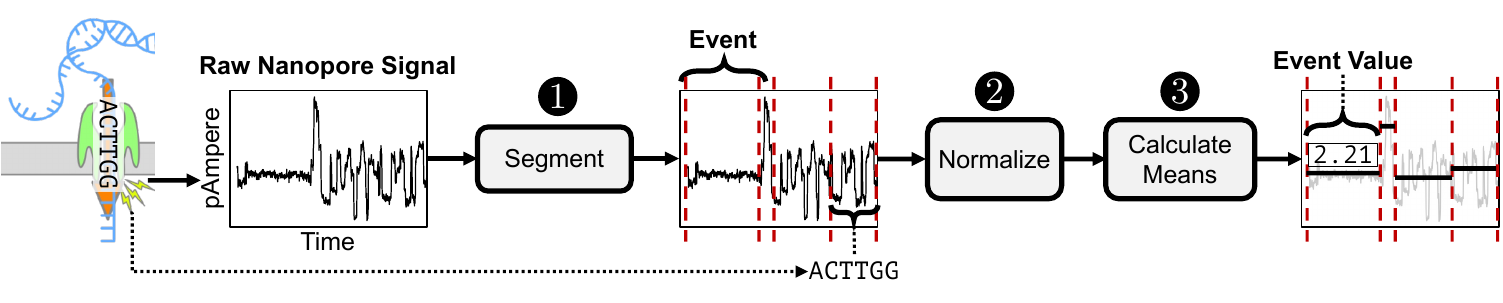}
  \caption{Main steps for raw signal analysis.}
  \label{rht:fig:rawsignal}
\end{figure}

Nanopore sequencing generates raw signal data in the form of time series~\cite{gamaarachchi_simulation_2024}, which reflect the changes in ionic current as nucleic acids translocate through the nanopore. Analyzing these time series data effectively addresses the variable translocation speed of the nucleic acid molecules, a critical aspect that influences signal noise. Figure~\ref{rht:fig:rawsignal} shows several steps for processing raw nanopore signals to reduce certain types of noise in raw nanopore signals.

\revb{First, to reduce noise caused by variable translocation speed during sequencing, the raw signal data is segmented using time-series statistical analysis~\cite{oxford_nanopore_technologies_scrappie_2019, firtina_rawhash_2023,zhang_real-time_2021}. Segmentation\revc{\circlednumber{1}} involves dividing the continuous signal into discrete parts or segments, called \emph{events}, based on detected abrupt changes in the signal intensity and pattern. To identify these abrupt changes, statistical methods such as the t-test~\cite{ruxton_unequal_2006} are typically used. Each event is assumed to correspond to the sequencing of a specific k-mer that transiently occupies the nanopore (see Section~\ref{bg:subsec:nanopore}), although these segmentation mechanisms have limitations in identifying particular types of noise related to translocation speed (i.e., skip and stay errors). The segmentation process is useful for identifying regions of the signal generated by specific k-mers to facilitate more accurate raw signal analysis without the translocation speed effect.}

\revc{Second, to account for the fact that raw signals generated under different sequencing conditions (e.g., different nanopores within the same flow cell) may appear at different scales, signal values are normalized (e.g., by calculating their standard scores~\cite{zhang_real-time_2021})\revc{\circlednumber{2}}.} Signal values may appear at different scales due to variations between nanopores and within the same nanopore~\cite{bhattacharya_molecular_2012, kawano_controlling_2009}. Normalization adjusts the signal values to a common scale to harmonize these differences and enable consistent subsequent analyses.

\revc{Third, the mean or median values of the segmented regions are calculated from the normalized signal values. This step\revc{\circlednumber{3}} helps reduce the effects of uninformative outliers and transient signal fluctuations within the segment.} Additional \revc{methods for} outlier detection can be employed~\cite{zhang_real-time_2021, firtina_rawhash_2023} to exclude anomalous data points that may result from noise. \revc{The value generated for each event is called \emph{event value}.}

\revc{Despite these initial processing steps, the resulting series of normalized signal values remains noisy due to various types of variations (see Section~\ref{bg:subsec:nanopore}) and potential skip and stay errors~\cite{shivakumar_sigmoni_2024}\mytodo{Cite more}. To further mitigate and eliminate this noise, various techniques have been developed, which we describe in Chapter~\ref{chap:related}.}

\revc{
\head{Benefits of Raw Signal Analysis}
Analyzing raw nanopore signals directly, rather than relying solely on basecalled sequences, offers several unique benefits that can enhance genomic analyses. 

First, raw nanopore signals \emph{capture subtle variations in the ionic current} that occur when modified nucleotides, such as methylated bases~\cite{\citemethraw}, pass through the nanopore~\cite{wan_beyond_2022}. These modifications can have profound biological implications, affecting gene expression and regulation~\cite{sigurpalsdottir_comparison_2024}\mytodo{Cite more}. While specialized basecallers can infer certain modifications by incorporating models trained on known modification patterns~\cite{\citemethbasecalled}, they may not detect novel or rare modifications not represented in the training data~\cite{yao_effective_2024}. Analyzing raw signals directly allows for a more comprehensive and potentially unbiased detection of nucleotide modifications.

Second, raw signal analysis can \emph{improve accuracy in challenging genomic regions}, such as homopolymeric tracts (e.g., estimating polyA tails~\cite{\citepolyaraw}) and repetitive sequences (e.g., short tandem repeats~\cite{\citestrraw}). Although modern basecallers~\cite{\citebasecallnanodnn} provide highly accurate basecalling, these basecallers may still introduce errors in these challenging regions due to difficulties in accurately determining the exact number of repeating units~\cite{sitarcik_warpstr_2023,rang_squiggle_2018}.

Third, analyzing raw signals can \emph{reduce computational requirements} in certain applications~\cite{firtina_rawhash_2023}. Basecalling involves complex deep learning models~\cite{\citebasecallnanodnn} that require substantial computational resources, such as GPUs~\cite{\citehwbasecgpu} or specialized hardware~\cite{\citehwbaseccust}. In resource-constrained environments, such as field portable sequencing setups~\cite{shih_efficient_2023,samarakoon_genopo_2020}, these requirements may not be practical~\cite{shih_efficient_2023}. By bypassing the basecalling step and analyzing raw signals directly with more lightweight algorithms, it is possible to reduce computational overhead and enable scalable analyses that would otherwise be infeasible.

Fourth, raw signal analysis can \emph{enable new directions in research} by leveraging the rich information contained in the signals. As new algorithms and techniques are developed for directly analyzing raw nanopore signals without basecalling~\cite{\citesignalanalysis,\citemethraw,\citepolyaraw,\citestrraw}, many applications that use basecalled sequences for the analysis (e.g., variant calling~\cite{\citevariantcallers} or \emph{de novo} assembly~\cite{\citeassembly}) can be performed using raw signals. These applications can utilize the rich information in raw signals to improve their accuracy while reducing the latency by avoiding the costly basecalling step. This advancement opens up possibilities for more accurate and comprehensive genomic studies, enhancing our understanding of genetics and molecular biology.

In summary, raw signal analysis offers unique advantages by providing access to richer information~\cite{wan_beyond_2022,ni_rna_2024,bao_squigglenet_2021}, improving accuracy in complex genomic regions, reducing computational requirements, and enabling new directions in genomic research. While some benefits can be partially achieved through advanced basecalling techniques, these may involve trade-offs such as loss of detail, increased computational demands, and potential inaccuracies. Therefore, analyzing raw nanopore signals directly can enhance genomic studies, particularly in applications where these factors are critical.
}

\subsubsection{Unique Opportunities: Real-time Analysis and Adaptive Sampling}
\label{bg:subsubsec:adaptive}

Computational techniques that can analyze the raw signals while they are generated at a speed that is as fast as the throughput of nanopore sequencing are called \emph{real-time analysis techniques}~\cite{loose_real-time_2016, firtina_rawhash_2023}. These signals can be analyzed in real-time with~\cite{\citebasecalledreal} or without~\cite{\citesignalanalysis} basecalling, as explained in the earlier parts of this section. There are unique benefits and challenges to performing real-time analysis.

\begin{figure}[tbh]
  \centering
  \includegraphics[width=0.8\columnwidth]{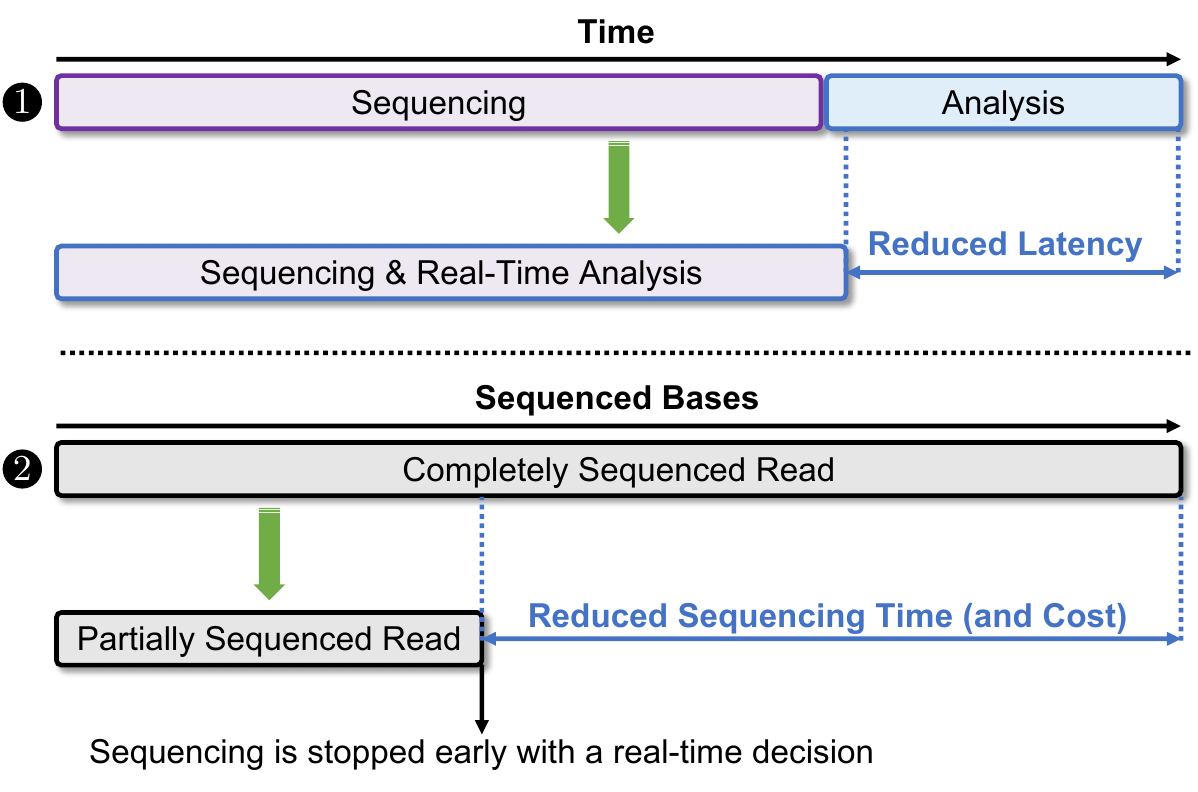}
  \caption{Two main benefits of real-time analysis with nanopore sequencing.}
  \label{rht:fig:real-time}
\end{figure}

\head{Benefits} Figure~\ref{rht:fig:real-time} shows the two unique benefits that real-time analysis offers.
First, real-time analysis allows for overlapping sequencing time with analysis time\revc{\circlednumber{1}}, as raw signals can be analyzed while they are being generated. This benefit enables reducing the overall latency of genome analysis.
Second, computational mechanisms can stop the sequencing of a molecule or the entire sequencing run early without sequencing the entire molecule\revc{\circlednumber{2}} or the sample using techniques known as \emph{adaptive sampling} with the Read Until~\cite{loose_real-time_2016} and Run Until~\cite{payne_readfish_2021} features. Adaptive sampling is useful for the selective sequencing of \emph{useful} molecules while minimizing the sequencing of \emph{useless} molecules, which can substantially reduce sequencing time and costs. To determine if a molecule is useful, computational mechanisms can start analyzing raw sequencing data with real-time analysis. Based on this analysis, the computational mechanism can determine if further sequencing of the nucleic acid molecule is necessary and can stop sequencing the molecule by \emph{ejecting} it from the nanopore~\cite{loose_real-time_2016}. To eject the molecule, the electric potential is temporarily reversed or altered in such a way that the molecule is pushed back out of the nanopore or drawn back into the \emph{cis} side \revc{(upper part of the membrane as explained in Section~\ref{bg:subsec:nanopore})} of the membrane~\cite{loose_real-time_2016, weilguny_dynamic_2023}. This process can fully remove the molecule from the nanopore and allows the nanopore to capture a new molecule for sequencing.

\head{Challenges} There are four key challenges in achieving effective adaptive sampling with real-time analysis.
First, the sequencing of unnecessary reads must be stopped as early as possible to minimize wasted time and costs. Specifically, failing to stop sequencing a read after a certain amount of time (i.e., sequencing a read longer than around 5 seconds) significantly increases the risk of blocking the nanopore~\cite{munro_icarust_2024}, rendering it unusable for the remainder of the sequencing run. The probability of blocking a nanopore usually increases with the increased sequenced length of a read, as longer molecules are more likely to be tangled during the ejection process. To prevent this, reads are not ejected after being sequenced for a certain period of time~\cite{payne_readfish_2021}.
Second, the speed of analysis must keep pace with the data generation speed (i.e., the throughput of the sequencer). This is critical to avoid data backlog and ensure timely decision-making~\cite{firtina_rawhash_2023}.
Third, the analysis techniques must effectively tolerate noise in raw nanopore signals to provide accurate results.
Fourth, real-time analysis, particularly in portable and resource-constrained environments, must be power-efficient and scalable. Low power consumption is crucial, as the analysis must handle large genomes and high-throughput data while operating within the limited power budgets of portable sequencing setups~\cite{shih_efficient_2023,samarakoon_genopo_2020}\mytodo{Cite more}. \revc{This necessitates the design of energy-efficient algorithms and hardware architectures that can support real-time, portable sequencing applications\revc{, which we discuss in Sections~\ref{bg:sec:accelerating} and~\ref{rel:subsec:basecalledreal}}.}

\vspace{0.175cm}
\subsection{Challenges in Analyzing Sequencing Data}
\label{bg:subsubsec:challengeshts}

\revc{High-throughput sequencing (HTS) technologies have significantly advanced our ability to sequence genomes rapidly and cost-effectively; however, they also present inherent limitations that pose challenges for genomic analysis.}

First, the reads generated by HTS technologies are produced without prior knowledge of their origin within the genome\revb{, except for specialized approaches that perform tagging and barcoding on the DNA to identify the origins of \emph{certain} sequences of nucleotides~\cite{fleischmann_whole-genome_1995, myers_whole-genome_2000, alkan_limitations_2011, nagarajan_sequence_2013, heather_sequence_2016}.} This lack of contextual information necessitates accurately and efficiently identifying each read’s origin. To achieve this, it is essential to compare the reads either 1)~against each other, in a process known as \revc{\emph{read overlapping}}~\cite{li_minimap_2016}\mytodo{Cite more}, or 2)~against a known representative of a species’ genome, referred to as a \emph{reference genome}. This sequence similarity identification is usually performed in a step called \emph{read mapping}. Given the large volume of sequencing data produced and the substantial size of certain reference genomes (e.g., a human genome), read mapping becomes challenging in terms of both scalability and accuracy~\cite{alser_technology_2021}.

Second, the reads produced by HTS technologies are prone to variations due to sequencing errors and natural mutations~\cite{alkan_limitations_2011, shendure_dna_2017}. \revc{Accurately detecting and distinguishing these variations is crucial for the integrity of genomic analysis. However, identifying these variations in the vast amount of data generated by HTS technologies is computationally intensive~\cite{alser_technology_2021}, requiring the development of scalable algorithms capable of handling large datasets while maintaining precision.}

Addressing these limitations requires the development of efficient algorithms and data structures that can perform rapid similarity identifications, scale to handle the vast amount of data generated, and accurately identify sequence variations while maintaining computational efficiency.

\section{Facilitating Practical Sequence Analysis with Read Mapping}
\label{bg:sec:mapping}

\begin{figure}[tbh]
  \centering
  \includegraphics[width=0.9\columnwidth]{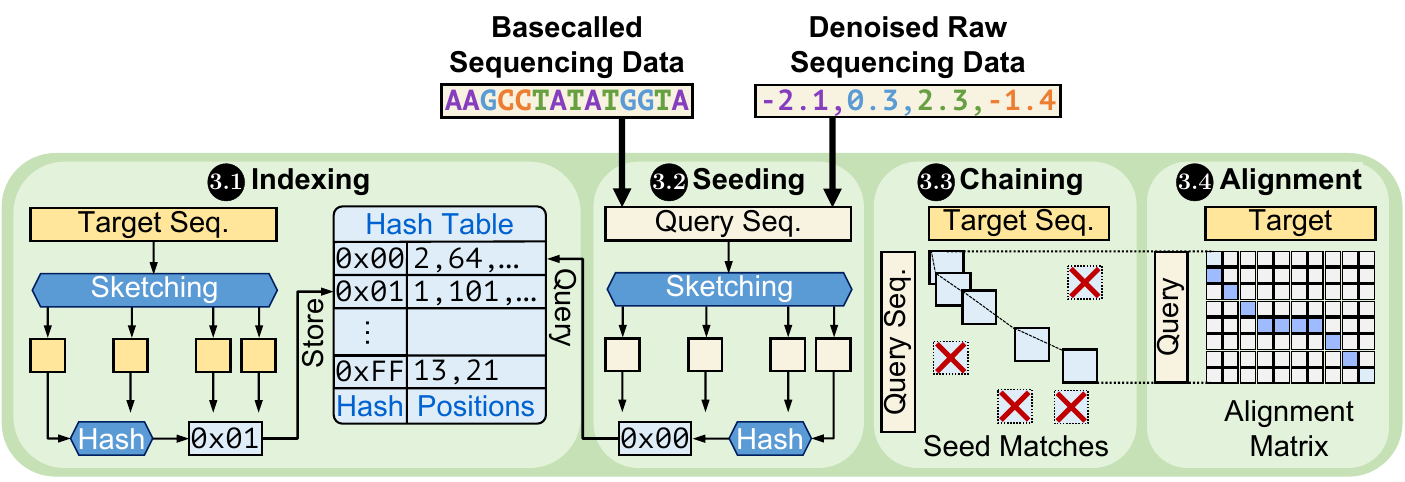}
  \caption{Main steps in read mapping.}
  \label{rht:fig:mapping}
\end{figure}

\revc{The goal of read mapping is to identify similarities and differences between genomic sequences, such as between 1)~a read (i.e., query sequence) and 2)~either another read or a reference sequence representing a species, known as a \emph{reference genome} (i.e., target sequence).} Due to genomic variants and sequencing errors, differences and similarities between these sequences (i.e., matches, substitutions, insertions, and deletions) are identified mainly using an approximate string matching (ASM) algorithm to generate an \emph{alignment score} that quantifies the degree of similarity between a pair of sequences (e.g., sequence segments from a reference genome and a read). This process is known as \emph{sequence alignment}.
To identify the degree of similarity, the minimum number of single-character edits (insertions, deletions, or substitutions) required to transform one sequence into another, known as \emph{edit distance}, is usually measured as well as the corresponding edits.
However, ASM algorithms~\cite{\citemappairalign} often have quadratic time and space complexity due to the dynamic programming nature of these algorithms~\cite{baichoo_computational_2017}. This makes the use of sequence alignment computationally challenging between a large number of sequence pairs.
To ease the identification of similarities within vast amounts of sequencing data, the read mapping process includes multiple steps designed to narrow down the search space and reduce computational overhead, as shown in Figure~\ref{rht:fig:mapping}. \revc{These steps include:
\begin{itemize}
    \item Constructing (i.e., indexing\revc{\circlednumber{3.1}}) and using (i.e., seeding\revc{\circlednumber{3.2}}) a database~\cite{\citemaphashtable,\citemapsuffix}. These steps use various sketching (i.e., sampling)~\cite{\citemapsketch} and hashing~\cite{\citemaphash} methods to selectively sample and hash sequencing data, which allows for rapid comparison. \revc{We describe 1)~indexing and seeding in Section~\ref{bg:subsubsec:seeding}, 2)~sketching in Section~\ref{bg:subsubsec:sketching} and 3)~hashing techniques in Section~\ref{bg:subsubsec:hashing}.}
    \item Pre-alignment filtering~\cite{\citemapfilter} and chaining~\cite{\citemapchain}\revc{\circlednumber{3.3}}, which discards dissimilar sequences early in the process to avoid unnecessary and expensive alignment computations. \revc{We describe these steps in Section~\ref{bg:subsubsec:filtering_chaining}.}
    \item Sequence alignment~\cite{\citemappairalign}\revc{\circlednumber{3.4}} to determine the alignment and the edit distance between sequence pairs. \revc{We describe this step in Section~\ref{bg:subsubsec:alignment}.}
\end{itemize}}

\subsection{Indexing and Seeding}
\label{bg:subsubsec:seeding}

To quickly find similar regions between genomic sequence pairs, read mapping typically utilizes data structures, such as a \revb{hash table~\cite{\citemaphashtable} or a suffix array-based structure~\cite{\citemapsuffix}}. These data structures enable efficient similarity identification between target and query sequence pairs within a pre-built database of the collection of target sequences (e.g., a reference genome). \revc{Such a database is called an \emph{index}~\cite{lee_fast_2015}, and the process of constructing this index is called \emph{indexing}.}

\revc{To narrow down the search space between target and query sequences, an index allows for quick matching of short sequence segments (i.e., \emph{seeds}). This process significantly reduces the need for computationally costly steps like sequence alignment by focusing on regions with matching seeds.}

\begin{figure}[tbh]
  \centering
  \includegraphics[width=0.7\linewidth]{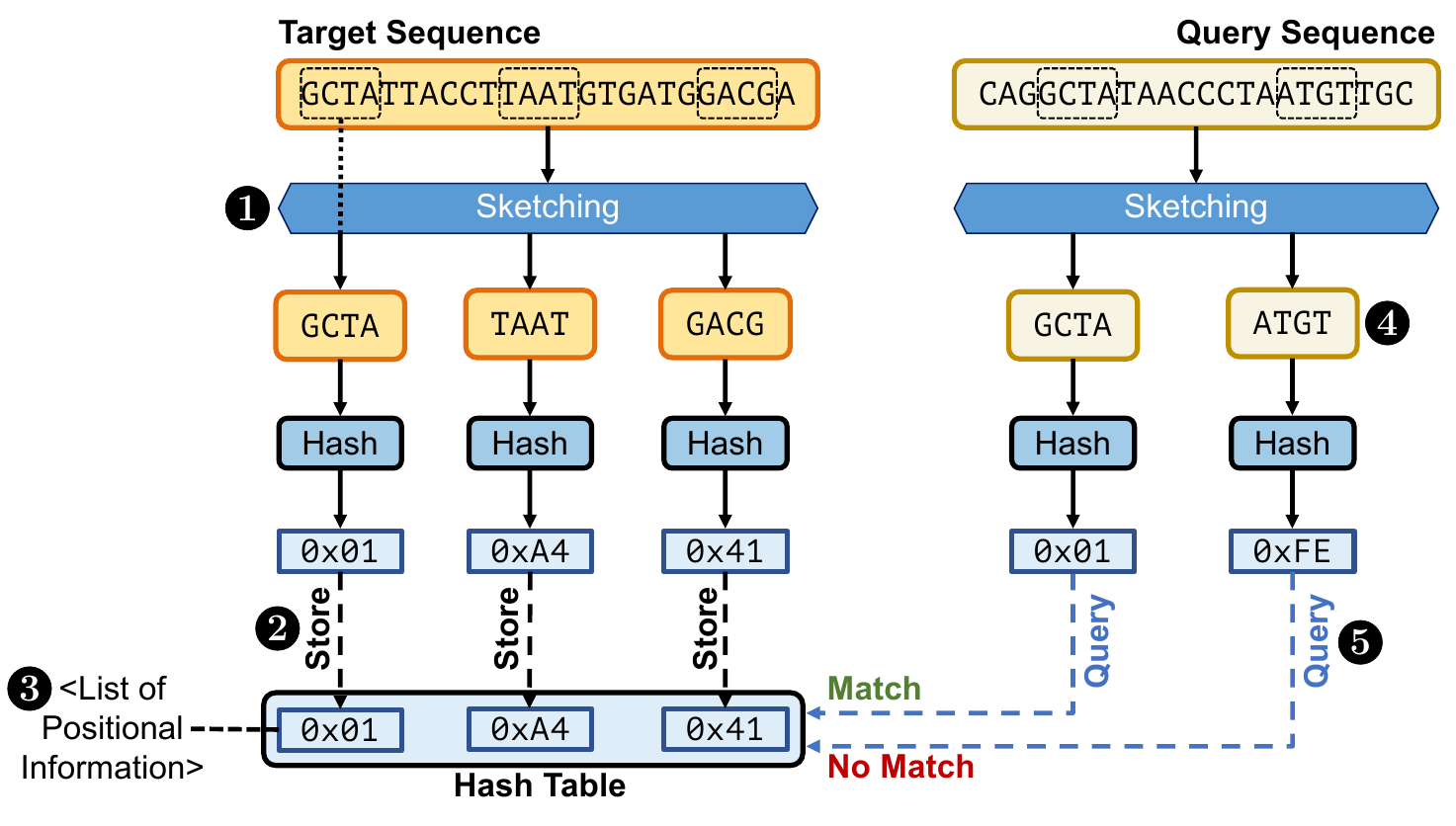}
  \caption{Steps in indexing (on the left side) and seeding (on the right side) to find matching sequence segments between target and query sequences using their hash values.}
  \label{fig:hashing}
\end{figure}

\revb{To find the matching seeds efficiently, a common approach is to match the hash values of seeds with a \emph{single lookup} (for each seed) using a hash table as an index. Figure~\ref{fig:hashing} shows an overview of how hash tables are used as an index to find seed matches between sequence pairs in two steps.
First, certain sequence segments are determined by a sampling mechanism on the entire \revc{target} sequence to store them in the hash table \revc{(\circlednumber{3.1} in Figure~\ref{rht:fig:mapping})}. Such sampling mechanisms are called sketching techniques\revc{\circlednumber{1}}, which we explain in \revc{Section~\ref{bg:subsubsec:sketching}}. The hash values of these sampled sequence segments are stored as the \emph{keys} of the hash table\revc{\circlednumber{2}}, while their positional information, such as the sequence ID (e.g., a chromosome name), is stored as the \emph{value} returned by the key\revc{\circlednumber{3}}. Since there can be multiple sequence segments with the same hash value (i.e., either due to repeated sequence segments or hash collisions), the value is returned as a list of such positional information that contains all seeds with the same hash value.
Second, certain sequence segments with the same sampling strategy are extracted from query sequences to use them as seeds\revc{\circlednumber{4}} \revc{(\circlednumber{3.2} in Figure~\ref{rht:fig:mapping})}. The hash values of these seeds are used to query the hash table\revc{\circlednumber{5}} to quickly find the positions where a seed from the query sequence appears in the target sequence with a single lookup. Such a query can quickly and substantially narrow down the search space from the entire target and query sequence to fewer regions (i.e., candidate regions) where seed matches appear.}

\subsubsection{Why Indexing and Hash-based Search}
\revd{The goal of indexing is to reduce the search space from an entire target and query sequences into smaller sets of sequences. This approach makes the analysis of an entire genome practical, as it avoids unnecessarily performing costly operations such as chaining and sequence alignment. For example, in a naive but the most optimal approach, it is possible to perform sequence alignment for each read (query sequence) in the entire reference genome space (target sequence). This approach guarantees finding the \emph{optimal} origin of a read in the entire reference genome with a \emph{minimal edit distance} because the optimal sequence alignment is performed in the \emph{entire} reference genome and read without reducing the search space. However, this naive approach becomes impractical, as the time complexity of sequence alignment is quadratic~\cite{smith_identification_1981, shaw_proving_2023} with respect to the lengths of query and target sequences, which can be billions of characters for a human reference genome. The lack of an indexing technique limits the application's scalability to usually smaller genomes such as viral genomes~\cite{dunn_squigglefilter_2021}, as the read needs to align to an entire reference genome rather than a few smaller regions in the genome. This practicality concern necessitates an \emph{approximate but practical} approach to quickly and accurately reduce the search space from an entire large sequence space into a few and smaller ones.

To alleviate the practical concerns, approximate indexing (and seeding) techniques have been widely applied in genome analysis since around 1990s~\cite{alser_technology_2021}. These approaches provide \emph{fast} solutions mainly to find \emph{exact matching short segments} between target and query sequences as explained in Section~\ref{bg:subsubsec:seeding}. For example, a hash-based search can find an exact matching region with a single hash value lookup in O(1) time complexity~\cite{alser_technology_2021}, depending on the implementation. Although these indexing and seeding techniques are fast, they are \emph{approximate} as they rely on heuristics such as finding several short exact (or near-exact) matching regions between sequence pairs~\cite{shaw_proving_2023}. Thus, these heuristics do \emph{not} provide optimal guarantees to find the region for a read in the entire reference genome with the smallest edit distance. However, such heuristics are extensively used in today's read mappers~\cite{\citemaphashtable,\citemapsuffix} by carefully designing certain parameters related to the length and frequency of exact matching regions, which provides biologically relevant and computationally practical solutions. These heuristics are mainly biologically relevant because it is expected to find exact (or near-exact) matches between sequence pairs that are likely to align with each other with a small edit distance, given their corresponding parameters are adjusted carefully (which is often a trade-off between performance and accuracy~\cite{alser_technology_2021}).}

\subsection{Sketching}
\label{bg:subsubsec:sketching}

\begin{figure}[tbh]
  \centering
  \includegraphics[width=0.8\linewidth]{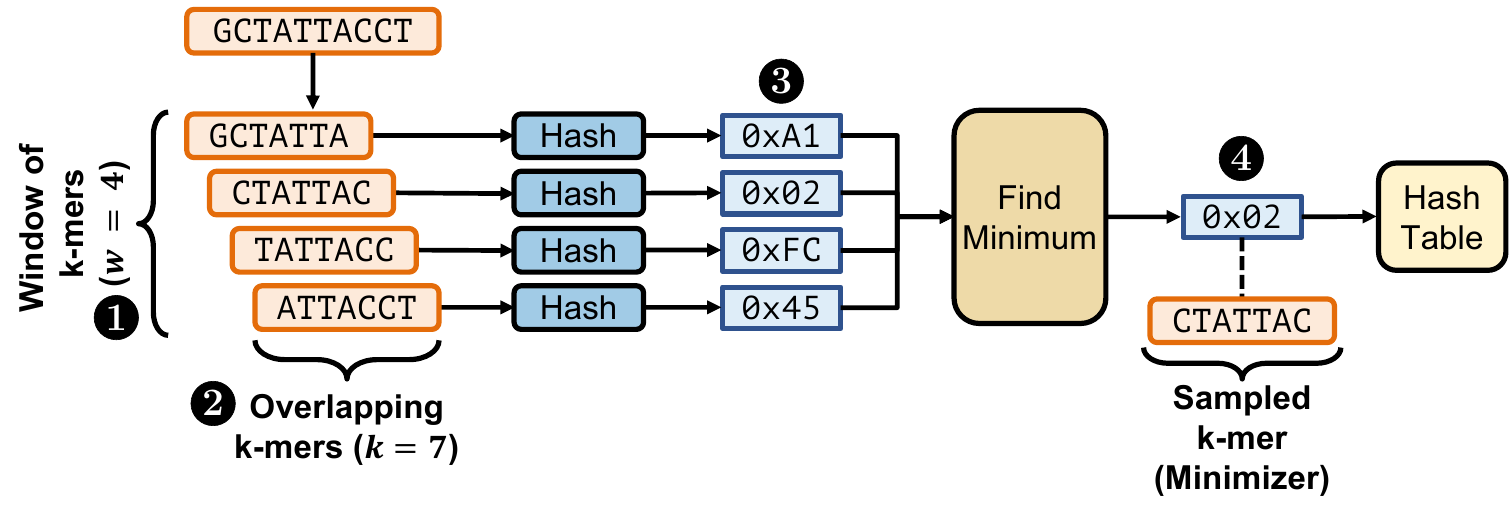}
  \caption{A sampling (i.e., sketching) mechanism, called \emph{minimizer sketching}.}
  \label{bg:fig:minimizer}
\end{figure}

Sketching \revc{(\circlednumber{3.1} and \circlednumber{3.2} in Figure~\ref{rht:fig:mapping})} aims to accurately represent an entire genomic sequence by sampling it, which enables reducing certain overheads (e.g., performance and memory overheads) of using the entire sequence while minimizing the useful information loss after sampling~\cite{firtina_blend_2023}. \revc{Although there are several sketching methods~\cite{\citemapsketch}, one commonly used method is \emph{minimizer sketching}~\cite{schleimer_winnowing_2003, roberts_reducing_2004, li_minimap2_2018}.} 
Figure~\ref{bg:fig:minimizer} shows the steps in minimizer sketching. First, subsequences of fixed length $k$, called k-mers, in a window of $w$\revc{\circlednumber{1}} overlapping k-mers\footnote{The term ``overlapping k-mers'' is commonly used interchangeably with ``adjacent k-mers'' or ``consecutive k-mers.'' All these terms refer to the same set of k-mers: a list of k-mers generated by shifting one character at a time along the sequence.} are extracted from a sequence\revc{\circlednumber{2}} to generate the hash values of these k-mers\revc{\circlednumber{3}}.
Second, among these k-mers, only one k-mer with the minimum hash value is sampled\revc{\circlednumber{4}} to represent the sequence segment that $w$-many overlapping k-mers cover. The window length can be tuned to increase the probability of matching k-mers between sequence pairs. The next window of k-mers is usually generated by removing the first k-mer and including the next consecutive k-mer of the sequence. Finding the minimizers using minimum values is effective since it builds on the assumption that many overlapping windows will share the same minimum value due to hitting the local minima in a large region~\cite{schleimer_winnowing_2003}. This enables finding the same local minima between similar sequence pairs even though these sequence pairs are not exactly the same.

\revb{Similar to the minimizer sketching, the MinHash sketching technique~\cite{broder_resemblance_1997} can be used~\cite{berlin_assembling_2015} to sample k-mers based on their minimum hash values. \revc{To do this, $n$ hash functions are used, generating $n$ hash values for each k-mer in a sequence.} For each hash function, only the k-mer with the minimum hash value is selected from the hash values generated by that function. Although \revc{(unlike minimizer sketching)} MinHash does not provide window-based sampling like minimizer sketching, it guarantees sampling a fixed number of k-mers, determined by the number of hash functions~\cite{berlin_assembling_2015}.

MinHash is particularly effective for matching sequences of similar lengths, as the number of hash functions is constant across all sequences. However, when sequence lengths vary significantly, MinHash can generate too many seeds for shorter sequences~\cite{li_minimap_2016}.

There are other alternatives to minimizers and MinHash, such as syncmers~\cite{edgar_syncmers_2021}, paired-minimizers~\cite{chin_human_2019, sahlin_effective_2021, sahlin_flexible_2022}, and HyperLogLog-based sketching~\cite{baker_dashing_2019,baker_dashing_2023}. These methods offer different trade-offs in terms of performance, accuracy, and search space size (e.g., comparisons between entire genomic sequences). We discuss some of these approaches in Chapter~\ref{chap:related}.}

\subsection{Hashing Techniques}
\label{bg:subsubsec:hashing}

\begin{figure}[tbh]
  \centering
  \includegraphics[width=0.7\linewidth]{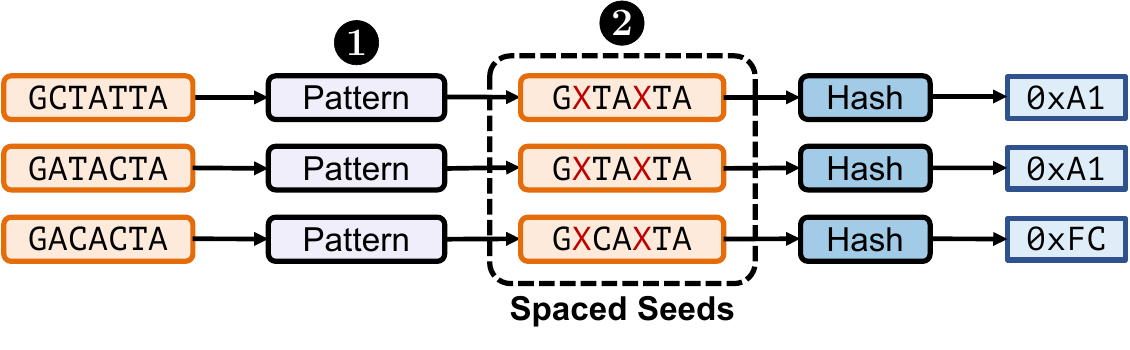}
  \caption{A spaced seeding technique. A pre-defined fixed pattern is applied to all three input sequences. Masked characters are highlighted by X in red. The first two input sequences generate the same output hash value since the characters where they differ from each other are masked, generating the same spaced seed and the same corresponding hash value of these seeds.}
  \label{bg:fig:spacedseeds}
\end{figure}

\revb{\revc{The main goal of using hashing techniques (\circlednumber{3.1} and \circlednumber{3.2} in Figure~\ref{rht:fig:mapping})} is to \emph{compress} the input values from a larger space into a smaller space. \revc{The characteristics of hashing techniques used in genome analysis} can broadly be defined in three ways. First, low-collision hash functions~\cite{\citemaplowcolhash} are designed to avoid assigning the same hash values to different input values, called \emph{collision}. \revc{Collisions of dissimilar seeds are incorrect seed matches (i.e., useless) and can lead to 1)~unnecessary execution of computationally costly steps in read mapping (e.g., chaining and alignment) and 2)~incorrect analysis.} Although low-collision hash functions are useful for avoiding incorrect seed matches, using these hash functions requires finding \emph{only} \revb{exactly matching} seeds. The exact matching requirement imposes challenges when determining the seed length. Longer seed lengths significantly decrease the probability of finding the \revb{exactly matching} seeds between sequences due to genetic variations and sequencing errors. Short seed lengths (e.g., 8-21 bases) result in matching a large number of seeds due to both the repetitive nature of most genomes and the high probability of finding the same short seed frequently in a long sequence of DNA letters~\cite{xin_accelerating_2013}.

Second, masking operations\cite{\citemapmaskedhash} aim to alter the content of the original seed sequence or its corresponding hash value, usually to provide better locality for similar seeds (e.g., increasing the chance for assigning the same hash value for similar sequences) and to enable the use of efficient similarity estimations~\cite{greenberg_lexichash_2023}. These approaches usually apply a pre-determined and fixed pattern (i.e., masks) to these sequences. Figure~\ref{bg:fig:spacedseeds} shows the use of such a fixed pattern\revc{\circlednumber{1}} where certain characters are marked $X$ \revc{(i.e., \emph{don't care characters})} to mask them\revc{\circlednumber{2}} \revc{(this technique is called \emph{spaced seeding}~\cite{\citemapmaskedhash})}. Although fixed patterns can specifically be useful for ignoring potential substitutions at particular positions between seeds, generating hash values from fixed patterns prevents ignoring differences (i.e.,  substitutions, insertions, and deletions) at arbitrary positions between sequences.} \emph{These substitutions and indels between seeds from basecalled sequences \revc{form a part of what we call} \emph{noise} in the scope of this thesis.}

\begin{figure}[tbh]
  \centering
  \includegraphics[width=0.7\linewidth]{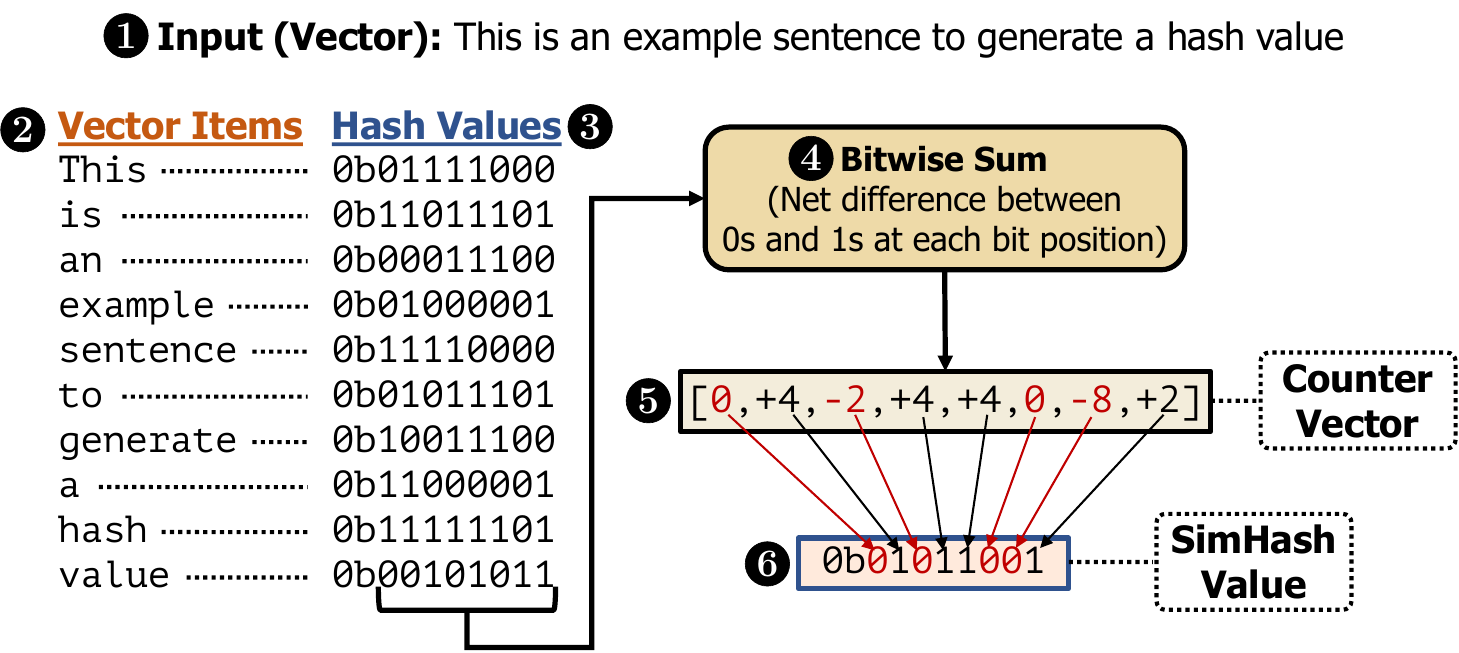}
  \caption{Generating a hash value for a vector of items using SimHash technique. Example input is a sentence (i.e., vector) where the hash values (shown in binary form) of these words (i.e., items) are used to generate the hash value for the entire sentence.}
  \label{bg:fig:simhash}
\end{figure}

\revc{Third, SimHash~\cite{charikar_similarity_2002, manku_detecting_2007} is a locality-sensitive hashing (LSH) technique~\cite{\citemaplsh} that enables generating the same hash values for highly similar vectors (e.g., seeds) even when some arbitrary items between two similar vectors differ (e.g., noise at arbitrary positions). Figure~\ref{bg:fig:simhash} shows the main steps to calculate the SimHash value for a given input vector (e.g., vector of words).
The input vector (e.g., a sentence) is provided as input\revc{\circlednumber{1}} to generate its hash value (i.e., SimHash value). To do so, the vector items are defined, such as all words in the sentence\revc{\circlednumber{2}}. Each item (e.g., word) of the vector is hashed using a low-collision hash function\revc{\circlednumber{3}}. These hash values are shown in their binary forms.
A bitwise sum is then computed across all the generated hash values\revc{\circlednumber{4}}. This is done by adding the corresponding bits across all the hash values at each bit position. The result is a counter vector\revc{\circlednumber{5}}, where each entry represents the net sum (difference between 1s and 0s) for that specific bit position.
The SimHash value is generated from the counter vector\revc{\circlednumber{6}}. For each position in the counter vector, if the sum is positive, the corresponding bit in the SimHash value is set to 1; if the sum is negative, the bit is set to 0. This results in a SimHash value for the given input vector. The SimHash technique enables generating the same hash values for highly similar vectors because a few different individual items (e.g., noise) may result in a counter vector with the same sign information, even though the exact count information may be different. This property allows SimHash to produce the same hash values despite minor variations between vectors.}

These properties of the SimHash technique enable estimating the cosine similarity between a pair of vectors~\cite{goemans_improved_1995} based on the Hamming distance~\cite{hamming_error_1950} of their hash values that SimHash generates (i.e., \emph{SimHash values})~\cite{charikar_similarity_2002, pratap_scaling_2020}.
To efficiently find the pairs of SimHash values with a small Hamming distance, the number of matching most significant bits between different permutations of these SimHash values are computed~\cite{manku_detecting_2007}. This \emph{permutation-based} approach enables exploiting the Hamming distance similarity properties of SimHash technique for various applications that find near-duplicate items~\cite{manku_detecting_2007, uddin_effectiveness_2011, sood_probabilistic_2011, feng_near-duplicate_2014, frobe_copycat_2021}.

\subsection{Filtering and Chaining}
\label{bg:subsubsec:filtering_chaining}

The computational cost of performing sequence alignment on a vast number of seed matches, where many sequences may not actually align, can be prohibitively expensive. This results in a significant waste of computational resources. To mitigate this, read mappers employ 1)~pre-alignment filtering~\cite{\citemapfilter} and 2)~chaining techniques~\cite{\citemapchain} to eliminate dissimilar sequence pairs before alignment and chain many useful seed matches into longer regions.

\subsubsection{Filtering}

Filtering techniques are used to quickly eliminate dissimilar sequences before they proceed to the computationally expensive alignment step. There are four main filtering techniques.

\head{Frequency Filters} 
To quickly eliminate high-frequency, non-informative k-mers that may lead to incorrect seed matches, frequency filters are used~\cite{li_minimap2_2018}. These frequency filters usually calculate the average number of occurrences of a seed in an index and set a frequency threshold based on this calculation to identify if a seed appears frequently. By focusing on less frequent, more informative seeds, frequency filters help reduce the computational burden and improve the accuracy of the subsequent alignment steps.

\head{Pigeonhole Principle} To quickly detect and discard dissimilar sequences from further analysis, the pigeonhole principle~\cite{xin_accelerating_2013, alser_gatekeeper_2017,alser_shouji_2019} assumes that if two sequences differ by up to \( E \) edits, they should still share \emph{at least one common subsequence} among any set of \( E+1 \) non-overlapping subsequences. If a sequence pair does not share a sufficient number of common subsequences, it can be filtered out by matching these subsequences, thereby avoiding unnecessary and costly sequence alignment.

\head{Base Counting} To recognize similar regions by quickly estimating the edit distance between two sequences, the base counting approach~\cite{hach_mrsfast-ultra_2014,rizk_gassst_2010} compares the counts of individual nucleotide bases (A, C, G, T) between sequence pairs. The sum of the absolute differences in base counts provides an upper bound on the similarity identification measured based on edit distance. If this sum exceeds a certain threshold, the sequences are identified as dissimilar, and further alignment is avoided.

\head{q-gram Filtering} To filter out dissimilar sequences by examining shared subsequences, the q-gram filtering approach~\cite{kim_grim-filter_2018} considers all overlapping substrings of length \( q \) (known as q-grams) within the sequence pairs. Since differences between sequences affect only a certain number of q-grams, the technique can estimate whether the sequences are sufficiently similar by quickly counting the number of shared q-grams. The sequence pair is filtered out if the number of shared q-grams is below a threshold.

\subsubsection{Chaining}
\label{bg:subsubsec:chaining}

\begin{figure}[tbh]
\centering
\includegraphics[width=0.8\columnwidth]{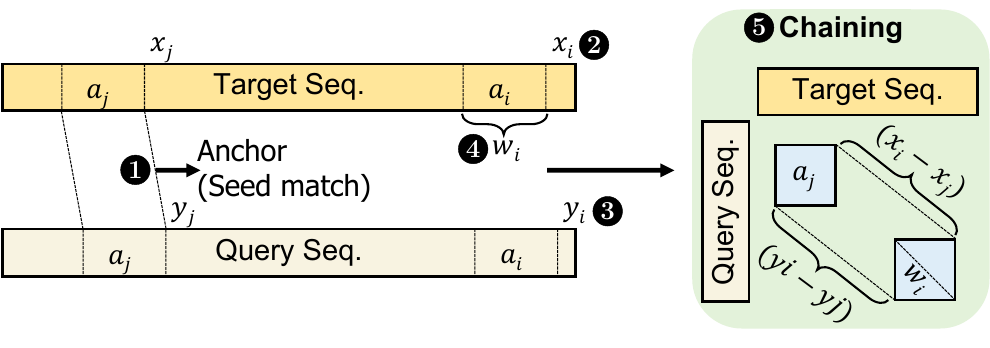}
\caption{Chaining anchors (seed matches) between target and query sequences. The chaining approach calculates the distance between a pair of anchors to identify if they should be included in the same chain.}
\label{fig:seeding_chaining}
\end{figure}

\revc{To identify the seed matches, known as \emph{anchors}, that appear in close proximity in both target and query sequences are linked together in a step called \emph{chaining}~\cite{\citemapchain} \revc{(\circlednumber{3.3} in Figure~\ref{rht:fig:mapping})}. Identifying co-linear chains of anchors is useful as it can eliminate 1)~the anchors that are not included in chains and 2)~the chain of anchors that are unlikely to provide a mapping between target and query sequence pairs, which further reduces the search space (after the filtering step) for the more computationally intensive sequence alignment step. Figure~\ref{fig:seeding_chaining} shows a pair of anchors used in the chaining step to identify if these anchors should be linked together in the same chain. We explain the chaining process in several steps.}

\revc{To explain the chaining process, we define each anchor\revc{\circlednumber{1}} as a 3-tuple $(x,y,w)$, where $x$ denotes the end position of the matching interval in the reference sequence\revc{\circlednumber{2}}, $y$ denotes the end position in the query sequence\revc{\circlednumber{3}}, and $w$ represents the length of the matching region (i.e., the span of the seed match)\revc{\circlednumber{4}}.} The goal of chaining is to calculate the optimal chain score $f(i)$ for each anchor $i$ given a predecessor anchor $j$, which is determined through a dynamic programming (DP) approach\revc{\circlednumber{5}} in Equation~\ref{bg:eq:chain}:

\begin{equation}\label{bg:eq:chain}
f(i) = \max\left\{\max_{i>j\ge 1} \left\{ f(j)+\alpha(j,i)-\beta(j,i) \right\}, w_i\right\}
\end{equation}

where $\alpha(j,i) = \min\left\{\min\{y_i-y_j, x_i-x_j\}, w_i\right\}$ represents the number of matching bases that are covered by anchor $i$ and not by predecessor anchor $j$, while the gap cost $\beta(j,i)$ penalizes gaps between anchors to ensure that only those anchors with minimal insertions or deletions chained together. The gap cost can be defined as:

\begin{equation}\label{eq:chain-gap}
\beta(j,i) = \gamma_c\left((y_i-y_j)-(x_i-x_j)\right)
\end{equation}

where the gap penalty function $\gamma_c(l)$ is given by:

\[
\gamma_c(l)=\left\{\begin{array}{ll}
0.01\cdot \bar{w}\cdot|l|+0.5\log_2|l| & (l\not=0) \\
0 & (l=0)
\end{array}\right.
\]

where $\bar{w}$ is the average seed length, and $l$ is the length of the gap between the two anchors.

To efficiently compute the near-optimal chain score across many anchors, a heuristic is applied to reduce computational complexity. Instead of evaluating all possible predecessors for each anchor, the algorithm can limit the number of comparisons to a fixed number of previous anchors. This heuristic significantly reduces the time complexity from $O(N^2)$ to $O(hN)$, where $N$ is the number of anchors.

\paragraph{Backtracking and Identifying Primary Chains}
\label{bg:subsubsec:backtracking_primary}

To identify the best chain among the several chains given a set of anchors, the backtracking step starts with an anchor with the highest score.
The process continues by following the chain of predecessors for each anchor and marking them \emph{used} until no further predecessor that is not used in another chain can be found. This backtracking step can identify multiple chains that represent similar regions between the sequence pairs.

This filtering and chaining process significantly improves the efficiency and accuracy of the read mapping process to enable the alignment of sequences in a computationally feasible manner, even when dealing with large genomic datasets.

\subsection{Sequence Alignment}
\label{bg:subsubsec:alignment}

To accurately identify the positions and types of differences and similarities between sequence pairs, approximate string matching (ASM) approaches are commonly used for \emph{sequence alignment} \revc{(\circlednumber{3.4} in Figure~\ref{rht:fig:mapping})}. This process aims to identify differences and similarities, such as matches, substitutions, insertions, and deletions, by calculating the \emph{edit distance} between sequence pairs—the minimum number of single-character edits required to transform one sequence into another. Widely used methods for performing sequence alignment are mainly based on the dynamic programming (DP) approach~\cite{\citemappairalign}.

\begin{figure}[tbh]
\centering
\includegraphics[width=1\columnwidth]{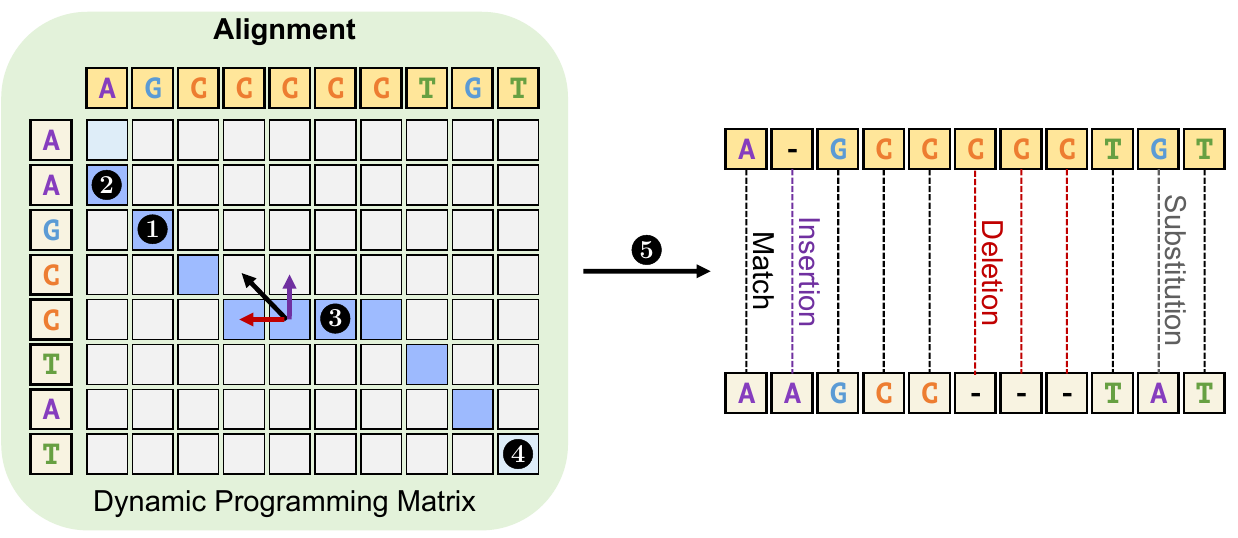}
\caption{Dynamic Programming (DP) Matrix used to identify edit operations: Match, Substitution, Insertion, and Deletion. Each cell’s value is calculated based on the values of adjacent cells, as indicated by the arrows. The optimal alignment path is highlighted in dark blue, while light blue cells indicate two subsequent anchors identified during the chaining step. The corresponding edit operations for the optimal alignment are shown on the right side of the DP Matrix.}
\label{fig:alignment}
\end{figure}

\revb{Figure~\ref{fig:alignment} shows how a DP matrix is constructed and used for sequence alignment. Each cell in the matrix represents a score for a potential edit operation between characters from the two sequences being aligned. The score in each cell is computed based on three main scenarios: 1)~aligning the current characters from both sequences (resulting in either a match or a substitution)\circlednumber{1}, 2)~aligning the current character from one sequence by \emph{opening} a gap in the other sequence (insertion or deletion)\circlednumber{2}, and \revc{3)~extending the open gap in the other sequence to handle insertions (or deletions) that appear in a row.\circlednumber{3}} The arrows in the matrix indicate how each cell's score is calculated based on the values from the adjacent cells.}

\revb{To minimize the number of edits \revc{between sequence pairs}, the alignment process usually penalizes operations that introduce \revc{and extend the existing open} gaps. This is typically achieved using the \emph{affine gap} scoring function~\cite{li_minimap2_2018}. The affine gap model applies two types of penalties: 1)~a \emph{gap opening penalty}, which usually incurs a higher cost when a new gap is introduced, and 2)~a \emph{gap extension penalty}, which mainly imposes a smaller incremental cost for extending an existing gap. These penalties help ensure that gaps are used judiciously in the alignment.

The optimal alignment path is identified by backtracking from the cell with the highest score, as illustrated with cells marked by blue colors starting from the lower-right corner\circlednumber{4} in Figure~\ref{fig:alignment}. This path shows the sequence of matches, substitutions, insertions, and deletions that results in the optimal alignment. The right side of Figure~\ref{fig:alignment} shows the corresponding edit operations for the aligned sequences. Each operation is marked as matches (black lines), insertions (purple lines), deletions (red lines), and substitutions (grey lines). The final alignment score, which results from this backtracking process, provides a quantitative measure of sequence similarity by accounting for the types and number of edits needed to align the sequences.

DP methods provide a rigorous approach to \revc{perform} sequence alignment. However, their computational complexity, typically \(O(m \cdot n)\)~\cite{baichoo_computational_2017} for sequences of length \(m\) and \(n\), can become a limitation for long reads or large reference genomes. To improve efficiency, filtering and chaining steps are often used before alignment to narrow down the search space, which can prevent using the alignment process unnecessarily.}

\revc{
\subsection{Necessity of Read Mapping}
A fine-grained mapping of reads (i.e., basecalled sequences and raw sequencing data) by finding the base positions (i.e., \emph{base-level read mapping}) where they map either within a corresponding genome or to another read is essential for various reasons. First, base-level mapping enables the detection of genetic variants such as single nucleotide polymorphisms (SNPs), insertions and deletions (indels), and structural variations by comparing the aligned reads to a reference genome, as we discuss in Section~\ref{bg:subsec:variant_calling}. Second, read mapping facilitates genome assembly and reconstruction by identifying overlapping regions between reads based on their exact positions, which is crucial for \emph{de novo} genome assembly to construct an organism's genome from scratch without a reference genome, as discussed in Section~\ref{bg:subsec:denovo}. Third, base-level mapping allows for accurate quantification of gene expression levels in RNA sequencing (RNA-Seq) experiments by mapping reads to specific genomic locations to measure transcript abundance, identify differentially expressed genes, and detect alternative splicing events~\cite{song_computational_2019,denti_asgal_2018}. Fourth, read mapping is essential for \emph{haplotype phasing} and detection of \emph{de novo} mutations to determine which genetic variants are inherited together on the same chromosome and to study the detection of new mutations that arise in a single generation (\emph{de novo} events)~\cite{cheng_haplotype-resolved_2021, zhang_haplotype-resolved_2021, stanojevic_telomere--telomere_2024}. Fifth, base-level mapping enables the detection of \emph{rare} or \emph{low-frequency variants}, especially in heterogeneous samples such as cancer tissues, which is crucial for understanding tumor heterogeneity and developing personalized treatment strategies~\cite{nabavi_identification_2022}. Read mapping is essential for many applications to achieve sensitive and fast analysis, even for scenarios where potential alternatives such as polymerase chain reaction (PCR)~\cite{kanagawa_bias_2003}\mytodo{Cite more} can fall short compared to read mapping.
}

\section{Further Steps in Sequence Analysis}
\label{bg:subsec:downstream}

\subsection{\emph{De novo} Assembly Construction}
\label{bg:subsec:denovo}

\revc{The goal of \emph{de novo} assembly is to reconstruct a genome from reads without the use of a reference genome. This is achieved by identifying similarities between read pairs, known as \emph{all-vs-all overlapping}, and stitching these reads together.} There are mainly two reasons for constructing \emph{de novo} assemblies. First, the reference genome may be unavailable or incomplete, particularly for non-model organisms or populations with high genetic diversity~\cite{chaisson_genetic_2015}. Second, analyzing an individual's genome without reference to a potentially divergent reference genome reduces biases introduced by reference-specific artifacts and assumptions~\cite{wong_towards_2020}.

\head{All-vs-all Overlapping} To stitch the reads together from their ends, the first step in reconstructing a genome from sequencing reads involves identifying overlapping regions among all pairs of reads~\cite{li_minimap_2016}. Typically, two reads are considered overlapping if they contain matching sequences from their ends (i.e., between the suffix of one read and the prefix of another). To identify these matching regions, read mapping steps discussed in Section~\ref{bg:sec:mapping} are mainly used.

\head{Assembling Overlapping Reads} Constructing the assembly from the paths generated by overlapping reads is challenging due to the presence of numerous redundant and incorrect~\cite{li_minimap_2016} overlapping pairs. To identify the underlying genome assembly, the overlap information between all read pairs is processed by constructing a graph from overlapping reads using a graph structure, such as an assembly graph (i.e., string graph)~\cite{myers_fragment_2005} or a de Bruijn graph~\cite{iqbal_novo_2012}.

\revc{These assembly graphs often contain errors~\cite{koren_canu_2017}, which are resolved through several steps collectively known as \emph{graph cleaning}.}
\revc{First, incorrect or incomplete overlap information can result in direct connections (i.e., edges) between two reads, even when a longer path (i.e., multiple nodes and edges) with multiple overlapping read pairs connects them. These direct edges, known as \emph{transitive edges}, are removed because they do not affect the overall connectivity of the graph~\cite{myers_fragment_2005}.}
Second, at least two sub-graphs with several nodes can exist in the graph such that they share the same start and end nodes. This creates alternative paths in the assembly graph called \emph{bubble}. These bubbles can be resolved by either collapsing them, keeping only one path~\cite{li_minimap_2016} or by presenting the alternative paths as different potential assemblies~\cite{cheng_haplotype-resolved_2022,cheng_haplotype-resolved_2021}. These bubbles, often caused by variations and sequencing errors, are useful for identifying different copies of chromosomes in diploid and polyploid genomes.
Third, the graph is traversed to identify paths that represent the original genome sequence, ensuring the assembled sequences are accurate and continuous. Additional steps can be used to simplify the graph further by identifying errors and artifacts in sequencing and during the overlapping process~\cite{li_minimap_2016}. \revc{The resulting graph typically includes paths that are \emph{trivial} to parse (i.e., without multiple edges from a node), representing the assembled genome by stitching the reads to each other as identified by the path in the graph.}

\subsection{Metagenomics Analysis}
\label{bg:subsec:metagenomics}

Metagenomics involves the study of genetic material recovered directly from environmental samples to allow for the analysis of microbial communities without the need for culturing individual species. This approach provides insights into the diversity and composition of microbial ecosystems.

\head{Read Classification} The initial step in metagenomics analysis is the classification of sequencing reads, where DNA or RNA sequences obtained from the environment are assigned to known taxonomic groups. \revc{This classification is usually performed by aligning reads to a reference database of known genomes~\cite{lapierre_metalign_2020}. Alternatively, k-mer-based matching can quickly identify the closest taxonomic matches based on sequence similarity~\cite{breitwieser_krakenuniq_2018,wood_improved_2019}.} The classification process enables the identification of the species present in the sample and helps map the genetic potential and functional roles of these organisms within their ecosystems.

\head{Relative Abundance Estimation} Following classification, estimating the relative abundance of each identified species is a common next step in metagenomics. This process involves quantifying the number of reads assigned to each taxonomic group and normalizing these counts to mitigate biases introduced by variable gene copy numbers, read length, and genomic size among different organisms. Relative abundance data are useful for understanding the structure of microbial communities and assessing how environmental changes or different conditions affect microbial composition and dynamics.

\head{Challenges} Despite its potential, metagenomics faces several challenges. First, the sheer diversity of microbial communities, coupled with the vast amount of data generated, poses computational and analytical challenges~\cite{mansouri_ghiasi_megis_2024}. \revc{The limited availability of complete reference genomes can hinder accurate taxonomic classification and functional prediction, especially for rapidly evolving microbial genomes~\cite{bexfield_metagenomics_2011}.} Third, the presence of DNA from dead cells can complicate interpretations of community activity and function~\cite{cangelosi_gerard_a_dead_2014}. 

\revc{Addressing these challenges is critical for advancing our understanding of microbial ecosystems. Improvements in computational methods, database completeness, and data interpretation can allow for more accurate and comprehensive insights into microbial diversity, dynamics, and functional roles, further expanding the potential of metagenomics in environmental and clinical research.}

\subsection{Variant Calling}
\label{bg:subsec:variant_calling}

The objective of variant calling is to identify genomic variants between an individual's genome and a reference genome~\cite{\citevariantcallers}. These variants are mainly categorized as single-nucleotide polymorphisms (SNPs), insertions, deletions, and larger structural variations (SVs). Accurate and efficient detection of these variants is vital for understanding of the genetic basis of diseases~\cite{lawrence_mutational_2013}, population genetics~\cite{poplin_scaling_2018}, evolutionary studies~\cite{kanehisa_toward_2019}, personalized medicine~\cite{dong_genome-wide_2019} and pharmacogenomics~\cite{sangkuhl_pharmacogenomics_2020}.

Variant calling involves processing the read mapping output and detecting variants. First, read mapping output is processed by sorting and optionally identifying duplicate information to minimize bias introduced during the \emph{polymerase chain reaction} (PCR) step of sample preparation~\cite{zverinova_variant_2022}. Second, mapped reads are analyzed to distinguish genuine variants from sequencing errors or misalignments using resource-intensive statistical techniques~\cite{nickerson_polyphred_1997, weckx_novosnp_2005, poplin_scaling_2018} or machine learning techniques~\cite{poplin_universal_2018}. \revc{Despite the high computational demands, variant calling plays a crucial role in advancing our understanding of genetic variations, driving research in genomics and personalized medicine. Ongoing improvements in variant calling accuracy, speed, and efficiency remain essential as the volume of sequencing data continues to grow.}

\section{Accelerating Genome Analysis}
\label{bg:sec:accelerating}

The goal of accelerating genome analysis is to reduce the time \revc{and energy} required to process sequencing data and produce actionable insights. This acceleration \revc{would in turn} reduce the overall latency \revc{and energy consumption} of \revc{end-to-end genome} analyses, \revc{while also improving throughput of analyses}~\cite{mutlu_accelerating_2023}.

We categorize the strategies for accelerating genome analysis into two main categories: 1)~acceleration with software (and algorithmic) optimizations \revc{as we discuss in Section~\ref{bg:subsec:accelerating:sw}}, and 2)~acceleration through hardware and software co-design \revc{as we discuss in Section~\ref{bg:subsec:accelerating:hw}}.

\revc{We explain how these strategies are used in many steps in the genome analysis pipeline in Section~\ref{bg:subsec:accelerating:pipeline}.}

\subsection{Acceleration with Software Optimizations}
\label{bg:subsec:accelerating:sw}

Software-based optimizations aim to improve the performance (i.e., speed) of genome analysis tools by modifying the existing or designing new algorithms and data structures. These optimizations often (\revc{but not always}) involve trade-offs between performance and accuracy, aiming to reduce computational load while maintaining acceptable \revc{(and ideally the same or better)} levels of accuracy. We describe several common strategies below.

\head{Reducing Computational and Space Complexity} Developing new algorithms that reduce the computational and space complexity of genome analysis tasks can directly accelerate the execution of an application~\cite{marco-sola_fast_2021, marco-sola_optimal_2023}\mytodo{Cite more}. This usually involves replacing either part of the entire execution flow of an application with different sets of algorithms. \revc{For example,} this can include replacing the computationally complex use of DNN approaches with simpler operations~\cite{firtina_rawhash_2023}\mytodo{Cite more} or eliminating the use of costly operations with DNN approaches~\cite{poplin_universal_2018, poplin_scaling_2018}\mytodo{Cite more}.

\head{\revb{Eliminating Unnecessary} Computation} \revc{Eliminating unnecessary computations by identifying and discarding irrelevant or low-quality data early in the pipeline significantly reduces computational overhead and accelerates overall performance.} We explain five common strategies to reduce the overall computations.

First, for DNN-based approaches (e.g., basecalling), a large number of parameters can be pruned without substantially reducing the accuracy to reduce computational overhead~\cite{singh_rubicon_2024}\mytodo{Cite more}.

Second, data resolution can be reduced (i.e., quantization), which is useful, \revc{for example}, for DNN-based works as they include a large number of complex operations with high precision~\cite{firtina_rawhash_2023,singh_rubicon_2024,shivakumar_sigmoni_2024}.

Third, certain filters or sampling techniques can quickly eliminate computations and data that are likely useless in later steps by using cheaper computations to identify such useless computations and data (e.g., sketching and pre-alignment filters techniques \revc{as discussed in Sections~\ref{bg:subsubsec:sketching} and~\ref{bg:subsubsec:filtering_chaining}, respectively})~\cite{cavlak_targetcall_2024,\citemapsketch,\citemapfilter}

Fourth, in many DP operations, such as in chaining~\cite{li_minimap2_2018}\mytodo{Cite more} and when using profile Hidden Markov Models (pHMMs)~\cite{firtina_aphmm_2024}, the connections to backward cells (or nodes) can usually be limited by how long a single connection can extend backward. This enables substantially reducing the \revc{high-overhead} data-dependent computations that are usually performed iteratively.

Fifth, redundant computations are common when re-running a certain application on slightly different data~\cite{\citemaplift}. This is, \revc{for example}, the case for read mapping when remapping reads from an older reference genome to a newer one~\cite{kim_airlift_2024}. In such cases, the similarities between the multiple versions of the data can be analyzed quickly to eliminate redundant computations by re-using the already-generated information from the earlier execution and identifying when it is necessary to perform additional computation on the newer data.

\head{Data Structure and Data Access Pattern Optimizations} Optimizing the choice of data structures and their access patterns~\cite{hach_mrsfast_2010,pan_kmerind_2016} is another key method for accelerating genome analysis. \revc{Efficient data structures, such as compressed matrices~\cite{guidi_bella_2021,ellis_dibella_2019}\mytodo{Cite more}, graphs~\cite{kanellopoulos_smash_2019}\mytodo{Cite more}, and specialized indexing schemes~\cite{senol_cali_segram_2022}\mytodo{Cite more}, can significantly reduce memory usage and accelerate data retrieval.}

\head{Exploiting Parallelism} Leveraging parallelism is a fundamental approach to accelerating genome analysis. Modern commodity processors offer multi-core and many-core architectures that can be utilized to execute multiple analysis tasks \revc{that are} \revb{multithreaded}. Single Instruction, Multiple Data (SIMD) operations~\cite{flynn_very_1966,xin_shifted_2015, firtina_blend_2023} can be used to perform the same operation on multiple data points simultaneously, making them highly effective for processing large genomic datasets. \revc{To significantly accelerate genome analysis tasks, massively parallel operations in genome analysis are exploited using 1)~Graphics Processing Units (GPUs)~\cite{\citehwgpuall}, as they are well-suited for handling the large-scale parallel computations required in genomics, and 2)~general distributed computing techniques~\cite{\citehwdistall}.}

\subsection{Acceleration with Hardware and Software Co-design}
\label{bg:subsec:accelerating:hw}

Hardware-software (HW/SW) co-design, where hardware and software \revc{(and algorithm)} are designed together \revc{to overcome} the \revc{different} limitations of and \revc{take advantage of the different} capabilities of underlying hardware and software, can offer substantial improvements in both performance and energy efficiency. \revc{As such, these HW/SW co-designed acceleration approaches have formed a thriving area of research~\cite{mutlu_accelerating_2023, alser_accelerating_2020, alser_molecules_2022}.} We discuss key challenges in software-only optimizations and potential hardware-software co-design solutions.

\head{Challenges in Pure Software Optimizations} Many software optimization strategies still involve significant data movement between computation and memory units, which can create bottlenecks due to limited bandwidth and \revc{high power requirements~\cite{alser_accelerating_2020, mansouri_ghiasi_genstore_2022, mansouri_ghiasi_megis_2024,mutlu_modern_2023,ghose_processing--memory_2019,mutlu_processing_2019,seshadri_simple_2017,singh_fpga-based_2021,ghose_processing--memory_2019-1,seshadri_-dram_2020,oliveira_accelerating_2022,oliveira_damov_2021}.}
Second, a substantial portion of the data processed in genome analysis pipelines can be useless~\cite{mao_genpip_2022,mansouri_ghiasi_genstore_2022,\citehwfiltall}, resulting in wasted compute cycles. Identifying and discarding useless computations earlier in the pipeline can save time and energy. However, even with software optimizations that identify useless computations, data must still be moved through different levels of memory hierarchy, causing unnecessary delays~\cite{mutlu_accelerating_2023,bera_constable_2024}.
Third, to accommodate a wide range of applications, general-purpose processors usually provide limited computational resources and speed to execute \revc{effectively} a \revc{particular operation~\cite{hameed_understanding_2010,hameed_understanding_2010-1}.} \revc{This makes it challenging to exploit the opportunities that a certain application provides (e.g., data access patterns with certain locality).}
Fourth, HTS technologies generate data at an ever-increasing rate, presenting a challenge to keep up with the output, especially in time-critical applications~\cite{alser_accelerating_2020, firtina_rawhash_2023}.

To address these challenges, several hardware design strategies can be employed. \revc{These strategies almost always necessitate co-design of software along with the hardware.}

\head{Designing Customized Hardware} Application-Specific Integrated Circuits (ASICs) and Field-Programmable Gate Arrays (FPGAs) (and customized system on chip designs) can be used to design customized hardware to accelerate specific computational tasks~\cite{\citehwcustall}. FPGAs and ASICs are particularly important for designing special processing units (PUs) that are highly performant and efficient hardware. These PUs are specifically designed based on the needs of the application, especially when these applications provide a certain data flow such that intermediate data generated after the execution of each PU is moved to another PU that performs either the same or a different operation. \revc{In such cases, these PUs can be connected using a design known as \emph{systolic arrays}~\cite{kung_systolic_1979,kung_why_1982}, which stream data efficiently between processing units, reducing the need for additional read and write operations to external memory.} \revc{Such specialized designs avoid the overheads present in general-purpose hardware.}

\head{Memory-Centric Design} Memory-centric approaches, such as Processing-in-Memory (PIM) (i.e., Processing Using Memory (PUM) and Processing Near Memory (PNM))~\cite{\citesafaripim}, can bring computation closer to the data to reduce the data movement \revb{overheads} in genome analysis~\cite{\citehwpimall}. PUM approaches can perform computations directly within memory cells (e.g., by using analog operations within DRAM), reducing the need to transfer data between memory and processing units \revb{and exploiting the inherent parallelism present in memory arrays~\cite{seshadri_ambit_2017,hajinazar_simdram_2021,seshadri_buddy-ram_2016,seshadri_rowclone_2013,seshadri_processing_2016,oliveira_damov_2021}.} PNM approaches place processing elements near memory to allow \revb{high-throughput} data access \revb{at} reduced latency \rev{and energy}~\cite{oliveira_damov_2021}. \revc{It is essential to understand the application's requirements to fully leverage PUM and PNM approaches, optimizing both computation~\cite{seshadri_ambit_2017,seshadri_rowclone_2013} and communication between memory-centric accelerators~\cite{oliveira_damov_2021,mutlu_lightning_2023}.}

\head{Storage-Centric Design} Storage-centric designs integrate computational capabilities directly into storage devices, such as Solid-State Drives (SSDs)~\cite{\citehwispall,kim_evanesco_2020}. To accelerate the execution of an application, in-storage processing can leverage the higher internal \revc{SSD} bandwidth to reduce data movement to external processing units by processing the data inside the SSD~\cite{\citehwfiltisp}. This can be achieved by utilizing the existing cores and DRAM within SSDs~\cite{zou_assasin_2022}, integrating specialized hardware inside SSDs~\cite{mansouri_ghiasi_genstore_2022}, and performing computations with flash chips in SSDs~\cite{park_flash-cosmos_2022}. Storage-centric computation provides many opportunities, especially when the data reuse of an application is low, and the application can benefit from the increased bandwidth inside SSDs.

\head{Extending Instruction Set Architecture with Specialized Instructions} Extending the Instruction Set Architecture (ISA) with specialized instructions can improve the performance of genome analysis tasks. By incorporating custom instructions tailored to specific genomic operations, such as DP operations in sequence alignment, it is possible to execute these tasks more efficiently, reducing the need for complex software routines and lowering the overall computational overhead~\cite{doblas_gmx_2023,pavon_quetzal_2024,liyanage_accelerating_2024,elster_nvidia_2022}. This approach involves designing custom logic and data handling mechanisms with special instructions to execute specific tasks directly within the processor, bypassing the limitations of general-purpose computation.

\head{\revb{Exploiting} Emerging Technologies} Emerging technologies such as analog computation using Resistive Random-Access Memory (ReRAM)~\cite{shahroodi_swordfish_2023,mao_genpip_2022,kaplan_resistive_2017,kaplan_rassa_2018,merlin_diper_2024}, data search using Content Addressable Memory (CAM) arrays~\cite{jahshan_dash-cam_2023,hanhan_edam_2022,harary_gcoc_2024,garzon_hamming_2022}, and optical computing~\cite{solli_analog_2015} provide opportunities for further acceleration. These technologies can potentially improve performance and energy efficiency for specific genome analysis tasks by leveraging novel computational paradigms.

By combining these hardware-centric approaches with software optimizations, highly efficient and scalable genome analysis pipelines capable of meeting the demands of modern sequencing technologies and applications can be created.

\subsection{Accelerating the Common Steps in a Genome Analysis Pipeline}
\label{bg:subsec:accelerating:pipeline}

\revb{Hardware and software co-designed approaches are commonly utilized to accelerate several steps in the common basecalled analysis of genomes. These include accelerating the steps in read mapping and variant calling, which can be integrated into real-time analysis to reduce the latency of genome analysis. We explain these acceleration efforts in this section. We explain the acceleration efforts for basecalling and raw signal analysis in Sections~\ref{rel:subsec:basecalledreal} and~\ref{rel:subsec:rawreal}, respectively.

\subsubsection{Accelerating Read Mapping}
\label{rel:subsec:accelerating:map}
Since read mapping is a crucial and computationally expensive step in many genome analysis pipelines, numerous works focus on accelerating it in various ways.
First, a significant fraction of sequence pairs does \emph{not} align, which leads to wasted computation and energy during alignment~\cite{kim_grim-filter_2018}. To avoid this useless computation, various works propose \emph{pre-alignment filtering}, another step in read mapping that can efficiently detect and eliminate highly dissimilar sequence pairs \emph{without} using alignment. Most pre-alignment filtering works~\cite{\citehwfiltall} provide algorithm-architecture co-design using FPGAs~\cite{\citehwfiltcust}, GPUs~\cite{\citehwfiltgpu}, and PIM~\cite{\citehwfiltpim} to substantially accelerate the entire read mapping process by exploiting massive parallelism, efficient bitwise operations, and specialized hardware logic for detecting similarities among a large number of sequences. Apart from accelerating the filtering process, many works co-design hardware and software for the chaining~\cite{\citehwchainall} \revc{(i.e., by using custom hardware design~\cite{\citehwchaincust}, GPUs~\cite{\citehwchaingpu} and PIM~\cite{\citehwchainpim}) and seeding steps~\cite{\citehwseedall} (i.e., by using custom hardware design~\cite{\citehwseedcust}, GPUs~\cite{\citehwseedgpu} and PIM~\cite{\citehwseedpim}) to accelerate the pre-alignment step in genome analysis.}

\revc{Second, \revc{various works~\cite{\citehwfiltisp,\citehwmetaisp} such as} GenStore~\cite{mansouri_ghiasi_genstore_2022}, show that a large amount of sequencing data unnecessarily moves from a storage system (e.g., the solid-state drive (SSD)) to memory during read mapping, significantly increasing latency and energy consumption. To eliminate this wasteful data movement, available computation capabilities and the high bandwidth within the SSD, as well as the specialized logic that can be integrated within a storage system, can be exploited to process sequencing data in the storage system without moving to the main memory or the CPU, thereby eliminating unnecessary data movement in the system.}

Third, numerous studies~\cite{\citehwalignall}, including GenASM~\cite{senol_cali_genasm_2020} and Darwin~\cite{turakhia_darwin_2018}, focus on accelerating the underlying ASM algorithm employed in sequence alignment through efficient algorithm-architecture co-design. They do so by exploiting systolic arrays~\cite{fei_fpgasw_2018,senol_cali_genasm_2020}, GPUs~\cite{\citehwaligngpu}, customized hardware designs (and system on chip)~\cite{\citehwaligncust}, and PIM~\cite{\citehwalignpim}. These works provide substantial speedups of up to multiple orders of magnitude compared to software baselines. Among these works, SeGraM~\cite{senol_cali_segram_2022} is the first to accelerate mapping sequences to graphs~\cite{\citehwstgall} that are used to reduce population bias and improve genome analysis accuracy by representing a large population (instead of a few individuals) within a single reference genome.

Fourth, several works focus on accelerating either all or some of the steps together~\cite{\citehwrmapall} using custom hardware accelerators~\cite{\citehwstgcust}, GPUs~\cite{\citehwrmapgpu} and PIM~\cite{\citehwrmappim}.

\subsubsection{Accelerating \emph{De novo} Assembly}
\label{rel:subsec:denovo_acceleration}

\revb{\emph{De novo} assembly pipeline commonly uses many steps in read mapping (i.e., seeding, sketching, and chaining). We discuss the acceleration efforts of these read mapping steps in Section~\ref{rel:subsec:accelerating:map}.

\revc{There are unique challenges to \emph{de novo} assembly, such as graph construction and cleaning (as discussed in Section~\ref{bg:subsec:denovo}) and k-mer counting~\cite{nisa_distributed-memory_2021}. Methods specifically designed for accelerating \emph{de novo} assembly~\cite{\citehwassall} utilize specialized algorithms and hardware accelerators to efficiently handle large-scale assembly with distributed computing techniques~\cite{\citehwassdist}, custom hardware accelerators~\cite{\citehwasscust}, GPUs~\cite{\citehwassgpu}, and PIM~\cite{\citehwasspim}. Most of these accelerators target accelerating k-mer counting due to the nature of embarrassingly parallel operations in k-mer counting that can be performed independently on many-core architectures.}}

\subsubsection{Accelerating Metagenomics Analysis}
\label{bg:subsec:metagenomics_acceleration}

Metagenomic analysis poses significant computational challenges due to the large volume of data involved~\cite{mansouri_ghiasi_megis_2024}. Efforts to optimize metagenomic analysis have focused on both software and hardware-based solutions.

\head{Software Optimization of Metagenomics}
Various software tools~\cite{\citemetagenomics} aim to improve the efficiency of metagenomic analysis by optimizing the use of databases. Tools like MetaAlign~\cite{lapierre_metalign_2020} and Centrifuge~\cite{kim_centrifuge_2016,song_centrifuger_2024} aim for high accuracy with computationally expensive approaches. However, these tools often incur significant computational and I/O costs due to the large sizes of these databases~\cite{mansouri_ghiasi_megis_2024}. To mitigate these issues, some tools apply sampling techniques to reduce database size, such as MetaPalette~\cite{koslicki_david_metapalette_2016} and MetaCache~\cite{kobus_metacache-gpu_2021}. While these techniques improve performance by reducing the amount of data that needs to be processed, they can lead to a loss in accuracy.

\head{Hardware Acceleration of Metagenomics}
\revb{Several studies~\cite{\citehwmetaall} have explored hardware acceleration to overcome the computational bottlenecks in metagenomics. GPUs have been widely used to speed up various stages of metagenomic analysis by offloading compute-intensive tasks~\cite{\citehwmetagpu}. FPGAs provide opportunities for acceleration, offering configurable hardware that can be tailored to specific tasks within the metagenomic pipeline~\cite{\citehwmetacust}. Processing-in-memory (PIM) techniques have been employed to address the I/O bottlenecks in metagenomic analysis~\cite{\citehwmetapim}. These hardware-accelerated approaches significantly reduce computation time and energy consumption but often still face challenges related to the I/O overhead associated with large metagenomic datasets. MegIS~\cite{mansouri_ghiasi_megis_2024} alleviates the data movement overhead in metagenomics analysis from the storage system~\cite{\citehwmetaisp} 1)~via its efficient and cooperative pipeline between the host and the SSD and 2)~by leveraging in-storage processing for metagenomics.}

\subsubsection{Accelerating Variant Calling}
Variant callers~\cite{\citevariantcallers} like GATK HaplotypeCaller~\cite{poplin_scaling_2018} use costly probabilistic calculations to analyze the likelihood of specific variants in large sequencing datasets. \revc{DeepVariant~\cite{poplin_universal_2018}, a DNN-based variant caller, processes read alignment data as images, a method that demands substantial GPU resources and memory due to the complexity of deep neural networks.} Reducing computational requirements through algorithmic optimizations, parallelization, and efficient data representation is crucial for faster, more accurate genetic variant analyses.

To accelerate variant calling, several works propose algorithm-architecture co-designs~\cite{\citehwvcall}. These include fast execution of Pair Hidden Markov Models (Pair HMMs) in FPGAs or ASICs~\cite{\citehwvccust}, reducing data movement overheads in GPUs~\cite{\citehwvcgpu}, and pipelining processing steps with tools like elPrep~\cite{herzeel_multithreaded_2021}, system-on-chip designs~\cite{wu_975-mw_2021}, and PIM~\cite{\citehwvcpim}.}

\section{Summary and Further Reading}
\label{bg:sec:summary}

\revc{Significant research efforts are essential across the entire sequence analysis pipeline. These efforts range from improvements in sequencing technologies \revc{(including device and circuit-level advancements)} to improvements in read mapping and downstream analysis (spanning algorithmic, software, and hardware optimizations). Each step, from efficiently processing raw sequencing data to accurately assembling and analyzing genomes, plays a critical role in advancing our understanding of genetic information and its practical applications.}

\revc{For further reading, we recommend that interested readers explore works on sequencing technologies~\cite{\citesbs,\citesmrt,\citesanger,\citenanopore}, algorithmic approaches~\cite{wood_improved_2019,song_centrifuger_2024,shen_kmcp_2023,kim_centrifuge_2016,boucher_prefix-free_2019,gagie_fully_2020,li_bwt_2024,charikar_similarity_2002,manku_detecting_2007,sinha_fruit-fly_2021,chen_using_2020,sharma_improving_2018,firtina_blend_2023,joudaki_fast_2021,ryali_bio-inspired_2020,firtina_genomic_2016, firtina_hercules_2018, firtina_apollo_2020}, and hardware accelerations~\cite{\citehwall} for genome analysis.}

\chapter{Related Work}
\label{chap:related}

Many prior works study noise in sequence analysis and develop techniques for understanding, tolerating, and reducing this noise. This section provides an overview of closely-related works \revc{and describes where the existing body of work falls short}.

\revc{In Section~\ref{rel:sec:fuzzy}, we describe the limitations of the existing seeding techniques to set the stage for how the work we discuss in Chapter~\ref{chap:blend} fills the gap to resolve the limitations we discuss.

In Section~\ref{rel:sec:realtime}, we discuss the existing works that perform real-time analysis with and without basecalling. We identify their challenges. In Chapters~\ref{chap:rh} and~\ref{chap:rht}, we discuss how our works overcome these challenges.}

\section{Tolerating Noise in Seeding – Fuzzy Seed Matching}
\label{rel:sec:fuzzy}

\revb{Commonly used read mappers and other related works that perform sequence analysis, such as minimap2~\cite{li_minimap2_2018}, MHAP~\cite{berlin_assembling_2015}, Winnowmap2~\cite{jain_long-read_2022, jain_weighted_2020}, re$M_{u}$val~\cite{deblasio_practical_2019}, and CAS~\cite{xin_context-aware_2020} use sampling techniques to choose a subset of k-mers from all k-mers of a read without significantly reducing their accuracy. For example, minimap2 uses the minimizer sketching technique~\cite{roberts_reducing_2004} to reduce overall \revc{storage (and memory)} space requirements while \revc{also} improving \revc{performance}. These works use low-collision hash functions that require exact matches of seeds generated after their sketching mechanisms. As discussed in Section~\ref{bg:subsubsec:hashing}, \revc{using low-collision hash functions} provides certain challenges when determining parameters that impact performance and accuracy. 
Although it is important to reduce the collision rate for assigning different hash values for dissimilar seeds for accuracy and performance reasons, choosing low-collision hash functions also makes it unlikely to assign the same hash value for similar seeds. Thus, seeds must exactly match to find matches between sequences with a single lookup.}

\revc{Mitigating \revc{this exact-matching requirement} allows similar seeds to share the same hash value, thereby significantly improving the performance and sensitivity of applications. These applications can then tolerate substitutions and indels at arbitrary positions during seed matching, a capability we refer to as \emph{fuzzy seed matching}.}

There are typically two directions for tolerating mismatches (i.e., substitutions and indels) when finding seed matches: 1)~mechanisms that tolerate substitutions at fixed positions~\cite{\citemapmaskedhash} and 2)~mechanisms that tolerate substitutions and indels~\cite{chin_human_2019,sahlin_effective_2021,sahlin_flexible_2022}. To our knowledge, our hash-based fuzzy seeding approach (explained in Chapter~\ref{chap:blend}) is the first work that can efficiently find fuzzy matches that can allow arbitrary mismatches between seeds.

\subsection{Spaced Seeds to Tolerate Substitutions}
\label{rel:subsec:spacedseeds}

\revc{To allow substitutions when matching k-mers, a common approach is to \emph{mask} certain characters of k-mers extracted from target and query sequences~\cite{\citemapmaskedhash}. These masks are used as \emph{don't care} characters such that the substitutions that appear at these positions are ignored. Predefined \emph{patterns} determine the \emph{fixed} masking positions for a given input k-mer.} Seeds generated from masked k-mers are known as \emph{spaced seeds}~\cite{ma_patternhunter_2002}.

Tools such as ZOOM!~\cite{lin_zoom_2008} and SHRiMP2~\cite{david_shrimp2_2011} use spaced seeds to improve sensitivity when mapping short reads (i.e., Illumina paired-end reads). S-conLSH~\cite{chakraborty_conlsh_2020, chakraborty_s-conlsh_2021} generates many spaced seeds from each k-mer using different masking patterns to improve the sensitivity when matching spaced seeds with locality-sensitive hashing techniques.
There have been recent improvements in determining the masking patterns to improve the sensitivity of spaced seeds~\cite{petrucci_iterative_2020, mallik_ales_2021}.
\revc{However, spaced seeds are limited to tolerating substitutions at certain positions because these techniques use 1)~fixed patterns that allow substitutions only at certain positions of k-mers and 2)~\emph{low-collision hashing} techniques used for finding \emph{only} \revb{exactly matching} spaced seeds.}

\subsection{Linked k-mers to Tolerate Substitutions and Indels}
\label{rel:subsec:linked}

\begin{figure}[tbh]
  \centering
  \includegraphics[width=0.8\linewidth]{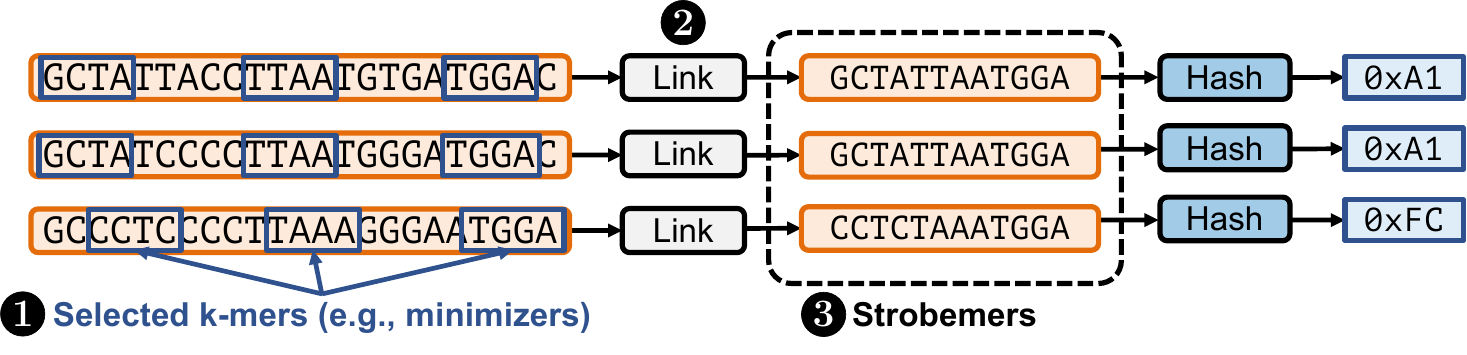}
  \caption{A simple example of the strobemers technique with three different input sequences. Sequences between selected k-mers are ignored to tolerate insertions and deletions. The first two input sequences generate the same hash values as output after generating their strobemer seeds even though these input sequences are not exactly matching.}
  \label{rel:fig:strobemers}
\end{figure}

Linking k-mers aims to allow both substitutions and indels when matching k-mers. A common approach is to select a few k-mers from an input sequence and link them to use these linked k-mers as seeds, such as paired-minimizers~\cite{chin_human_2019} and strobemers~\cite{sahlin_effective_2021, sahlin_flexible_2022} as Figure~\ref{rel:fig:strobemers} shows. These approaches can ignore large gaps between the linked k-mers~\circlednumber{$1$}. \revc{For example, the strobemer technique links~\circlednumber{$2$} a subset of selected k-mers of a sequence (e.g., by finding minimizers of the input sequence) to generate a strobemer sequence~\circlednumber{$3$}, which is used as a seed.}
Strobemers enable masking some characters within sequences without requiring a fixed pattern, unlike spaced k-mers. This makes strobemers a more sensitive approach for detecting indels with varying lengths as well as substitutions. However, the nature of the hash function used in strobemers requires exact matches of \emph{all} concatenated k-mers in strobemer sequences when matching seeds. Such an exact match requirement introduces challenges for further improving the sensitivity of strobemers for detecting indels and substitutions between sequences.

\revc{We find that the existing approaches used for tolerating substitutions and indels require exact matches of the resulting seeds that these mechanisms generate (i.e., spaced seeds or strobemer seeds). Due to the limited parameter choices when finding only \revb{exactly matching} seeds, the selected seeds either 1)~increase the use of the computationally costly steps after seeding or 2)~provide limited sensitivity. In Chapter~\ref{chap:blend}, we explain our mechanism to find fuzzy seed matches while enabling highly similar seeds to mismatch at arbitrary positions.}

\section{Real-Time Analysis During Sequencing}
\label{rel:sec:realtime}

\revb{Our works that can analyze raw nanopore signals without basecalling (as we discuss in Chapters~\ref{chap:rh},~\ref{chap:rht}, and~\ref{chap:rs}) aim to provide fast, scalable, accurate, and real-time genome analysis. There are several prior works that perform real-time analysis using 1)~the Sequencing by Synthesis (SBS) technology with Illumina sequencers~\cite{\citesbsrealtime}, 2)~basecallers for nanopore sequencing data to perform basecalled sequence analysis~\cite{\citebasecalledreal}, and 3)~raw nanopore signal analysis without basecalling~\cite{\citesignalanalysis}. \revc{Specialized hardware accelerators are commonly employed to perform real-time analysis~\cite{\citehwbasecall,\citehwrawall}.}

\revc{To show the real-time analysis capabilities of other sequencing technologies, we discuss the works that can perform real-time analysis using the Sequencing by Synthesis (SBS) technology with Illumina sequencers~\cite{\citesbsrealtime} in Section~\ref{rel:subsec:realillumina}.

We explain the works that perform real-time analysis with basecalling in Section~\ref{rel:subsec:basecalledreal}. We argue these works are not scalable as they rely on computationally costly basecallers.

In Section~\ref{rel:subsec:rawreal}, we show that there is a large body of recent work that focuses on the real-time analysis of raw nanopore signals without basecalling. We 1)~describe the two prior works most related to our works we describe in Chapters~\ref{chap:rh},~\ref{chap:rht} and~\ref{chap:rs} and 2)~explain the limitations of these prior works.}

\subsection{Real-time Analysis with SBS}
\label{rel:subsec:realillumina}

There are \revc{multiple} works~\cite{\citesbsrealtime} that analyze the raw sequencing data that SBS generates, specifically from Illumina sequencers, by basecalling the raw sequencing data from SBS followed by read mapping to reduce the overall genome analysis latency. HiLive~\cite{lindner_hilive_2017} and HiLive2~\cite{loka_reliable_2019} perform basecalling after a certain amount of sequencing cycles to enable accurate basecalling, followed by read mapping. Real-time basecalling and mapping can be used for many use cases, such as k-mer based taxonomic classification in LiveKraken~\cite{tausch_livekrakenreal-time_2018}, variant calling~\cite{loka_reliable_2019}, molecular diagnosis for rapid treatment~\cite{stranneheim_rapid_2014}, pathogen identification~\cite{zhang_optimized_2022, tausch_patholivereal-time_2022}, and privacy-preserving analysis~\cite{loka_prilive_2018}.

Our works described in Chapters~\ref{chap:rh},~\ref{chap:rht}, and~\ref{chap:rs} are different from these works as \revc{our works} focus on analyzing raw nanopore sequencing data to exploit the unique opportunities that nanopore sequencing provides.}

\subsection{Basecalled Real-time Analysis with Nanopore Sequencing}
\label{rel:subsec:basecalledreal}

\revc{Various} works~\cite{\citebasecalledreal} perform real-time basecalling and further analysis using the raw nanopore signals generated during sequencing. Among these works,
ReadFish~\cite{payne_readfish_2021}, ReadBouncer~\cite{ulrich_readbouncer_2022}, and RUBRIC~\cite{edwards_real-time_2019} perform basecalling followed by read mapping. SPUMONI~\cite{ahmed_pan-genomic_2021} and SPUMONI 2~\cite{ahmed_spumoni_2023} use basecalled sequences to perform binary classification between target and non-target species in a sample containing a large collection of genomes. To achieve this classification with a space-efficient mechanism, these tools utilize \revc{the} r-index \revc{data structure}~\cite{gagie_optimal-time_2018,boucher_prefix-free_2019,gagie_fully_2020} for indexing multiple genomes. Coriolis~\cite{mikalsen_coriolis_2023} enables performing real-time analysis for metagenomics classification in mobile devices after performing basecalling. 

\revc{However, all of these works predominantly rely on GPUs~\cite{\citehwbasecgpu} for performing real-time basecalling during sequencing due to the high computational demands of DNN-based basecallers. This reliance presents significant challenges for achieving scalable, energy-efficient, and portable on-site analysis, particularly in resource-constrained environments.} \revc{Additionally, DNN-based basecallers} are mainly optimized to use long chunks of raw nanopore signal produced after a few seconds. \revc{Since stopping the sequencing decisions must made quickly in adaptive sampling (as we discuss in Section~\ref{bg:subsubsec:adaptive}),} using raw signals generated within a second or less than a second with these basecallers provides less accurate basecalling and analysis~\cite{kovaka_targeted_2021, zhang_real-time_2021}.

\subsubsection{Accelerating Basecalling}
Basecallers developed for nanopore sequencing mainly use deep neural networks (DNNs)~\cite{\citebasecallnanodnn} to achieve high accuracy.
\revc{However, the use of complex deep learning models makes basecalling slow and memory-hungry~\cite{mao_genpip_2022, singh_rubicon_2024}.
To accelerate basecalling, various algorithm-architecture co-design techniques have been proposed~\cite{\citehwbasecall}.
First, to provide faster and more energy-efficient computations, multiple works use custom logic design to accelerate the basecalling operations on reconfigurable (i.e., FPGAs) or specialized (i.e., ASICs) hardware~\cite{\citehwbaseccust}.
Second, to perform a large number of arithmetic operations that are independent of each other (e.g., multiplications of many single-precision floating point operations), a large body of DNN-based basecallers exploit the massive parallelism that GPUs provide~\cite{\citehwbasecgpu}. To optimize these GPU-based basecallers, various works aim to reduce the unnecessary computations by 1)~reducing the DNN parameters and precision~\cite{singh_rubicon_2024} or 2)~introducing pre-basecalling filters~\cite{cavlak_targetcall_2024}.}
\revb{Third, PIM approaches~\cite{\citehwbasecpim} can substantially accelerate DNN operations \revc{by performing 1)~analog computation in-memory, 2)~minimizing the data movement latency, and 3)~exploiting large internal parallelism in memory arrays.}}

\subsection{Real-time Raw Nanopore Signal Analysis}
\label{rel:subsec:rawreal}

\revc{A number of} works perform real-time genome analysis of raw nanopore signals by utilizing adaptive sampling~\cite{\citesignalanalysis}.

SquiggleNet~\cite{bao_squigglenet_2021}, DeepSelectNet~\cite{senanayake_deepselectnet_2023}, and RawMap~\cite{sadasivan_rapid_2023} use machine learning techniques to classify raw nanopore signals to certain species without performing read mapping. Sigmoni~\cite{shivakumar_sigmoni_2024} uses a similar strategy used in SPUMONI~\cite{ahmed_pan-genomic_2021} and SPUMONI 2~\cite{ahmed_spumoni_2023} for classification without basecalling.

\unc~\cite{kovaka_targeted_2021} and \sig~\cite{zhang_real-time_2021} are the most relevant works to our raw nanopore signal analysis works (explained in Chapters~\ref{chap:rh},~\ref{chap:rht}, and~\ref{chap:rs}). These works map raw nanopore signals to a reference genome \emph{without} using computationally costly basecallers.

\revc{To map raw signals to a reference genome, \unc~\cite{kovaka_targeted_2021} calculates the probability of k-mers that each raw signal segment (i.e., event) can represent using a k-mer model that provides the expected event values for each possible k-mer.} \unc identifies the sequence of matching k-mers between the most probable k-mers of events and a reference genome using an FM-index~\cite{ferragina_opportunistic_2000}. \revc{However, as genome size increases, this probabilistic model struggles to accurately identify matching regions due to the sheer number of potential matches~\cite{kovaka_targeted_2021}. While \unc remains highly accurate for small genomes (e.g., \emph{E. coli} and \emph{Yeast}), its accuracy declines significantly when applied to larger genomes, such as human genomes (as we demonstrate in Chapters~\ref{chap:rh} and~\ref{chap:rht}).}

\revc{To accurately map raw nanopore signals to genomes larger than \emph{Yeast}, \sig~\cite{zhang_real-time_2021} calculates the Euclidean distance between raw signals and a reference genome after converting the reference genome into its expected signal representation by using k-mer models. Distance calculation enables identifying a segment of raw signals that are close in value to a particular segment in the reference genome. Identifying segments with a small distance between them can tolerate noise in raw signals, as a pair raw signal segment generated from the same identical content does not exactly match due to noise (as we discuss in Section~\ref{bg:subsec:nanopore}) but expected to be close in value~\cite{zhang_real-time_2021}. To calculate Euclidean distance between signal segments, \sig creates a vector from each $n$ consecutive event values (i.e., $n$-dimensional vector space) from the reference genome (i.e., the indexing step) and measures the Euclidean distance between these vectors and the vectors generated from raw nanopore signals (i.e., the mapping step) using a k-d tree structure~\cite{bentley_multidimensional_1975}. \revc{While calculating the distance between raw signals and a reference genome can provide accurate mapping positions, this distance calculation is computationally expensive and suffers from the \emph{curse of dimensionality}~\cite{bellman_dynamic_1966}. This issue makes it increasingly difficult to efficiently scale the number of events within a single vector, limiting accuracy for large genomes.} Increasing the dimension of vectors is needed to reduce the number of mapping positions in the reference genome as the genome size increases, which can substantially impact the overall runtime and accuracy of mapping.}

\revc{We find that none of these earlier works can provide accurate and fast analysis for mapping raw nanopore signals to large genomes such as a human genome in real-time. Our work we discuss in Chapter~\ref{chap:rh} provides the first mechanism to map raw nanopore signals in real-time accurately and quickly to large genomes without basecalling.}

\subsubsection{Accelerating Raw Signal Analysis}
\revc{To accelerate the costly computations that are commonly used in the raw signal analysis (e.g., alignment using dynamic time warping~\cite{bellman_adaptive_1959}), several works~\cite{\citehwrawall} propose hardware-algorithm co-design.} SquiggleFilter~\cite{dunn_squigglefilter_2021} provides an ASIC accelerator that quickly filters non-related raw electrical signals before basecalling for viral detection. HARU~\cite{shih_efficient_2023} is an FPGA accelerator that accelerates real-time selective genome sequencing on resource-constrained devices for detecting viral genomes. There are various other works that use GPUs\rev{~\cite{sadasivan_accelerated_2024, gamaarachchi_gpu_2020}} to accelerate raw signal analysis. All of these works mainly focus on accelerating the costly signal alignment algorithm known as dynamic time warping (DTW).

\revc{These works are orthogonal to the works we describe in Chapters~\ref{chap:rh} and~\ref{chap:rht}, as the alignment step can be performed separately after quickly and accurately identifying the mapping positions for a raw signal.}

\chapter[\blend: A Noise-tolerant Mechanism to Find Fuzzy Seed Matches]{Tolerating Arbitrary Noise with Hash-based Fuzzy Seeding}
\label{chap:blend}

In this chapter, we 1)~explain the types of noise we find in seed matching that impact the sensitivity of sequence analysis and 2)~introduce a mechanism that utilizes a noise-tolerant hashing mechanism to find fuzzy seed matching with single hash value lookups between seeds.

\section{Background and Motivation}
\label{blend:sec:bg}

High-throughput sequencing (HTS) technologies have revolutionized the field of genomics due to their ability to produce millions of nucleotide sequences at a relatively low cost~\cite{shendure_dna_2017}.
Although HTS technologies are key enablers of almost \emph{all} genomics studies~\cite{aynaud_multiplexed_2021, logsdon_long-read_2020, mantere_long-read_2019, friedman_genome-wide_2019, merker_long-read_2018, alkan_genome_2011}, HTS technology-provided data comes with two key shortcomings.
\revc{First, HTS technologies produce short genome fragments called \emph{reads}, which cover only a small region of a genome and contain from around one hundred to a million bases, depending on the technology~\cite{shendure_dna_2017}.}
Second, HTS technologies can misinterpret signals during sequencing and thus provide reads that contain \emph{sequencing errors}~\cite{goodwin_coming_2016}.
The average frequency of sequencing errors in a read highly varies from 0.1\% up to 15\% depending on the HTS technology~\cite{stoler_sequencing_2021, zhang_comprehensive_2020, hon_highly_2020, ma_analysis_2019, senol_cali_nanopore_2019}.
To address the shortcomings of HTS technologies, various computational approaches must be taken to process the reads into meaningful information accurately and efficiently. 
These include 1)~read mapping~\cite{li_minimap_2016, li_minimap2_2018, canzar_short_2017, kim_airlift_2024, kim_fastremap_2022}, 2)~\emph{de novo} assembly~\cite{ekim_minimizer-space_2021, cheng_haplotype-resolved_2021, robertson_novo_2010}, 3)~read classification in metagenomic studies~\cite{meyer_critical_2022, lapierre_metalign_2020, wood_improved_2019}, 4)~correcting sequencing errors~\cite{firtina_apollo_2020, vaser_fast_2017, loman_complete_2015}.

\revc{At the core of these computational approaches, identifying sequence similarities through seed matching is essential to overcome the limitations of HTS technologies.} However, identifying the similarities across \emph{all} pairs of sequences is not practical due to the costly algorithms used to calculate the distance between two sequences, such as sequence alignment algorithms using dynamic programming (DP) approaches~\cite{alser_technology_2021, alser_molecules_2022}. To practically identify similarities, it is essential to avoid calculating the distance between dissimilar sequence pairs.
A common heuristic is to find matching \emph{short} sequence segments, called \emph{seeds}, between sequence pairs by using a hash table~\cite{\citemaphashtable}. Sequences that have no or few seed matches are quickly filtered out from performing costly sequence alignment. There are several techniques that generate seeds from sequences, known as \emph{seeding techniques}.
To find the matching seeds efficiently, a common approach is to match the hash values of seeds with a \emph{single lookup} using a hash table that contains the hash values of all seeds of interest.
The use of seeds drastically reduces the search space from all possible sequence pairs to similar sequence pairs to facilitate efficient distance calculations over many sequence pairs~\cite{baichoo_computational_2017, roberts_reducing_2004, schleimer_winnowing_2003}.

\subsection{Motivation: Need for Tolerating Arbitrary Noise in Seeding}
\label{blend:subsec:bg_motiv}

To our knowledge, there is no work that can \emph{efficiently} find fuzzy matches of seeds \emph{without} requiring 1)~\emph{exact matches} of all k-mers (i.e., any k-mer can mismatch) and 2)~imposing high performance and memory space overheads. In this work, we observe that existing works have such a limitation mainly because they employ hash functions with low-collision rates when generating the hash values of seeds. Although it is important to reduce the collision rate for assigning different hash values for dissimilar seeds for accuracy and performance reasons, the choice of hash functions also makes it unlikely to assign the same hash value for similar seeds. Thus, seeds \emph{must} exactly match to find matches between sequences with a single lookup. \revc{Mitigating this exact-matching requirement allows similar seeds to share the same hash value, significantly improving performance and sensitivity. This enables applications to tolerate substitutions and indels at arbitrary positions when matching seeds.}

\subsection{Overview of Fuzzy Seed Matching in BLEND}
\label{blend:subsec:bg_fuzzy}

\textbf{Our goal} in this work is to enable finding \emph{fuzzy} matches of seeds as well as \revb{exactly matching} seeds between sequences (e.g., reads) with a single lookup of hash values of these seeds.
To this end, we propose \emph{\blend}, the \emph{first} efficient and accurate mechanism that can identify both \revb{exactly matching} and highly similar seeds with a single lookup of their hash values.
\revc{The \textbf{key idea} in \blend is to assign the same hash value to highly similar seeds. To achieve this, \blend~1)~leverages the SimHash technique~\cite{charikar_similarity_2002, manku_detecting_2007}, and 2)~introduces mechanisms that adapt hash-based seeding techniques to SimHash for finding fuzzy seed matches with a single hash lookup.} This provides us with two key benefits.
First, \blend can generate the same hash value for highly similar seeds \emph{without} imposing exact matches of seeds, unlike existing seeding mechanisms that use hash functions with low-collision rates.
Second, \blend enables finding fuzzy seed matches with a single lookup of a hash value rather than 1)~using various permutations to find the longest prefix matches~\cite{lederman_random-permutations-based_2013} or 2)~matching many hash values for calculating costly similarity scores (e.g., Jaccard similarity~\cite{jaccard_nouvelles_1908}) that the conventional locality-sensitive hashing-based methods use, such as MHAP~\cite{berlin_assembling_2015} or S-conLSH~\cite{chakraborty_conlsh_2020, chakraborty_s-conlsh_2021}. These two ideas ensure that \blend can efficiently find both 1)~all \revb{exactly matching} seeds that a seeding technique finds using a conventional hash function with a low-collision rate and 2)~approximate seed matches that these conventional hashing mechanisms cannot find with a single lookup of a hash value.

\begin{figure}[tbh]
  \centering
  \includegraphics[width=0.9\linewidth]{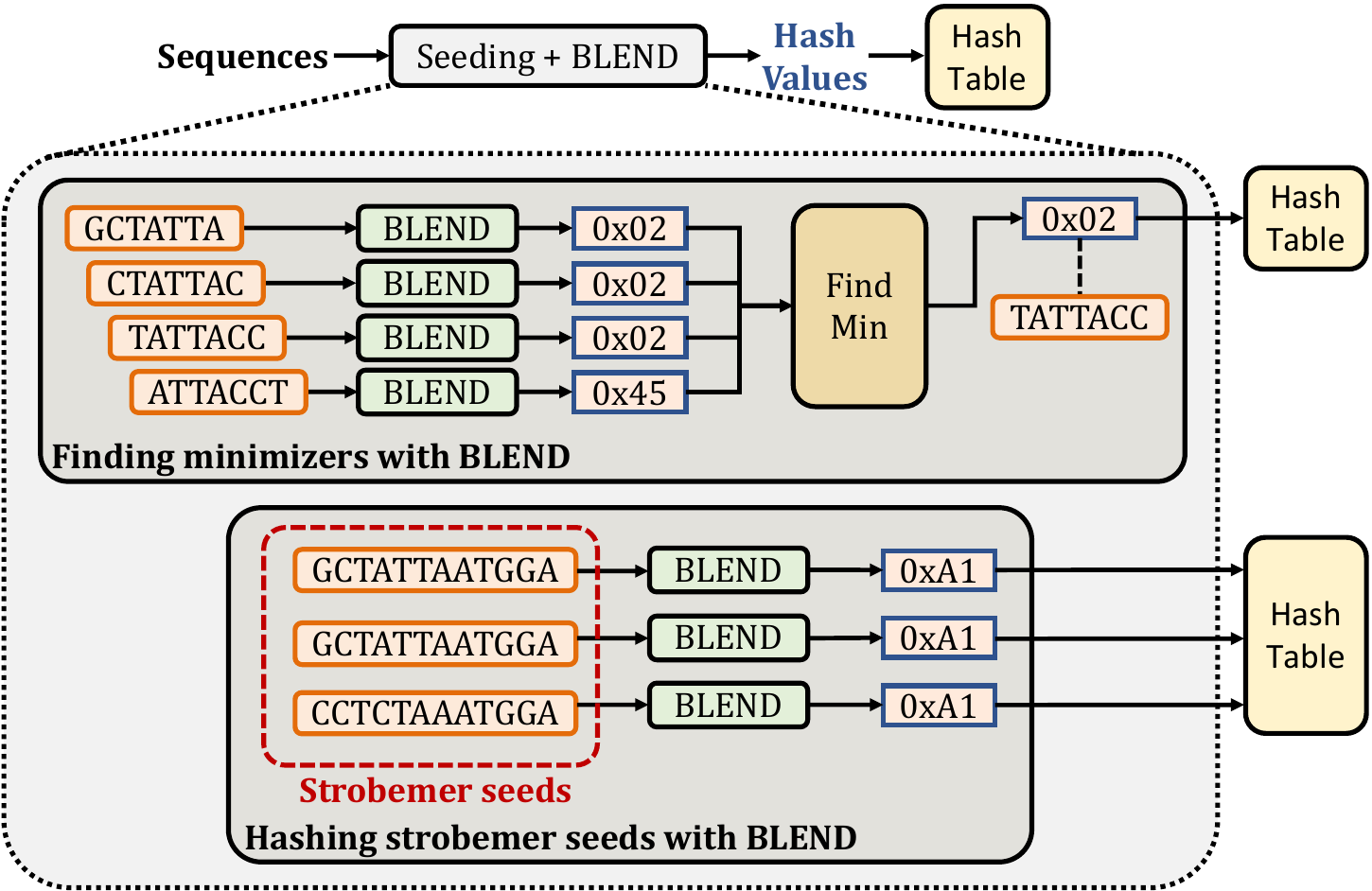}
  \caption{Replacing the hash functions in seeding techniques with \blend.}
  \label{blend:fig:seeding_and_blend}
\end{figure}

Figure~\ref{blend:fig:seeding_and_blend} shows two examples of how \blend can replace the conventional hash functions that the seeding techniques use. \revc{The \textbf{key challenge} is to accurately and efficiently define the vector items derived from seeds for the SimHash technique. To address this, \blend provides two mechanisms for converting seeds into vectors of items: 1)~\texttt{\blend-I} and 2)~\texttt{\blend-S}.} To perform a sensitive detection of substitutions, \texttt{\blend-I} uses all overlapping smaller k-mers of a potential seed sequence as the items of a vector for generating the hash value with SimHash. To allow mismatches between the linked k-mers that strobemers and similar seeding mechanisms use, \texttt{\blend-S} uses only the linked k-mers as the vector with SimHash. We envision that \blend can be integrated with any seeding technique that uses hash values for matching seeds with a single lookup by replacing their hash function with \blend and using the proper mechanism for converting seeds into a vector of items.

\subsection{Overview of Experimental Studies}
\label{blend_subsec:bg_exp}

Using erroneous (ONT and PacBio CLR), highly accurate (PacBio HiFi), and short (Illumina) reads, we experimentally show the benefits of \blend on two important applications in genomics: 1)~read overlapping and 2)~read mapping. \revc{First, read overlapping identifies overlaps between all pairs of reads based on seed matches. These overlaps are crucial for generating assemblies of the sequenced genome~\cite{pop_comparative_2004, li_minimap_2016}.} We compare \blend with minimap2 and MHAP by finding overlapping reads. We then generate the assemblies from the overlapping reads to compare the qualities of these assemblies. Second, read mapping uses seeds to find similar portions between a reference genome and a read before performing the read alignment. Aligning a read to a reference genome shows the edit operations (i.e., match, substitution, insertion, and deletions) to make the read identical to the portion of the reference genome, which is useful for downstream analysis (e.g., variant calling~\cite{mckenna_genome_2010}). We compare \blend with minimap2, LRA~\cite{ren_lra_2021}, Winnowmap2, S-conLSH, and Strobealign by mapping long and paired-end short reads to their reference genomes. \revc{We evaluate the impact of long read mapping on downstream analysis by identifying structural variants (SVs) and assessing their detection accuracy.}

\section{\blend Mechanism} 
\label{blend:sec:methods}

\revc{We propose \textbf{\emph{\blend}}, a mechanism that efficiently finds highly similar (fuzzy) seed matches using a single hash value lookup. \blend allows for assigning identical hash values to highly similar seeds and integrates with hash-based seeding approaches (e.g., minimizers or strobemers) to find fuzzy seed matches with a single lookup.}

\begin{figure}[tbh]
  \centering
  \includegraphics[width=0.7\linewidth]{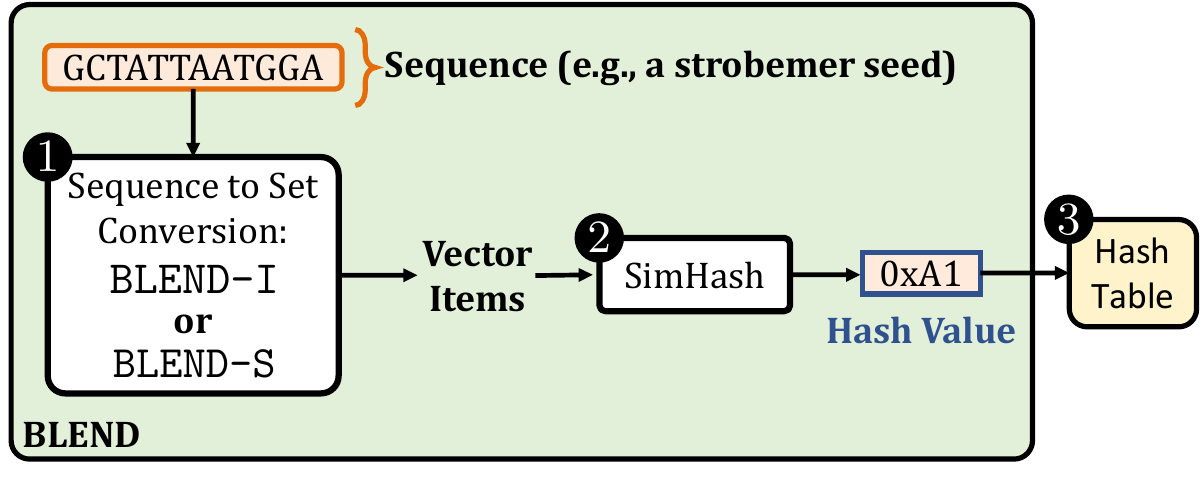}
  \caption{Overview of \blend.\circlednumber{$1$}~\blend uses \texttt{\blend-I} or \texttt{\blend-S} for converting a sequence into its vector of items.\circlednumber{$2$}~\blend generates the hash value of the input sequence using its vector of items with the SimHash technique.\circlednumber{$3$}~\blend uses hash tables for finding fuzzy seed matches with a single lookup of the hash values that \blend generates.}
  \label{blend:fig:blend-overview}
\end{figure}

Figure~\ref{blend:fig:blend-overview} shows the overview of steps to find fuzzy seed matches with a single lookup in three steps.
First, \blend starts with converting the input sequence it receives from a seeding technique (e.g., a strobemer sequence in Figure~\ref{blend:fig:seeding_and_blend}) to its vector representation as the SimHash technique generates the hash value of the vector using its items\circlednumber{$1$}. To enable effective and efficient integration of seeds with the SimHash technique, \blend proposes two mechanisms for identifying the items of the vector of the input sequence: 1)~\texttt{\blend-I} and 2)~\texttt{\blend-S}.
Second, after identifying the items of the vector, \blend uses this vector with the SimHash technique to generate the hash value for the input sequence\circlednumber{$2$}. \blend uses the SimHash technique as it allows for generating the same hash value for highly similar vectors. 
Third, \blend uses the hash tables with the hash values it generates to enable finding fuzzy seed matches with a single lookup of their hash values\circlednumber{$3$}.

\subsection{Sequence to vector Conversion}\label{blend:subsec:seed}
\begin{figure}[tbh]
  \centering
  \includegraphics[width=0.8\linewidth]{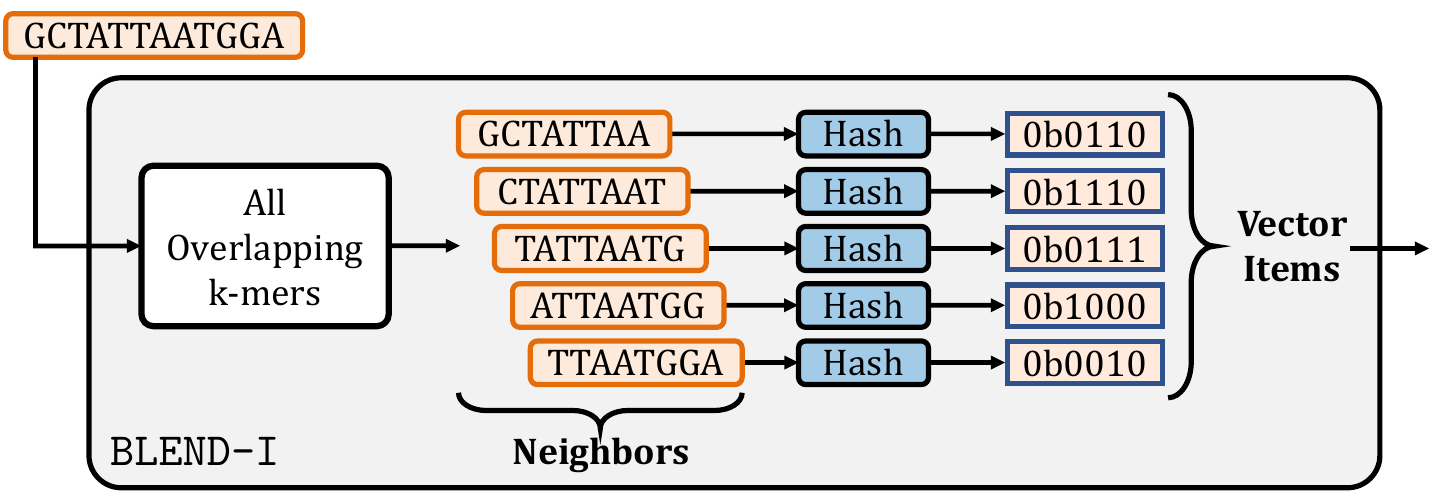}
  \caption{Overview of \texttt{\blend-I}. \texttt{\blend-I} uses the hash values of all the overlapping k-mers of an input sequence as the vector items.}
  \label{blend:fig:blend-i}
\end{figure}

\begin{figure}[tbh]
  \centering
  \includegraphics[width=0.8\linewidth]{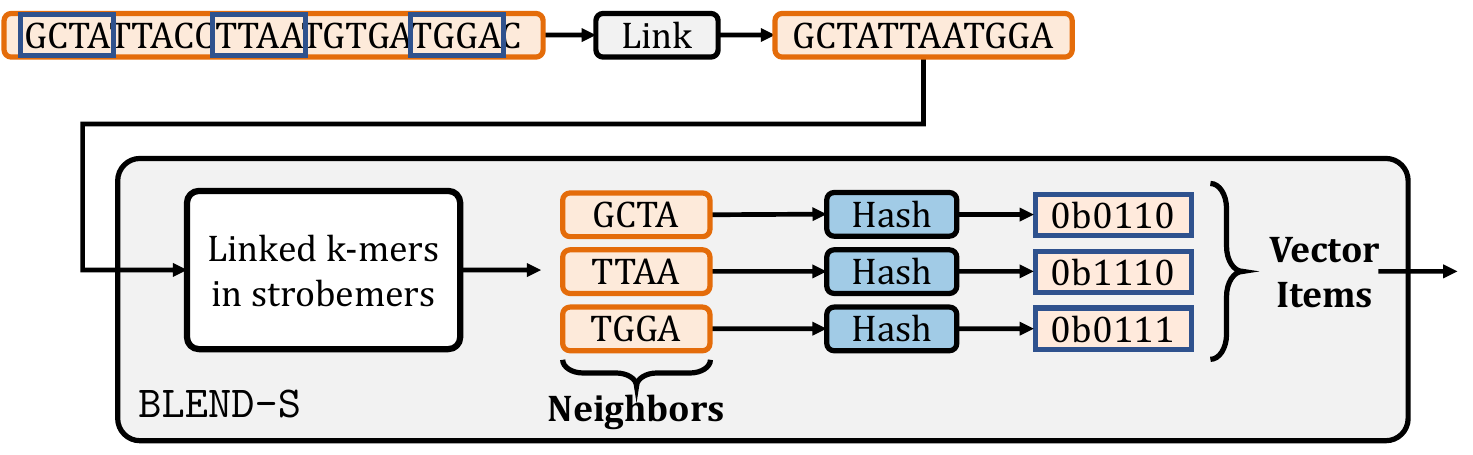}
  \caption{Overview of \texttt{\blend-S}. \texttt{\blend-S} uses the hash values of only the k-mers selected by the strobemer seeding mechanism.}
  \label{blend:fig:blend-s}
\end{figure}

Our goal is to convert the input sequences that \blend receives from any seeding technique (Figure~\ref{blend:fig:seeding_and_blend}) to their proper vector representations so that \blend can use the items of vectors for generating the hash values of input sequences with the SimHash technique.
To achieve effective and efficient conversion of sequences into their vector representations in different scenarios, \blend provides two mechanisms: 1)~\texttt{\blend-I} and 2)~\texttt{\blend-S}, as we show in Figures~\ref{blend:fig:blend-i} and~\ref{blend:fig:blend-s}, respectively.

The goal of the first mechanism, \texttt{\blend-I}, is to provide high sensitivity for a single character change in the input sequences that seeding mechanisms provide when generating their hash values such that two sequences are likely to have the same hash value if they differ by a few characters. \texttt{\blend-I} has three steps. First, \texttt{\blend-I} extracts \emph{all} the overlapping k-mers of an input sequence, as shown in Figure~\ref{blend:fig:blend-i}. For simplicity, we use the \emph{neighbors} term to refer to all the k-mers that \texttt{\blend-I} extracts from an input sequence. Second, \texttt{\blend-I} generates the hash values of these k-mers using any hash function. Third, \texttt{\blend-I} uses the hash values of the k-mers as the vector items of the input sequence for SimHash. Although \texttt{\blend-I} can be integrated with any seeding mechanism, we integrate it with the minimizer seeding mechanism, as shown in Figure~\ref{blend:fig:seeding_and_blend} as proof of work.

The goal of the second mechanism, \texttt{\blend-S}, is to allow indels and substitutions when matching the sequences such that two sequences are likely to have the same hash value if these sequences differ by a few k-mers. \texttt{\blend-S} has three steps. First, \texttt{\blend-S} uses \emph{only} the selected k-mers that the strobemer-like seeding mechanisms find and link~\cite{sahlin_effective_2021} as neighbors, as shown in Figure~\ref{blend:fig:blend-s}. \revc{\texttt{\blend-S} allows for mismatches between linked k-mers in strobemer sequences because a single character difference does not propagate to other linked k-mers. This is unlike \texttt{\blend-I}, where a single character difference affects multiple overlapping k-mers.} To ensure the correctness of strobemer seeds when matching them based on their hash values, \texttt{\blend-S} uses \emph{only} the selected k-mers from the same strand. Second, \texttt{\blend-S} generates the hash values of these linked k-mers using any hash function. Third, \texttt{\blend-S} uses the hash values of all such selected k-mers as the vector items of the input sequence for SimHash.

\subsection{Integrating the SimHash Technique}\label{blend:subsec:seedhash}

Our goal is to enable efficient comparisons of equivalence or high similarity between seeds with a single lookup by generating the same hash value for highly similar or equivalent seeds.
To enable generating the same hash value for these seeds, \blend uses the SimHash technique~\cite{charikar_similarity_2002}. The SimHash technique takes a vector of items and generates a hash value for the vector using its items. The key benefit of the SimHash technique is that it allows generating the same hash value for highly similar vectors while enabling any \emph{arbitrary} items to mismatch between vectors. To exploit the key benefit of the SimHash technique, \blend efficiently and effectively integrates the SimHash technique with the vector items that \texttt{\blend-I} or \texttt{\blend-S} determine. \blend uses these vector items for generating the hash values of seeds such that highly similar seeds can have the same hash value to enable finding fuzzy seed matches with a single lookup of their hash values.

\begin{figure}[tbh]
  \centering
  \includegraphics[width=0.8\linewidth]{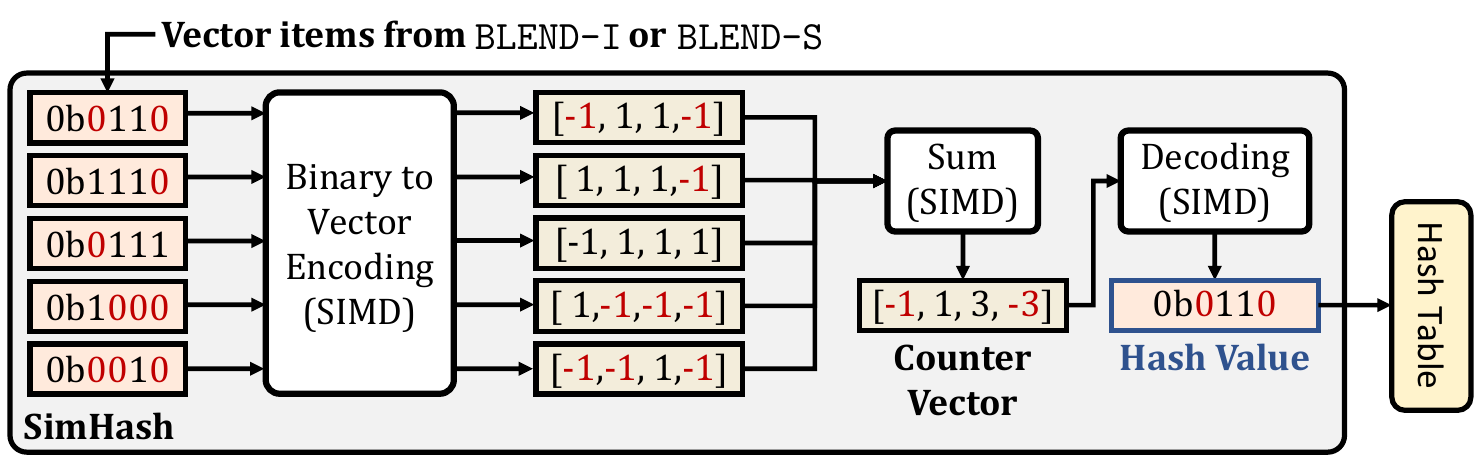}
  \caption{The overview of the steps in the SimHash technique for calculating the hash value of a given vector of items. The vector items are the hash values represented in their binary form. Binary to Vector Encoding converts these vector items to their corresponding bit vector representations. Sum performs the vector additions and stores the result in a separate vector that we call the \emph{counter vector}. Decoding generates the hash value of the vector based on the values in the counter vector. \blend uses SIMD operations for these three steps, as indicated by SIMD. We highlight in red how $0$ bits are converted and propagated in the SimHash technique.}
  \label{blend:fig:simhash}
\end{figure}

\blend employs the SimHash technique in three steps: 1)~encoding the input vector items as bit vectors, 2)~performing additions for the bit vector, resulting in a counter vector, and 3)~decoding the counter vector to generate the hash value for the input vector that \texttt{\blend-I} or \texttt{\blend-S} determine, as we show in Figure~\ref{blend:fig:simhash}. To enable efficient computations between vectors, \blend uses SIMD operations when performing all these three steps. We provide the details of our SIMD implementation in Section~\ref{blend:suppsec:simd_implementation}.

First, the goal of the \emph{binary to vector encoding} step is to transform all the hash values of vector items from the binary form into their corresponding bit vector representations so that \blend can efficiently perform the bitwise arithmetic operations that the SimHash technique uses in the vector space.
For each hash value in the vector item, the encoding can be done in two steps. \revc{The first step creates a bit vector with $n$ elements corresponding to an $n$-bit hash value. Initially, all elements in the bit vector are set to $1$.} For each bit position $t$ of the hash value, the second step assigns $-1$ to the $t^{th}$ element in the bit vector if the bit at position $t$ is $0$, as we highlight in Figure~\ref{blend:fig:simhash} with red colors of $0$ bits and their corresponding $-1$ values in the vector space. For each hash value in vector items, the resulting bit vector includes $1$ for the positions where the corresponding bit of a hash value is $1$ and $-1$ for the positions where the bit is $0$.

Second, the goal of the vector addition operation is to determine the bit positions where the number of $1$ bits is greater than the number of $0$ bits among the vector items, which we call determining the \emph{majority} bits. The key insight in determining these majority bits is that highly similar vectors are likely to result in \emph{similar} majority results because a few differences between two similar vectors are unlikely to change the majority bits at each position, given that there is a sufficiently large number of items involved in this majority calculation. To efficiently determine the majority of bits at each position, \blend counts the number of $1$ and $0$ bits at a position by using the bit vectors it generates in the vector encoding step, as shown with the addition step (Sum) in Figure~\ref{blend:fig:simhash}. \revc{The vector addition step adds $+1$ or $-1$ values between bit vector elements and stores the results in a separate \emph{counter} vector.} The values in this counter vector show the majority of bits at each position of the vector items. Since \blend assigns $-1$ for $0$ bits and $1$ for $1$ bits, the majority of bits at a position is either 1)~$1$ if the corresponding value in the counter vector is greater than $0$ or 2)~$0$ if the values are less than or equal to $0$.

Third, to generate the hash value of a vector, \blend uses the majority of bits that it determines by calculating the counter vector. To this end, \blend decodes the counter vector into a hash value in its binary form, as shown in Figure~\ref{blend:fig:simhash} with the decoding step. The decoding operation is a simple conditional operation where each bit of the final hash value is determined based on its corresponding value at the same position in the counter vector. \blend assigns the bit either 1) $1$ if the value at the corresponding position of the counter vector is greater than $0$ or 2) $0$ if otherwise. Thus, each bit of the final hash value shows the majority voting result of input vector items of a seed. We use this final hash value for the input sequence that the seeding techniques provide because highly similar sequences are likely to have many characters or k-mers in common, which essentially leads to generating \emph{similar} vector items by using \texttt{\blend-I} or \texttt{\blend-S}. Properly identifying the vector items of similar sequences enables \blend to find similar majority voting results with the SimHash technique, which can lead to generating the same final hash value for similar sequences. This enables \blend to find fuzzy seed matches with a single lookup using these hash values. We provide a step-by-step example of generating the hash values for two different seeds in Supplementary Section~\ref{blend:suppsec:real} and Supplementary Tables~\ref{blend:supptab:kmers7-15-1-32}-~\ref{blend:supptab:kmers15-7-2-16}.

\subsection{SIMD Implementation of the SimHash Technique}\label{blend:suppsec:simd_implementation}

\begin{figure}[htb]
\centering
\includegraphics[width=0.9\linewidth]{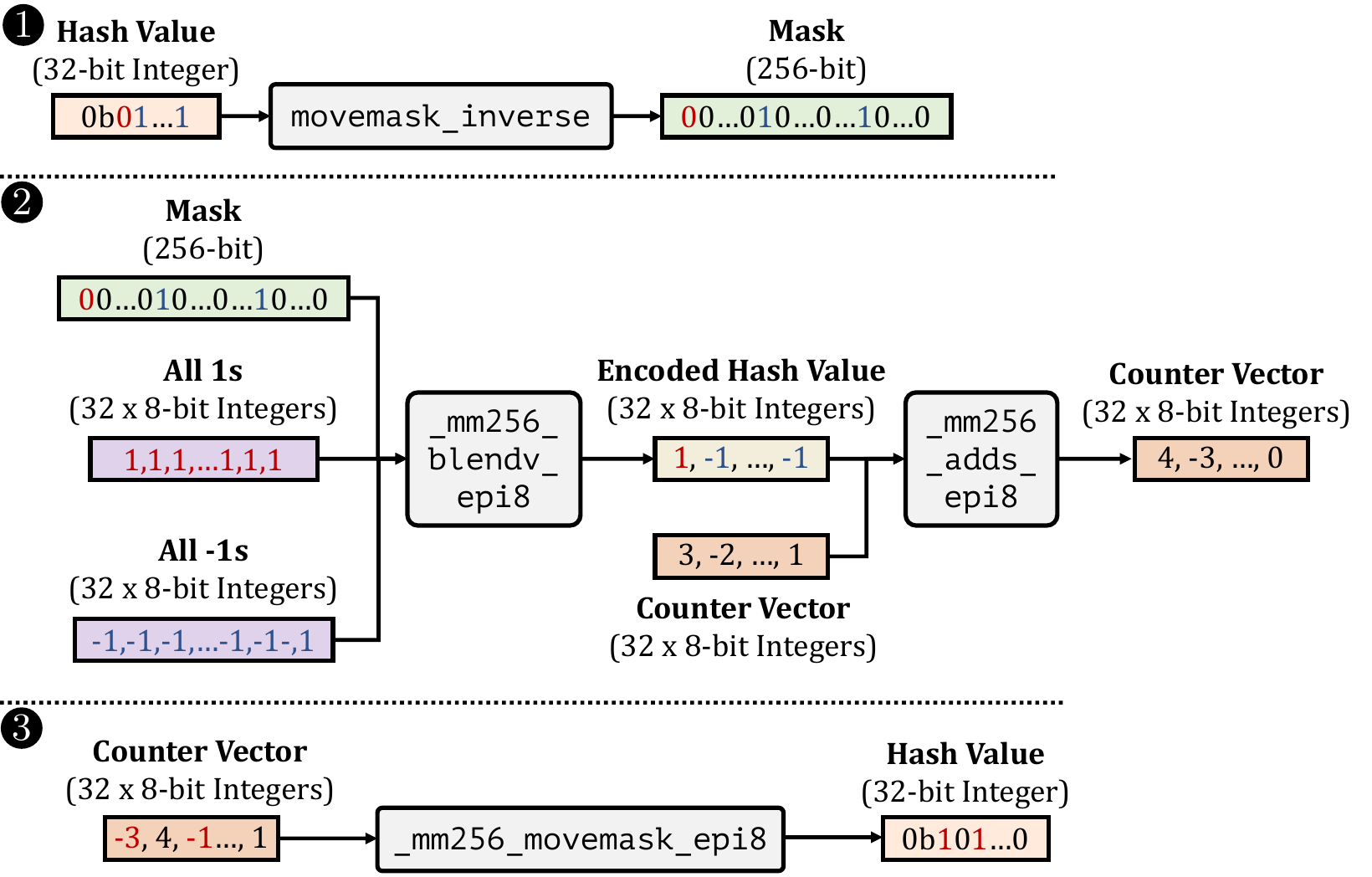}
\caption{SIMD execution flow when generating the hash value of a seed from its vector items that \texttt{\blend-I} or \texttt{\blend-S} identifies. Colors highlight the propagation of bits and values to the outputs of functions.}
\label{blend:suppfig:simd1}
\end{figure}

Figure~\ref{blend:suppfig:simd1} shows the high-level execution flow when calculating the hash value of a seed from its vector items that \texttt{\blend-I} or \texttt{\blend-S} identifies, as explained in Sections~\ref{blend:subsec:seed} and~\ref{blend:subsec:seedhash}. To efficiently perform the bitwise operations in the SimHash technique, \blend utilizes the SIMD operations in three steps.

\begin{figure}[htb]
\centering
\includegraphics[width=0.7\linewidth]{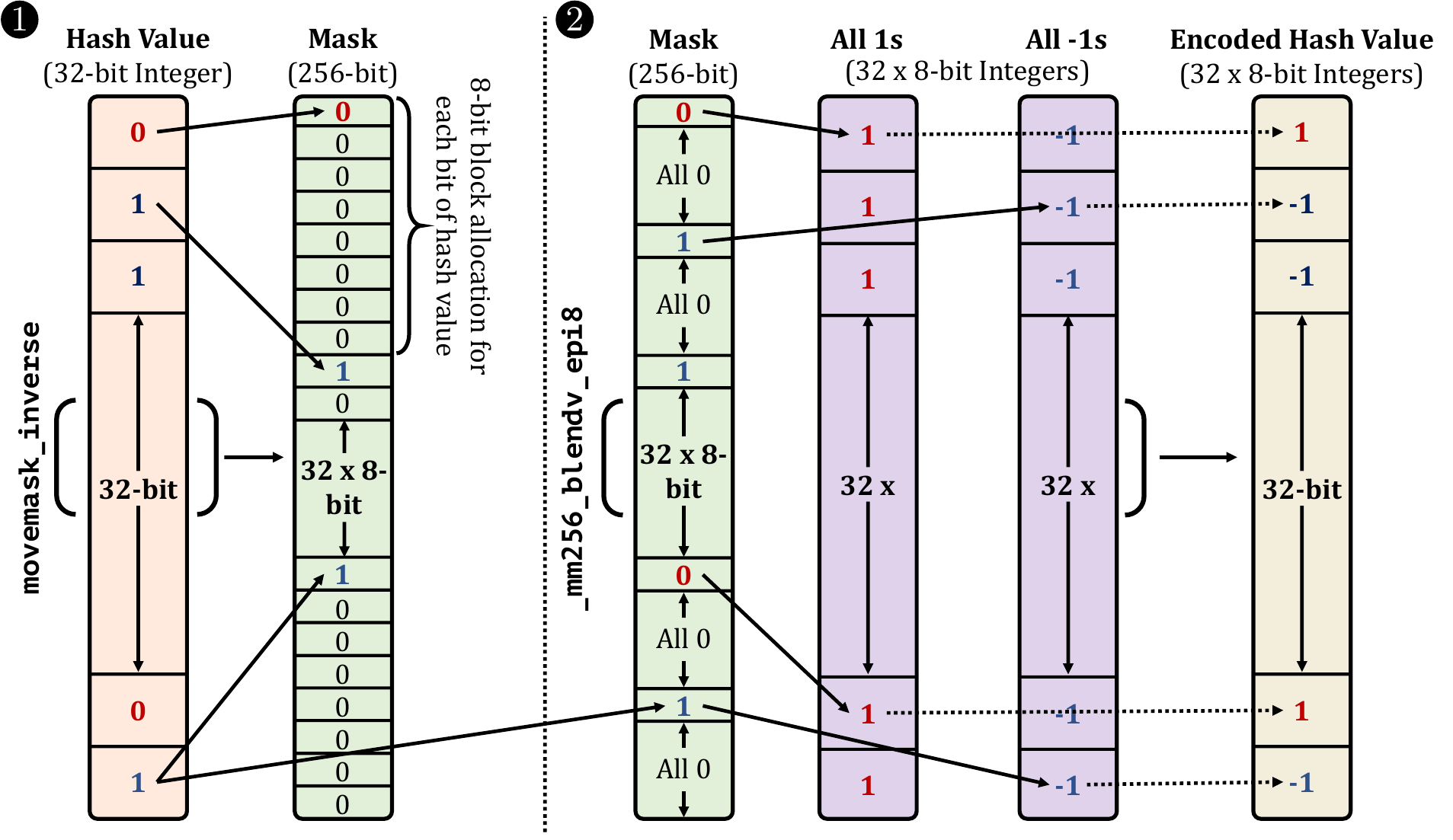}
\caption{Details of the \texttt{movemask\_inverse} and \texttt{\_mm256\_blendv\_epi8} executions. Colors and arrows highlight the propagation of bits and values to the outputs of functions.}
\label{blend:suppfig:simd2}
\end{figure}

First, for each \emph{hash value} in vector items, \blend creates its corresponding \emph{mask} using the \texttt{movemask\_inverse} function, as shown in~\circlednumber{$1$}. For each bit position $t$ of the hash value, the \texttt{movemask\_inverse} function assigns the bit at position $t$ of the hash value to the bit position $t*8+7$ of a 256-bit SIMD register (i.e., the most significant bit positions of each 8-bit block), which \blend uses it as a \emph{mask} in the next steps. We assume 0-based indexing and the mask register is initially 0. Figure~\ref{blend:suppfig:simd2} shows how each bit in hash value propagates to the mask register in~\circlednumber{$1$}. The \texttt{movemask\_inverse} is an in-house implementation that performs the reverse behavior of the \texttt{\_mm256\_movemask\_epi8}\footnote{\url{https://software.intel.com/sites/landingpage/IntrinsicsGuide/\#text=\_mm256\_movemask\_epi8&ig\_expand=4874}} SIMD function. Our function efficiently utilizes several other SIMD functions to perform the reverse behavior of \texttt{\_mm256\_movemask\_epi8}.

Second, for each mask created in the first step, \blend updates the values in the counter vector (explained in Section~\ref{blend:subsec:seedhash}), as shown in~\circlednumber{$2$}. To encode the hash value into its bit vector representation, \blend uses the \texttt{\_mm256\_blendv\_epi8}\footnote{\url{https://software.intel.com/sites/landingpage/IntrinsicsGuide/\#text=\_mm256\_blendv\_epi8&ig\_expand=515}} SIMD function with 1)~the mask register \blend creates in the first step, 2)~two 256-bit wide SIMD registers that include $32 \times$ 8-bit integers. For the first register, all 8-bit values are initialized to 1, and for the second register, all 8-bit values are initialized to -1. The \texttt{\_mm256\_blendv\_epi8} function generates a new 256-bit register with 8-bit integers where each 8-bit block is copied from either the first or the second register based on the most significant value in the mask register. If the most significant value in the mask register is 0, the corresponding 8-bit block in the first register is copied. Otherwise, the 8-bit block in the second register is copied. We show in detail how the values in these registers propagate to the resulting 256-bit register in Figure~\ref{blend:suppfig:simd2} in~\circlednumber{$2$}. \blend, then, performs addition using the \texttt{\_mm256\_adds\_epi8}\footnote{\url{https://software.intel.com/sites/landingpage/IntrinsicsGuide/\#text=\_mm256\_adds\_epi8&ig\_expand=220}} function between the register that the \texttt{\_mm256\_blendv\_epi8} function generates and the 256-bit \emph{counter vector} that includes $32 \times$ 8-bit integers. We assume that all the 8-bit values in the counter vector are initially 0. \blend keeps updating the counter vector as it iterates through the vector items (i.e., hash values). The resulting value is written back to the counter vector to use it in the next iterations with the next vector item. We note that the current design encodes 1 bits as -1 and 0 bits as 1, which is the opposite case of our explanation in Section~\ref{blend:subsec:seedhash}. Although we perform our encoding in an opposite way, \blend generates the final hash values as originally explained. The reason for such a design change is due to the behavior of the function we use in the third step.

Third, \blend converts the final result in the counter vector to its corresponding 32-bit hash value of a vector (explained in the decoding step of Section~\ref{blend:subsec:seedhash}), as shown in~\circlednumber{$3$}. \blend uses the \texttt{\_mm256\_movemask\_epi8} function that takes a 256-bit register of 8-bit blocks and assigns the corresponding bit accordingly in a 32-bit value. The behavior of this function is essentially the reverse behavior of the \texttt{movemask\_inverse} function that we explain in the first step (i.e., reversing the arrows in Figure~\ref{blend:suppfig:simd2} \circlednumber{$1$} can simulate the \texttt{\_mm256\_movemask\_epi8} function). Thus, it assigns 1 to the bit position $t$ of a hash value if the bit at position $t*8+7$ (i.e., the most significant bit of an 8-bit integer) is 1. Since the most significant is 1 \emph{only} for the negative values according to the signed integer value convention, this function creates the opposite behavior of our decoding step we explain in Section~\ref{blend:subsec:seedhash}. We resolve this issue by performing the encoding in an opposite way in the second step, where the resulting counter vector includes negative values when the majority of bits at a bit position is 1. Thus, the final hash value contains the same bits as explained in our main paper.

Although we omit the details here, \blend avoids performing redundant computations when calculating the hash values of each input sequence, as the vector items between each of these input sequences are likely to be shared. For example, minimap2 generates a hash value for each k-mer in a sequence and selects the k-mer with the minimum hash value in a window of k-mers as the minimizer. Assuming that the vector items of each k-mer are l-mers, each k-mer differs by two l-mers at most with its next overlapping k-mer: one different l-mer contains the leading character of the first k-mer, and the other contains the trailing character of the next k-mer. Since there are two l-mer changes at most, \blend calculates only the difference between subsequent k-mers. Thus, \blend keeps a buffer in a first-in, first-out fashion such that the corresponding encoded hash value of the l-mer that is missing in the next k-mer is subtracted from the counter vector and popped from the queue while performing an addition \emph{only} for the l-mer that is missing from the previous k-mer and pushing it into the queue.

We perform our operations on 256-bit wide SIMD registers. \blend works on 8-bit integer blocks assigned for each bit in a hash value. Since our registers are 256-bit wide, \blend uses 32-bit hash values when calculating the SimHash value of a seed. Our implementation allows working on up to 64-bit hash values by dividing the most and least significant 32 bits into two 32-bit hash values. Each 32-bit hash value can independently follow the three steps we show in Figure~\ref{blend:suppfig:simd1}, and the final 64-bit value can be generated by the shift operations between the final 32-bit hash values. Although our approach is scalable to allow hash values with a larger number of bits, the current implementation does not support such flexible scaling and works on up to 64-bits.

\subsection{Using a Hash Table for Quick Fuzzy Seed Matching}\label{blend:subsec:storeseed}

Our goal is to enable an efficient lookup of the hash values of seeds to find fuzzy seed matches with a single lookup. To this end, \blend uses hash tables in two steps. First, \blend stores the hash values of all the seeds of target sequences (e.g., a reference genome) in a hash table, usually known as the \emph{indexing} step. Keys of the hash table are hash values of seeds, and the value that a key returns is a \emph{list} of metadata information (i.e., seed length, position in the target sequence, and the unique name of the target sequence). \blend keeps minimal metadata information for each seed sufficient to locate seeds in target sequences. Since similar or equivalent seeds can share the same hash value, \blend stores these seeds using the same hash value in the hash table. Thus, a query to the hash table returns all fuzzy seed matches with the same hash value.

Second, \blend iterates over all query sequences (e.g., reads) and uses the hash table from the indexing step to find fuzzy seed matches between query and target sequences. The query to the hash table returns the list of seeds of the target sequences that have the same hash value as the seed of a query sequence. Thus, the list of seeds that the hash table returns is the list of fuzzy seed matches for a seed of a query sequence as they share the same hash value. \blend can find fuzzy seed matches with a single lookup using the hash values it generates for the seeds from both query and target sequences.

\blend finds fuzzy seed matches mainly for two important genomics applications: read overlapping and read mapping. For these applications, \blend stores all the list of fuzzy seed matches between query and target sequences to perform \emph{chaining} among fuzzy seed matches that fall in the same target sequence (overlapping reads) optionally, followed by alignment (read mapping) as described in minimap2~\cite{li_minimap2_2018}.

\section{Results}
\label{blend:sec:evaluation}

\subsection{Evaluation Methodology}
\label{blend:subsec:evaluation-methodology}
We replace the mechanism in minimap2 that generates hash values for seeds with \blend to find fuzzy seed matches when performing end-to-end read overlapping and read mapping. We also incorporate the \texttt{\blend-I} and \texttt{\blend-S} mechanisms in the implementation and provide the user to choose either of these mechanisms when using \blend. We provide a set of default parameters we optimize based on sequencing technology and the application to perform (e.g., read overlapping). We explain the details of the \blend parameters in Supplementary Table~\ref{blend:supptab:pardef} and the parameter configurations we use for each tool and dataset in Supplementary Tables~\ref{blend:supptab:ovpars} and~\ref{blend:supptab:mappars}. We determine these default parameters empirically by testing the performance and accuracy of \blend with different values for some parameters (i.e., k-mer length, number of k-mers to include in a seed, and the window length) as shown in Supplementary Table~\ref{blend:supptab:parameter_exploration}. We show the trade-offs between the seeding mechanisms \texttt{\blend-I} and \texttt{\blend-S} in Supplementary Figures~\ref{blend:suppfig:overlap_perf-blend} and ~\ref{blend:suppfig:read_mapping_perf-blend} and Supplementary Tables~\ref{blend:supptab:overlap_assembly-blend} -~\ref{blend:supptab:mapping_accuracy-blend} regarding their performance and accuracy.

For our evaluation, we use real and simulated read datasets as well as their corresponding reference genomes. We list the details of these datasets in Table~\ref{blend:tab:dataset}. To evaluate \blend in several common scenarios in read overlapping and read mapping, we classify our datasets into three categories: 1)~highly accurate long reads (i.e., PacBio HiFi), 2)~erroneous long reads (i.e., PacBio CLR and Oxford Nanopore Technologies), and 3)~short reads (i.e., Illumina). We use PBSIM2~\cite{ono_pbsim2_2021} to simulate the erroneous PacBio and Oxford Nanopore Technologies (ONT) reads from the Yeast genome.
To use realistic depth of coverage, we use SeqKit~\cite{shen_seqkit_2016} to down-sample the original \emph{E. coli}, and \emph{D. ananassae} reads to $100\times$ and $50\times$ sequencing depth of coverage, respectively.

\begin{table}[htb]
\centering
\caption{Details of datasets used in evaluation.}
\begin{tabular}{@{}llrrll@{}}\toprule
\textbf{Organism} & \textbf{Library} & \textbf{Reads (\#)} & \textbf{Seq.} & \textbf{SRA} & \textbf{Reference}\\
\textbf{} & \textbf{} & \textbf{} & \textbf{Depth} & \textbf{Accession} & \textbf{Genome}\\\midrule
\emph{Human CHM13} & PacBio HiFi & 3,167,477 & 16 & SRR1129212{2-3} & T2T-CHM13 (v1.1) \\
                   & ONT$^{*}$ & 10,380,693 & 30 & Simulated R9.5 & T2T-CHM13 (v2.0) \\\midrule
\emph{Human HG002} & PacBio HiFi & 11,714,594 & 52 & SRR1038224{4-9} & GRCh37 \\\midrule
\emph{D. ananassae} & PacBio HiFi & 1,195,370 & 50 & SRR11442117 & \cite{tvedte_comparison_2021} \\\midrule
\emph{Yeast} & PacBio CLR$^{*}$ & 270,849 & 200 & Simulated P6-C4 & GCA\_000146045.2\\
             & ONT$^{*}$ & 135,296 & 100 & Simulated R9.5 & GCA\_000146045.2\\
             & Illumina MiSeq & 3,318,467 & 80 & ERR1938683 & GCA\_000146045.2\\\midrule
\emph{E. coli} & PacBio HiFi & 38,703 & 100 & SRR11434954 & \cite{tvedte_comparison_2021}\\
               & PacBio CLR & 76,279 & 112 & SRR1509640 & GCA\_000732965.1\\\bottomrule
\multicolumn{6}{l}{\footnotesize $^{*}$ We use PBSIM2 to generate the simulated PacBio and ONT reads.}\\
\multicolumn{6}{l}{We show the simulated chemistry under the SRA Accession column.}\\
\end{tabular}

\label{blend:tab:dataset}
\end{table}

We evaluate \blend based on two use cases: 1)~read overlapping and 2)~read mapping to a reference genome. For read overlapping, we perform \emph{all-vs-all overlapping} to find all pairs of overlapping reads within the same dataset (i.e., the target and query sequences are the same set of sequences). To calculate the overlap statistics, we report the overall number of overlaps, the average length of overlaps, and the number of seed matches per overlap. To evaluate the quality of overlapping reads based on the accuracy of the assemblies we generate from overlaps, we use miniasm~\cite{li_minimap_2016}. We use miniasm because it does not perform error correction when generating \emph{de novo} assemblies, which allows us to directly assess the quality of overlaps without using additional approaches that externally improve the accuracy of assemblies. We use \texttt{mhap2paf.pl} package as provided by miniasm to convert the output of MHAP to the format miniasm requires (i.e., PAF). We use QUAST~\cite{gurevich_quast_2013} to measure statistics related to the contiguity, length, and accuracy of \emph{de novo} assemblies, such as the overall assembly length, largest contig, NG50, and NGA50 statistics (i.e., statistics related to the length of the shortest contig at the half of the overall reference genome length), k-mer completeness (i.e., amount of shared k-mers between the reference genome and an assembly), number of mismatches per 100Kb, and GC content (i.e., the ratio of G and C bases in an assembly). We use dnadiff~\cite{marcais_mummer4_2018} to measure the accuracy of \emph{de novo} assemblies based on 1)~the average identity of an assembly when compared to its reference genome and 2)~the fraction of overall bases in a reference genome that align to a given assembly (i.e., genome fraction). We compare \blend to minimap2~\cite{li_minimap2_2018} and MHAP~\cite{berlin_assembling_2015} for read overlapping. \revc{For human genomes, MHAP either 1)~requires more memory than available in our system (i.e., 1TB) or 2)~produces an output so large that miniasm cannot generate the assembly due to memory constraints.}

For read mapping, we map all reads in a dataset (i.e., query sequences) to their corresponding reference genome (i.e., target sequence). We evaluate read mapping in terms of accuracy, quality, and the effect of read mapping on downstream analysis by calling structural variants. We compare \blend with minimap2, LRA~\cite{ren_lra_2021}, Winnowmap2~\cite{jain_long-read_2022, jain_weighted_2020}, S-conLSH~\cite{chakraborty_conlsh_2020, chakraborty_s-conlsh_2021}, and Strobealign~\cite{sahlin_flexible_2022}. We do not evaluate 1)~LRA, Winnowmap2, and S-conLSH for short reads as these tools do not support mapping paired-end short reads, 2)~Strobealign for long reads as it is a short read aligner, 3)~S-conLSH for the \emph{D. ananassae} as S-conLSH crashes due to a segmentation fault when mapping reads to the \emph{D. ananassae} reference genome, and 4)~S-conLSH for mapping HG002 reads as its output cannot be converted into a sorted BAM file, which is required for variant calling. We do not evaluate the read mapping accuracy of LRA and S-conLSH because 1)~LRA generates a CIGAR string with characters that the \texttt{paftools mapeval} tool cannot parse to calculate alignment positions, and 2)~S-conLSH due to its poor accuracy results we observe in our preliminary analysis.

\textbf{Read mapping accuracy.} We measure 1)~the overall read mapping error rate and 2)~the distribution of the read mapping error rate with respect to the fraction of mapped reads. To generate these results, we use the tools in \texttt{paftools} provided by minimap2 in two steps. First, the \texttt{paftools pbsim2fq} tool annotates the read IDs with their true mapping information that PBSIM2 generates. The \texttt{paftools mapeval} tool calculates the error rate of read mapping tools by comparing the mapping regions that the read mapping tools find with their true mapping regions annotated in read IDs. The error rate shows the ratio of reads mapped to incorrect regions over the entire mapped reads.

\textbf{Read mapping quality.} We measure 1)~the breadth of coverage (i.e., percentage of bases in a reference genome covered by at least one read), 2)~the average depth of coverage (i.e., the average number of read alignments per base in a reference genome), 3)~mapping rate (i.e., number of aligned reads) and 4)~rate of properly paired reads for paired-end mapping. To measure the breadth and depth of coverage of read mapping, we use BEDTools~\cite{quinlan_bedtools_2010} and Mosdepth~\cite{pedersen_mosdepth_2018}, respectively. To measure the mapping rate and properly paired reads, we use BAMUtil~\cite{jun_efficient_2015}.

\textbf{Downstream analysis.} We use sniffles2~\cite{sedlazeck_accurate_2018, smolka_comprehensive_2022} to call structural variants (SVs) from the HG002 long read mappings. We use Truvari~\cite{english_truvari_2022} to compare the resulting SVs with the benchmarking SV set (i.e., the \emph{Tier 1} set) released by the Genome in a Bottle (GIAB) consortium~\cite{zook_robust_2020} in terms of their true positives ($TP$), false positives ($FP$), false negatives ($FN$), precision ($P = TP/(TP+FP)$), recall ($R = TP/(TP + FN)$) and the $F_1$ scores ($F_{1} = 2 \times (P \times R)/(P+R)$). False positives show the number of the called SVs missing in the benchmarking set. False negatives show the number of SVs in the benchmarking set missing from the called SV set. The Tier 1 set includes 12,745 sequence-resolved SVs that include the \texttt{PASS} filter tag. GIAB provides the high-confidence regions of these SVs with low errors. We follow the benchmarking strategy that GIAB suggests~\cite{zook_robust_2020}, where we compare the SVs with the \texttt{PASS} filter tag within the high-confidence regions.

For both use cases, we use the \texttt{time} command in Linux to evaluate the performance and peak memory footprints. We provide the average speedups and memory overhead of \blend compared to each tool, while dataset-specific results are shown in our corresponding figures. When applicable, we use the default parameters of all the tools suggested for certain use cases and sequencing technologies (e.g., mapping HiFi reads in minimap2). Since minimap2 and MHAP do not provide default parameters for read overlapping using HiFi reads, we use the parameters that HiCanu~\cite{nurk_hicanu_2020} uses for overlapping HiFi reads with minimap2 and MHAP. We provide the details regarding the parameters and versions we use for each tool in Supplementary Tables~\ref{blend:supptab:ovpars}, \ref{blend:supptab:mappars}, and \ref{blend:supptab:version}. When applicable in read overlapping, we use the same window and the seed length parameters that \blend uses in minimap2 and show the performance and accuracy results in Supplementary Figure~\ref{blend:suppfig:overlap_perf-eq} and Supplementary Table~\ref{blend:supptab:overlap_assembly-eq}. For read mapping, the comparable default parameters in \blend are already the same as in minimap2.

\subsection{Empirical Analysis of Fuzzy Seed Matching}\label{blend:subsec:fuzzy_matching}
\revc{We evaluate the effectiveness of fuzzy seed matching by identifying non-identical seeds that share the same hash value (i.e., collisions) using both the low-collision hash function of minimap2 (\texttt{hash64}) and \blend.}

\subsubsection{Finding minimizer collisions.}\label{blend:subsubsec:mincollisions}
\begin{figure*}[htb]
\centering
\includegraphics[width=\linewidth]{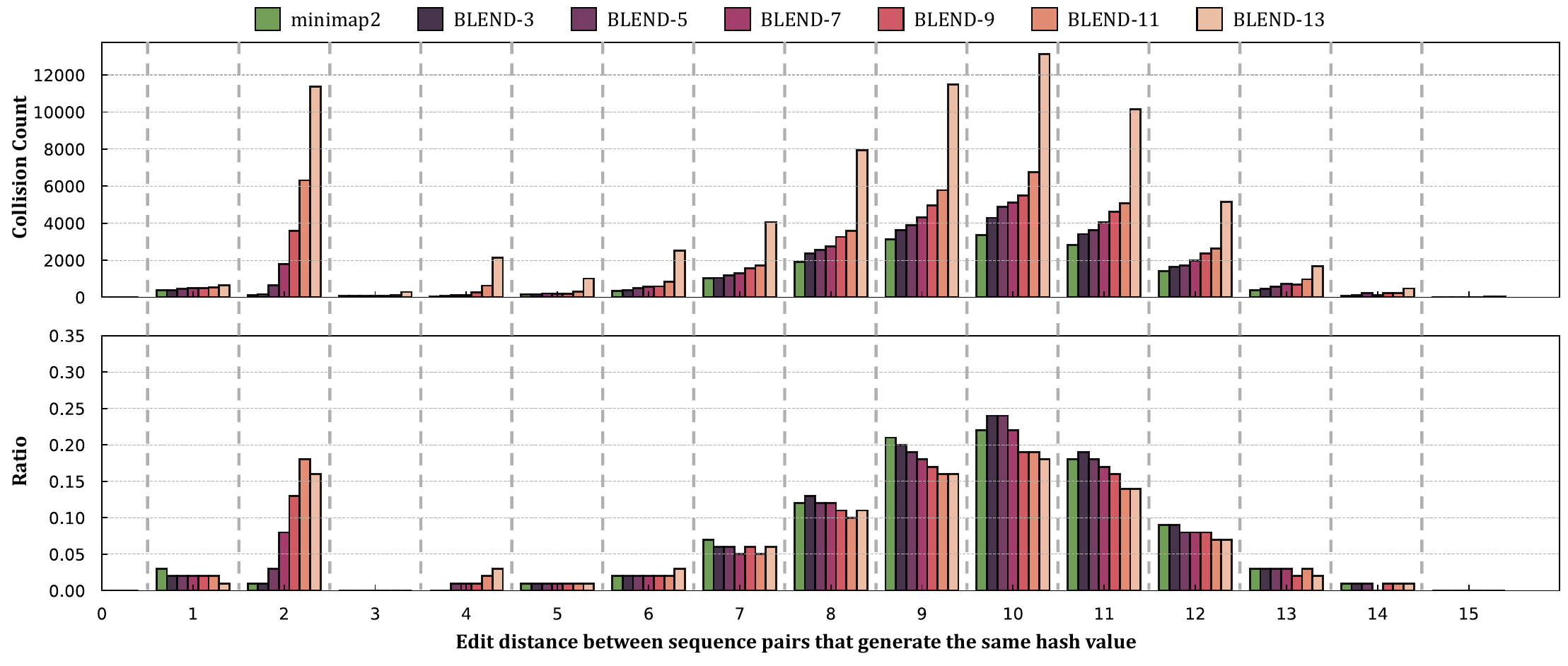}
\caption{Fuzzy seed matching statistics. Collision count shows the number of non-identical seeds that generate the same hash value and the edit distance between these sequences. Ratio is the proportion of collisions between non-identical sequences at a certain edit distance over all collisions. \blend-$n$ shows the number of neighbors ($n$) that \blend uses.}
\label{blend:fig:fuzzy_seed_matching}
\end{figure*}

\revc{Our goal is to evaluate the effects of using a low-collision hash function and the \blend mechanism on hash collisions between non-identical minimizers.} We find minimizers using 1)~the low collision hash function that minimap2 uses (i.e., \texttt{hash64}) and 2)~the SimHash technique~\cite{charikar_similarity_2002, manku_detecting_2007} we use in \blend. For \blend, we use the \texttt{\blend-I} technique to directly compare the minimizers found using \blend and minimap2. We keep the seed length constant, 16. For \blend, we use various numbers of immediately overlapping k-mers that \blend extracts from seed sequences (i.e., \emph{neighbors}), as explained in Section~\ref{blend:subsec:seed}. To keep the seed length ($|S|$) constant with a varying number of neighbors ($n$), we calculate the k-mer length ($k$) we extract from seeds as follows: $|S| = n+k-1$ where $|S|$ and $n$ are known. For each tool and configuration, we report the overall number of minimizers we find, the number of minimizer pairs that generate the same hash value (i.e., \emph{collision}), the ratio of collisions to all minimizers, and the average edit distance between the minimizer pairs that have the same hash value. We make our resulting dataset that includes the statistics shown in Figure~\ref{blend:fig:fuzzy_seed_matching} and Table~\ref{blend:tab:min_fuzzy_seed_matching} available at Zenodo\footnote{\url{https://doi.org/10.5281/zenodo.7317896}}.

\begin{table}[htb]
\centering
\caption{Fuzzy seed matching statistics of minimizer seeds that we find using minimap2 and \blend. The number of overlapping k-mers that \blend extracts from seed sequences (i.e., neighbors or \emph{n}) are annotated as \blend-n}
\resizebox{\linewidth}{!}{
\begin{tabular}{@{}lrrrr@{}}\toprule
\textbf{Tool} & \textbf{Number of}  & \textbf{Number of} & \textbf{Collision/Minimizer} & \textbf{Avg. Edit Distance} \\	
 	          & \textbf{Minimizers} & \textbf{Collisions}       & \textbf{Ratio}               & \textbf{Between Minimizers} \\
 	          &                     &                               &                              & \textbf{With Collision}     \\\midrule
minimap2      & 903,043             & 15,306                        & 0.016949                     & 9.327061                    \\\midrule
\blend-3   & 1,014,173           & 18,224                        & 0.017969                     & 9.393437                    \\
\blend-5   & 1,090,468           & 20,659                        & 0.018945                     & 9.213660                    \\
\blend-7   & 1,140,254           & 23,591                        & 0.020689                     & 8.874698                    \\
\blend-9   & 1,173,198           & 28,411                        & 0.024217                     & 8.495301                    \\
\blend-11  & 1,186,687           & 35,500                        & 0.029915                     & 8.067549                    \\
\blend-13  & 1,197,966           & 72,078                        & 0.060167                     & 8.075918                    \\\bottomrule
\end{tabular}

}
\label{blend:tab:min_fuzzy_seed_matching}
\end{table}

Table~\ref{blend:tab:min_fuzzy_seed_matching} shows the overall statistics of the fuzzy seed matching, and Figure~\ref{blend:fig:fuzzy_seed_matching} shows the edit distance between non-identical seeds with hash collision when using minimap2 and \blend. We make three key observations. 
First, \blend significantly increases the ratio of hash collisions between highly similar minimizer pairs (e.g., edit distance less than 3) compared to using a low-collision hash function in minimap2. This result shows that \blend favors increasing the collisions for highly similar seeds (i.e., fuzzy seed matching) than uniformly increasing the number of collisions by keeping the same ratio across all edit distance values.
Second, the number of collisions that minimap2 and \blend find are similar to each other for the minimizer pairs that have a large edit distance between them (e.g., larger than 6). The only exception to this observation is \blend-13, which substantially increases all collisions for any edit distance due to using many small k-mers (i.e., thirteen 4-mers) when generating the hash values of 16-character long seeds. We note that the number of collisions is significantly higher when the edit distance between minimizers is $2$ compared to the collisions with edit distance $1$. We argue that this may be due to the distribution of the edit distances between minimizer pairs where there may be significantly a large number of minimizer pairs with edit distance $2$ than $1$.
Third, increasing the number of neighbors can effectively reduce the average edit distance between fuzzy seed matches with the cost of increasing the overall number of minimizer seeds, as shown in Table~\ref{blend:tab:min_fuzzy_seed_matching}. We conclude that \blend can effectively find highly similar seeds with the same hash value as it increases the ratio of collisions between similar seeds while providing a collision ratio similar to minimap2 for dissimilar seeds.

\subsubsection{Identifying similar sequences.}
Our goal is to find non-identical k-mer matches with the same hash value (i.e., fuzzy k-mer matches) between highly similar sequence pairs. To this end, we prepare a dataset that includes 25-character long sequences in four steps. First, we extract 25-character long non-overlapping sequences from the \emph{E. coli} reference genome~\cite{tvedte_comparison_2021} shown in Table~\ref{blend:tab:dataset}, which we call \emph{sampled sequences} for simplicity. To evenly sample these sequences, each sampled sequence is separated by 75 characters from the previous sampled sequence. Second, our goal is to find \emph{all} sequences in the reference genome that are \emph{similar} and non-identical to the sampled sequences. To achieve this, we use bowtie~\cite{langmead_aligning_2010} and find \emph{all} sequences in the \emph{E. coli} reference genome that the sampled sequences align with at least one mismatch and, at most, three mismatches (i.e., at least $\sim 88\%$ similarity). Third, we extract the sequences from the reference regions that the sampled sequences align, which we call \emph{aligned sequences}. Fourth, we prepare our dataset that contains 1,077 FASTA files and 4,130 25-character long sequences overall. Each FASTA file includes 1)~a sampled sequence that has at least one alignment in the reference genome based on our mismatching criteria and 2)~all aligned sequences that the sampled sequence is aligned to.

To find the non-identical k-mers with the same hash value in each FASTA file, we generate the hash values of all overlapping 16-mers of all sequences in a FASTA file. We use the low-collision hash function that minimap2 uses (i.e., \texttt{hash64}) and the \texttt{\blend-I} technique in \blend to generate these hash values. For \blend, we use various numbers of neighbors when generating the hash values of 16-mers (see Section~\ref{blend:subsubsec:mincollisions} for the relation between the number of neighbors and the seed length, which is 16 in our evaluation). In Table~\ref{blend:tab:seq_fuzzy_seed_matching}, we report the number of sequences in our dataset, the number of sequences that have at least one non-identical k-mer pair with the same hash value (i.e., collisions), the ratio of collisions to the overall number of sequences, and the average edit distance between k-mers with collision. We make our dataset available at Zenodo\footnote{\url{https://doi.org/10.5281/zenodo.7319786}}, which includes 1,077 FASTA files and the resulting files that we generate the numbers we show in Table~\ref{blend:tab:seq_fuzzy_seed_matching}. These resulting files include the non-identical k-mers with the same hash value, the sequence pairs that we extract these k-mers from, and the edit distances with these k-mers.

\begin{table}[htb]
\centering
\caption{Fuzzy k-mer matching statistics of sequences that we find using minimap2 and \blend. The number of overlapping k-mers that \blend extracts from seed sequences (i.e., neighbors or \emph{n}) are annotated as \blend-n}
\resizebox{\linewidth}{!}{
\begin{tabular}{@{}lrrrr@{}}\toprule
\textbf{Tool} & \textbf{Number of}  & \textbf{Number of Sequences} & \textbf{Collision/Sequence} & \textbf{Avg. Edit Distance} \\	
 	          & \textbf{Sequences}  & \textbf{with Collision}   & \textbf{Ratio}              & \textbf{Between K-mers}  \\
 	          &                     &                           &                             & \textbf{With Collision}     \\\midrule
minimap2      & 4,130               & 0                         & 0                           & N/A                         \\\midrule
\blend-3   & 4,130               & 0                         & 0                           & N/A                         \\
\blend-5   & 4,130               & 11                        & 0.00263663                  & 1.45455                     \\
\blend-7   & 4,130               & 50                        & 0.0119847                   & 1.5                         \\
\blend-9   & 4,130               & 77                        & 0.0184564                   & 2.01299                     \\
\blend-11  & 4,130               & 273                       & 0.0654362                   & 2.80952                     \\
\blend-13  & 4,130               & 329                       & 0.0788591                   & 2.20669                     \\\bottomrule
\end{tabular}

}
\label{blend:tab:seq_fuzzy_seed_matching}
\end{table}

Table~\ref{blend:tab:seq_fuzzy_seed_matching} shows the number and portion of similar sequence pairs that we can find using \emph{only} fuzzy k-mer matches. We make two key observations.
First, \blend is the only mechanism that can identify similar sequences from their fuzzy k-mer matches since low-collision hash functions cannot increase the collision rates for high similarity matches. Second, \blend can identify a larger number of similar sequence pairs with an increasing number of neighbors. For the number of neighbors larger than 5, the percentage of these similar sequence pairs that \blend can identify ranges from $1.2\%$ to $7.9\%$ of the overall number of sequences we use in our dataset. We conclude that \blend enables finding similar sequence pairs from fuzzy k-mer matches that low-collision hash functions cannot find.

\subsection{Use Case 1: Read Overlapping}\label{blend:subsec:overlap}
\subsubsection{Performance.}\label{blend:subsubsec:overlap-performance}

\begin{figure}[htb]
\centering
\includegraphics[width=0.9\linewidth]{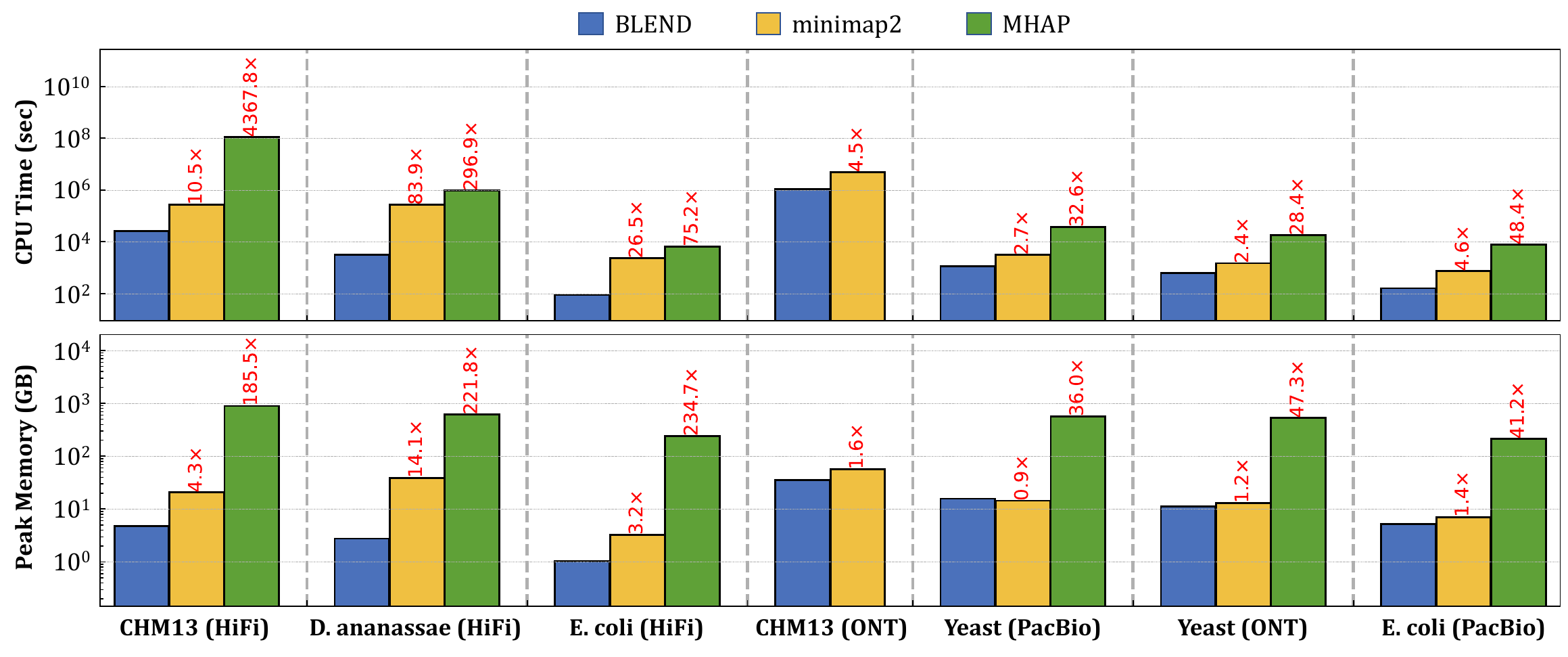}
\caption{CPU time and peak memory footprint comparisons of read overlapping.}
\label{blend:fig:overlap_perf}
\end{figure}

Figure~\ref{blend:fig:overlap_perf} shows the CPU time and peak memory footprint comparisons for read overlapping. We make the following five observations.
First, \blend provides an average speedup of \blendavgovpM and \blendavgovpMH while reducing the memory footprint by \blendavgovmM and \blendavgovmMH compared to minimap2 and MHAP, respectively. \blend is significantly more performant and provides less memory overheads than MHAP because MHAP generates many hash values for seeds regardless of the length of the sequences, while \blend allows sampling the number of seeds based on the sequence length with the windowing guarantees of minimizers and strobemer seeds.
Second, when considering only HiFi reads, \blend provides significant speedups by \blendavgovpHM and \blendavgovpHMH while reducing the memory footprint by \blendavgovmHM and \blendavgovmHMH compared to minimap2 and MHAP, respectively.
HiFi reads allow \blend to increase the window length (i.e., $w=200$) when finding the minimizer k-mer of a seed, which improves the performance and reduces the memory overhead without reducing the accuracy. This is possible mainly because \blend can find \emph{both} fuzzy and exact seed matches, which enables \blend to find \emph{unique} fuzzy seed matches that minimap2 \emph{cannot} find due to its exact-matching seed requirement.
Third, we find that \blend requires less than 16GB of memory space for almost all the datasets, making it largely possible to find overlapping reads even with a personal computer with relatively small memory space.
\blend has a lower memory footprint because 1)~\blend uses as many seeds as the number of minimizer k-mers per sequence to benefit from the reduced storage requirements that minimizer k-mers provide, and 2)~the window length is larger than minimap2 as \blend can tolerate increasing this window length with the fuzzy seed matches without reducing the accuracy.
Fourth, when using erroneous reads (i.e., PacBio CLR and ONT), \blend performs better than other tools with memory overheads similar to minimap2. The set of parameters we use for erroneous reads prevents \blend from using large windows (i.e., $w=10$ instead of $w=200$) without reducing the accuracy of read overlapping. Smaller window lengths generate more seeds, which increases the memory space requirements.
Fifth, we use the same parameters (i.e., the seed length and the window length) with minimap2 that \blend uses to observe the benefits that \blend provides with PacBio CLR and ONT datasets. We cannot perform the same experiment for the HiFi datasets because \blend uses strobemer seeds of length $31$, which minimap2 cannot support due to its minimizer seeds and the maximum seed length limitation in its implementation (i.e., max. $28$). We use \emph{minimap2-Eq} to refer to the version of minimap2, where it uses the parameters equivalent to the \blend parameters for a given dataset in terms of the seed and window lengths.
We show in Supplementary Figure~\ref{blend:suppfig:overlap_perf-eq} that minimap2-Eq performs, on average, $\sim5\%$ better than \blend with similar memory space requirements when using the same set of parameters with the \texttt{\blend-I} technique. Minimap2-Eq provides worse accuracy than \blend when generating the ONT assemblies, as shown in Supplementary Table~\ref{blend:supptab:overlap_assembly-eq}, while the erroneous PacBio assemblies are more accurate with minimap2-Eq. The main benefit of \blend is to provide overall higher accuracy than both the baseline minimap2 and minimap-Eq, which we can achieve by finding unique fuzzy seed matches that minimap2 cannot find.
We conclude that \blend is significantly more memory-efficient and faster than other tools to find overlaps, especially when using HiFi reads with its ability to sample many seeds using large values of $w$ without reducing the accuracy.

\subsubsection{Overlap Statistics.}\label{blend:subsubsec:overlap-statistics}
\begin{figure}[htb]
\centering
  \includegraphics[width=0.7\linewidth]{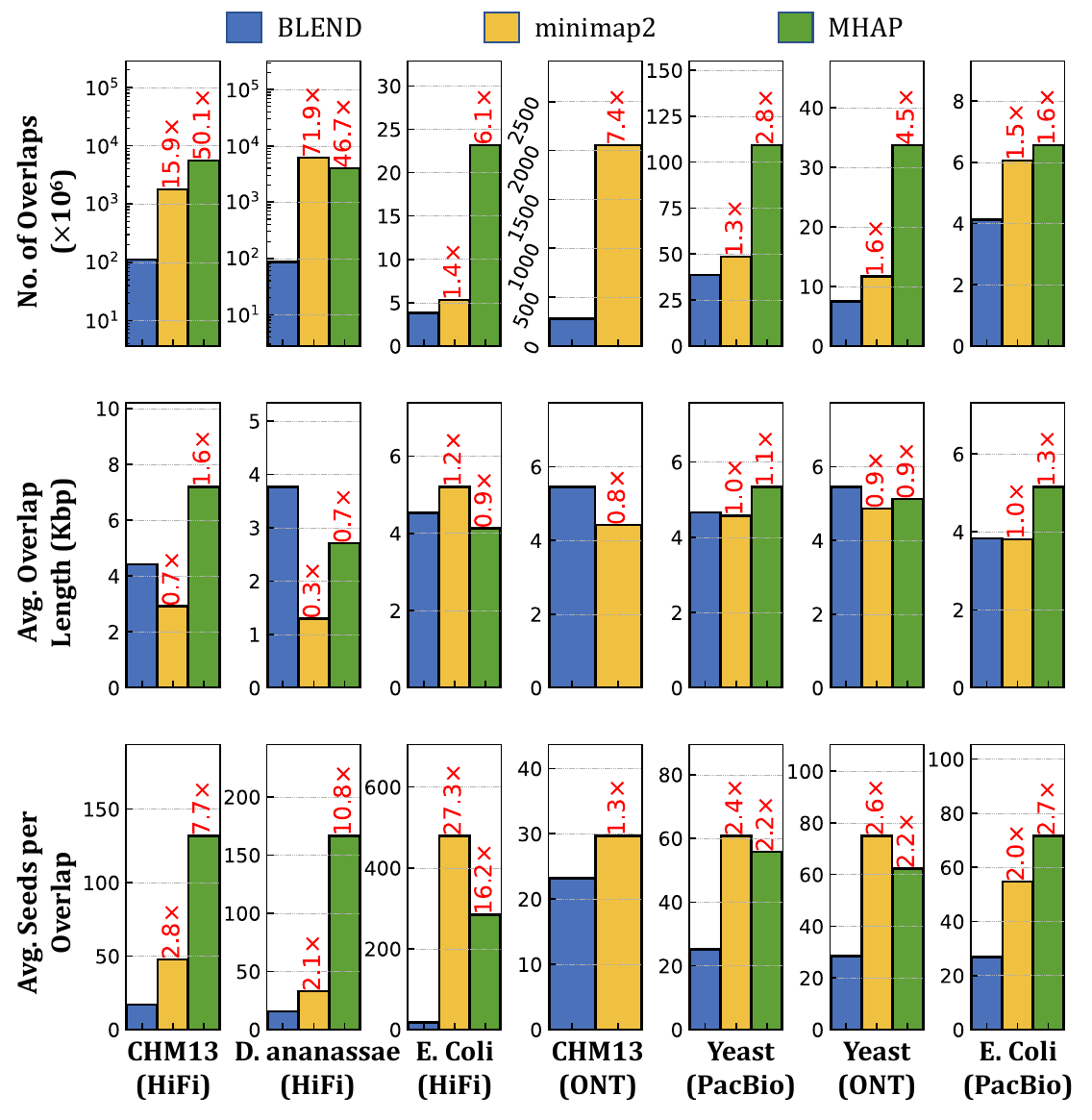}
    \caption{Average number and length of overlaps, and average number of seeds used to find a single overlap.}
    \label{blend:fig:overlap_stats}
\end{figure}

Figure~\ref{blend:fig:overlap_stats} shows the overall number of overlaps, the average length of overlaps, and the average number of seed matches that each tool finds to identify the overlaps between reads. \revc{The total number of overlaps combined with the average number of seed matches per overlap determines the total number of seeds found by each method.} We make the following four key observations.
First, we observe that \blend finds overlaps longer than minimap2 and MHAP can find in most cases. \blend can 1)~uniquely find the fuzzy seed matches that the exact-matching-based tools cannot find and 2)~perform chaining on these fuzzy seed matches to increase the length of overlap using many fuzzy seed matches that are relatively close to each other. Finding more distinct seeds and chaining these seeds enable \blend to find longer overlaps than other tools. Although these unique features of \blend can lead to chaining longer overlaps, we also note that \blend may not be able to find very short overlaps due to the larger window lengths it uses, which can also contribute to increasing the average length of overlaps.
Second, \blend uses significantly fewer seed matches per overlap than other tools, up to \blendovmaxseed, to find these longer overlaps. This is mainly because \blend needs much fewer seeds per overlap as it uses 1)~larger window lengths than minimap2 and 2)~provides windowing guarantees, unlike MHAP.
Third, finding fewer seed matches per overlap leads to 1)~finding fewer overlaps than minimap2 and MHAP find and 2)~reporting fewer seed matches overall. These overlaps that \blend cannot find are mainly because of the strict parameters that minimap2 and MHAP use due to their exact seed matching limitation (e.g., smaller window lengths). \blend can increase the window length while producing more accurate and complete assemblies than minimap2 and MHAP (Table~\ref{blend:tab:overlap_assembly}). This suggests that minimap2 and MHAP find redundant overlaps and seed matches that have no significant benefits in generating accurate and complete assemblies from these overlaps. 
Fourth, the sequencing depth of coverage has a larger impact on the number of overlaps that \blend can find compared to the impact on minimap2 and MHAP. We observe this trend when comparing the number of overlaps found using the PacBio ($200\times$ coverage) and ONT ($100\times$ coverage) reads of the Yeast genome. The gap between the number of overlaps found by \blend and other tools increases as the sequencing coverage decreases. This suggests that \blend can be less robust to the sequencing depth of coverage. Such a trend does not impact the accuracy of the assemblies that we generate using the \blend overlaps, while it provides lower NGA50 and NG50 values as shown in Table~\ref{blend:tab:overlap_assembly}.
We conclude that the performance and memory-efficiency improvements in read overlapping are proportional to the reduction in the seed matches that \blend uses to find overlapping reads. Thus, finding fewer non-redundant seed matches can dramatically improve the performance and memory space usage without reducing the accuracy.

\subsubsection{Assembly Quality Assessment.}\label{blend:subsubsec:overlap-assembly}

Our goal is to assess the quality of assemblies generated using the overlapping reads found by \blend, minimap2, and MHAP. Table~\ref{blend:tab:overlap_assembly} shows the statistics related to the accuracy of assemblies (i.e., the six statistics on the leftmost part of the table) and the statistics related to assembly length and contiguity (i.e., the four statistics on the rightmost part of the table) when compared to their respective reference genomes. We make the following five key observations based on the accuracy results of assemblies.
First, \revc{we observe that assemblies generated using the overlapping reads found by \blend are more accurate in terms of average identity and k-mer completeness compared to those generated by minimap2 and MHAP.} These results show that the assemblies we generate using the \blend overlaps are more similar to their corresponding reference genome. \blend can find unique and accurate overlaps using fuzzy seed matches that lead to more accurate \emph{de novo} assemblies than the minimap2 and MHAP overlaps due to their lack of support for fuzzy seed matching.
Second, we observe that assemblies generated using \blend overlaps usually cover a larger fraction of the reference genome than minimap2 and MHAP overlaps.
Third, although the average identity and genome fraction results seem mixed for the PacBio CLR and ONT reads such that \blend is best in terms of either average identity or genome fraction, we believe these two statistics should be considered together (e.g., by multiplying both results). This is because a highly accurate but much smaller fraction of the assembly can align to a reference genome, giving the best results for the average identity. We observe that this is the case for the \emph{D. ananassae} and \emph{Yeast} (PacBio CLR) genomes such that MHAP provides a very high average identity only for the much smaller fraction of the assemblies than the assemblies generated using \blend and minimap2 overlaps. Thus, when we combine average identity and genome fraction results, we observe that \blend consistently provides the best results for all the datasets.
Fourth, \blend usually provides the best results in terms of the aligned length and the number of mismatches per 100Kb. In some cases, QUAST cannot generate these statistics for the MHAP results as a small portion of the assemblies aligns the reference genome when the MHAP overlaps are used.
Fifth, we find that assemblies generated from \blend overlaps are less biased than minimap2 and MHAP overlaps, based on the average GC content results that are mostly closer to their corresponding reference genomes. We conclude that \blend overlaps yield assemblies with higher accuracy and less bias than the assemblies that the minimap2 and MHAP overlaps generate in most cases.

\begin{table}[htb]
\centering
\caption{Assembly quality comparisons.}
\resizebox{\linewidth}{!}{
\begin{tabular}{@{}clrrrrrrrrrr@{}}\toprule
\textbf{Dataset} 	  & \textbf{Tool} & \textbf{Average}	   & \textbf{Genome}		& \textbf{K-mer}	   & \textbf{Aligned}	   & \textbf{Mismatch per} & \textbf{Average} & \textbf{Assembly}	  & \textbf{Largest} 	  & \textbf{NGA50}    & \textbf{NG50}	  \\
					  & 			  & \textbf{Identity (\%)} & \textbf{Fraction (\%)} & \textbf{Compl. (\%)} & \textbf{Length (Mbp)} & \textbf{100Kbp (\#)}  & \textbf{GC (\%)} & \textbf{Length (Mbp)} & \textbf{Contig (Mbp)} & \textbf{(Kbp)}    & \textbf{(Kbp)}    \\\midrule
\emph{CHM13} 	  	  & \blend 	  & \textbf{99.8526}	   & \textbf{98.4847}		& \textbf{90.15}	   & 3,092.54 			   & \textbf{22.02}	  	   & \textbf{40.78}	  & \textbf{3,095.21} 	  & 22.8397 			  & 5,442.25 		  & 5,442.31 		  \\
(HiFi)				  & minimap2 	  & 99.7421				   & 97.1493 				& 83.05 			   & \textbf{3,094.79} 	   & 55.96 			  	   & 40.71 			  & 3,100.97 			  & \textbf{47.1387} 	  & \textbf{7,133.43} & \textbf{7,134.31} \\
					  & MHAP 		  & N/A					   & N/A 					& N/A 				   & N/A 				   & N/A 				   & N/A 			  & N/A 				  & N/A 				  & N/A 		      & N/A 		      \\
					  & Reference 	  & 100 				   & 100 					& 100 				   & 3,054.83 			   & 0.00 				   & 40.85 			  & 3,054.83 			  & 248.387 			  & 154,260 	      & 154,260 	   	  \\\midrule
\emph{D. ananassae}   & \blend 	  & \textbf{99.7856}	   & \textbf{97.2308}		& \textbf{86.43}	   & 240.391 			   & \textbf{143.13} 	   & \textbf{41.75}	  & \textbf{247.153} 	  & \textbf{6.23256} 	  & \textbf{792.407}  & \textbf{798.913}  \\
(HiFi)				  & minimap2 	  & 99.7044 			   & 96.3190 				& 72.33 			   & \textbf{289.453} 	   & 191.53	  			   & 41.68 			  & 298.28 			  	  & 4.43396 			  & 273.398 		  & 278.775 		  \\
					  & MHAP 		  & 99.5551 			   & 0.7276 				& 0.21 				   & 2.29 				   & 239.76 			   & 42.07 			  & 2.34951 			  & 0.028586 			  & N/A 		   	  & N/A 		   	  \\
					  & Reference 	  & 100 				   & 100 					& 100 				   & 213.805 			   & 0.00 				   & 41.81 			  & 213.818 			  & 30.6728 			  & 26,427.4 		  & 26,427.4 		  \\\midrule
\emph{E. coli} 		  & \blend 	  & \textbf{99.8320} 	   & \textbf{99.8801} 		& \textbf{87.91}	   & \textbf{5.12155} 	   & \textbf{3.77}	  	   & \textbf{50.53}	  & 5.12155 			  & \textbf{3.41699}	  & \textbf{3,416.99} & \textbf{3,416.99} \\
(HiFi)				  & minimap2 	  & 99.7064 			   & 99.8748 				& 79.27 			   & 5.09249 			   & 19.71 			  	   & 50.47 			  & \textbf{5.09436} 	  & 3.08849 			  & 3,087.05 		  & 3,087.05 		  \\
					  & MHAP 		  & N/A 				   & N/A 					& N/A 				   & N/A 				   & N/A 				   & N/A 			  & N/A 				  & N/A 				  & N/A 		   	  & N/A 		   	  \\
					  & Reference 	  & 100 				   & 100 					& 100 				   & 5.04628 			   & 0.00 				   & 50.52 			  & 5.04628 			  & 4.94446 			  & 4,944.46 		  & 4,944.46 		  \\\midrule
\emph{CHM13} 	  	  & \blend 	  & N/A	   				   & N/A					& \textbf{29.26}	   & \textbf{2,891.28} 	   & \textbf{4,077.53}	   & \textbf{41.32}	  & 2,897.87 	  		  & 25.2071 			  & 5,061.52 		  & 5,178.59 		  \\
(ONT)				  & minimap2 	  & N/A	   				   & N/A					& 28.32 			   & 2,860.26 	   		   & 4,660.73 			   & 41.36 			  & \textbf{2,908.55} 	  & \textbf{66.7564}	  & \textbf{13,189.2} & \textbf{13,820.3} \\
					  & Reference 	  & 100 				   & 100 					& 100 			   	   & 3,117.29 			   & 0.00 				   & 40.75 			  & 3,117.29 			  & 248.387 			  & 150,617 	   	  & 150,617 	      \\\midrule
\emph{Yeast} 		  & \blend 	  & 89.1677 	   		   & \textbf{97.0854} 		& \textbf{33.81}	   & \textbf{12.3938} 	   & 2,672.37 	   		   & 38.84	  		  & 12.4176 			  & 1.54807 			  & 635.966 		  & 636.669 		  \\
(PacBio)			  & minimap2 	  & 88.9002 			   & 96.9709 				& 33.38 			   & 12.0128 			   & 2,684.38 			   & 38.85 			  & \textbf{12.3325} 	  & \textbf{1.56078}	  & \textbf{810.046}  & \textbf{828.212}  \\
					  & MHAP 		  & \textbf{89.2182} 	   & 88.5928 				& 32.39 			   & 10.9039 			   & \textbf{2,552.05} 	   & \textbf{38.81}   & 10.9896 			  & 1.02375 			  & 85.081 		   	  & 436.285 		  \\
					  & Reference 	  & 100 				   & 100 					& 100 				   & 12.1571 			   & 0.00 				   & 38.15 			  & 12.1571 			  & 1.53193 			  & 924.431 		  & 924.431 		  \\\midrule
\emph{Yeast} 	      & \blend 	  & \textbf{89.6889} 	   & 99.2974 				& \textbf{35.95}	   & \textbf{12.3222} 	   & \textbf{2,529.47} 	   & \textbf{38.64}	  & \textbf{12.3225} 	  & 1.10582 			  & 793.046 		  & 793.046 		  \\
(ONT)			      & minimap2 	  & 88.9393 			   & \textbf{99.6878}		& 34.84 			   & 12.304 			   & 2,782.59 			   & 38.74 			  & 12.3725 			  & \textbf{1.56005}	  & \textbf{796.718}  & \textbf{941.588}  \\
					  & MHAP 		  & 89.1970 			   & 89.2785 				& 33.58 			   & 10.8302 			   & 2,647.19 			   & 38.84 			  & 10.9201 			  & 1.44328 			  & 118.886 		  & 618.908 		  \\
					  & Reference 	  & 100 				   & 100 					& 100 				   & 12.1571 			   & 0.00 				   & 38.15 			  & 12.1571 			  & 1.53193 			  & 924.431 		  & 924.431 		  \\\midrule
\emph{E. coli} 		  & \blend 	  & \textbf{88.5806} 	   & \textbf{96.5238} 		& \textbf{32.32}	   & \textbf{5.90024} 	   & \textbf{1,857.56}	   & \textbf{49.81}	  & 6.21598 			  & 2.40671	  			  & \textbf{769.981}  & 2,060.4 	 	  \\
(PacBio)			  & minimap2 	  & 88.1365 			   & 92.7603 				& 30.74 			   & 5.37728 			   & 2,005.72 			   & 49.66 			  & \textbf{6.02707} 	  & \textbf{3.77098} 	  & 367.442    	   	  & \textbf{3,770.98} \\
					  & MHAP 		  & 88.4883 			   & 90.5533 				& 31.32 			   & 5.75159 			   & 1,999.48 			   & 49.69 			  & 6.26216 			  & 1.04286 			  & 110.535    	   	  & 456.01 		   	  \\
					  & Reference 	  & 100 				   & 100 					& 100 				   & 5.6394 			   & 0.00 				   & 50.43 			  & 5.6394 				  & 5.54732 			  & 5,547.32 		  & 5,547.32 		  \\\bottomrule
\multicolumn{12}{l}{\footnotesize Best results are highlighted with \textbf{bold} text. For most metrics, the best results are the ones closest to the corresponding value of the reference genome.}\\
\multicolumn{12}{l}{\footnotesize The best results for \emph{Aligned Length} are determined by the highest number within each dataset. We do not highlight the reference results as the best results.}\\
\multicolumn{12}{l}{\footnotesize N/A indicates that we could not generate the corresponding result because tool, QUAST, or dnadiff failed to generate the statistic.} \\
\end{tabular}

}
\label{blend:tab:overlap_assembly}
\end{table}

Table~\ref{blend:tab:overlap_assembly} shows the results related to assembly length and contiguity on its rightmost part. We make the following three observations.
First, we show that \blend yields assemblies with better contiguity when using HiFi reads based on the largest NG50, NGA50, and contig length results compared to minimap2 with the exception of the human genome.
Second, minimap2 provides better contiguity for the human genomes and erroneous reads.
Third, the overall length of all assemblies is mostly closer to the reference genome assembly.
We conclude that minimap2 provides better contiguity for the assemblies from erroneous and human reads while \blend is usually better suited for using the HiFi reads.

\subsection{Use Case 2: Read Mapping}\label{blend:subsec:mapping}
\subsubsection{Performance.}\label{blend:subsubsec:mapping-perf}

\begin{figure}[htb]
\centering
  \includegraphics[width=\linewidth]{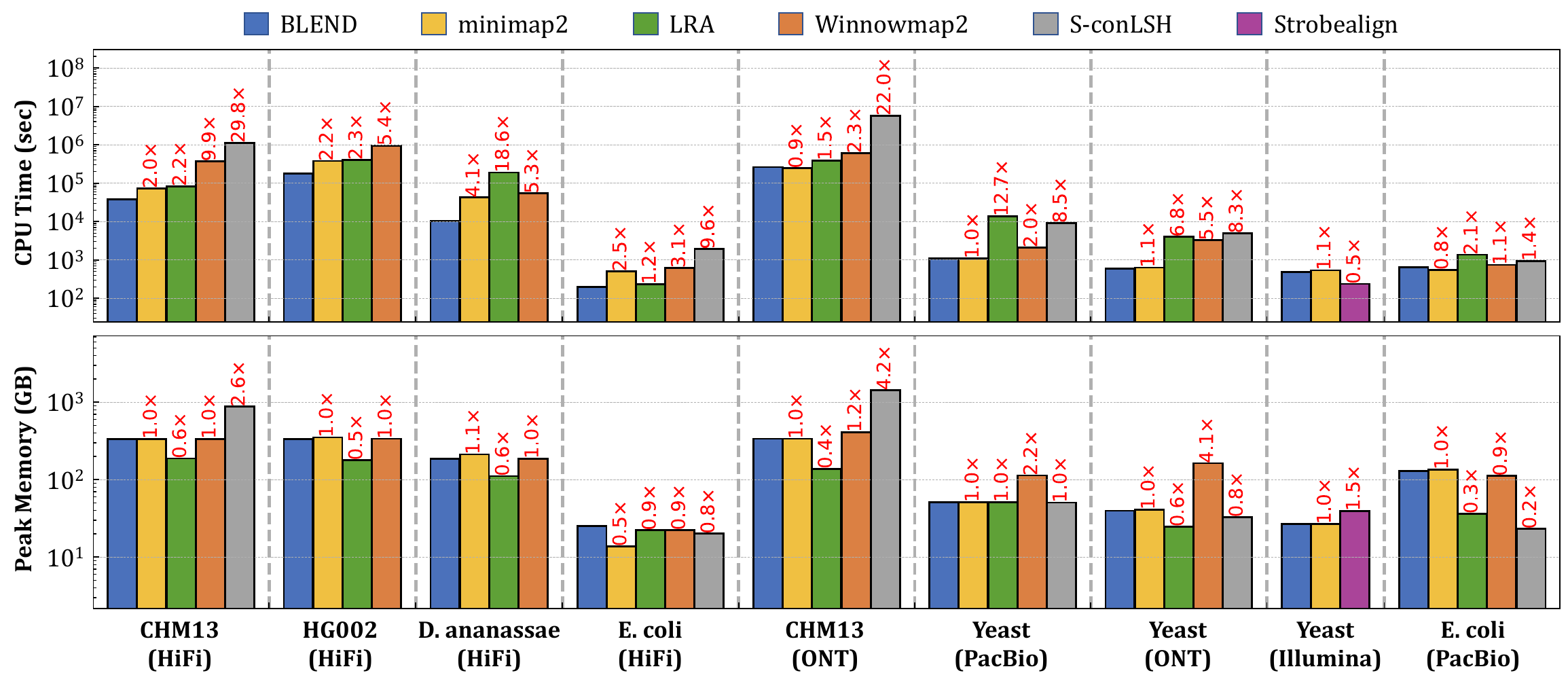}
    \caption{CPU time and peak memory footprint comparisons of read mapping.}
    \label{blend:fig:mapping_perf}
\end{figure}

Figure~\ref{blend:fig:mapping_perf} shows the CPU time and the peak memory footprint comparisons when performing read mapping to the corresponding reference genomes. We make the following four key observations.
First, we observe that \blend provides an average speedup of \blendavgrmpM, \blendavgrmpL, \blendavgrmpW, and \blendavgrmpS over minimap2, LRA, Winnowmap2, and S-conLSH, respectively. \revc{While \blend outperforms most of these tools, the speedups observed are generally lower than those in read overlapping.} Read mapping includes an additional computationally costly step that read overlapping skips, which is the read alignment. The extra overhead of read alignment slightly hinders the benefits that \blend provides that we observe in read overlapping.
Second, we find that LRA and minimap2 require \blendavgrmmL and \blendavgrmmM of the memory space that \blend uses, while Winnowmap2 and S-conLSH have a larger memory footprint by \blendavgrmmW and \blendavgrmmS, respectively. \blend cannot provide similar reductions in the memory overhead that we observe in read overlapping due to the narrower window length ($w=50$ instead of $w=200$) it uses to find the minimizer k-mers for HiFi reads. Using a narrow window length generates more seeds to store in a hash table, which proportionally increases the peak memory space requirements.
Third, \blend provides performance and memory usage similar to minimap2 when mapping the erroneous ONT and PacBio reads because \blend uses the same parameters as minimap2 for these reads (i.e., same $w$ and seed length).
Fourth, Strobealign is the best-performing tool for mapping short reads with the cost of larger memory overhead.
We conclude that \blend, on average, 1)~performs better than all tools for mapping long reads and 2)~provides a memory footprint similar to or better than minimap2, Winnowmap2, S-conLSH, and Strobealign, while LRA is the most memory-efficient tool.

\subsubsection{Read Mapping Accuracy.}\label{blend:subsubsec:mapping-accuracy}

\begin{figure*}[htb]
\centering
  \includegraphics[width=\linewidth]{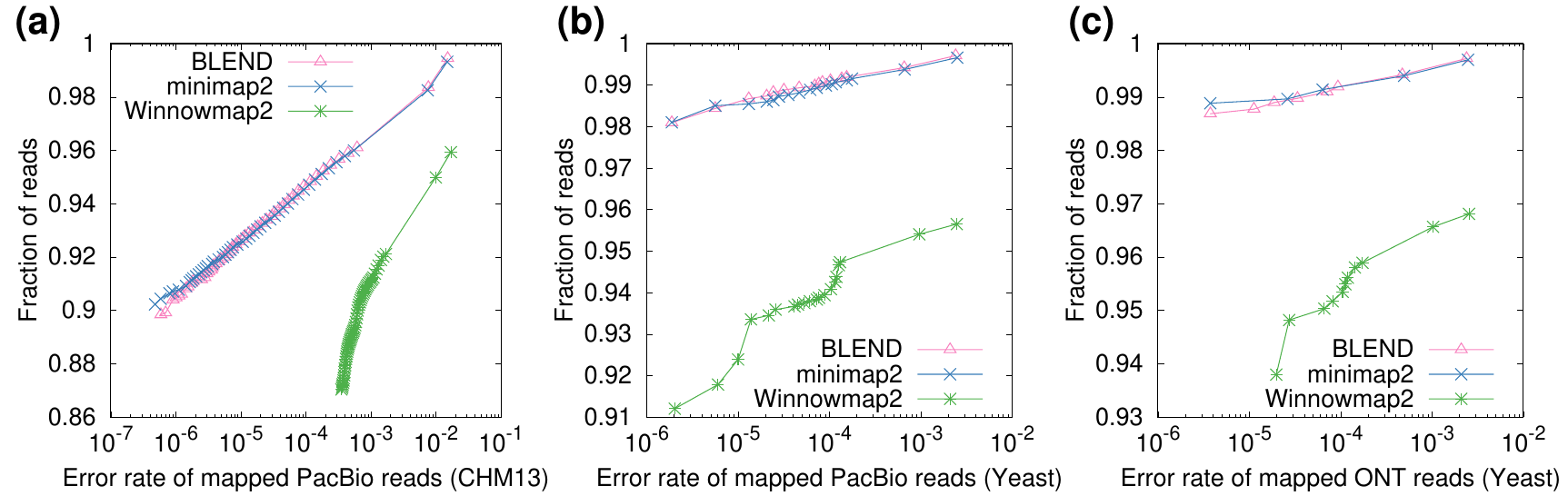}
    \caption{Fraction of simulated reads with an average mapping error rate. Reads are binned by their mapping quality scores. There is a bin for each mapping quality score as reported by the read mapper, and bins are sorted based on their mapping quality scores in descending order. For each tool, the $n^{th}$ data point from the left side of the x-axis shows the rate of incorrectly mapped reads among the reads in the first $n$ bins. We show the number of reads in these bins in terms of the fraction of the overall number of reads in the dataset. The data point with the largest fraction shows the average mapping error rate of all mapped reads.}
    \label{blend:fig:mapping_accuracy}
\end{figure*}

Table~\ref{blend:tab:mapping_accuracy} and Figure~\ref{blend:fig:mapping_accuracy} show the overall read mapping accuracy and fraction of mapped reads with their average mapping accuracy, respectively. We make two observations. First, we observe that \blend generates the most accurate read mapping in most cases, while minimap2 provides the most accurate read mapping for the human genome. These two tools are on par in terms of their read mapping accuracy and the fraction of mapped reads. Second, although Winnowmap2 provides more accurate read mapping than minimap2 for the PacBio reads from the Yeast genome, Winnowmap2 always maps a smaller fraction of reads than those \blend and minimap2 map. We conclude that although the results are mixed, \blend is the only tool that generates either the most or the second-most accurate read mapping in all datasets, providing the overall best accuracy results.

\begin{table}[htb]
\centering
\caption{Read mapping accuracy comparisons.}
\begin{tabular}{@{}lrrr@{}}\toprule
\textbf{Dataset} 		& \multicolumn{3}{c}{\textbf{Overall Error Rate (\%)}} \\\cmidrule{2-4}
				 		& \blend          & minimap2           & Winnowmap2 \\\midrule
\emph{CHM13} (ONT) 	    & 1.5168427 	     & \textbf{1.4914009} & 1.7001222  \\\midrule
\emph{Yeast} (PacBio) 	& \textbf{0.2403134} & 0.2504307          & 0.2474206  \\\midrule
\emph{Yeast} (ONT) 		& \textbf{0.2386617} & 0.2468770          & 0.2534777  \\\bottomrule
\multicolumn{4}{l}{\footnotesize Best results are highlighted with \textbf{bold} text.} \\
\end{tabular}

\label{blend:tab:mapping_accuracy}
\end{table}

\subsubsection{Read Mapping Quality.}\label{blend:subsubsec:mapping-quality}

Our goal is to assess the quality of read mappings in terms of four metrics: average depth of coverage, breadth of coverage, number of aligned reads, and the ratio of the paired-end reads that are properly paired in mapping. Table~\ref{blend:tab:mapping_quality} shows the quality of read mappings based on these metrics when using \blend, minimap2, LRA, Winnowmap2, and Strobealign. We exclude S-conLSH from the read mapping quality comparisons as we cannot convert its SAM output to BAM format to properly index the BAM file due to issues with its SAM output format. We make five observations.

\begin{table}[htb]
\centering
\caption{Read mapping quality comparisons.}
\resizebox{0.6\linewidth}{!}{
\begin{tabular}{@{}clrrrr@{}}\toprule
\textbf{Dataset} 		& \textbf{Tool} & \textbf{Average} 			 & \textbf{Breadth of} & \textbf{Aligned} 	 & \textbf{Properly}\\
				 		& 				& \textbf{Depth of} 		 & \textbf{Coverage}   & \textbf{Reads} 	 & \textbf{Paired}\\
				 		& 				& \textbf{Cov. (${\times}$)} & \textbf{(\%)} 	   & \textbf{(\#)} 		 & \textbf{(\%)}\\\midrule
\emph{CHM13} 	        & \blend 	& \textbf{16.58} 			 & \textbf{99.991} 	   & 3,171,916           & NA \\
(HiFi)					& minimap2 		& \textbf{16.58} 			 & \textbf{99.991} 	   & \textbf{3,172,261}  & NA \\
						& LRA 			& 16.37 					 & 99.064 			   & 3,137,631 			 & NA \\
						& Winnowmap2 	& \textbf{16.58} 			 & 99.990 	           & 3,171,313 			 & NA \\\midrule
\emph{HG002} 	        & \blend 	& 51.25 			         & 92.245 	           & 11,424,762          & NA \\
(HiFi)					& minimap2 		& 53.08 			         & 92.242 	           & 12,407,589          & NA \\
						& LRA 			& 52.48 					 & \textbf{92.275} 	   & \textbf{13,015,195} & NA \\
						& Winnowmap2 	& \textbf{53.81} 			 & 92.248 	           & 12,547,868 	     & NA \\\midrule
\emph{D. ananassae} 	& \blend 	& 57.37 					 & 99.662 			   & 1,223,388 			 & NA \\
(HiFi)					& minimap2 		& \textbf{57.57} 			 & \textbf{99.665} 	   & 1,245,931 			 & NA \\
						& LRA 			& 57.06 					 & 99.599 			   & 1,235,098 			 & NA \\
						& Winnowmap2 	& 57.40 			 		 & 99.663 	   		   & \textbf{1,249,575}  & NA \\\midrule
\emph{E. coli} 			& \blend 	& \textbf{99.14} 			 & 99.897 			   & 39,048 			 & NA\\
(HiFi)					& minimap2 		& \textbf{99.14} 			 & 99.897 			   & \textbf{39,065} 	 & NA\\
						& LRA 			& 99.10 			 		 & 99.897 			   & 39,063 			 & NA\\
						& Winnowmap2 	& \textbf{99.14} 			 & 99.897 			   & 39,036 			 & NA\\\midrule
\emph{CHM13} 	        & \blend 	& \textbf{29.34} 			 & \textbf{99.999} 	   & \textbf{10,322,767} & NA \\
(ONT)			        & minimap2 		& 29.33 			         & \textbf{99.999} 	   & 10,310,182          & NA \\
						& LRA 			& 28.84 					 & 99.948 			   & 9,999,432 			 & NA \\
						& Winnowmap2 	& 28.98 			         & 99.936 	           & 9,958,402 			 & NA \\\midrule
\emph{Yeast} 	        & \blend 	& \textbf{195.87} 			 & \textbf{99.980} 	   & \textbf{270,064}    & NA\\
(PacBio)		        & minimap2 		& 195.86 			         & \textbf{99.980} 	   & 269,935 	         & NA\\
						& LRA 			& 194.65 					 & 99.967 			   & 267,399 			 & NA\\
						& Winnowmap2 	& 192.35 			 		 & 99.977	           & 259,073 			 & NA\\\midrule
\emph{Yeast} 		    & \blend 	& \textbf{97.88} 			 & \textbf{99.964} 	   & \textbf{134,919} 	 & NA\\
(ONT)			        & minimap2 		& \textbf{97.88} 			 & \textbf{99.964} 	   & 134,885 	         & NA\\
						& LRA 			& 97.25 					 & 99.952 			   & 132,862 			 & NA\\
						& Winnowmap2 	& 97.04 			 		 & 99.963 	   		   & 130,978 			 & NA\\\midrule
\emph{Yeast}            & \blend 	& \textbf{79.92} 			 & \textbf{99.975} 	   & 6,493,730           & 95.88 \\
(Illumina)				& minimap2 		& 79.91 					 & 99.974 			   & 6,492,994 			 & 95.89 \\
				        & Strobealign   & \textbf{79.92} 			 & 99.970 			   & \textbf{6,498,380}  & \textbf{97.59} \\\midrule
\emph{E. coli} 			& \blend 	& \textbf{97.51} 			 & 100 			       & 83,924 			 & NA\\
(PacBio)			    & minimap2 		& 97.29 			         & 100 			       & \textbf{85,326} 	 & NA\\
						& LRA 			& 93.61 			 		 & 100 			       & 80,802 			 & NA\\
						& Winnowmap2 	& 89.78 			         & 100 			       & 69,884 			 & NA\\\bottomrule
\multicolumn{6}{l}{\footnotesize Best results are highlighted with \textbf{bold} text.} \\
\multicolumn{6}{l}{\footnotesize Properly paired rate is only available for paired-end Illumina reads.} \\
\end{tabular}

}
\label{blend:tab:mapping_quality}
\end{table}

First, all tools cover a large portion of the reference genomes based on the breadth of coverage of the reference genomes. Although LRA provides the lowest breadth of coverage in most cases compared to the other tools, it also provides the best breadth of coverage after mapping the human HG002 reads. This result shows that these tools are less biased in mapping reads to particular regions with their high breadth of coverage, and the best tool for covering the largest portion of the genome depends on the dataset.

Second, both \blend and minimap2 map an almost complete set of reads to the reference genome for all the datasets, while Winnowmap2 suffers from a slightly lower number of aligned reads when mapping erroneous PacBio CLR and ONT reads. The only exception to this observation is the HG002 dataset, where \blend provides a smaller number of aligned reads compared to other tools, while \blend provides the same breadth of coverage as minimap2. We investigate if such a smaller number of aligned reads leads to a coverage bias genome-wide in Supplementary Figures~\ref{blend:suppfig:genomedepth-1}, \ref{blend:suppfig:genomedepth-13}, and \ref{blend:suppfig:genomedepth_diff}. We find that the distribution of the depth of coverage of \blend is mostly similar to minimap2. There are a few regions in the reference genome where minimap2 provides substantially higher coverage than \blend provides, as we show in Supplementary Figure~\ref{blend:suppfig:genomedepth_diff}, which causes \blend to align a smaller number of reads than minimap2 aligns. Since these regions are still covered by both \blend and minimap2 with different depths of coverage, these two tools generate the same breadth of coverage without leading to no significant coverage bias genome-wide.

Third, we find that all the tools generate read mappings with a depth of coverage significantly close to their sequencing depth of coverage. This shows that almost all reads map to the reference genome evenly. 
Fourth, Strobealign generates the largest number of 1)~short reads mappings to the reference genome and 2)~properly paired reads compared to \blend and minimap2. Strobealign can map more reads using less time (Figure~\ref{blend:fig:mapping_perf}, which makes its throughput much higher than \blend and minimap2.
Fifth, although Strobealign can map more reads, it covers the smallest portion of the reference genome based on the breadth of coverage compared to \blend and minimap2. This suggests that Strobealign provides a higher depth of coverage at certain regions of the reference genome than \blend and minimap2 while leaving larger gaps in the reference genome.
We conclude that the read mapping qualities of \blend, minimap2, and Winnowmap2 are highly similar, while LRA provides slightly worse results. It is worth noting that \blend provides a better breadth of coverage than minimap2 provides in most cases while using the same parameters in read mapping. \blend does this by finding unique fuzzy seed matches that the other tools cannot find due to their \revb{exactly matching} seed requirements.

\subsubsection{Downstream Analysis.}\label{blend:subsubsec:sv_calling}
To evaluate the effect of read mapping on downstream analysis, we call SVs from the HG002 long read mappings that \blend, minimap2, LRA, and Winnowmap2 generate. Table~\ref{blend:tab:sv_calling} shows the benchmarking results. We make two key observations. First, we find that \blend provides the best overall accuracy in downstream analysis based on the best $F_1$ score compared to other tools. This is because \blend provides the best true positive and false negative numbers while providing the second-best false positive numbers after LRA. These two best values overall contribute to achieving the best recall and second-best precision that is on par with the precision LRA provides. Second, although LRA generates the second-best $F_1$ score, it provides the worst recall results due to the largest number of false negatives. We conclude that \blend is consistently either the best or second-best in terms of the metrics we show in Table~\ref{blend:tab:sv_calling}, which leads to providing the best overall $F_1$ accuracy in structural variant calling.

\begin{table}[htb]
\centering
\caption{Benchmarking the structural variant (SV) calling results.}
\begin{tabular}{@{}lrrrrrr@{}}\toprule
              & \multicolumn{6}{c}{HG002 SVs (High-confidence Tier 1 SV Set)} \\\cmidrule{2-7}
\textbf{Tool} & \textbf{TP (\#)} &\textbf{FP (\#)} & \textbf{FN (\#)} & \textbf{Precision} & \textbf{Recall} & \textbf{$F_1$}  \\\midrule
\blend	  & \textbf{9,229}   & 855             & \textbf{412}     & 0.9152 		       & \textbf{0.9573} & \textbf{0.9358} \\
minimap2	  & 9,222            & 915             & 419              & 0.9097 			   & 0.9565		     & 0.9326          \\
LRA			  & 9,155            & \textbf{830}    & 486              & \textbf{0.9169}    & 0.9496			 & 0.9329          \\
Winnowmap2	  & 9,170            & 1029            & 471              & 0.8991 			   & 0.9511		     & 0.9244          \\\bottomrule
\multicolumn{6}{l}{\footnotesize Best results are highlighted with \textbf{bold} text.} \\
\end{tabular}

\label{blend:tab:sv_calling}
\end{table}

\section{Discussion} \label{blend:sec:discussion}
We demonstrate that there are usually too many redundant \emph{short} and \emph{exactly matching} seeds used to find overlaps between sequences, as shown in Figure~\ref{blend:fig:overlap_stats}. These redundant seeds usually exacerbate the performance and peak memory space requirement problems that read overlapping and read mapping suffer from as the number of chaining and alignment operations proportionally increases with the number of seed matches between sequences~\cite{li_minimap2_2018}. Such redundant computations have been one of the main limitations against developing population-scale genomics analysis due to the high runtime of a single high-coverage genome analysis.

There has been a clear interest in using long or fuzzy seed matches because of their potential to find similarities between target and query sequences efficiently and accurately~\cite{alser_technology_2021}. To achieve this, earlier works mainly focus on either 1)~chaining the exact k-mer matches by tolerating the gaps between them to increase the seed region or 2)~linking multiple consecutive minimizer k-mers such as strobemer seeds. Chaining algorithms are becoming a bottleneck in read mappers as the complexity of chaining is determined by the number of seed matches~\cite{guo_hardware_2019}. Linking multiple minimizer k-mers enables tolerating indels when finding the matches of short sequence segments between genomic sequence pairs, but these seeds (e.g., strobemer seeds) should still exactly match due to the nature of the hash functions used to generate the hash values of seeds. This requires the seeding techniques to generate exactly the same seed to find either \revb{exactly matching} or approximate matches of short sequence segments.
We state that any arbitrary k-mer in the seeds should be tolerated to mismatch to improve the sensitivity of any seeding technique, which has the potential for finding more matching regions while using fewer seeds. Thus, we believe \blend solves the main limitation of earlier works such that it can generate the same hash value for similar seeds to find fuzzy seed matches with a single lookup while improving the performance, memory overhead, and accuracy of the applications that use seeds.

We hope that \blend advances the field and inspires future work in several ways, some of which we list next. 
First, we observe that \blend is \emph{most effective} when using high coverage and highly accurate long reads. Thus, \blend is already ready to scale for longer and more accurate sequencing reads. 
Second, the vector operations are suitable for hardware acceleration to improve the performance of \blend further. Such an acceleration is mainly useful when a massive amount of k-mers in a seed are used to generate the hash value for a seed, as these calculations can be done in parallel. We already provide the SIMD implementation to calculate the hash values \blend. We encourage implementing our mechanism for the applications that use seeds to find sequence similarity using processing-in-memory and near-data processing~\cite{senol_cali_segram_2022, mansouri_ghiasi_genstore_2022, shahroodi_demeter_2022, diab_high-throughput_2022, khalifa_filtpim_2021, khatamifard_genvom_2021, senol_cali_genasm_2020, chen_parc_2020, kaplan_bioseal_2020, laguna_seed-and-vote_2020, angizi_pim-aligner_2020, nag_gencache_2019, kim_grim-filter_2018}, GPUs~\cite{sadasivan_accelerating_2023, a_zeni_logan_2020, goenka_segalign_2020}, and FPGAs and ASICs~\cite{singh_fpga-based_2021, chen_high-throughput_2021, yan_accel-align_2021, fujiki_seedex_2020, alser_sneakysnake_2020, turakhia_darwin_2018} to exploit the massive amount of embarrassingly parallel bitwise operations in \blend to find fuzzy seed matches.
Third, we believe it is possible to apply the hashing technique we use in \blend for many seeding techniques with a proper design. We already show we can apply SimHash in regular minimizer k-mers or strobemers. Strobemers can be generated using k-mer sampling strategies other than minimizer k-mers, which are based on syncmers and random selection of k-mers (i.e., randstrobes)~\cite{sahlin_flexible_2022}. It is worth exploring and rethinking the hash functions used in these seeding techniques.
Fourth, potential machine learning applications can be used to generate more sensitive hash values for fuzzy seed matching based on learning-to-hash approaches~\cite{wang_survey_2018} and recent improvements on SimHash for identifying nearest neighbors in machine learning and bioinformatics~\cite{sharma_improving_2018, chen_using_2020, sinha_fruit-fly_2021}.

\subsection{Limitations}
We identify two main limitations of our work that requires further improvements. First, \blend may generate the same hash values for $1\%-8\%$ of all the similar sequence pairs in a dataset, as we show in Table~\ref{blend:tab:seq_fuzzy_seed_matching}. These $1\%-8\%$ of similar sequence pairs that cannot be found using low-collision hash functions can be significant in improving the accuracy and performance of some genomics applications. However, such a percentage may also be considered low for other use cases. We observe that increasing the number of neighbors ($n$) can increase the percentage of similar sequence pairs that \blend can find with the cost of causing more collisions for dissimilar sequence pairs. A newer generation of the SimHash-like hash functions such as DenseFly~\cite{sharma_improving_2018} or FlyHash~\cite{dasgupta_neural_2017} has the potential to improve the rate of similar sequence pairs with the same hash value. Second, the advantage of \blend is mainly observed when using highly accurate and long reads with high sequencing depth of coverage in read overlapping and downstream analysis, while the improvements are lower in other datasets. Although \blend scales better as the sequencing technologies become cheaper and generate longer and highly accurate reads, it is also essential to further improve its accuracy and performance for existing read datasets with erroneous long reads and short reads. This requires further optimizations in the parameter settings for erroneous long reads and short reads. We leave these two limitations as future work along with the other potential future works that we discuss earlier.

\section{Summary}
\label{blend:sec:summary}

We propose \blend, a mechanism that can efficiently find fuzzy seed matches between sequences to improve the performance, memory space efficiency, and accuracy of two important applications significantly: 1)~read overlapping and 2)~read mapping. Based on the experiments we perform using real and simulated datasets, we make six key observations.
First, for read mapping, \blend provides an average speedup of \blendavgovpM and \blendavgovpMH while reducing the peak memory footprint by \blendavgovmM and \blendavgovmMH compared to minimap2 and MHAP.
Second, we observe that \blend finds longer overlaps, in general, while using significantly fewer seed matches by up to \blendovmaxseed to find these overlaps. 
Third, we find that we can usually generate more \emph{accurate} assemblies when using the overlaps that \blend finds than those found by minimap2 and MHAP.
Fourth, for read mapping, we find that \blend, on average, provides speedup by 1) \blendavgrmpM, \blendavgrmpL, \blendavgrmpW, and \blendavgrmpS compared to minimap2, LRA, Winnowmap2, and S-conLSH, respectively. Fifth, Strobealign performs best for short read mapping, while \blend provides better memory space usage than Strobealign.
Sixth, we observe that \blend, minimap2, and Winnowmap2 provide both high quality and better accuracy in read mapping in all datasets, while \blend and LRA provide the best SV calling results in terms of downstream analysis accuracy. We conclude that \blend can use fewer fuzzy seed matches to significantly improve the performance and reduce the memory overhead of read overlapping without losing accuracy, while \blend, on average, provides better performance and a similar memory footprint in read mapping without reducing the read mapping quality and accuracy.
\revb{We hope that \blend will lead to the design of more accurate genome analysis techniques with the ability to find fuzzy seed matches as well as exact matching seeds.}

\clearpage
\setsuppbasednumbering
\section{Supplementary Materials}
\subsection{A Real Example of Generating the Hash Values of Seeds $S_k$ and $S_l$} \label{blend:suppsec:real}

Our goal is to show how the k-mer length $k$ and the number of k-mers to include in a seed, $n$, affect the final hash value. To this end, we use the following two seeds as found in the Yeast reference genome: $S_k:$ \texttt{CGGATGCTACAGTATATACCA} and $S_l:$ \texttt{ATGCTACAGTATATACCATCT}. Both seeds are 21-character long. We use two different parameter settings when generating the hash values of these seeds. The first setting uses $k=7$ as the k-mer length and $n=15$ as the number of immediately overlapping k-mers to include in a seed so that we can generate the 21-character long seeds $S_l$ and $S_k$. The second setting uses $k=15$ as the k-mer length and $n=7$ as the number of k-mers to include in a seed. We use the  \texttt{hash64} hash function as provided in the minimap2 implementation to generate the hash values of the k-mers of seeds.

In Supplementary Tables~\ref{blend:supptab:kmers7-15-1-32} - \ref{blend:supptab:kmers15-7-2-16} we show k-mers, the hash values of the k-mers in their binary form, and the gradual change in the counter vectors used to calculate the hash values for seeds $S_k$ and $S_l$. We update the counter vectors based on the bits in the hash values of each k-mer. Finally, we show the hash values of $S_k$ and $S_l$ in the last rows of each table. In Supplementary Tables~\ref{blend:supptab:kmers7-15-1-32}-~\ref{blend:supptab:kmers7-15-2-16}, we use $k=7$ as the k-mer length and $n=15$ as the number of immediately overlapping k-mers to include in a seed. In Supplementary Tables~\ref{blend:supptab:kmers15-7-1-32}-~\ref{blend:supptab:kmers15-7-2-16}, we use $k=15$ as the k-mer length and $n=7$ as the number of k-mers to include in a seed.

We make two key observations. First, we observe that the hash values of $S_k$ and $S_l$ are equal ($B(S_k) = B(S_l) = $\texttt{0b11000100 01101100 11101001 10110100}) when we use a short k-mer with high number of neighbors even though these two seeds differ by 3 k-mers. Second, the hash values of these two seeds are not equal when we use fewer neighbors with larger k-mers. For $S_k$ we find the hash value $B(S_k) = $\texttt{0b01101000 01000001 01110100 11000000} and for $S_l$ we find $B(S_l) = $\texttt{0b00101101 10110000 01111100 01010011}. We note that the bit positions with large values in their corresponding counter vectors are less likely to differ between two seeds when the seeds have a large number of k-mers in common. This motivates as to design more intelligent hash functions that are aware of the values in the counter vectors to increase the chance of generating the hash value for similar seeds.

\begin{table*}[hbt]
\centering
\caption{Hash Values of the k-mers of seed $S_k$: \texttt{CGGATGCTACAGTATATACCA} for $k=7$ and $n=15$. We show the most significant 16 bits of the counter vector $C(S_k)$. Last row shows the most significant 16 bits of the hash value of the seed.}
\resizebox{\linewidth}{!}{
\begin{tabular}{@{}llrrrrrrrrrrrrrrrr@{}}\toprule
\textbf{K-mer} & \textbf{Hash Value} & C[31] & C[30] & C[29] & C[28] & C[27] & C[26] & C[25] & C[24] & C[23] & C[22] & C[21] & C[20] & C[19] & C[18] & C[17] & C[16]\\\midrule
\texttt{CGGATGC} & \texttt{0b 10100000 01111111 10000110 10110101} & 1 & -1 & 1 & -1 & -1 & -1 & -1 & -1 & -1 & 1  & 1  & 1  & 1  & 1  & 1  & 1 \\
\texttt{GGATGCT} & \texttt{0b 10101101 11110000 01110100 11010000} & 2 & -2 & 2 & -2 & 0  & 0  & -2 & 0  & 0  & 2  & 2  & 2  & 0  & 0  & 0  & 0 \\
\texttt{GATGCTA} & \texttt{0b 01000010 01001011 11011001 10011011} & 1 & -1 & 1 & -3 & -1 & -1 & -1 & -1 & -1 & 3  & 1  & 1  & 1  & -1 & 1  & 1 \\
\texttt{ATGCTAC} & \texttt{0b 11001100 01110101 01010011 00100110} & 2 &  0 & 0 & -4 & 0  & 0  & -2 & -2 & -2 & 4  & 2  & 2  & 0  & 0  & 0  & 2 \\
\texttt{TGCTACA} & \texttt{0b 10110001 01101010 10101001 10100111} & 3 & -1 & 1 & -3 & -1 & -1 & -3 & -1 & -3 & 5  & 3  & 1  & 1  & -1 & 1  & 1 \\
\texttt{GCTACAG} & \texttt{0b 11000101 11010111 11010101 00100101} & 4 & 0 & 0  & -4 & -2 & 0  & -4 & 0  & -2 & 6  & 2  & 2  & 0  & 0  & 2  & 2 \\
\texttt{CTACAGT} & \texttt{0b 11001001 01111101 01001010 10110101} & 5 & 1 & -1 & -5 & -1 & -1 & -5 & 1  & -3 & 7  & 3  & 3  & 1  & 1  & 1  & 3 \\
\texttt{TACAGTA} & \texttt{0b 00101011 01101111 11111000 11111000} & 4 & 0 & 0  & -6 & 0  & -2 & -4 & 2  & -4 & 8  & 4  & 2  & 2  & 2  & 2  & 4 \\
\texttt{ACAGTAT} & \texttt{0b 11100100 01001110 01110101 00011010} & 5 & 1 & 1  & -7 & -1 & -1 & -5 & 1  & -5 & 9  & 3  & 1  & 3  & 3  & 3  & 3 \\
\texttt{CAGTATA} & \texttt{0b 10010010 00001101 00100011 10110100} & 6 & 0 & 0  & -6 & -2 & -2 & -4 & 0  & -6 & 8  & 2  & 0  & 4  & 4  & 2  & 4 \\
\texttt{AGTATAT} & \texttt{0b 01110100 00110000 10101100 00000000} & 5 & 1 & 1  & -5 & -3 & -1 & -5 & -1 & -7 & 7  & 3  & 1  & 3  & 3  & 1  & 3 \\
\texttt{GTATATA} & \texttt{0b 11001111 10110000 11001001 10010110} & 6 & 2 & 0  & -6 & -2 & 0  & -4 & 0  & -6 & 6  & 4  & 2  & 2  & 2  & 0  & 2 \\
\texttt{TATATAC} & \texttt{0b 10000001 00001000 00101111 01111111} & 7 & 1 & -1 & -7 & -3 & -1 & -5 & 1  & -7 & 5  & 3  & 1  & 3  & 1  & -1 & 1 \\
\texttt{ATATACC} & \texttt{0b 11001100 11100000 00101000 11011010} & 8 & 2 & -2 & -8 & -2 & 0  & -6 & 0  & -6 & 6  & 4  & 0  & 2  & 0  & -2 & 0 \\
\texttt{TATACCA} & \texttt{0b 00110100 00000100 11110100 10010100} & 7 & 1 & -1 & -7 & -3 & 1  & -7 & -1 & -7 & 5  & 3  & -1 & 1  & 1  & -3 & -1 \\\cmidrule{1-18}
& & B[31] & B[30] & B[29] & B[28] & B[27] & B[26] & B[25] & B[24] & B[23] & B[22] & B[21] & B[20] & B[19] & B[18] & B[17] & B[16]\\\midrule
& & 1 & 1 & 0 & 0 & 0 & 1 & 0 & 0 & 0 & 1 & 1 & 0 & 1 & 1 & 0 & 0 \\\bottomrule\\
\multicolumn{18}{l}{\footnotesize Best results are highlighted with \textbf{bold} text.} \\
\end{tabular}}
\label{blend:supptab:kmers7-15-1-32}
\end{table*}

\begin{table*}[hbt]
\centering
\caption{Hash Values of the k-mers of seed $S_k$: \texttt{CGGATGCTACAGTATATACCA} for $k=7$ and $n=15$. We show the least significant 16 bits of the counter vector $C(S_k)$. Last row shows the least significant 16 bits of the hash value of the seed.}
\resizebox{\linewidth}{!}{
\begin{tabular}{@{}llrrrrrrrrrrrrrrrr@{}}\toprule
\textbf{K-mer} & \textbf{Hash Value} & C[15] & C[14] & C[13] & C[12] & C[11] & C[10] & C[9] & C[8] & C[7] & C[6] & C[5] & C[4] & C[3] & C[2] & C[1] & C[0]\\\midrule
\texttt{CGGATGC} & \texttt{0b 10100000 01111111 10000110 10110101} & 1  & -1 & -1 & -1 & -1 & 1  & 1  & -1 & 1  & -1 & 1  & 1  & -1 & 1  & -1 & 1 \\
\texttt{GGATGCT} & \texttt{0b 10101101 11110000 01110100 11010000} & 0  & 0  & 0  & 0  & -2 & 2  & 0  & -2 & 2  & 0  & 0  & 2  & -2 & 0  & -2 & 0 \\
\texttt{GATGCTA} & \texttt{0b 01000010 01001011 11011001 10011011} & 1  & 1  & -1 & 1  & -1 & 1  & -1 & -1 & 3  & -1 & -1 & 3  & -1 & -1 & -1 & 1 \\
\texttt{ATGCTAC} & \texttt{0b 11001100 01110101 01010011 00100110} & 0  & 2  & -2 & 2  & -2 & 0  & 0  & 0  & 2  & -2 & 0  & 2  & -2 & 0  & 0  & 0 \\
\texttt{TGCTACA} & \texttt{0b 10110001 01101010 10101001 10100111} & 1  & 1  & -1 & 1  & -1 & -1 & -1 & 1  & 3  & -3 & 1  & 1  & -3 & 1  & 1  & 1 \\
\texttt{GCTACAG} & \texttt{0b 11000101 11010111 11010101 00100101} & 2  & 2  & -2 & 2  & -2 & 0  & -2 & 2  & 2  & -4 & 2  & 0  & -4 & 2  & 0  & 2 \\
\texttt{CTACAGT} & \texttt{0b 11001001 01111101 01001010 10110101} & 1  & 3  & -3 & 1  & -1 & -1 & -1 & 1  & 3  & -5 & 3  & 1  & -5 & 3  & -1 & 3 \\
\texttt{TACAGTA} & \texttt{0b 00101011 01101111 11111000 11111000} & 2  & 4  & -2 & 2  & 0  & -2 & -2 & 0  & 4  & -4 & 4  & 2  & -4 & 2  & -2 & 2 \\
\texttt{ACAGTAT} & \texttt{0b 11100100 01001110 01110101 00011010} & 1  & 5  & -1 & 3  & -1 & -1 & -3 & 1  & 3  & -5 & 3  & 3  & -3 & 1  & -1 & 1 \\
\texttt{CAGTATA} & \texttt{0b 10010010 00001101 00100011 10110100} & 0  & 4  & 0  & 2  & -2 & -2 & -2 & 2  & 4  & -6 & 4  & 4  & -4 & 2  & -2 & 0 \\
\texttt{AGTATAT} & \texttt{0b 01110100 00110000 10101100 00000000} & 1  & 3  & 1  & 1  & -1 & -1 & -3 & 1  & 3  & -7 & 3  & 3  & -5 & 1  & -3 & -1 \\
\texttt{GTATATA} & \texttt{0b 11001111 10110000 11001001 10010110} & 2  & 4  & 0  & 0  & 0  & -2 & -4 & 2  & 4  & -8 & 2  & 4  & -6 & 2  & -2 & -2 \\
\texttt{TATATAC} & \texttt{0b 10000001 00001000 00101111 01111111} & 1  & 3  & 1  & -1 & 1  & -1 & -3 & 3  & 3  & -7 & 3  & 5  & -5 & 3  & -1 & -1 \\
\texttt{ATATACC} & \texttt{0b 11001100 11100000 00101000 11011010} & 0  & 2  & 2  & -2 & 2  & -2 & -4 & 2  & 4  & -6 & 2  & 6  & -4 & 2  & 0  & -2 \\
\texttt{TATACCA} & \texttt{0b 00110100 00000100 11110100 10010100} & 1  & 3  & 3  & -1 & 1  & -1 & -5 & 1  & 5  & -7 & 1  & 7  & -5 & 3  & -1 & -3 \\\cmidrule{1-18}
& & B[15] & B[14] & B[13] & B[12] & B[11] & B[10] & B[9] & B[8] & B[7] & B[6] & B[5] & B[4] & B[3] & B[2] & B[1] & B[0]\\\midrule
& & 1 & 1 & 1 & 0 & 1 & 0 & 0 & 1 & 1 & 0 & 1 & 1 & 0 & 1 & 0 & 0 \\\bottomrule\\
\multicolumn{18}{l}{\footnotesize Best results are highlighted with \textbf{bold} text.} \\
\end{tabular}}
\label{blend:supptab:kmers7-15-1-16}
\end{table*}

\begin{table*}[hbt]
\centering
\caption{Hash Values of the k-mers of seed $S_l$: \texttt{ATGCTACAGTATATACCATCT} for $k=7$ and $n=15$. We show the most significant 16 bits of the counter vector $C(S_l)$. Last row shows the most significant 16 bits of the hash value of the seed.}
\resizebox{\linewidth}{!}{
\begin{tabular}{@{}llrrrrrrrrrrrrrrrr@{}}\toprule
\textbf{K-mer} & \textbf{Hash Value} & C[31] & C[30] & C[29] & C[28] & C[27] & C[26] & C[25] & C[24] & C[23] & C[22] & C[21] & C[20] & C[19] & C[18] & C[17] & C[16]\\\midrule
\texttt{ATGCTAC} & \texttt{0b 11001100 01110101 01010011 00100110} & 1  & 1  & -1 & -1 & 1  & 1  & -1 & -1 & -1 & 1  & 1  & 1  & -1 & 1  & -1 & 1 \\
\texttt{TGCTACA} & \texttt{0b 10110001 01101010 10101001 10100111} & 2  & 0  & 0  & 0  & 0  & 0  & -2 & 0  & -2 & 2  & 2  & 0  & 0  & 0  & 0  & 0 \\
\texttt{GCTACAG} & \texttt{0b 11000101 11010111 11010101 00100101} & 3  & 1  & -1 & -1 & -1 & 1  & -3 & 1  & -1 & 3  & 1  & 1  & -1 & 1  & 1  & 1 \\
\texttt{CTACAGT} & \texttt{0b 11001001 01111101 01001010 10110101} & 4  & 2  & -2 & -2 & 0  & 0  & -4 & 2  & -2 & 4  & 2  & 2  & 0  & 2  & 0  & 2 \\
\texttt{TACAGTA} & \texttt{0b 00101011 01101111 11111000 11111000} & 3  & 1  & -1 & -3 & 1  & -1 & -3 & 3  & -3 & 5  & 3  & 1  & 1  & 3  & 1  & 3 \\
\texttt{ACAGTAT} & \texttt{0b 11100100 01001110 01110101 00011010} & 4  & 2  & 0  & -4 & 0  & 0  & -4 & 2  & -4 & 6  & 2  & 0  & 2  & 4  & 2  & 2 \\
\texttt{CAGTATA} & \texttt{0b 10010010 00001101 00100011 10110100} & 5  & 1  & -1 & -3 & -1 & -1 & -3 & 1  & -5 & 5  & 1  & -1 & 3  & 5  & 1  & 3 \\
\texttt{AGTATAT} & \texttt{0b 01110100 00110000 10101100 00000000} & 4  & 2  & 0  & -2 & -2 & 0  & -4 & 0  & -6 & 4  & 2  & 0  & 2  & 4  & 0  & 2 \\
\texttt{GTATATA} & \texttt{0b 11001111 10110000 11001001 10010110} & 5  & 3  & -1 & -3 & -1 & 1  & -3 & 1  & -5 & 3  & 3  & 1  & 1  & 3  & -1 & 1 \\
\texttt{TATATAC} & \texttt{0b 10000001 00001000 00101111 01111111} & 6  & 2  & -2 & -4 & -2 & 0  & -4 & 2  & -6 & 2  & 2  & 0  & 2  & 2  & -2 & 0 \\
\texttt{ATATACC} & \texttt{0b 11001100 11100000 00101000 11011010} & 7  & 3  & -3 & -5 & -1 & 1  & -5 & 1  & -5 & 3  & 3  & -1 & 1  & 1  & -3 & -1 \\
\texttt{TATACCA} & \texttt{0b 00110100 00000100 11110100 10010100} & 6  & 2  & -2 & -4 & -2 & 2  & -6 & 0  & -6 & 2  & 2  & -2 & 0  & 2  & -4 & -2 \\
\texttt{ATACCAT} & \texttt{0b 00000111 10111111 11010101 01001100} & 5  & 1  & -3 & -5 & -3 & 3  & -5 & 1  & -5 & 1  & 3  & -1 & 1  & 3  & -3 & -1 \\
\texttt{TACCATC} & \texttt{0b 01010110 11100111 00100010 11001101} & 4  & 2  & -4 & -4 & -4 & 4  & -4 & 0  & -4 & 2  & 4  & -2 & 0  & 4  & -2 & 0 \\
\texttt{ACCATCT} & \texttt{0b 00010010 11001000 11001010 11100111} & 3  & 1  & -5 & -3 & -5 & 3  & -3 & -1 & -3 & 3  & 3  & -3 & 1  & 3  & -3 & -1 \\\cmidrule{1-18}
& & B[31] & B[30] & B[29] & B[28] & B[27] & B[26] & B[25] & B[24] & B[23] & B[22] & B[21] & B[20] & B[19] & B[18] & B[17] & B[16]\\\midrule
& & 1 & 1 & 0 & 0 & 0 & 1 & 0 & 0 & 0 & 1 & 1 & 0 & 1 & 1 & 0 & 0 \\\bottomrule\\
\multicolumn{18}{l}{\footnotesize Best results are highlighted with \textbf{bold} text.} \\
\end{tabular}}
\label{blend:supptab:kmers7-15-2-32}
\end{table*}

\clearpage

\begin{table*}[hbt]
\centering
\caption{Hash Values of the k-mers of seed $S_l$: \texttt{ATGCTACAGTATATACCATCT} for $k=7$ and $n=15$. We show the least significant 16 bits of the counter vector $C(S_l)$. Last row shows the least significant 16 bits of the hash value of the seed.}
\resizebox{\linewidth}{!}{
\begin{tabular}{@{}llrrrrrrrrrrrrrrrr@{}}\toprule
\textbf{K-mer} & \textbf{Hash Value} & C[15] & C[14] & C[13] & C[12] & C[11] & C[10] & C[9] & C[8] & C[7] & C[6] & C[5] & C[4] & C[3] & C[2] & C[1] & C[0]\\\midrule
\texttt{ATGCTAC} & \texttt{0b 11001100 01110101 01010011 00100110} & -1 & 1  & -1 & 1  & -1 & -1 & 1  & 1  & -1 & -1 & 1  & -1 & -1 & 1  & 1  & -1  \\
\texttt{TGCTACA} & \texttt{0b 10110001 01101010 10101001 10100111} & 0  & 0  & 0  & 0  & 0  & -2 & 0  & 2  & 0  & -2 & 2  & -2 & -2 & 2  & 2  & 0  \\
\texttt{GCTACAG} & \texttt{0b 11000101 11010111 11010101 00100101} & 1  & 1  & -1 & 1  & -1 & -1 & -1 & 3  & -1 & -3 & 3  & -3 & -3 & 3  & 1  & 1  \\
\texttt{CTACAGT} & \texttt{0b 11001001 01111101 01001010 10110101} & 0  & 2  & -2 & 0  & 0  & -2 & 0  & 2  & 0  & -4 & 4  & -2 & -4 & 4  & 0  & 2  \\
\texttt{TACAGTA} & \texttt{0b 00101011 01101111 11111000 11111000} & 1  & 3  & -1 & 1  & 1  & -3 & -1 & 1  & 1  & -3 & 5  & -1 & -3 & 3  & -1 & 1  \\
\texttt{ACAGTAT} & \texttt{0b 11100100 01001110 01110101 00011010} & 0  & 4  & 0  & 2  & 0  & -2 & -2 & 2  & 0  & -4 & 4  & 0  & -2 & 2  & 0  & 0  \\
\texttt{CAGTATA} & \texttt{0b 10010010 00001101 00100011 10110100} & -1 & 3  & 1  & 1  & -1 & -3 & -1 & 3  & 1  & -5 & 5  & 1  & -3 & 3  & -1 & -1  \\
\texttt{AGTATAT} & \texttt{0b 01110100 00110000 10101100 00000000} & 0  & 2  & 2  & 0  & 0  & -2 & -2 & 2  & 0  & -6 & 4  & 0  & -4 & 2  & -2 & -2  \\
\texttt{GTATATA} & \texttt{0b 11001111 10110000 11001001 10010110} & 1  & 3  & 1  & -1 & 1  & -3 & -3 & 3  & 1  & -7 & 3  & 1  & -5 & 3  & -1 & -3 \\
\texttt{TATATAC} & \texttt{0b 10000001 00001000 00101111 01111111} & 0  & 2  & 2  & -2 & 2  & -2 & -2 & 4  & 0  & -6 & 4  & 2  & -4 & 4  & 0  & -2  \\
\texttt{ATATACC} & \texttt{0b 11001100 11100000 00101000 11011010} & -1 & 1  & 3  & -3 & 3  & -3 & -3 & 3  & 1  & -5 & 3  & 3  & -3 & 3  & 1  & -3  \\
\texttt{TATACCA} & \texttt{0b 00110100 00000100 11110100 10010100} & 0  & 2  & 4  & -2 & 2  & -2 & -4 & 2  & 2  & -6 & 2  & 4  & -4 & 4  & 0  & -4  \\
\texttt{ATACCAT} & \texttt{0b 00000111 10111111 11010101 01001100} & 1  & 3  & 3  & -1 & 1  & -1 & -5 & 3  & 1  & -5 & 1  & 3  & -3 & 5  & -1 & -5  \\
\texttt{TACCATC} & \texttt{0b 01010110 11100111 00100010 11001101} & 0  & 2  & 4  & -2 & 0  & -2 & -4 & 2  & 2  & -4 & 0  & 2  & -2 & 6  & -2 & -4 \\
\texttt{ACCATCT} & \texttt{0b 00010010 11001000 11001010 11100111} & 1  & 3  & 3  & -3 & 1  & -3 & -3 & 1  & 3  & -3 & 1  & 1  & -3 & 7  & -1 & -3 \\\cmidrule{1-18}
& & B[15] & B[14] & B[13] & B[12] & B[11] & B[10] & B[9] & B[8] & B[7] & B[6] & B[5] & B[4] & B[3] & B[2] & B[1] & B[0]\\\midrule
& & 1 & 1 & 1 & 0 & 1 & 0 & 0 & 1 & 1 & 0 & 1 & 1 & 0 & 1 & 0 & 0 \\\bottomrule\\
\multicolumn{18}{l}{\footnotesize Best results are highlighted with \textbf{bold} text.}
\end{tabular}}
\label{blend:supptab:kmers7-15-2-16}
\end{table*}

\begin{table*}[hbt]
\centering
\caption{Hash Values of the k-mers of seed $S_k$: \texttt{CGGATGCTACAGTATATACCA} for $k=15$ and $n=7$. We show the most significant 16 bits of the counter vector $C(S_k)$. Last row shows the most significant 16 bits of the hash value of the seed.}
\resizebox{\linewidth}{!}{
\begin{tabular}{@{}llrrrrrrrrrrrrrrrr@{}}\toprule
\textbf{K-mer} & \textbf{Hash Value} & C[31] & C[30] & C[29] & C[28] & C[27] & C[26] & C[25] & C[24] & C[23] & C[22] & C[21] & C[20] & C[19] & C[18] & C[17] & C[16]\\\midrule
\texttt{CGGATGCTACAGTAT} & \texttt{0b 01001010 11101011 00100110 11001101} & -1 & 1  & -1 & -1 & 1  & -1 & 1  & -1 & 1  & 1  & 1  & -1 & 1  & -1 & 1  & 1  \\
\texttt{GGATGCTACAGTATA} & \texttt{0b 01101100 01000011 11111000 11000000} & -2 & 2  & 0  & -2 & 2  & 0  & 0  & -2 & 0  & 2  & 0  & -2 & 0  & -2 & 2  & 2  \\
\texttt{GATGCTACAGTATAT} & \texttt{0b 01011000 01000101 00110001 11011000} & -3 & 3  & -1 & -1 & 3  & -1 & -1 & -3 & -1 & 3  & -1 & -3 & -1 & -1 & 1  & 3  \\
\texttt{ATGCTACAGTATATA} & \texttt{0b 11100001 01110100 01100010 01000010} & -2 & 4  & 0  & -2 & 2  & -2 & -2 & -2 & -2 & 4  & 0  & -2 & -2 & 0  & 0  & 2  \\
\texttt{TGCTACAGTATATAC} & \texttt{0b 10111100 10011010 00111111 01011011} & -1 & 3  & 1  & -1 & 3  & -1 & -3 & -3 & -1 & 3  & -1 & -1 & -1 & -1 & 1  & 1  \\
\texttt{GCTACAGTATATACC} & \texttt{0b 11101010 01001100 01000100 11100001} & 0  & 2  & 2  & -2 & 4  & -2 & -2 & -4 & -2 & 4  & -2 & -2 & 0  & 0  & 0  & 0  \\
\texttt{CTACAGTATATACCA} & \texttt{0b 00101001 10010001 11111100 01010000} & -1 & 1  & 3  & -3 & 5  & -3 & -3 & -3 & -1 & 3  & -3 & -1 & -1 & -1 & -1 & 1  \\\cmidrule{1-18}
\multicolumn{2}{l}{Seed \texttt{CGGATGCTACAGTATATACCA}} & B[31] & B[30] & B[29] & B[28] & B[27] & B[26] & B[25] & B[24] & B[23] & B[22] & B[21] & B[20] & B[19] & B[18] & B[17] & B[16]\\\midrule
& & 0 & 1 & 1 & 0 & 1 & 0 & 0 & 0 & 0 & 1 & 0 & 0 & 0 & 0 & 0 & 1 \\\bottomrule\\
\multicolumn{18}{l}{\footnotesize Best results are highlighted with \textbf{bold} text.}
\end{tabular}}
\label{blend:supptab:kmers15-7-1-32}
\end{table*}

\begin{table*}[hbt]
\centering
\caption{Hash Values of the k-mers of seed $S_k$: \texttt{CGGATGCTACAGTATATACCA} for $k=15$ and $n=7$. We show the least significant 16 bits of the counter vector $C(S_k)$. Last row shows the least significant 16 bits of the hash value of the seed.}
\resizebox{\linewidth}{!}{
\begin{tabular}{@{}llrrrrrrrrrrrrrrrr@{}}\toprule
\textbf{K-mer} & \textbf{Hash Value} & C[15] & C[14] & C[13] & C[12] & C[11] & C[10] & C[9] & C[8] & C[7] & C[6] & C[5] & C[4] & C[3] & C[2] & C[1] & C[0]\\\midrule
\texttt{CGGATGCTACAGTAT} & \texttt{0b 01001010 11101011 00100110 11001101} & -1 & -1 & 1  & -1 & -1 & 1  & 1  & -1 & 1  & 1  & -1 & -1 & 1  & 1  & -1 & 1  \\
\texttt{GGATGCTACAGTATA} & \texttt{0b 01101100 01000011 11111000 11000000} & 0  & 0  & 2  & 0  & 0  & 0  & 0  & -2 & 2  & 2  & -2 & -2 & 0  & 0  & -2 & 0  \\
\texttt{GATGCTACAGTATAT} & \texttt{0b 01011000 01000101 00110001 11011000} & -1 & -1 & 3  & 1  & -1 & -1 & -1 & -1 & 3  & 3  & -3 & -1 & 1  & -1 & -3 & -1 \\
\texttt{ATGCTACAGTATATA} & \texttt{0b 11100001 01110100 01100010 01000010} & -2 & 0  & 4  & 0  & -2 & -2 & 0  & -2 & 2  & 4  & -4 & -2 & 0  & -2 & -2 & -2 \\
\texttt{TGCTACAGTATATAC} & \texttt{0b 10111100 10011010 00111111 01011011} & -3 & -1 & 5  & 1  & -1 & -1 & 1  & -1 & 1  & 5  & -5 & -1 & 1  & -3 & -1 & -1 \\
\texttt{GCTACAGTATATACC} & \texttt{0b 11101010 01001100 01000100 11100001} & -4 & 0  & 4  & 0  & -2 & 0  & 0  & -2 & 2  & 6  & -4 & -2 & 0  & -4 & -2 & 0  \\
\texttt{CTACAGTATATACCA} & \texttt{0b 00101001 10010001 11111100 01010000} & -3 & 1  & 5  & 1  & -1 & 1  & -1 & -3 & 1  & 7  & -5 & -1 & -1 & -5 & -3 & -1 \\\cmidrule{1-18}
\multicolumn{2}{l}{Seed \texttt{CGGATGCTACAGTATATACCA}} & B[15] & B[14] & B[13] & B[12] & B[11] & B[10] & B[9] & B[8] & B[7] & B[6] & B[5] & B[4] & B[3] & B[2] & B[1] & B[0]\\\midrule
& & 0 & 1 & 1 & 1 & 0 & 1 & 0 & 0 & 1 & 1 & 0 & 0 & 0 & 0 & 0 & 0 \\\bottomrule\\
\multicolumn{18}{l}{\footnotesize Best results are highlighted with \textbf{bold} text.}
\end{tabular}}
\label{blend:supptab:kmers15-7-1-16}
\end{table*}

\begin{table*}[hbt]
\centering
\caption{Hash Values of the k-mers of seed $S_l$: \texttt{ATGCTACAGTATATACCATCT} for $k=15$ and $n=7$. We show the most significant 16 bits of the counter vector $C(S_l)$. Last row shows the most significant 16 bits of the hash value of the seed.}
\resizebox{\linewidth}{!}{
\begin{tabular}{@{}llrrrrrrrrrrrrrrrr@{}}\toprule
\textbf{K-mer} & \textbf{Hash Value} & C[31] & C[30] & C[29] & C[28] & C[27] & C[26] & C[25] & C[24] & C[23] & C[22] & C[21] & C[20] & C[19] & C[18] & C[17] & C[16]\\\midrule
\texttt{ATGCTACAGTATATA} & \texttt{0b 11100001 01110100 01100010 01000010} & 1  & 1  & 1  & -1 & -1 & -1 & -1 & 1  & -1 & 1  & 1  & 1  & -1 & 1  & -1 & -1  \\
\texttt{TGCTACAGTATATAC} & \texttt{0b 10111100 10011010 00111111 01011011} & 2  & 0  & 2  & 0  & 0  & 0  & -2 & 0  & 0  & 0  & 0  & 2  & 0  & 0  & 0  & -2  \\
\texttt{GCTACAGTATATACC} & \texttt{0b 11101010 01001100 01000100 11100001} & 3  & 1  & 3  & -1 & 1  & -1 & -1 & -1 & -1 & 1  & -1 & 1  & 1  & 1  & -1 & -3  \\
\texttt{CTACAGTATATACCA} & \texttt{0b 00101001 10010001 11111100 01010000} & 2  & 0  & 4  & -2 & 2  & -2 & -2 & 0  & 0  & 0  & -2 & 2  & 0  & 0  & -2 & -2  \\
\texttt{TACAGTATATACCAT} & \texttt{0b 00001110 00100000 11011100 11110110} & 1  & -1 & 3  & -3 & 3  & -1 & -1 & -1 & -1 & -1 & -1 & 1  & -1 & -1 & -3 & -3  \\
\texttt{ACAGTATATACCATC} & \texttt{0b 00101111 10111010 00010000 11011111} & 0  & -2 & 4  & -4 & 4  & 0  & 0  & 0  & 0  & -2 & 0  & 2  & 0  & -2 & -2 & -4  \\
\texttt{CAGTATATACCATCT} & \texttt{0b 01100101 11100111 10111011 00111011} & -1 & -1 & 5  & -5 & 5  & 1  & -1 & 1  & 1  & -1 & 1  & 1  & -1 & -1 & -1 & -3 \\\cmidrule{1-18}
\multicolumn{2}{l}{Seed \texttt{ATGCTACAGTATATACCATCT}} & B[31] & B[30] & B[29] & B[28] & B[27] & B[26] & B[25] & B[24] & B[23] & B[22] & B[21] & B[20] & B[19] & B[18] & B[17] & B[16]\\\midrule
& & 0 & 0 & 1 & 0 & 1 & 1 & 0 & 1 & 1 & 0 & 1 & 1 & 0 & 0 & 0 & 0 \\\bottomrule\\
\multicolumn{18}{l}{\footnotesize Best results are highlighted with \textbf{bold} text.}
\end{tabular}}
\label{blend:supptab:kmers15-7-2-32}
\end{table*}

\begin{table*}[hbt]
\centering
\caption{Hash Values of the k-mers of seed $S_l$: \texttt{ATGCTACAGTATATACCATCT} for $k=15$ and $n=7$. We show the least significant 16 bits of the counter vector $C(S_l)$. Last row shows the least significant 16 bits of the hash value of the seed.}
\resizebox{\linewidth}{!}{
\begin{tabular}{@{}llrrrrrrrrrrrrrrrr@{}}\toprule
\textbf{K-mer} & \textbf{Hash Value} & C[15] & C[14] & C[13] & C[12] & C[11] & C[10] & C[9] & C[8] & C[7] & C[6] & C[5] & C[4] & C[3] & C[2] & C[1] & C[0]\\\midrule
\texttt{ATGCTACAGTATATA} & \texttt{0b 11100001 01110100 01100010 01000010} & -1 & 1  & 1  & -1 & -1 & -1 & 1  & -1 & -1 & 1  & -1 & -1 & -1 & -1 & 1  & -1 \\
\texttt{TGCTACAGTATATAC} & \texttt{0b 10111100 10011010 00111111 01011011} & -2 & 0  & 2  & 0  & 0  & 0  & 2  & 0  & -2 & 2  & -2 & 0  & 0  & -2 & 2  & 0  \\
\texttt{GCTACAGTATATACC} & \texttt{0b 11101010 01001100 01000100 11100001} & -3 & 1  & 1  & -1 & -1 & 1  & 1  & -1 & -1 & 3  & -1 & -1 & -1 & -3 & 1  & 1  \\
\texttt{CTACAGTATATACCA} & \texttt{0b 00101001 10010001 11111100 01010000} & -2 & 2  & 2  & 0  & 0  & 2  & 0  & -2 & -2 & 4  & -2 & 0  & -2 & -4 & 0  & 0  \\
\texttt{TACAGTATATACCAT} & \texttt{0b 00001110 00100000 11011100 11110110} & -1 & 3  & 1  & 1  & 1  & 3  & -1 & -3 & -1 & 5  & -1 & 1  & -3 & -3 & 1  & -1  \\
\texttt{ACAGTATATACCATC} & \texttt{0b 00101111 10111010 00010000 11011111} & -2 & 2  & 0  & 2  & 0  & 2  & -2 & -4 & 0  & 6  & -2 & 2  & -2 & -2 & 2  & 0  \\
\texttt{CAGTATATACCATCT} & \texttt{0b 01100101 11100111 10111011 00111011} & -1 & 1  & 1  & 3  & 1  & 1  & -1 & -3 & -1 & 5  & -1 & 3  & -1 & -3 & 3  & 1 \\\cmidrule{1-18}
\multicolumn{2}{l}{Seed \texttt{ATGCTACAGTATATACCATCT}} & B[15] & B[14] & B[13] & B[12] & B[11] & B[10] & B[9] & B[8] & B[7] & B[6] & B[5] & B[4] & B[3] & B[2] & B[1] & B[0]\\\midrule
& & 0 & 1 & 1 & 1 & 1 & 1 & 0 & 0 & 0 & 1 & 0 & 1 & 0 & 0 & 1 & 1 \\\bottomrule\\
\multicolumn{18}{l}{\footnotesize Best results are highlighted with \textbf{bold} text.}
\end{tabular}}
\label{blend:supptab:kmers15-7-2-16}
\end{table*}

\clearpage

\subsection{Parameter Exploration}
\subsubsection{The trade-off between \texttt{\blend-I} and \texttt{\blend-S}}\label{blend:suppsec:tradeoff-blend}

\begin{figure}[tbh]
\centering
\includegraphics[width=\linewidth]{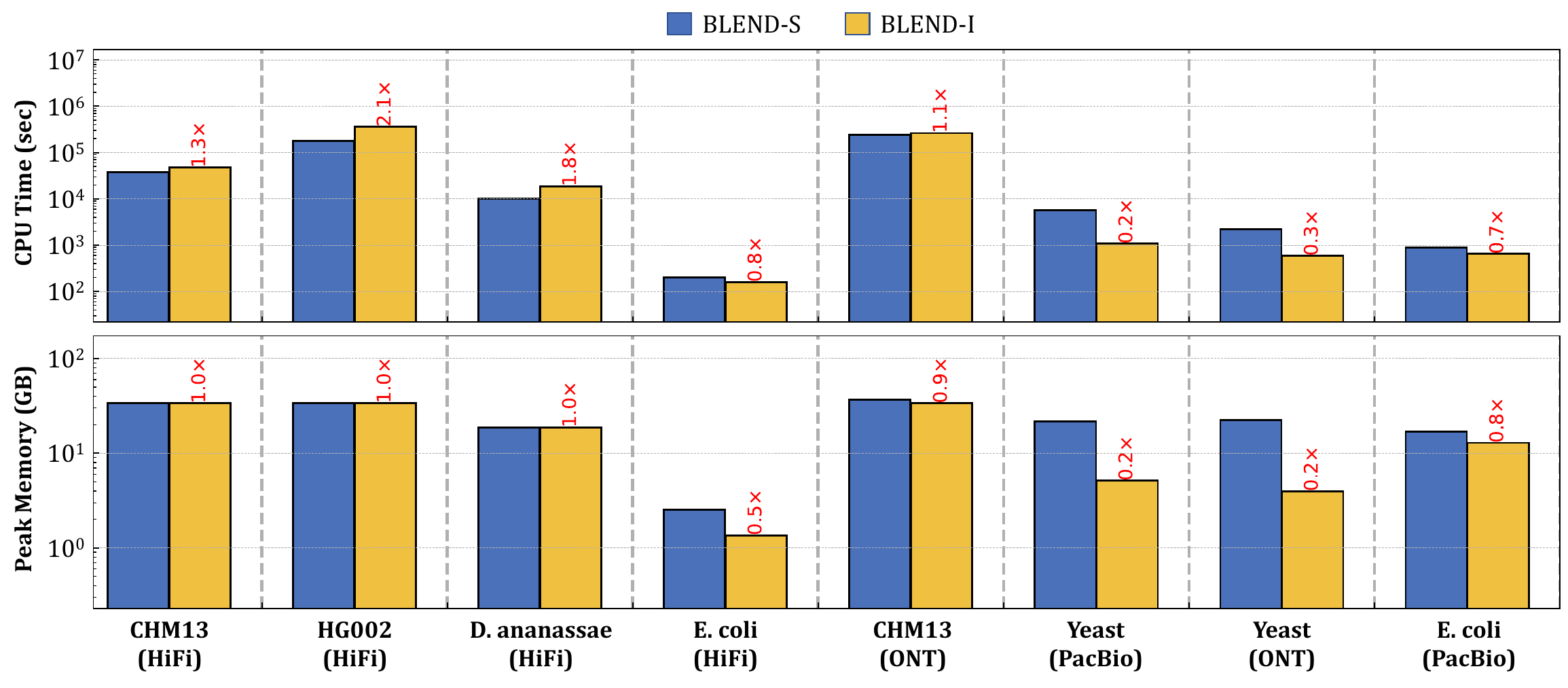}
\caption{CPU time and peak memory footprint comparisons of read mapping.}
\label{blend:suppfig:read_mapping_perf-blend}
\end{figure}

Our goal is to show the performance and accuracy trade-offs between the seeding techniques that \blend supports: \texttt{\blend-I} and \texttt{\blend-S}. In Supplementary Figures~\ref{blend:suppfig:overlap_perf-blend} and ~\ref{blend:suppfig:read_mapping_perf-blend}, we show the performance and peak memory usage comparisons when using \texttt{\blend-I} and \texttt{\blend-S} as the seeding technique by keeping all the other relevant parameters identical (e.g., number of k-mers to include in a seed $n$, window length $w$). In Supplementary Table~\ref{blend:supptab:overlap_assembly-blend}, we show the assembly quality comparisons in terms of the accuracy and contiguity of the assemblies that we generate using the overlaps that \texttt{\blend-I} and \texttt{\blend-S} find. In Supplementary Tables~\ref{blend:supptab:mapping_quality-blend} and ~\ref{blend:supptab:mapping_accuracy-blend}, we show the read mapping quality and accuracy results using these two seeding techniques, respectively.

We also show the values for different parameters we test with \blend in Supplementary Table~\ref{blend:supptab:parameter_exploration}. We determine the default parameters of \blend empirically based on the combination of best performance, memory overhead, and accuracy results.

\begin{figure}[tbh]
\centering
\includegraphics[width=\linewidth]{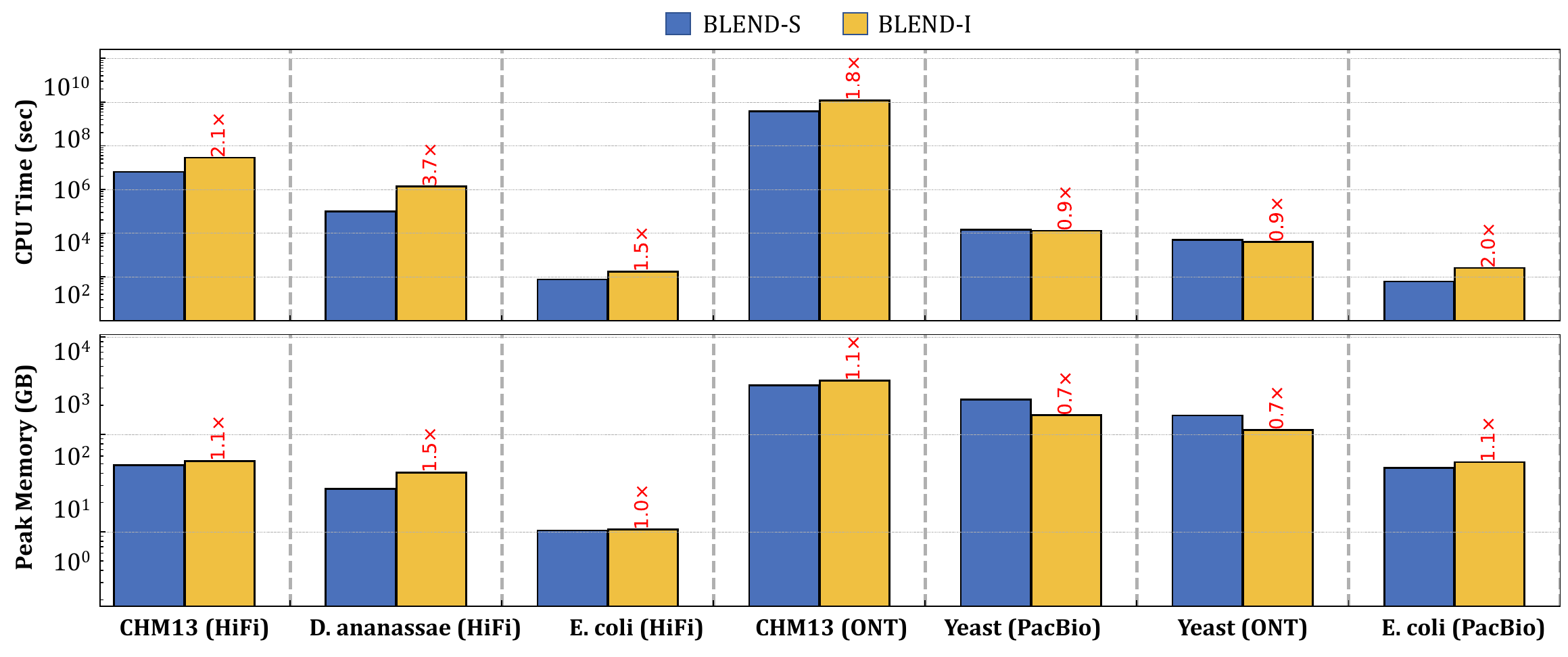}
\caption{CPU time and peak memory footprint comparisons of read overlapping.}
\label{blend:suppfig:overlap_perf-blend}
\end{figure}

\begin{table}[tbh]
\centering
\caption{Assembly quality comparisons between \texttt{\blend-I} and \texttt{\blend-S}.}
\resizebox{\linewidth}{!}{
\begin{tabular}{@{}clrrrrrrrrrr@{}}\toprule
\textbf{Dataset} 	  & \textbf{Tool} 		 & \textbf{Average}	      & \textbf{Genome}		    & \textbf{K-mer}	   & \textbf{Aligned}	   & \textbf{Mismatch per} & \textbf{Average} & \textbf{Assembly}	  & \textbf{Largest} 	  & \textbf{NGA50}    & \textbf{NG50}	  \\
					  & 			  		 & \textbf{Identity (\%)} & \textbf{Fraction (\%)}  & \textbf{Compl. (\%)} & \textbf{Length (Mbp)} & \textbf{100Kbp (\#)}  & \textbf{GC (\%)} & \textbf{Length (Mbp)} & \textbf{Contig (Mbp)} & \textbf{(Kbp)}    & \textbf{(Kbp)}    \\\midrule
\emph{CHM13} 	  	  & \texttt{\blend-I} & 99.7535	              & 96.7203		            & 83.65	               & 3,054.49 			   & 48.49	  	           & \textbf{40.79}	  & \textbf{3,059.29} 	  & \textbf{41.8342} 	  & \textbf{8,507.53} & \textbf{8,508.92} \\
(HiFi)				  & \texttt{\blend-S} & \textbf{99.8526}	      & \textbf{98.4847}		& \textbf{90.15}	   & \textbf{3,092.54} 	   & \textbf{22.02}	  	   & 40.78	          & 3,095.21 	          & 22.8397 			  & 5,442.25 		  & 5,442.31 		  \\
					  & Reference 	  		 & 100 				      & 100 					& 100 				   & 3,054.83 			   & 0.00 				   & 40.85 			  & 3,054.83 			  & 248.387 			  & 154,260           & 154,260 	   	  \\\midrule
\emph{D. ananassae}   & \texttt{\blend-I} & 99.6890	              & 97.2290		            & 77.85	               & \textbf{270.218} 	   & 233.18 	           & 41.95	          & 280.388 	          & 5.01099 	          & 356.745           & 356.745           \\
(HiFi)				  & \texttt{\blend-S} & \textbf{99.7856}	      & \textbf{97.2308}		& \textbf{86.43}	   & 240.391 			   & \textbf{143.13} 	   & \textbf{41.75}	  & \textbf{247.153} 	  & \textbf{6.23256} 	  & \textbf{792.407}  & \textbf{798.913}  \\
					  & Reference 	  		 & 100 				      & 100 					& 100 				   & 213.805 			   & 0.00 				   & 41.81 			  & 213.818 			  & 30.6728 			  & 26,427.4 		  & 26,427.4 		  \\\midrule
\emph{E. coli} 		  & \texttt{\blend-I} & 99.6902                & \textbf{99.8824} 		& 79.36	               & 5.04157 	           & 17.92	  	           & \textbf{50.52}	  & \textbf{5.04263} 	  & \textbf{4.94601}	  & \textbf{4,025.48} & \textbf{4,946.01} \\
(HiFi)				  & \texttt{\blend-S} & \textbf{99.8320} 	  & 99.8801 		        & \textbf{87.91}	   & \textbf{5.12155} 	   & \textbf{3.77}	  	   & 50.53	          & 5.12155 			  & 3.41699	              & 3,416.99          & 3,416.99          \\
					  & Reference 	  		 & 100 				      & 100 					& 100 				   & 5.04628 			   & 0.00 				   & 50.52 			  & 5.04628 			  & 4.94446 			  & 4,944.46 		  & 4,944.46 		  \\\midrule
\emph{CHM13} 	  	  & \texttt{\blend-I} & N/A	   			      & N/A					    & \textbf{29.26}	   & \textbf{2,891.28} 	   & 4,077.53	           & \textbf{41.32}	  & \textbf{2,897.87} 	  & \textbf{25.2071} 	  & \textbf{5,061.52} & \textbf{5,178.59} \\
(ONT)				  & \texttt{\blend-S} & N/A	   			      & N/A					    & 0 			       & 0.010546 	   		   & \textbf{3,250.70} 	   & 51.30 			  & 0.010548 	          & 0.010548	          & 0                 & 0                 \\
					  & Reference 	  		 & 100 				      & 100 					& 100 			   	   & 3,117.29 			   & 0.00 				   & 40.75 			  & 3,117.29 			  & 248.387 			  & 150,617 	   	  & 150,617 	      \\\midrule
\emph{Yeast} 		  & \texttt{\blend-I} & 89.1677 	   		      & \textbf{97.0854} 		& \textbf{33.81}	   & 12.3938 	           & \textbf{2,672.37} 	   & 38.84	  		  & \textbf{12.4176} 	  & \textbf{1.54807} 	  & \textbf{635.966}  & \textbf{636.669}  \\
(PacBio)			  & \texttt{\blend-S} & \textbf{90.3347}       & 83.8814 				& 33.17 			   & \textbf{22.9473} 	   & 4,795.58 			   & \textbf{38.71}   & 22.9523 	          & 0.265118	          & 114.125           & 116.143           \\
					  & Reference 	  		 & 100 				      & 100 					& 100 				   & 12.1571 			   & 0.00 				   & 38.15 			  & 12.1571 			  & 1.53193 			  & 924.431 		  & 924.431 		  \\\midrule
\emph{Yeast} 	      & \texttt{\blend-I} & 89.6889                & \textbf{99.2974} 	    & \textbf{35.95}	   & \textbf{12.3222} 	   & 2,529.47 	           & 38.64	          & \textbf{12.3225} 	  & \textbf{1.10582} 	  & \textbf{793.046}  & \textbf{793.046}  \\
(ONT)			      & \texttt{\blend-S} & \textbf{91.0865} 	  & 7.9798		            & 4.90 			       & 0.898565 			   & \textbf{2,006.91} 	   & \textbf{38.35}   & 0.899654 			  & 0.043321	          & 0                 & 0                 \\
					  & Reference 	  		 & 100 				      & 100 					& 100 				   & 12.1571 			   & 0.00 				   & 38.15 			  & 12.1571 			  & 1.53193 			  & 924.431 		  & 924.431 		  \\\midrule
\emph{E. coli} 		  & \texttt{\blend-I} & 88.5806                & \textbf{96.5238} 		& \textbf{32.32}	   & \textbf{5.90024} 	   & 1,857.56	           & \textbf{49.81}	  & \textbf{6.21598} 	  & \textbf{2.40671}	  & \textbf{769.981}  & \textbf{2,060.4}  \\
(PacBio)			  & \texttt{\blend-S} & \textbf{90.3551} 	  & 36.6230 				& 17.07 			   & 2.10137 			   & \textbf{1,299.50} 	   & 48.91 			  & 2.10704 	          & 0.095505 	          & 0    	   	      & 0                 \\
					  & Reference 	  		 & 100 				      & 100 					& 100 				   & 5.6394 			   & 0.00 				   & 50.43 			  & 5.6394 				  & 5.54732 			  & 5,547.32 		  & 5,547.32 		  \\\bottomrule
\multicolumn{12}{l}{\footnotesize Best results are highlighted with \textbf{bold} text. For most metrics, the best results are the ones closest to the corresponding value of the reference genome.}\\
\multicolumn{12}{l}{\footnotesize The best results for \emph{Aligned Length} are determined by the highest number within each dataset. We do not highlight the reference results as the best results.}\\
\multicolumn{12}{l}{\footnotesize N/A indicates that we could not generate the corresponding result because tool, QUAST, or dnadiff failed to generate the statistic.} \\
\end{tabular}

}
\label{blend:supptab:overlap_assembly-blend}
\end{table}

\clearpage

\begin{table}[tbh]
\centering
\caption{Read mapping quality comparisons between \texttt{\blend-I} and \texttt{\blend-S}.}
\resizebox{0.7\linewidth}{!}{
\begin{tabular}{@{}llrrrr@{}}\toprule
\textbf{Dataset} 		& \textbf{Tool} 		& \textbf{Average} 			 & \textbf{Breadth of} & \textbf{Aligned} 	 & \textbf{Properly}\\
				 		& 						& \textbf{Depth of} 		 & \textbf{Coverage}   & \textbf{Reads} 	 & \textbf{Paired}\\
				 		& 						& \textbf{Cov. (${\times}$)} & \textbf{(\%)} 	   & \textbf{(\#)} 		 & \textbf{(\%)}\\\midrule
\emph{CHM13} 	        & \texttt{\blend-I} 	& 16.58 			 		 & 99.991 	   		   & \textbf{3,172,305}  & NA \\
(HiFi)					& \texttt{\blend-S} 	& 16.58 			 		 & 99.991 	   		   & 3,171,916  		 & NA \\\midrule
\emph{HG002} 	        & \texttt{\blend-I} 	& \textbf{51.25} 			 & \textbf{92.245} 	   & 6,813,886           & NA \\
(HiFi)					& \texttt{\blend-S} 	& 11.24 			         & 13.860 	           & \textbf{11,424,762} & NA \\\midrule
\emph{D. ananassae} 	& \texttt{\blend-I} 	& \textbf{57.51} 			 & 99.650 			   & \textbf{1,249,666}  & NA \\
(HiFi)					& \texttt{\blend-S} 	& 57.37 					 & \textbf{99.662} 	   & 1,223,388 			 & NA \\\midrule
\emph{E. coli} 			& \texttt{\blend-I} 	& 99.14 			 		 & 99.897 			   & \textbf{39,064} 	 & NA \\
(HiFi)					& \texttt{\blend-S} 	& 99.14 			 		 & 99.897 			   & 39,048 			 & NA \\\midrule
\emph{CHM13} 	        & \texttt{\blend-I} 	& \textbf{29.34} 			 & \textbf{99.999} 	   & \textbf{10,322,767} & NA \\
(ONT)			        & \texttt{\blend-S} 	& 17.51 			 		 & 99.700 	   		   & 5,760,401 			 & NA \\\midrule
\emph{Yeast} 	        & \texttt{\blend-I} 	& \textbf{195.87} 			 & \textbf{99.980} 	   & \textbf{270,064}    & NA \\
(PacBio)		        & \texttt{\blend-S} 	& 142.31 			 		 & 99.975 	   		   & 179,039    		 & NA \\\midrule
\emph{Yeast} 		    & \texttt{\blend-I} 	& \textbf{97.88} 			 & \textbf{99.964} 	   & \textbf{134,919} 	 & NA \\
(ONT)			        & \texttt{\blend-S} 	& 59.57 			 		 & 99.906 	   		   & 75,110 	 		 & NA \\\midrule
\emph{E. coli} 			& \texttt{\blend-I} 	& \textbf{97.51} 			 & 100 			       & \textbf{83,924}     & NA \\
(PacBio)			    & \texttt{\blend-S} 	& 56.87 			 		 & 100 			       & 40,694 			 & NA \\\bottomrule
\multicolumn{6}{l}{\footnotesize Best results are highlighted with \textbf{bold} text.} \\
\multicolumn{6}{l}{\footnotesize Properly paired rate is only available for paired-end Illumina reads.} \\
\end{tabular}

}
\label{blend:supptab:mapping_quality-blend}
\end{table}

\begin{table}[tbh]
\centering
\caption{Read mapping accuracy comparisons between \texttt{\blend-I} and \texttt{\blend-S}.}
\resizebox{0.4\linewidth}{!}{
\begin{tabular}{@{}lrrr@{}}\toprule
\textbf{Dataset} 		& \multicolumn{2}{c}{\textbf{Overall Error Rate (\%)}} \\\cmidrule{2-4}
				 		& \texttt{\blend-I} & \texttt{\blend-S} \\\midrule
\emph{CHM13} (ONT) 	    & \textbf{1.5168427}   & 5.996888    		  \\\midrule
\emph{Yeast} (PacBio) 	& \textbf{0.2403134}   & 0.6959378            \\\midrule
\emph{Yeast} (ONT) 		& \textbf{0.2386617}   & 0.6284117            \\\bottomrule
\multicolumn{4}{l}{\footnotesize Best results are highlighted with \textbf{bold} text.} \\
\end{tabular}

}
\label{blend:supptab:mapping_accuracy-blend}
\end{table}

\begin{table}[tbh]
\centering
\caption{Performance, memory, and accuracy comparisons using different parameter settings in \blend.}
\resizebox{\linewidth}{!}{
\begin{tabular}{@{}lrrrrrrr@{}}\toprule
\textbf{Tool} 	   & \textbf{K-mer} 	   & \textbf{\# of k-mers} 	  & \textbf{Window} 	  & \textbf{CPU Time}  & \textbf{Peak}		  & \textbf{Average} 	   & \textbf{Genome}	    \\
 				   & \textbf{Length ($k$)} & \textbf{in a Seed ($n$)} & \textbf{Length ($w$)} & \textbf{(seconds)} & \textbf{Memory (KB)} & \textbf{Identity (\%)} & \textbf{Fraction (\%)} \\\midrule
\blend 		   & 9					   & 11						  & 200					  &	62.38			   & 1,115,384			  & 99.7255			  	   & 99.8502	  			\\
\blend 		   & 9					   & 13						  & 200					  &	58.13			   & 994,120				  & 99.7294			  	   & 99.7808	  			\\
\blend 		   & 9					   & 15						  & 200					  &	49.79			   & 1,030,148			  & 99.7411			  	   & 99.7619	  			\\
\blend 		   & 9					   & 17						  & 200					  &	45.03			   & 960,080				  & 99.7302			  	   & 99.7460	  			\\
\blend 		   & 9					   & 21						  & 200					  &	36.84			   & 976,456				  & 99.7257			  	   & 99.6640	  			\\\cmidrule{1-8}
\blend 		   & 15					   & 5						  & 200					  &	83.05			   & 1,168,612			  & 99.6735			  	   & 99.7625	  			\\
\blend 		   & 15					   & 7						  & 200					  &	74.93			   & 1,137,360			  & 99.7009			  	   & 99.5874	  			\\
\blend 		   & 15					   & 11						  & 200					  &	58.09			   & 1,051,912			  & 99.7149			  	   & 99.1166	  			\\\cmidrule{1-8}
\blend 		   & 19					   & 5						  & 200					  &	77.16			   & 1,130,604			  & 99.7312			  	   & 99.8802	  			\\
\blend 		   & 19					   & 7						  & 200					  &	50.50			   & 1,078,596			  & 99.7880			  	   & 99.8424	  			\\
\blend 		   & 19					   & 11						  & 200					  &	46.26			   & 977,060				  & 99.8078			  	   & 99.6438	  			\\\cmidrule{1-8}
\blend 		   & 21					   & 5						  & 200					  &	67.85			   & 1,116,684			  & 99.7472			  	   & 99.8835	  			\\
\blend 		   & 21					   & 7						  & 200					  &	61.63			   & 1,042,724			  & 99.7969			  	   & 99.8605	  			\\
\blend 		   & 21					   & 11						  & 200					  &	42.35			   & 969,184				  & 99.8340			  	   & 99.7515	  			\\\cmidrule{1-8}
\blend 		   & 25					   & 5						  & 200					  &	65.61			   & 1,057,804			  & 99.7769			  	   & 99.8818	  			\\
\blend 		   & 25					   & 7						  & 200					  &	54.88			   & 1,029,888			  & 99.8320			  	   & 99.8801	  			\\
\blend 		   & 25					   & 11						  & 200					  &	37.01			   & 936,260				  & 99.8646			  	   & 99.8001	  			\\
\blend 		   & 25					   & 15						  & 200					  &	29.83			   & 866,208				  & 99.8838			  	   & 99.7307	  			\\
\blend 		   & 25					   & 17						  & 200					  &	29.59			   & 826,456				  & 99.8784			  	   & 99.7521	  			\\
\blend 		   & 25					   & 21						  & 200					  &	26.09			   & 791,736				  & 99.8774			  	   & 99.6955	  			\\\cmidrule{1-8}
\blend 		   & 9					   & 11						  & 50					  &	263.82			   & 1,786,516			  & 99.7013			  	   & 99.8612	  			\\
\blend 		   & 9					   & 13						  & 50					  &	411.24			   & 1,805,800			  & 99.6995			  	   & 99.8573	  			\\
\blend 		   & 9					   & 15						  & 50					  &	271.00			   & 1,729,784			  & 99.6798			  	   & 99.8517	  			\\
\blend 		   & 9					   & 17						  & 50					  &	238.52			   & 1,690,912			  & 99.6690			  	   & 99.8083	  			\\
\blend 		   & 9					   & 21						  & 50					  &	206.76			   & 1,725,168			  & 99.6496			  	   & 99.8150	  			\\\cmidrule{1-8}
\blend 		   & 15					   & 5						  & 50					  &	330.84			   & 1,785,456			  & 99.6634			  	   & 99.8604	  			\\
\blend 		   & 15					   & 7						  & 50					  &	337.95			   & 1,812,052			  & 99.6280			  	   & 99.8177	  			\\
\blend 		   & 15					   & 11						  & 50					  &	236.82			   & 1,803,816			  & 99.5831			  	   & 99.6893	  			\\\cmidrule{1-8}
\blend 		   & 19					   & 5						  & 50					  &	328.67			   & 1,692,248			  & 99.7077			  	   & 99.8794	  			\\
\blend 		   & 19					   & 7						  & 50					  &	295.57			   & 1,713,940			  & 99.7188			  	   & 99.8579	  			\\
\blend 		   & 19					   & 11						  & 50					  &	201.79			   & 1,700,412			  & 99.7015			  	   & 99.8578	  			\\\cmidrule{1-8}
\blend 		   & 21					   & 5						  & 50					  &	378.58			   & 1,625,388			  & 99.7120			  	   & 99.8832	  			\\
\blend 		   & 21					   & 7						  & 50					  &	278.56			   & 1,695,476			  & 99.7333			  	   & 99.8832	  			\\
\blend 		   & 21					   & 11						  & 50					  &	189.33			   & 1,694,820			  & 99.7623			  	   & 99.8594	  			\\\cmidrule{1-8}
\blend 		   & 25					   & 5						  & 50					  &	323.69			   & 1,685,304			  & 99.7272			  	   & 99.8831	  			\\
\blend 		   & 25					   & 7						  & 50					  &	211.78			   & 1,647,984			  & 99.7722			  	   & 99.8831	  			\\
\blend 		   & 25					   & 11						  & 50					  &	170.60			   & 1,683,736			  & 99.8094			  	   & 99.8866	  			\\
\blend 		   & 25					   & 15						  & 50					  &	142.42			   & 1,622,452			  & 99.8170			  	   & 99.8576	  			\\
\blend 		   & 25					   & 17						  & 50					  &	103.96			   & 1,590,776			  & 99.8073			  	   & 99.8206	  			\\
\blend 		   & 25					   & 21						  & 50					  &	109.62			   & 1,548,228			  & 99.7792			  	   & 99.7880	  			\\\cmidrule{1-8}
\blend 		   & 9					   & 11						  & 20					  &	837.50			   & 2,769,552			  & 99.6916			  	   & 99.8784	  			\\
\blend 		   & 9					   & 13						  & 20					  &	813.50			   & 2,765,480			  & 99.6834			  	   & 99.8785	  			\\
\blend 		   & 9					   & 15						  & 20					  &	764.91			   & 2,795,848			  & 99.6797			  	   & 99.8756	  			\\
\blend 		   & 9					   & 17						  & 20					  &	739.52			   & 2,823,188			  & 99.6801			  	   & 99.8802	  			\\\cmidrule{1-8}
\multicolumn{8}{l}{\footnotesize We use the \emph{E.coli} dataset for all these runs} \\
\end{tabular}

\label{blend:supptab:parameter_exploration}
}
\end{table}

\clearpage

\subsubsection{The trade-off between \blend and minimap2}\label{blend:suppsec:tradeoff-eq}
Our goal is to compare \blend and minimap2 using the same set of parameters that \blend uses when generating its results. To achieve this, we control the following two conditions. First, we ensure that we use the same seeding technique that minimap2 uses. To this end, we use the \texttt{\blend-I} seeding technique, which uses minimizers as seeds. We should note that \texttt{\blend-I} does not always provide the best results in terms of performance or accuracy for the HiFi reads as the default seeding technique is \texttt{\blend-S} for HiFi datasets in \blend.

Second, we use the same seed length when we compare \blend with minimap2. In minimap2, the seed length is the same as the k-mer length as minimap2 finds the minimizer k-mers from the hash values of k-mers. The seed length in \texttt{\blend-I} is determined by \emph{both} the k-mer length and the number of k-mers that we include in a seed (i.e., $n$). For example, \blend uses the \texttt{\blend-I} seeding technique with the k-mer length $k=19$ and the number of neighbors $n=5$ for the PacBio reads. Combining immediately overlapping $5$-many $19$-mers generates seeds with length $19+5-1=23$. Thus, \texttt{\blend-I} uses seeds of length $23$ based on these parameters. Supplementary Table~\ref{blend:supptab:pardef} shows the seed length calculation for both \texttt{\blend-I} and \texttt{\blend-S}. We calculate the seed lengths for the datasets where \blend uses \texttt{\blend-I} as the default option (i.e., the PacBio and ONT datasets) in read overlapping. We note that \blend uses the same seed length and window length as in minimap2 for mapping long reads. Thus, we do not report the read mapping results in this section, which are already reported in the main paper when comparing \blend with minimap2. To run minimap2 with the same parameter conditions, we apply the same seed length and the window length that \blend uses to minimap2 using the $k$ and $w$ parameters, respectively. We show these parameters in Supplementary Table~\ref{blend:supptab:ovpars} (minimap-Eq). In the results we show below, minimap-Eq indicates the runs of minimap2 when using the same set of parameters that \blend uses with the \texttt{\blend-I} technique.

In Supplementary Figure~\ref{blend:suppfig:overlap_perf-eq}, we show the performance and peak memory comparisons when using \blend with the \texttt{\blend-I} seeding technique, minimap2, and minimap2-Eq. In Supplementary Table~\ref{blend:supptab:overlap_assembly-eq}, we show the assembly quality comparisons in terms of the accuracy and contiguity of the assemblies that we generate using the overlaps that each tool finds.

\begin{figure}[tbh]
\centering
\includegraphics[width=0.7\linewidth]{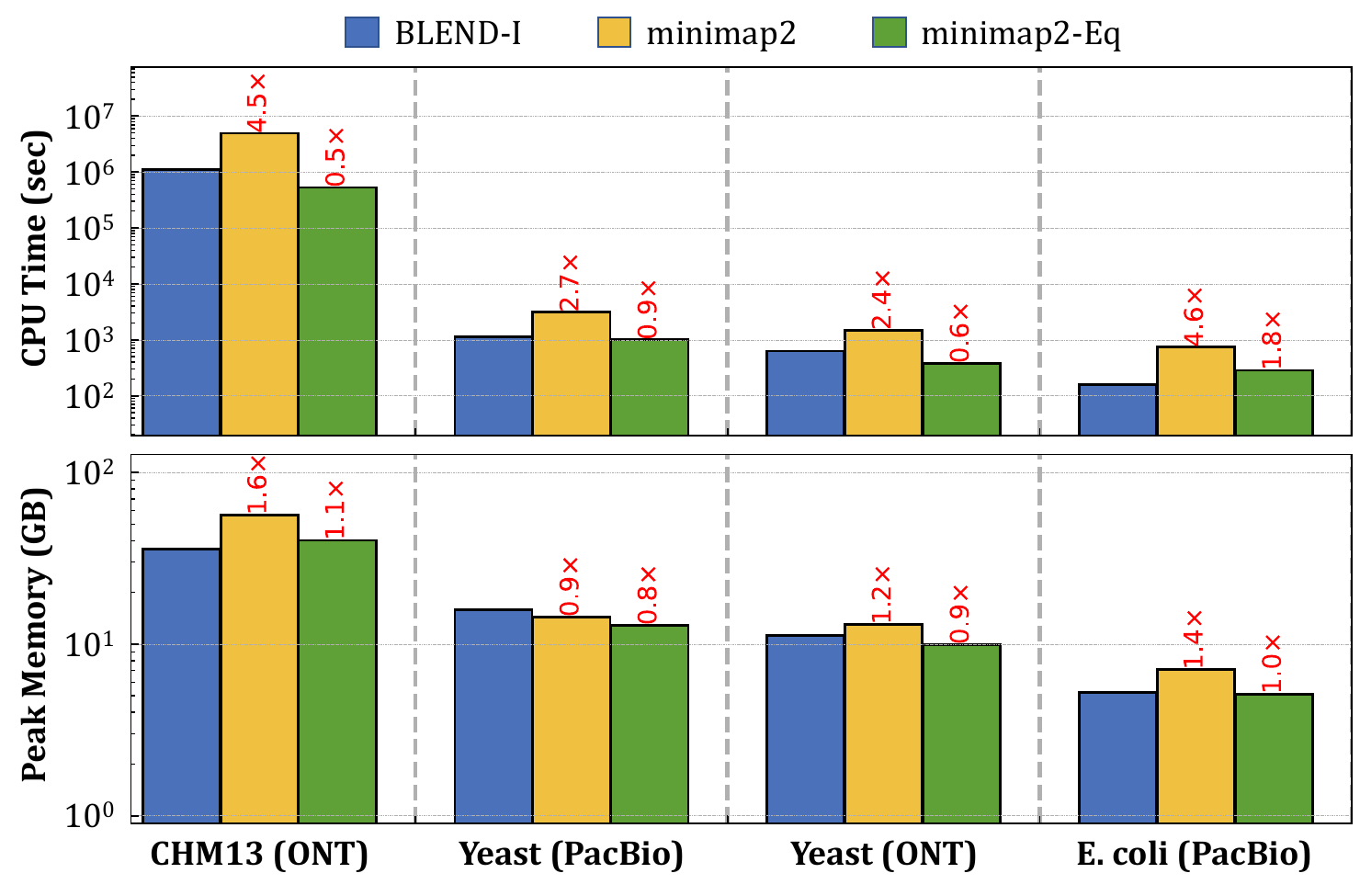}
\caption{CPU time and peak memory footprint comparisons of read overlapping.}
\label{blend:suppfig:overlap_perf-eq}
\end{figure}

\begin{table}[tbh]
\centering
\caption{Assembly quality comparisons when using the parameters equivalent to \texttt{\blend-I}.}
\resizebox{\linewidth}{!}{
\begin{tabular}{@{}clrrrrrrrrrr@{}}\toprule
\textbf{Dataset} 	  & \textbf{Tool} 		 & \textbf{Average}	      & \textbf{Genome}		    & \textbf{K-mer}	   & \textbf{Aligned}	   & \textbf{Mismatch per} & \textbf{Average} & \textbf{Assembly}	  & \textbf{Largest} 	  & \textbf{NGA50}    & \textbf{NG50}	  \\
					  & 			  		 & \textbf{Identity (\%)} & \textbf{Fraction (\%)}  & \textbf{Compl. (\%)} & \textbf{Length (Mbp)} & \textbf{100Kbp (\#)}  & \textbf{GC (\%)} & \textbf{Length (Mbp)} & \textbf{Contig (Mbp)} & \textbf{(Kbp)}    & \textbf{(Kbp)}    \\\midrule
\emph{CHM13} 	  	  & \texttt{\blend-I} & N/A	   			      & N/A					    & 29.26	   			   & 2,891.28 	   		   & 4,077.53	           & \textbf{41.32}	  & 2,897.87 	  		  & 25.2071 	  		  & 5,061.52 		  & 5,178.59 		  \\
(ONT)				  & minimap2 	  		 & N/A	   				  & N/A						& 28.32 			   & 2,860.26 	   		   & 4,660.73 			   & 41.36 			  & \textbf{2,908.55} 	  & \textbf{66.7564}	  & \textbf{13,189.2} & \textbf{13,820.3} \\
					  & minimap2-Eq 		 & N/A	   			      & N/A					    & \textbf{29.32} 	   & \textbf{3,117.29} 	   & \textbf{4,025.22} 	   & \textbf{41.32}   & 2,882.94 	          & 24.6651	          	  & 3,634.05          & 3,653.47          \\
					  & Reference 	  		 & 100 				      & 100 					& 100 			   	   & 3,117.29 			   & 0.00 				   & 40.75 			  & 3,117.29 			  & 248.387 			  & 150,617 	   	  & 150,617 	      \\\cmidrule{1-12}
\emph{Yeast} 		  & \texttt{\blend-I} & 89.1677 	   		      & 97.0854 				& 33.81	   			   & \textbf{12.3938} 	   & 2,672.37 	   		   & 38.84	  		  & 12.4176 	  		  & 1.54807 	  		  & 635.966  		  & 636.669  		  \\
(PacBio)			  & minimap2 	  		 & 88.9002 			   	  & 96.9709 				& 33.38 			   & 12.0128 			   & 2,684.38 			   & 38.85 			  & \textbf{12.3325} 	  & \textbf{1.56078}	  & \textbf{810.046}  & \textbf{828.212}  \\
					  & minimap2-Eq 		 & \textbf{89.2166}       & \textbf{97.2674} 		& \textbf{33.93} 	   & 12.3886 	   		   & \textbf{2,653.08} 	   & \textbf{38.82}   & 12.4241 	          & 1.53435	          	  & 643.136           & 781.136           \\
					  & Reference 	  		 & 100 				      & 100 					& 100 				   & 12.1571 			   & 0.00 				   & 38.15 			  & 12.1571 			  & 1.53193 			  & 924.431 		  & 924.431 		  \\\cmidrule{1-12}
\emph{Yeast} 	      & \texttt{\blend-I} & \textbf{89.6889}       & 99.2974 	    		& \textbf{35.95}	   & \textbf{12.3222} 	   & 2,529.47 	           & \textbf{38.64}	  & \textbf{12.3225} 	  & 1.10582 	  		  & 793.046  		  & 793.046  		  \\
(ONT)			      & minimap2 	  		 & 88.9393 			   	  & \textbf{99.6878}		& 34.84 			   & 12.304 			   & 2,782.59 			   & 38.74 			  & 12.3725 			  & \textbf{1.56005}	  & \textbf{796.718}  & \textbf{941.588}  \\
					  & minimap2-Eq 		 & 89.6653 	  			  & 97.3273		            & 35.62 			   & 11.826 			   & \textbf{2,465.87} 	   & \textbf{38.64}   & 11.8282 			  & 1.07367	          	  & 605.201           & 677.415  		  \\
					  & Reference 	  		 & 100 				      & 100 					& 100 				   & 12.1571 			   & 0.00 				   & 38.15 			  & 12.1571 			  & 1.53193 			  & 924.431 		  & 924.431 		  \\\cmidrule{1-12}
\emph{E. coli} 		  & \texttt{\blend-I} & 88.5806                & 96.5238 				& 32.32	   			   & \textbf{5.90024} 	   & 1,857.56	           & \textbf{49.81}	  & 6.21598 	  		  & 2.40671	  			  & 769.981  		  & 2,060.4  		  \\
(PacBio)			  & minimap2 	  		 & 88.1365 			   	  & 92.7603 				& 30.74 			   & 5.37728 			   & 2,005.72 			   & 49.66 			  & \textbf{6.02707} 	  & 3.77098 	  		  & 367.442    	   	  & 3,770.98 		  \\
					  & minimap2-Eq 		 & \textbf{88.6371} 	  & \textbf{96.8540} 		& \textbf{32.33} 	   & 5.82218 			   & \textbf{1,816.29} 	   & 49.76 			  & 6.05821 	          & \textbf{3.77318} 	  & \textbf{1,119.04} & \textbf{3,773.18} \\
					  & Reference 	  		 & 100 				      & 100 					& 100 				   & 5.6394 			   & 0.00 				   & 50.43 			  & 5.6394 				  & 5.54732 			  & 5,547.32 		  & 5,547.32 		  \\\bottomrule
\multicolumn{12}{l}{\footnotesize Best results are highlighted with \textbf{bold} text. For most metrics, the best results are the ones closest to the corresponding value of the reference genome.}\\
\multicolumn{12}{l}{\footnotesize The best results for \emph{Aligned Length} are determined by the highest number within each dataset. We do not highlight the reference results as the best results.}\\
\multicolumn{12}{l}{\footnotesize N/A indicates that we could not generate the corresponding result because tool, QUAST, or dnadiff failed to generate the statistic.} \\
\end{tabular}

}
\label{blend:supptab:overlap_assembly-eq}
\end{table}

\clearpage

\subsection{The Genome-wide Coverage Comparison}\label{blend:suppsec:genomecov}

\begin{figure}[tbh]
\centering
\includegraphics[width=\linewidth]{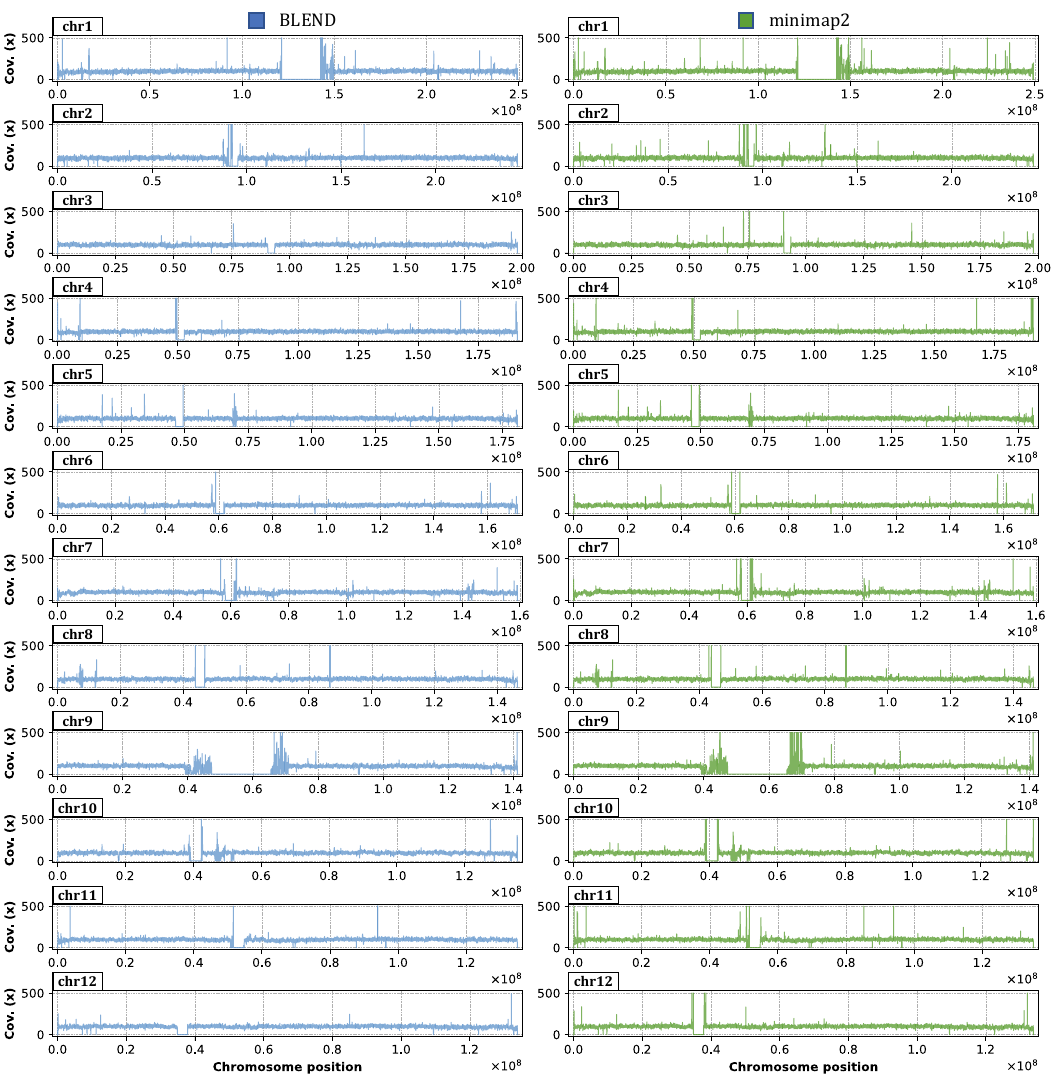}
\caption{Depth of coverage at each position (binned) of the GRCh37 reference genome (chromosomes 1 to 12) after mapping the HG002 reads using \blend and minimap2. We label the chromosomes on the top left corner of each plot.}
\label{blend:suppfig:genomedepth-1}
\end{figure}

\begin{figure}[tbh]
\centering
\includegraphics[width=\linewidth]{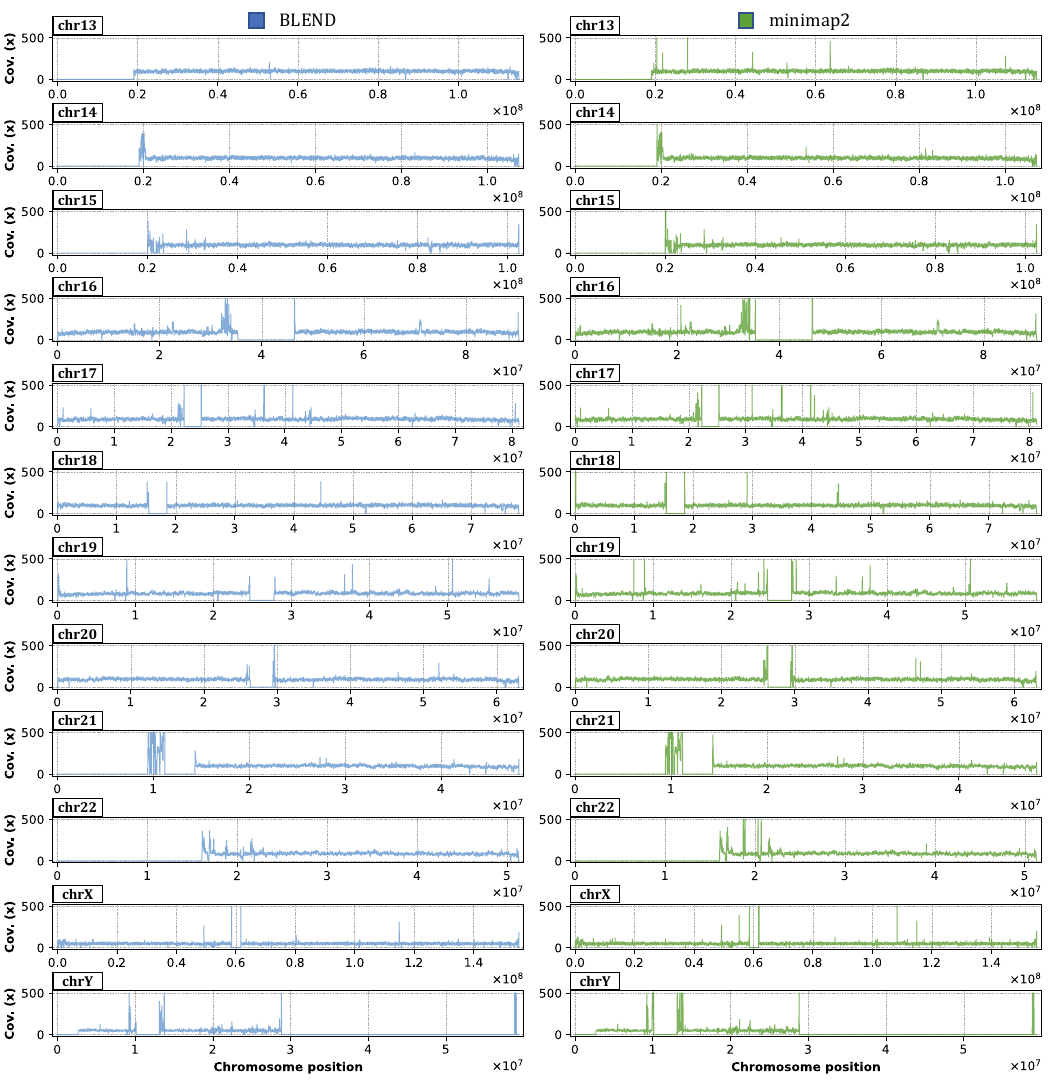}
\caption{Depth of coverage at each position (binned) of the GRCh37 reference genome (chromosomes 13 to Y) after mapping the HG002 reads using \blend and minimap2. We label the chromosomes on the top left corner of each plot.}
\label{blend:suppfig:genomedepth-13}
\end{figure}

We map the HG002 reads to the human reference genome (GRCh37) using \blend and minimap2. Supplementary Figures~\ref{blend:suppfig:genomedepth-1} and \ref{blend:suppfig:genomedepth-13} show the depth of mapping coverage at each position of the reference genome chromosomes for \blend and minimap2 on the left and right sides of the figures, respectively. To calculate the position-wise depth of coverage, we use the \texttt{multiBamSummary} tool from the deepTools2 package~\cite{ramirez_deeptools2_2016}. The \texttt{multiBamSummary} tool divides the reference genome into consecutive bins of equal size (10,000 bases) to calculate the genome-wide coverage in fine granularity. For positions where the coverage is higher than $500\times$, we set the coverage to $500\times$ for visibility reasons as there are only a negligible amount of such regions where either \blend or minimap2 exceeds this threshold without the other one exceeding it.

\begin{figure}[tbh]
\centering
\includegraphics[width=\linewidth]{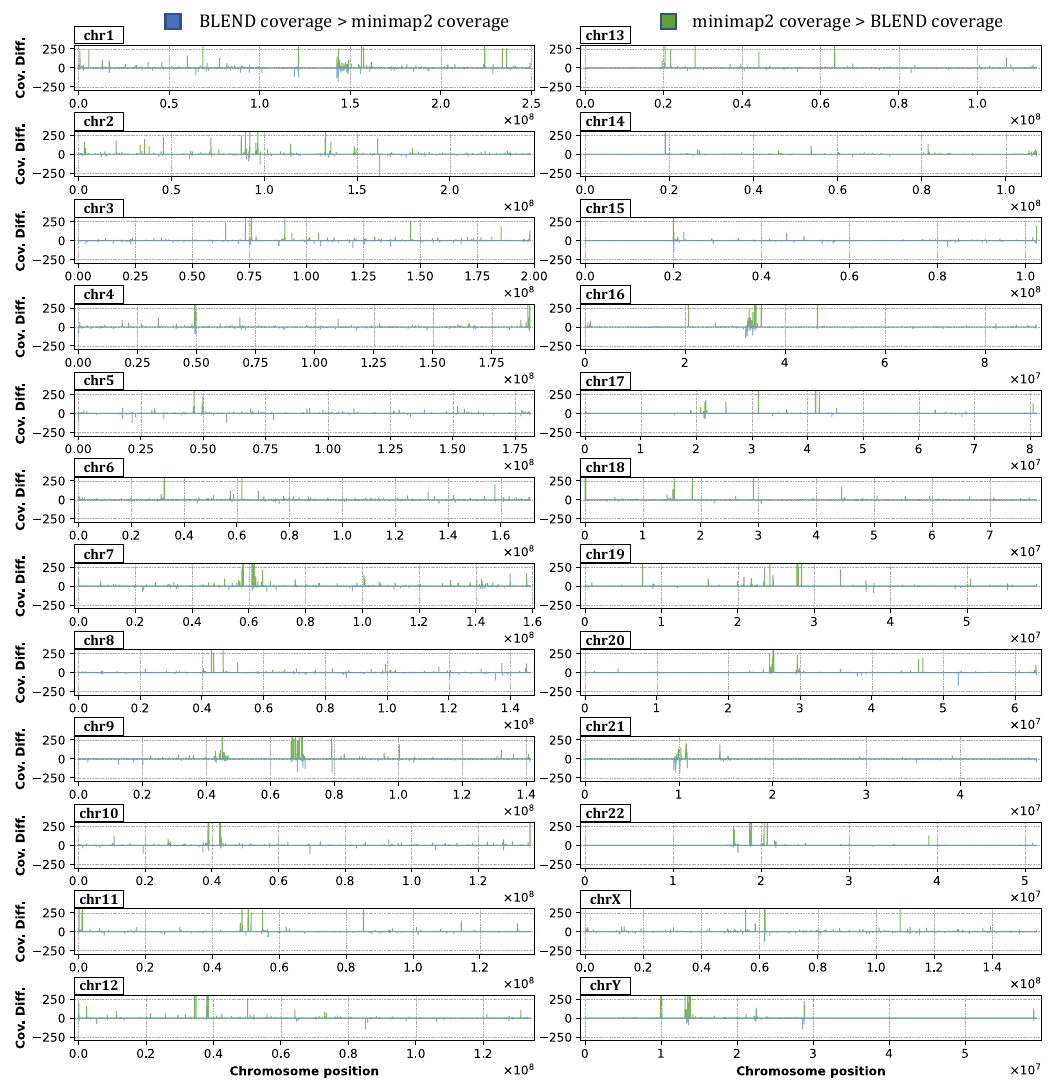}
\caption{Difference between the depth of coverage of minimap2 and \blend. Positive values show the positions where minimap2 has higher coverage and negative values show the positions where \blend has higher coverage. We label the chromosomes on the top left corner of each plot.}
\label{blend:suppfig:genomedepth_diff}
\end{figure}

To find the positions where the depth of coverage significantly differs between \blend and minimap2, we subtract the minimap2 coverage from the \blend coverage for each chromosome position that we show in Figures~\ref{blend:suppfig:genomedepth-1} and \ref{blend:suppfig:genomedepth-13}. We show the coverage differences in Figure~\ref{blend:suppfig:genomedepth_diff}, where the positive values show the positions that minimap2 has a higher depth of coverage than \blend, and negative values show the positions that \blend has a higher coverage.

\clearpage

\subsection{Parameters and Tool Versions}\label{blend:suppsec:parameters}

In Supplementary Table~\ref{blend:supptab:version}, we show the version numbers of each tool. When calculating the performance and peak memory usage, we use the time command from Linux and append the following command to the beginning of each of our runs: \texttt{/usr/bin/time -vp}.

Supplementary Table~\ref{blend:supptab:pardef} shows the parameters we use in \blend and their definition. Since \blend uses the minimap2 implementation as a baseline, the rest of the parameters we do not show in Supplementary Table~\ref{blend:supptab:pardef} can be found on the manual page of minimap2\footnote{\url{https://lh3.github.io/minimap2/minimap2.html}}. In Supplementary Table~\ref{blend:supptab:ovpars}, we show the parameters we use with \blend, minimap2, and MHAP~\protect\cite{berlin_assembling_2015} for read overlapping. Since there are no default parameters for minimap2 and MHAP when using the HiFi reads, we used the parameters as suggested by the HiCanu tool~\protect\cite{nurk_hicanu_2020}. We found these parameters in the source code of Canu. For minimap2 and MHAP, the HiFi parameters are found in the GitHub pages\footnote{\url{https://github.com/marbl/canu/blob/404540a944664cfab00617f4f4fa37be451b34e0/src/pipelines/canu/OverlapMMap.pm\#L63-L65}}$^,$\footnote{\url{https://github.com/marbl/canu/blob/404540a944664cfab00617f4f4fa37be451b34e0/src/pipelines/canu/OverlapMhap.pm\#L100-L131}}, respectively. In Supplementary Table~\ref{blend:supptab:ovpars}, \emph{minimap-Eq} shows the parameters that are equivalent to the parameters we use with \blend without the fuzzy seed matching capability.

In Supplementary Table~\ref{blend:supptab:mappars}, we show the parameters we use with \blend, minimap2~\protect\cite{li_minimap2_2018}, LRA~\protect\cite{ren_lra_2021}, Winnowmap2~\protect\cite{jain_long-read_2022, jain_weighted_2020}, S-conLSH~\protect\cite{chakraborty_conlsh_2020, chakraborty_s-conlsh_2021}, and Strobealign~\protect\cite{sahlin_flexible_2022} for read mapping.

\begin{table}[tbh]
\centering
\caption{Versions of each tool.}\label{blend:supptab:version}
\resizebox{\linewidth}{!}{
\begin{tabular}{@{}lll@{}}\toprule
\textbf{Tool} & \textbf{Version} & \textbf{GitHub or Conda Link to the Version} \\\midrule
\blend & 1.0 & \url{https://github.com/CMU-SAFARI/BLEND}\\\midrule
minimap2 & 2.24 & \url{https://github.com/lh3/minimap2/releases/tag/v2.24}\\\midrule
MHAP & 2.1.3 & \url{https://anaconda.org/bioconda/mhap/2.1.3/download/noarch/mhap-2.1.3-hdfd78af_1.tar.bz2}\\\midrule
LRA & 1.3.2 & \url{https://anaconda.org/bioconda/lra/1.3.2/download/linux-64/lra-1.3.2-ha140323_0.tar.bz2}\\\midrule
Winnowmap2 & 2.03 & \url{https://anaconda.org/bioconda/Winnowmap/2.03/download/linux-64/Winnowmap2-2.03-h2e03b76_0.tar.bz2}\\\midrule
S-conLSH & 2.0 & \url{https://github.com/anganachakraborty/S-conLSH-2.0/tree/292fbe0405f10b3ab63fc3a86cba2807597b582e} \\\midrule
Strobealign & 0.7.1 & \url{https://anaconda.org/bioconda/strobealign/0.7.1/download/linux-64/strobealign-0.7.1-hd03093a_1.tar.bz2} \\\bottomrule
\end{tabular}

}
\end{table}

\begin{table}[tbh]
\centering
\caption{Definition of parameters in \blend}
\resizebox{\linewidth}{!}{
\begin{tabular}{@{}ll@{}}\toprule
\textbf{Parameter} & \textbf{Definition} \\\midrule
--strobemers & Use the \texttt{\blend-S} mechanism when generating the list of k-mers of a seed\\\midrule
--immediate & Use the \texttt{\blend-I} mechanism when generating the list of k-mers of a seed\\\midrule
-H & Use homopolymer-compressed k-mers\\\midrule
-w INT & Window size used when finding minimizers.\\\midrule
-k INT & k-mer size used when generating the list of k-mers of a seed\\\midrule
--neighbors INT & Number of k-mers included in the list of seeds.\\
                & Combination of both -k ($k$) and --neighbors ($n$) determines the seed length.\\
                & Seed length in \texttt{\blend-S} is calculated as: $k \times n$\\
                & Seed length in \texttt{\blend-I} is calculated as: $k+(n-1)$\\\midrule
--fixed-bits INT & Bit length of hash values that \blend generates for each seed.\\
                 & Setting it to $2 \times k$ is the default behavior.\\\midrule
-t INT & Number of CPU threads to use.\\\midrule
-x STR & Preset for setting the default parameters given the use case (STR)\\
-x map-ont & Preset for mapping ONT reads. It uses the following parameters:\\
           & --immediate -w 10 -k 9 --neighbors 7 --fixed-bits 30\\
-x map-pb & Preset for mapping erroneous PacBio reads. It uses the following parameters:\\
           & --immediate -H -w 10 -k 13 --neighbors 7 --fixed-bits 32\\
-x map-hifi & Preset for mapping accurate long (HiFi) reads. It uses the following parameters:\\
           & --strobemers -w 50 -k 19 --neighbors 5 --fixed-bits 38\\
-x sr & Preset for mapping short reads. It uses the following parameters:\\
           & --immediate -w 11 -k 21 --neighbors 5 --fixed-bits 32\\
-x ava-ont & Preset for overlapping ONT reads. It uses the following parameters:\\
           & --immediate -w 10 -k 15 --neighbors 5 --fixed-bits 30\\
-x ava-pb & Preset for overlapping erroneous PacBio reads. It uses the following parameters:\\
           & --immediate -H -w 10 -k 19 --neighbors 5 --fixed-bits 38\\
-x ava-hifi & Preset for overlapping accurate long (HiFi) reads. It uses the following parameters:\\
           & --strobemers -w 200 -k 25 --neighbors 7 --fixed-bits 50\\\bottomrule
\end{tabular}

}
\label{blend:supptab:pardef}
\end{table}

\begin{table}[tbh]
\centering
\caption{Parameters$^{*}$ we use in our evaluation for each tool and dataset in read overlapping.}
\resizebox{\linewidth}{!}{
\begin{tabular}{@{}lll@{}}\toprule
\textbf{Tool} & \textbf{Dataset} & \textbf{Parameters} \\\midrule
\blend & \emph{CHM13 (HiFi)} & -x ava-hifi -t 32 \\
\blend & \emph{D. ananassae (HiFi)} & -x ava-hifi -t 32\\
\blend & \emph{E. coli (HiFi)} & -x ava-hifi -t 32\\
\blend & \emph{CHM13 (ONT)} & -x ava-ont -t 32\\
\blend & \emph{Yeast (PacBio)} & -x ava-pb -t 32\\
\blend & \emph{Yeast (ONT)} & -x ava-ont -t 32\\
\blend & \emph{E. coli (PacBio)} & -x ava-pb -t 32\\\midrule
minimap2 & \emph{CHM13 (HiFi)} & -x ava-pb -Hk21 -w14 -t 32 \\
minimap2 & \emph{D. ananassae (HiFi)} & -x ava-pb -Hk21 -w14 -t 32 \\
minimap2 & \emph{E. coli (HiFi)} & -x ava-pb -Hk21 -w14 -t 32\\
minimap2 & \emph{CHM13 (ONT)} & -x ava-ont -t 32\\
minimap2 & \emph{Yeast (PacBio)} & -x ava-pb -t 32\\
minimap2 & \emph{Yeast (ONT)} & -x ava-ont -t 32\\
minimap2 & \emph{E. coli (PacBio)} & -x ava-pb -t 32\\\midrule
minimap2-Eq & \emph{CHM13 (ONT)} & -x ava-ont -k19 -w10 -t 32\\
minimap2-Eq & \emph{Yeast (PacBio)} & -x ava-pb -k23 -w10 -t 32\\
minimap2-Eq & \emph{Yeast (ONT)} & -x ava-ont -k19 -w10 -t 32\\
minimap2-Eq & \emph{E. coli (PacBio)} & -x ava-pb -k23 -w10 -t 32\\\midrule
MHAP & \emph{CHM13 (HiFi)} & --store-full-id --ordered-kmer-size 18 --num-hashes 128 --num-min-matches 5\\
& & --ordered-sketch-size 1000 --threshold 0.95 --num-threads 32\\
MHAP & \emph{D. ananassae (HiFi)} & --store-full-id --ordered-kmer-size 18 --num-hashes 128 --num-min-matches 5\\
& & --ordered-sketch-size 1000 --threshold 0.95 --num-threads 32\\
MHAP & \emph{E. coli (HiFi)} & --store-full-id --ordered-kmer-size 18 --num-hashes 128 --num-min-matches 5\\
& & --ordered-sketch-size 1000 --threshold 0.95 --num-threads 32\\
MHAP & \emph{Yeast (PacBio)} & --store-full-id --num-threads 32\\
MHAP & \emph{Yeast (ONT)} & --store-full-id --num-threads 32\\
MHAP & \emph{E. coli (PacBio)} & --store-full-id --num-threads 32\\\bottomrule
\multicolumn{3}{l}{\footnotesize $^{*}$ For the definitions of the parameters we use in \blend, please see Supplementary Table~\ref{blend:supptab:pardef}}\\
\end{tabular}

}
\label{blend:supptab:ovpars}
\end{table}

\begin{table}[tbh]
\centering
\caption{Parameters we use in our evaluation for each tool and dataset in read mapping.}\label{blend:supptab:mappars}
\resizebox{0.6\linewidth}{!}{
\begin{tabular}{@{}lll@{}}\toprule
\textbf{Tool} & \textbf{Dataset} & \textbf{Parameters} \\\midrule
\blend & \emph{CHM13 (HiFi)} & -ax map-hifi -t 32 --secondary=no\\
\blend & \emph{HG002 (HiFi)} & -ax map-hifi -t 32 --secondary=no\\
\blend & \emph{D. ananassae (HiFi)} & -ax map-hifi -t 32 --secondary=no \\
\blend & \emph{E. coli (HiFi)} & -ax map-hifi -t 32 --secondary=no \\
\blend & \emph{CHM13 (ONT)} & -ax map-ont -t 32 --secondary=no\\
\blend & \emph{Yeast (PacBio)} & -ax map-pb -t 32 --secondary=no\\
\blend & \emph{Yeast (ONT)} & -ax map-ont -t 32 --secondary=no\\
\blend & \emph{Yeast (Illumina)} & -ax sr -t 32 \\
\blend & \emph{E. coli (PacBio)} & -ax map-pb -t 32 --secondary=no\\\midrule
minimap2 & \emph{CHM13 (HiFi)} & -ax map-hifi -t 32 --secondary=no\\
minimap2 & \emph{HG002 (HiFi)} & -ax map-hifi -t 32 --secondary=no\\
minimap2 & \emph{D. ananassae (HiFi)} & -ax map-hifi -t 32 --secondary=no\\
minimap2 & \emph{E. coli (HiFi)} & -ax map-hifi -t 32 --secondary=no\\
minimap2 & \emph{CHM13 (ONT)} & -ax map-ont -t 32 --secondary=no\\
minimap2 & \emph{Yeast (PacBio)} & -ax map-pb -t 32 --secondary=no\\
minimap2 & \emph{Yeast (ONT)} & -ax map-ont -t 32 --secondary=no\\
minimap2 & \emph{Yeast (Illumina)} & -ax sr -t 32 \\
minimap2 & \emph{E. coli (PacBio)} & -ax map-pb -t 32 --secondary=no\\\midrule
Winnowmap2 & \emph{CHM13 (HiFi)} & meryl count k=15\\
 & & meryl print greater-than distinct=0.9998\\
 & & -ax map-pb -t 32\\
Winnowmap2 & \emph{HG002 (HiFi)} & meryl count k=15\\
 & & meryl print greater-than distinct=0.9998\\
 & & -ax map-pb -t 32\\
Winnowmap2 & \emph{D. ananassae (HiFi)} & meryl count k=15\\
& &  meryl print greater-than distinct=0.9998\\
& & -ax map-pb -t 32\\
Winnowmap2 & \emph{E. coli (HiFi)} & meryl count k=15\\
& &  meryl print greater-than distinct=0.9998\\
& & -ax map-pb -t 32\\
Winnowmap2 & \emph{CHM13 (ONT)} & meryl count k=15\\
& &  meryl print greater-than distinct=0.9998\\
& & -ax map-ont -t 32\\
Winnowmap2 & \emph{Yeast (PacBio)} & meryl count k=15\\
& &  meryl print greater-than distinct=0.9998\\
& & -ax map-pb-clr -t 32\\
Winnowmap2 & \emph{Yeast (ONT)} & meryl count k=15\\
& &  meryl print greater-than distinct=0.9998\\
& & -ax map-ont -t 32\\
Winnowmap2 & \emph{E. coli (PacBio)} & meryl count k=15\\
& &  meryl print greater-than distinct=0.9998\\
& & -ax map-pb-clr -t 32\\\midrule
LRA & \emph{CHM13 (HiFi)} & align -CCS -t 32 -p s \\
LRA & \emph{HG002 (HiFi)} & align -CCS -t 32 -p s \\
LRA & \emph{D. ananassae (HiFi)} & align -CCS -t 32 -p s \\
LRA & \emph{E. coli (HiFi)} & align -CCS -t 32 -p s \\
LRA & \emph{CHM13 (ONT)} & align -ONT -t 32 -p s \\
LRA & \emph{Yeast (PacBio)} & align -CLR -t 32 -p s \\
LRA & \emph{Yeast (ONT)} & align -ONT -t 32 -p s \\
LRA & \emph{E. coli (PacBio)} & align -CLR -t 32 -p s \\\midrule
S-conLSH & \emph{CHM13 (HiFi)} & --threads 32 --align 1\\
S-conLSH & \emph{E. coli (HiFi)} & --threads 32 --align 1\\
S-conLSH & \emph{CHM13 (ONT)} & --threads 32 --align 1\\
S-conLSH & \emph{Yeast (PacBio)} & --threads 32 --align 1\\
S-conLSH & \emph{Yeast (ONT)} & --threads 32 --align 1\\
S-conLSH & \emph{E. coli (PacBio)} & --threads 32 --align 1\\\midrule
Strobealign & \emph{Yeast (Illumina)} & -t 32\\\bottomrule
\multicolumn{3}{l}{\footnotesize $^{*}$ For the definitions of the parameters we use in \blend, please see Supplementary Table~\ref{blend:supptab:pardef}}\\
\end{tabular}

}
\end{table}

\clearpage

\setchapterbasednumbering

\chapter[\rh: Enabling Scalable Hash-based Search of Raw Nanopore Signals]{Enabling Hash-based Search\\of Raw Nanopore Signals\\by Reducing Noise}
\label{chap:rh}

In the previous chapter, we present a mechanism for generating the same hash value when the basecalled seeds are highly similar to each other to improve the sensitivity of sequence analysis by tolerating noise in seeds (i.e., differences between seeds). Our goal in this chapter is to provide a similar mechanism for raw nanopore signals such that \textbf{noise in these signals are effectively reduced} to enable generating the same hash values between similar raw signals.

\section{Background and Motivation}
\label{rh:sec:bg}

High-throughput sequencing (HTS) devices can generate a large amount of genomic data at a relatively low cost. HTS can be used to analyze a wide range of samples, from small amounts of DNA or RNA to entire genomes. Oxford Nanopore Technologies (ONT) is one of the most widely-used HTS technologies that can sequence long genomic regions, called \emph{reads}, with up to a few million bases. ONT devices use the nanopore sequencing technique, which involves passing a single DNA or RNA strand through a tiny pore, \emph{nanopore} or channel, at an average speed of 450 bases per second~\cite{kovaka_targeted_2021} and measuring the electrical current as the strand passes through. Nanopore sequencing enables two key features. First, nanopores provide the electrical raw signals in \emph{real-time} as the DNA strand passes through a nanopore. Second, nanopore sequencing provides a functionality, known as \emph{\ru}~\cite{loose_real-time_2016}, that can partially sequence DNA strands without fully sequencing them. These two features of nanopores provide opportunities for 1)~real-time genome analysis and 2)~significantly reducing sequencing time and cost.

Real-time analysis of nanopore raw signals using \ru can reduce the sequencing time and cost per read by terminating the sequencing whenever sequencing the full read is not necessary. The freed-up nanopore can then be used to sequence a different read. A purely computational mechanism can send a signal to \emph{eject} a read from a nanopore by reversing the voltage if the partial sequencing of a read meets certain conditions for particular genome analysis, such as 1)~reaching a desired coverage for a species in a sample~\cite{payne_readfish_2021} or 2)~identifying that a read does not originate from a certain genome of interest (i.e., a target region)~\cite{kovaka_targeted_2021, zhang_real-time_2021} and hence, does not need to be fully sequenced. By terminating the sequencing of reads that do not correspond to the target region, the sequencer can spend time and resources on higher coverage sequencing of the reads that correspond to the target. This process is referred to as \emph{nanopore adaptive sampling}. By providing high coverage at target regions and avoiding unessential sequencing of reads outside those regions, this approach can improve the quality of sequencing and the downstream analysis utilizing the obtained data.

\subsection{Motivation: Need for Scalable Mechanisms Raw Nanopore Signal Analysis}
\label{rh:subsec:bg_motiv}

To effectively utilize adaptive sampling in nanopore sequencing, it is crucial to have computational methods that can accurately analyze the raw output signals from nanopores in real-time. These methods must provide 1)~low latency and 2)~throughput matching or exceeding that of the sequencer~\cite{dunn_squigglefilter_2021, kovaka_targeted_2021, zhang_real-time_2021}. Several works propose adaptive sampling methods for real-time analysis of raw nanopore signals~\cite{edwards_real-time_2019, dunn_squigglefilter_2021, bao_squigglenet_2021, kovaka_targeted_2021, zhang_real-time_2021, payne_readfish_2021, shih_efficient_2023, ulrich_readbouncer_2022, senanayake_deepselectnet_2023, sadasivan_rapid_2023}. However, these works have three key limitations. First, most techniques mainly use powerful computational resources, such as GPUs~\cite{payne_readfish_2021, bao_squigglenet_2021}, or specialized hardware~\cite{dunn_squigglefilter_2021, shih_efficient_2023} due to the use of computationally-intensive algorithms such as basecalling as we explain in detail in Chapter~\ref{chap:related}. This can make real-time genome analysis challenging for portable and low-cost nanopore-based sequencers, such as the ONT Flongle or MinION, which are not typically equipped with such resources. Therefore these techniques introduce challenges for using them in resource-constrained environments. Second, the sheer size of genomic data at the scale of large genomes (e.g., human genome) makes it challenging to process the data in real-time. This is because such large genomes require efficient and accurate similarity identification across a large number of regions. This renders many current methods~\cite{kovaka_targeted_2021, zhang_real-time_2021} inaccurate or useless for large genomes as they cannot either provide accurate results or match the throughput of nanopores for these genomes. Third, machine learning models used in past works~\cite{edwards_real-time_2019, payne_readfish_2021, bao_squigglenet_2021, ulrich_readbouncer_2022, senanayake_deepselectnet_2023} to analyze raw nanopore signals often require retraining or reconfiguring the model to improve accuracy for a certain experiment, which can be a barrier to flexibly and easily performing real-time analysis without retraining or reconfiguring these models. To our knowledge, there is no work that can efficiently and accurately perform real-time analysis of raw nanopore signals on a large scale (e.g., whole-genome analysis for human) without requiring powerful computational resources, which can easily and flexibly be applied to a wide range of applications that could benefit from real-time nanopore raw signal analysis.

\subsection{Challenge: Efficiently Handling Noise in Raw Nanopore Signal Analysis}
\label{rh:subsec:bg_challenge}

Our \textbf{goal} is to enable efficient and accurate real-time genome analysis for large genomes. To this end, we propose \textbf{\emph{\rh}}, the \emph{first} mechanism that can efficiently and accurately perform real-time analysis of raw nanopore signals for large genomes in resource-contained environments without requiring computationally-intensive algorithms such as basecalling. Our \textbf{key idea} is to encode regions of the raw nanopore signal into hash values such that similar signal regions can efficiently be identified by matching their hash values, facilitating efficient similarity identification between signals. However, enabling accurate hashing-based similarity identification in the raw signal domain is challenging because raw signals corresponding to the same DNA content are unlikely to have exactly the same signal amplitudes. This is because the raw signals generated by nanopores can vary each time the same DNA fragment is sequenced due to several factors impacting nanopores during sequencing, such as variations in the properties of the nanopores or the conditions in which the sequencing is performed~\cite{david_nanocall_2017}. Although the similarity identification of raw signals is possible via calculating the Euclidean distance between a sequence of signals in a multi-dimensional space~\cite{zhang_real-time_2021}, such an approach can become impractical when dealing with larger sequences as the number of dimensions increases with the length of the sequences. This increase in dimensionality can lead to computational complexity and the curse of dimensionality, making it expensive and impractical.

\subsection{Overview of Noise Reduction To Enable Hash-based Search in RawHash}
\label{rh:subsec:bg_rh}

To address these challenges, \rh provides three \textbf{key mechanisms} for efficient signal encoding and similarity identification. First, \rh encodes signal values that have a wider range of values into a smaller set of values using a quantization technique, such that signal values within a certain range are assigned to the same encoded value. This helps to alleviate the probability of having varying signal values for the same DNA content and enables \rh to directly match these values using a hashing technique. Second, \rh concatenates the quantized values of \emph{multiple} consecutive signals and generates a single hash value for them. The hashing mechanism enables \rh to efficiently identify similar signal regions of these consecutive signal values by directly matching their corresponding hash values. Representing many consecutive signals with a single hash value increases the size of the regions examined during similarity identification without suffering from the curse of dimensionality. Using larger regions can substantially reduce the number of possible matching regions that need to be examined. \rh is the \emph{first} work that can accurately use hash values in the raw signal domain, which enables using efficient data structures commonly-used used in the sequence domain (e.g., hash tables in minimap2~\cite{li_minimap2_2018}). Third, \rh uses an existing algorithm, known as chaining~\cite{li_minimap2_2018}, to find the colinear matches of hash values between signals to identify similar signal regions. These efficient and accurate mechanisms enable \rh to perform real-time genome analysis for large genomes.

While our proposed three key mechanisms have the potential to be used for various purposes in raw signal similarity identification, we design \rh as a tool for mapping nanopore raw signals to their corresponding reference genomes in real-time. \rhcap operates the mapping in two steps 1)~\emph{indexing} and 2)~\emph{mapping}. First, in the indexing step, \rh 1)~converts the reference genome sequence into \emph{expected} signal values by simulating the expected behavior of nanopores based on a previously-known model, 2)~generates the hash values from these signals, and 3)~stores the hash values in a hash table for efficient matching. Second, in the mapping step, \rh 1)~generates the hash values from the raw signals in a streaming fashion, 2)~queries the hash table from the indexing step with these hash values to find the matching regions in the reference genome with the same hash value, and 3)~performs chaining to find the similar region between the reference genome and the raw signal of a read.

\subsection{The \emph{Sequence Until} Mechanism: Dynamically Stopping the Entire Sequencing Run}
\label{rh:subsec:bg_sequenceuntil}

We propose a novel mechanism that can stop the entire sequencing run for certain applications (e.g., relative abundance estimation) when an accurate decision can be made without sequencing the entire sample, which we call \emph{Sequence Until}. \emph{Sequence Until} utilizes \emph{Run Until} to \emph{fully} stop the \emph{entire} sequencing of all subsequent reads after sequencing a certain amount of reads that is sufficient to make an accurate relative abundance estimation. Avoiding the redundant sequencing of further reads that are unlikely to substantially change the relative abundance estimation has the potential to significantly reduce the sequencing time and cost. To utilize \emph{Sequence Until}, \rh integrates a confidence calculation mechanism that evaluates the relative abundance estimations in real-time and fully stops the entire sequencing run if using more reads does not change its estimation. To stop the entire sequencing run for further reads, Run Until can be used to stop the entire sequencing run, which can enable the better utilization of nanopores. We find that \emph{Sequence Until} can be applied to other mechanisms (e.g., \unc) that can perform real-time relative abundance estimations. Prior work~\cite{weilguny_dynamic_2023} proposes a technique to terminate the sequencing process when species in the sample reach a certain coverage depth. The key difference of \emph{Sequence Until} is that it reduces the cost of sequencing for relative abundance estimation and is based on our adaptive, accurate, and low-cost confidence calculation during real-time abundance estimation.

\section{\rhcap Mechanism} \label{rh:sec:methods}
We propose \textbf{\rh}, a mechanism that can efficiently and accurately identify similarities between raw nanopore signals of a read and a large reference genome in real-time (i.e., while the read is sequenced). The raw nanopore signal of each read is a series of electrical current measurements as a strand of DNA passes through a nanopore. The reference genome is a set of strings over the alphabet \texttt{A,C,G,T}. \rhcap provides the mechanisms for generating hash values from both a raw nanopore signal and a reference genome such that similar regions between the two can be efficiently and accurately found by matching their hash values.

\subsection{Overview} \label{rh:subsec:overview}
\begin{figure}[tbh]
  \centering
  \includegraphics[width=0.6\columnwidth]{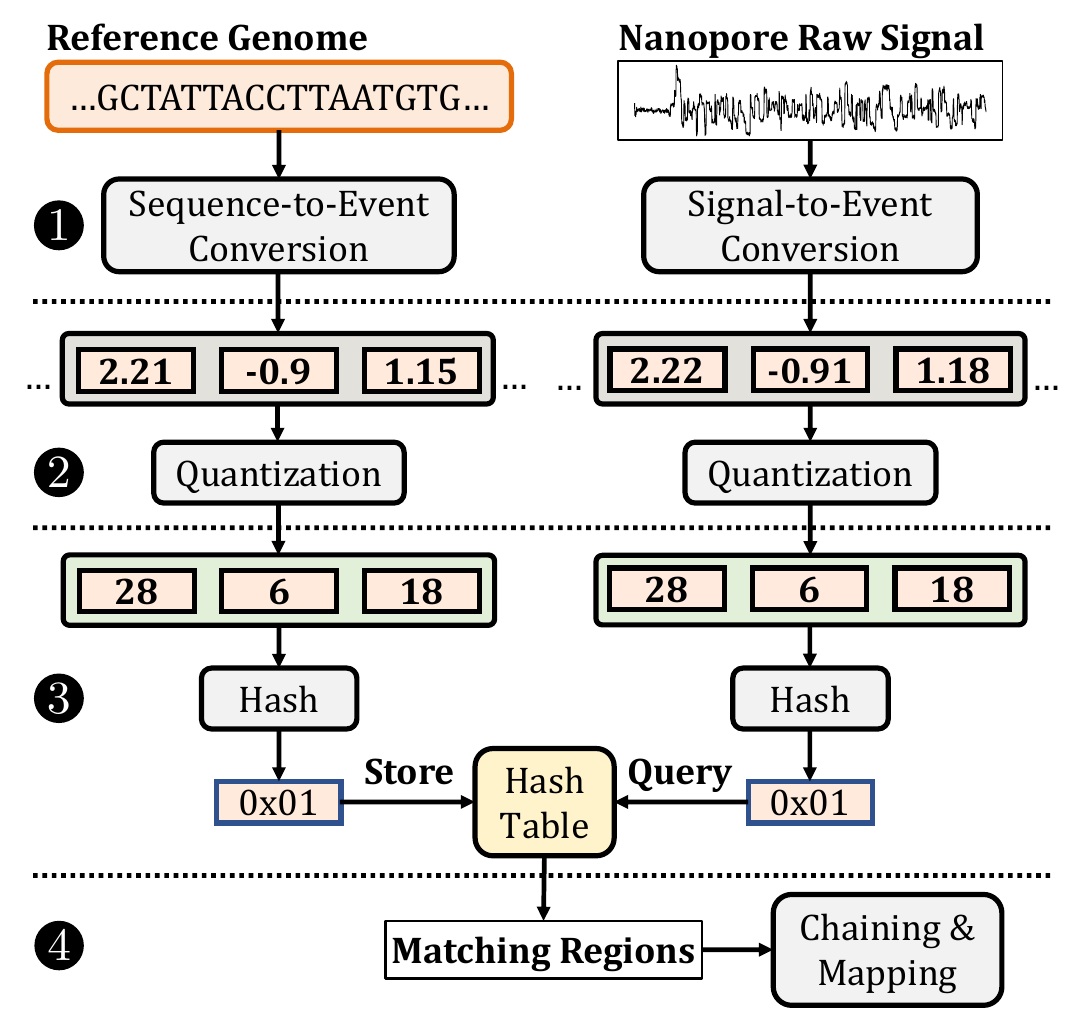}
  \caption{Overview of \rh.}
  \label{rh:fig:overview}
\end{figure}

Figure~\ref{rh:fig:overview} shows the overview of how \rh identifies similarities between raw nanopore signals of a read and a reference genome in four steps. First, \rh pre-processes both 1)~the raw nanopore signal and 2)~the reference genome into values that are comparable to each other. For raw signals, \rh segments the raw signal into non-overlapping regions such that each region is expected to contain a certain amount of signal values that are generated from reading a fixed number $k$ of DNA bases. Each such region is called an \emph{event}~\cite{david_nanocall_2017}. Each event is usually represented with a value derived from the signal values in the segment. For the reference genome, \rh translates each substring of length $k$ (called a \emph{k-mer}) into their \emph{expected} event values based on the nanopore model.

The event values from the reference genome are not directly comparable to the event values from raw nanopore signals due to variability in the current measurements in nanopores generating slightly different event values for the same k-mer~\cite{david_nanocall_2017}. To generate the same values from slightly different events that may contain the same k-mer information, the second step of \rh \emph{quantizes} the event values from a larger set of values into a smaller set. The quantization technique ensures that the event values within a certain range are likely to be assigned to the same quantized value such that the effect of signal variation is alleviated, i.e., the same k-mer is likely assigned the same quantized value.

Due to the nature of nanopores, each event usually represents a very small k-mer of length around $k$=6 bases, depending on the nanopore model~\cite{zhang_real-time_2021}. Such a short k-mer is likely to exist in a large number of locations in the reference genome, making it challenging to efficiently identify the correct one. To make the events more unique (i.e., such that they exist only in a small number of locations in the reference genome), the third step of \rh combines multiple consecutive quantized events into a single hash value. These hash values can then be used to efficiently identify similar regions between raw signals and the reference genome by matching the hash values generated from their events using efficient data structures such as hash tables.

Fourth, to map a raw nanopore signal of a read to a reference genome, \rh uses a chaining algorithm~\cite{zhang_real-time_2021, li_minimap2_2018} that find colinear matching hash values generated from regions that are close to each other both in the reference genome and the raw nanopore signal.

\subsection{Event Generation} \label{rh:subsec:eventgeneration}

Our goal is to translate a reference genome sequence and a raw nanopore signal into comparable values. To this end, \rh converts 1)~each k-mer of the reference genome and 2)~each segmented region of the raw signal into its corresponding event.

\begin{figure}[tbh]
  \centering
  \includegraphics[width=0.8\columnwidth]{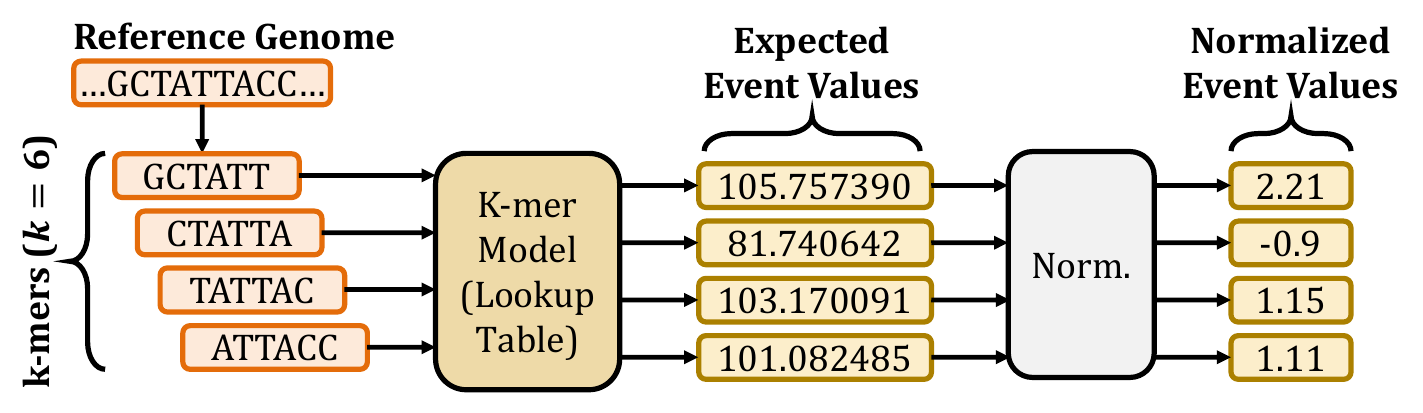}
  \caption{Converting sequences to event values based on the k-mer model of a nanopore.}
  \label{rh:fig:seqtoev}
\end{figure}

\head{Sequence-to-Event Conversion} To convert a reference genome sequence into a form that can be compared with raw nanopore signals, \rh converts the reference genome sequence into event values in three steps, as shown in Figure~\ref{rh:fig:seqtoev}.

First, \rh extracts all k-mers from the reference genome sequence, where $k$ depends on the nanopore. The \emph{k-mer model} of a nanopore\footnote{For many nanopore models, ONT provides the k-mer model including recent R10 and R10.4. These models can also be generated by users~\cite{simpson_detecting_2017}.} includes the information about the \emph{expected} k-mer length of an event and the \emph{expected} average event value for each k-mer based on certain variables affecting the signal outcome of the nanopore's current measurements. 

Second, \rh queries the k-mer model for each k-mer of the reference genome to convert k-mers into their expected event values. Although the k-mer model of a nanopore provides an extensive set of information for each possible k-mer, \rh uses only the mean values of events that provide an average value for the signals in the same event since these mean values provide a sufficient level of meaningful information for comparison with the raw nanopore signals.

Third, \rh normalizes the event values from the same reference genome sequence (e.g., entire chromosome sequence or a contig) by calculating the standard scores (i.e., z-scores) of these events. \rhcap uses these normalized values as event values since the same normalization step is taken for raw signals to avoid certain variables that may affect the range of raw signal amplitudes during sequencing~\cite{zhang_real-time_2021, kovaka_targeted_2021}.

\begin{figure}[tbh]
  \centering
  \includegraphics[width=\columnwidth]{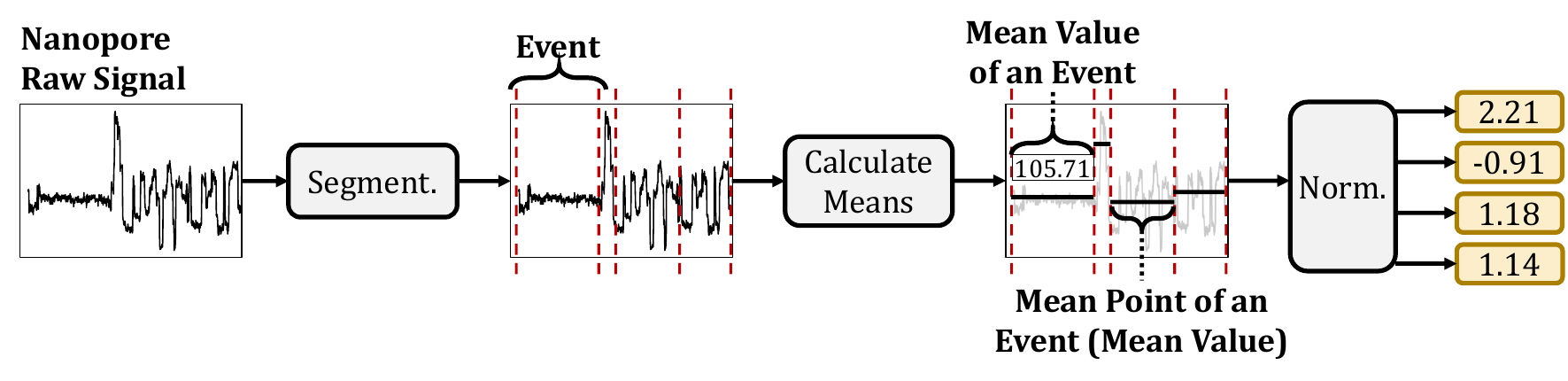}
  \caption{Detecting events from raw signals.}
  \label{rh:fig:sigtoev}
\end{figure}

\head{Signal-To-Event Conversion} Our goal is to accurately convert the series of raw nanopore signals into a set of values where each value corresponds to certain DNA sequences of fixed length $k$, k-mers, and consecutive values differ by one base. To achieve this, \rh converts the raw signals into their corresponding values in three steps, as shown in Figure~\ref{rh:fig:sigtoev}. First, to accurately identify the distinct regions in the raw signal that correspond to a certain k-mer from DNA, \rh performs a segmentation step as described in a basecalling tool, Scrappie, and used by earlier works \unc and \sig. The segmentation step aims to eliminate the factors that affect the speed of the DNA molecules passing through a nanopore, as the speed affects the number of signal measurements taken for a certain amount of bases in DNA. To perform the segmentation step, \rh identifies the boundaries in the signal where the signal value changes significantly compared to the certain amount of previously measured signal values, which indicates a base change in the nanopore. Such boundaries are computed using a statistical test, known as \emph{Welch's t-test}~\cite{ruxton_unequal_2006}, over a rolling window of consecutive signals. \rhcap performs this t-test for multiple windows of different lengths to avoid the variables that cause a change in the number of current measurements due to the varying speed of DNA through a nanopore, known as \emph{skip} and \emph{stay} errors~\cite{david_nanocall_2017}. Signals that fall within the same segment (i.e., between the same measured boundaries) are usually called \emph{events} since each event contains the signals from a reading of a fixed amount of DNA bases, k-mers.

Second, since the number of signals that each event includes is not constant across different events due to the stay and skip errors, \rh generates a single value for each event to quickly avoid these potential errors and other factors that cause variations from reading the same amount of DNA bases. To this end, \rh measures the mean value of the signals that fall within the same segment and uses this mean value for an event.

Third, since the amplitudes of the signal measurements may significantly vary when reading k-mers at different times, \rh normalizes the mean event values using the event values generated from the nanopore within the same certain time interval in a streaming fashion. Although this time interval parameter can be modified in our tool, the default configuration of \rh processes the events of signals generated by the nanopore within one second. For normalization, \rh uses the same z-score calculation that it uses for normalizing the event values generated from reference sequences as described earlier. \rhcap uses these normalized values as event values when comparing with the event values from reference sequences.

\subsection{Quantization of Events} \label{rh:subsec:quantization}

\begin{figure}[tbh]
  \centering
  \includegraphics[width=0.9\linewidth]{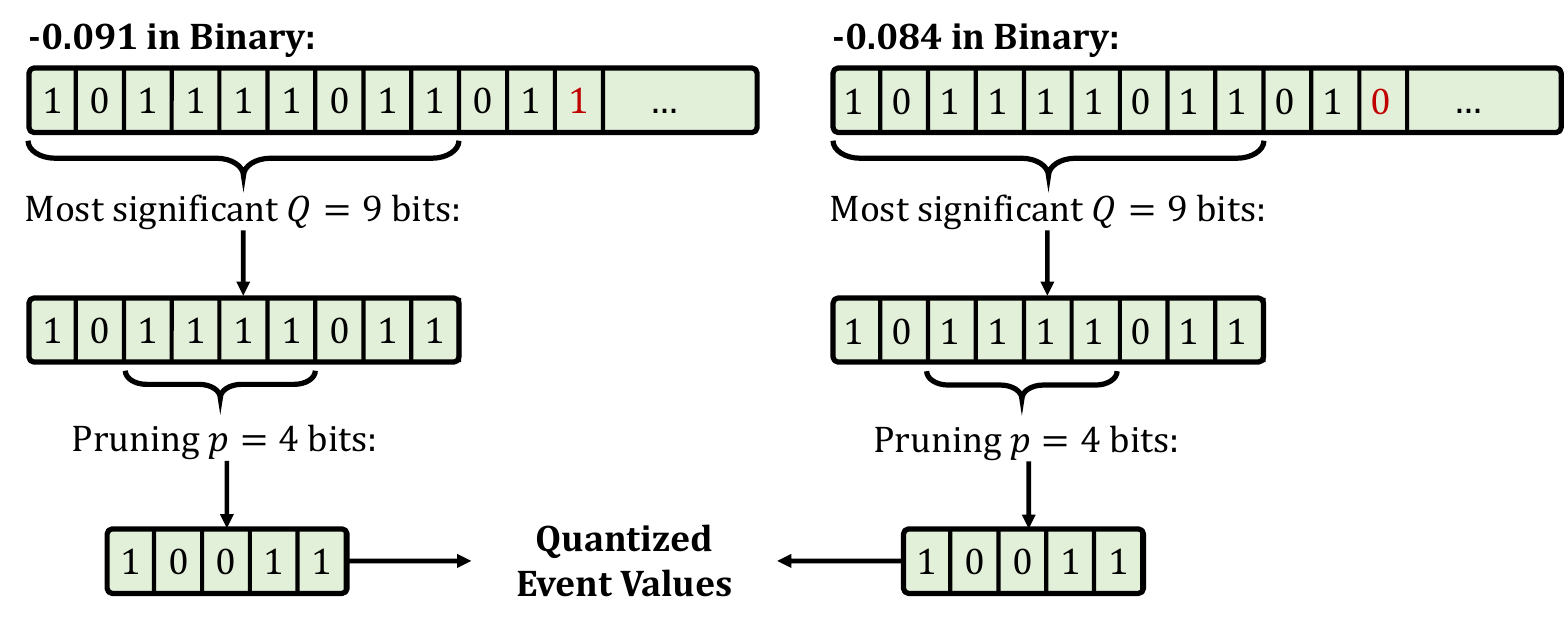}
  \caption{Quantization of two event values.}
  \label{rh:fig:quantization}
\end{figure}

Our goal is to avoid the effects of generating different event values when reading the same k-mer content from nanopores so that we can identify k-mer matches by directly matching events. Although the segmentation and normalization steps explained in Section~\ref{rh:subsec:eventgeneration} can avoid the potential sequencing errors, such as stay and skip errors and significant changes in the current readings at different times, these approaches still do not guarantee to generate \emph{exactly} the same event values when reading the \emph{same} k-mer content. This is because \emph{slight} changes in the normalized event values may occur when reading the same DNA content due to the high sensitivity and stochasticity of nanopores~\cite{david_nanocall_2017}. Thus, it is challenging to generate the same event value for the same k-mer content after the segmentation and normalization steps. Since these event values generated from reading the same k-mer content are expected to be close to each other~\cite{zhang_real-time_2021}, we propose a quantization mechanism that encodes event values so that events with close mean values can have the same quantized value in two steps as shown in Figure~\ref{rh:fig:quantization}.

First, to increase the probability of assigning the same value for similar event values, \rh trims the least significant fractional part of mean values by using only the most significant $Q$ bits of these mean event values from their binary format, which we represent as $E[1,Q]$ for simplicity where $E$ is the event value and $E[1,Q]$ gives the most significant $Q$ bits of $E$. We assume that the mean event values are represented by the standard single-precision floating-point format with the sign, exponent, and fraction bits. This enables \rh to reduce the wide range of floating-point numbers into a smaller range \emph{without} significantly losing from the accuracy such that event values closer to each other can be represented by the same value in the smaller range of values. We can perform this trimming technique without significant sensitivity loss because we observe that these normalized event values mostly use at most six digits from the fractional part of their values, leaving a large number of fractional bits useless.

Second, to avoid using redundant bits that may carry little or no information in the most significant $Q$ bits of an event value, \rh prunes $p$ bits after the most significant two bits of $E[1,Q]$ such that $2+p < Q$ and the resulting \emph{quantized value} is $E[1,2]E[3+p,Q]$. For simplicity, we show the quantized value of $E$ as $E_{Q,p}$. By ignoring these $p$ bits, we effectively pack $Q$ bits into $Q-p$ bits without losing significant information from event values. We can perform such a pruning operation because we observe that the normalized event values are usually in the range $[-3,3]$ such that these $p$ bits provide little information in distinguishing different event values due to the small range of values. We note that these $Q$ and $p$ values are parameters to \rh and can empirically be adjusted based on the required sensitivity and quantization efficiency. This quantization technique enables \rh to assign the same quantized values for a pair of \emph{close} event values, $E$ and $F$, that may be generated from reading the same k-mer such that $E_{Q,p} = F_{Q,p}$ where $|E-F| < \epsilon$ and $\epsilon$ is small enough for two events to represent the same k-mer content. \rhcap always uses the most significant two bits as these two bits consistently carry the most significant information of the normalized event values, including the sign bit.

\subsection{Generating the Hash Values} \label{rh:subsec:hashvalues}

\begin{figure}[tbh]
  \centering
  \includegraphics[width=0.6\columnwidth]{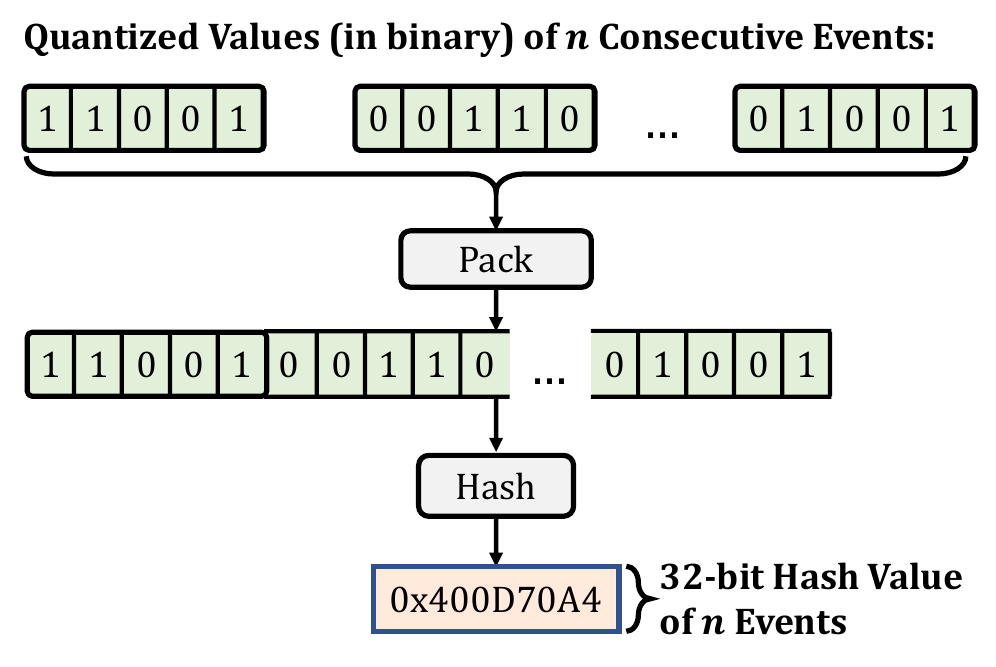}
  \caption{Generating a hash value from $n$ consecutive quantized event values.}
  \label{rh:fig:hashing}
\end{figure}

Our goal is to generate values for \emph{large} regions of raw nanopore signals and reference sequences such that these values can be used to efficiently and accurately identify similarities between raw signals and a reference genome. To this end, \rh generates hash values using quantized values of events in two steps, as shown in Figure~\ref{rh:fig:hashing}. First, to avoid finding a large number of matches, \rh uses the quantized values of $n$ consecutive events to pack them in $n \times (Q-p)$ bits while preserving the order information of these consecutive events. \rhcap uses several consecutive events in a single hash value because matching a single event is likely to generate a larger number of matches for larger genomes as a single event usually corresponds to a k-mer of 6 to 9 bases depending on the nanopore model~\cite{david_nanocall_2017}. It is essential to use several consecutive events to reduce the number of matching regions between raw signals and the reference genome by increasing the region that these consecutive events span.

Second, to efficiently and accurately find matches between large regions of raw signals and a reference genome using a constrained space, \rh uses a low collision hash function to generate a 32-bit hash value from $n \times (Q-p)$ bits of $n$ consecutive quantized event values. Since $n \times (Q-p)$ can be larger than 32, using such a hash function is likely to increase the collision rate for dissimilar regions. To avoid inaccurate similarity identifications due to these incorrect collisions, \rh requires several matches of hash values within close proximity for similarity identification, which we explain next.

\subsection{Seeding and Mapping} \label{rh:subsec:mapping}

To efficiently identify similarities, \rh uses hash values generated from raw nanopore signals and the reference genome in two steps. First, \rh efficiently identifies matching regions between raw nanopore signals and a reference genome by matching their hash values. These hash values used for matching are usually known as \emph{seeds}. Matching seeds enable efficiently finding similar regions between raw nanopore signals and a reference genome. Second, \rh uses the chaining algorithm proposed in Sigmap~\cite{zhang_real-time_2021} to identify the \emph{best} colinear matching seeds that are close to each other in both raw nanopore signal and a reference genome. The region that the \emph{best} chain of seed matches cover is the \emph{mapping position} that \rh identifies as a similar region.

The chaining algorithm is useful for two reasons. First, the chaining algorithm can tolerate mismatches and indels as it allows gaps between seed matches, which enables finding similar regions with many seed matches without requiring the entire region to match exactly. To investigate this, we show the average gap length between a pair of anchors that \rh finds when mapping raw nanopore signals in various datasets in Table~\ref{rh:tab:profilinggap}. Read and reference anchors show the average gap length between a pair of anchors found in reads and the reference genome, respectively. We find that the chaining algorithm can tolerate a large number of mismatches and indels. Second, incorrect seed matches due to collisions or our quantization mechanism that may generate the same quantized value for distinctly dissimilar events are likely to be filtered in the chaining step due to the difficulty of finding colinear seed matches in highly dissimilar regions. We note that we modify the original chaining algorithm in Sigmap by disabling the distance coefficient as \rh does not calculate the distance between seed matches.

\begin{table}[tbh]
\centering
\caption{The average gap length between a pair of anchors in reads and a reference genome.}
\begin{tabular}{@{}lrrrrr@{}}\toprule
\textbf{Tool}     & \textbf{\emph{SARS-CoV-2}} & \textbf{\emph{E. coli}} & \textbf{\emph{Yeast}} & \textbf{\emph{Green Algae}} & \textbf{\emph{Human}} \\\midrule
Read Anchors      & 25.37                      & 42.92                   & 79.31                 & 148.58                      & 200.06\\
Reference Anchors & 20.06                      & 33.39                   & 67.55                 & 127.03                      & 165.44\\\bottomrule
\end{tabular}

\label{rh:tab:profilinggap}
\end{table}

To efficiently map raw signals to a reference genome, \rh provides efficient data structures. To this end, \rh uses hash tables to store the hash values generated from reference genomes (i.e., the indexing step) and efficiently query the same hash table with the hash values generated from the raw signal as the read is sequenced from a nanopore to find positions in the reference genome with matching hash values. \rhcap uses the events in chunks (i.e., collection of events generated within a certain time interval) to find seed matches and perform chaining in a streaming fashion such that the chaining computation from previous chunks (i.e., seed matches) is transferred to the next chunk if the mapping is unsuccessful for the current chunk.

\subsection{The \emph{Sequence Until} Mechanism} \label{rh:subsec:sequenceuntil}

\revb{Our goal with the \emph{Sequence Until} mechanism is to reduce time and cost spent during sequencing by dynamically stopping the \emph{entire} sequencing run if further sequencing is unnecessary for an accurate analysis. To do so, \emph{Sequence Until} aims to continuously determine if further sequencing is unlikely to change the analysis outcome. Although \emph{Sequence Until} is not limited to a particular analysis type, we exemplify the workflow of \emph{Sequence Until} by focusing on the relative abundance estimation use case. Figure~\ref{rh:fig:sequenceuntil} shows the main steps in \emph{Sequnce Until}.

First, \emph{Sequence Until} continuously generates relative abundance estimations as new reads are sequenced. After every $n$ reads\circlednumber{1}, \emph{Sequence Until} performs an estimation of the relative abundance, producing a series of estimations as sequencing progresses.

Second, \emph{Sequence Until} maintains the last $t$ estimation results\circlednumber{2} to assess the consistency of the abundance estimations. By focusing on a rolling window of the most recent estimations, \emph{Sequence Until} can effectively monitor the ongoing trends in the relative abundance estimations.

Third, \emph{Sequence Until} employs a cross-correlation technique to detect outliers among the last $t$ relative abundance estimations\circlednumber{3}. Cross-correlation helps to identify whether recently sequenced reads significantly cause deviations from the general relative abundance estimation pattern, indicating that additional sequencing might still provide valuable information. The presence of outliers suggests that the sequencing process should continue, as further sequencing might still refine the abundance estimations.

Fourth, when no outliers are detected among the last $t$ estimations, \emph{Sequence Until} assumes that the abundance estimations have reached consistency. In such cases, the \emph{Sequence Until} mechanism signals that the sequencing can be fully stopped (by using the Run Until functionality in nanopore sequencers)\circlednumber{4}, as further reads are unlikely to alter the relative abundance results substantially. This complete halt of sequencing helps to avoid redundant sequencing, thereby reducing the overall sequencing time and cost.

The \emph{Sequence Until} mechanism is particularly beneficial for applications where the goal is to estimate the relative abundance of species accurately and efficiently. By dynamically adjusting the sequencing run based on real-time analysis, \emph{Sequence Until} enables reducing unnecessary sequencing. This mechanism can be integrated into various sequencing workflows and analysis tools other than \rh, including those utilizing other real-time relative abundance estimation tools~\cite{shivakumar_sigmoni_2024}.

\begin{figure}[tbh]
\centering
\includegraphics[width=0.9\columnwidth]{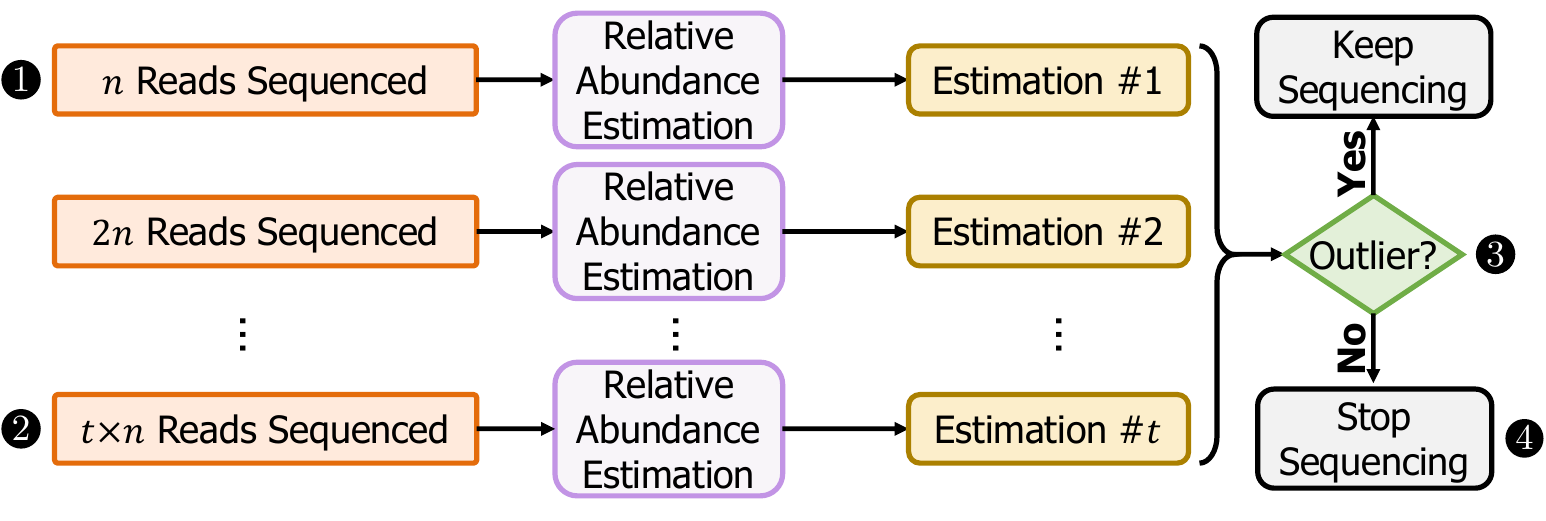}
\caption{Overview of the Sequence Until mechanism.}
\label{rh:fig:sequenceuntil}
\end{figure}
}

\section{Results} \label{rh:sec:results}
\subsection{Evaluation Methodology} \label{rh:subsec:evaluation}

We implement \rh as a tool for mapping raw nanopore signals to a reference genome. Similar to regular read mapping tools, \rh has two steps to complete the mapping process: 1)~indexing the reference genome and 2)~mapping raw signals. Although indexing is usually a one-time task that can be performed prior to the mapping step, the indexing of \rh can be performed relatively quickly within a few minutes for large genomes (Table~\ref{rh:tab:indexing_resources}). \rhcap provides the mapping information using a standard pairwise mapping format (PAF). In our implementation, we provide an extensive set of parameters that allow configuring several options to fit \rh for many other applications and nanopore models that we do not evaluate, such as configuring details about the nanopore model (e.g., number of bases per second), number of events that can be included in a single hash value, range of bits to quantize, enabling seeding techniques such as minimizers and fuzzy seed matching. We also provide a default set of parameters that we empirically choose for each common application of real-time genome analysis. These default parameters are set to accurately and efficiently analyze 1)~very small (e.g., viral) genomes, 2)~small and mid-sized genomes (i.e., genomes with less than a few hundred million bases), 3)~large genomes (e.g., genomes with a few billion bases such as a human genome). We show the details regarding these parameter selections and the versions of tools in Supplementary Tables~\ref{rh:tab:parameters}, \ref{rh:tab:presets}, and \ref{rh:tab:versions}.

We evaluate \rh in terms of its performance, peak memory usage, accuracy, and estimated benefits in sequencing time and cost compared to two state-of-the-art tools \unc and \sig. For performance, we evaluate the throughput and overall runtime of each tool in terms of the number of bases they can process per second. Throughput determines if the tool is at least as fast as the speed of DNA passing through a nanopore. For many nanopore models (e.g., R9.4), a DNA strand passes through a pore at around 450 bases per second~\cite{kovaka_targeted_2021, zhang_real-time_2021}. It is essential to provide a throughput higher than the throughput of the nanopore to enable real-time genome analysis. To calculate the throughput, we use the tool that UNCALLED provides, \texttt{UNCALLED pafstats}, which measures the throughput of the tool from the number of bases that the tool processes and the time it takes to process those bases. Although theoretically, it is not possible to exceed the throughput of a nanopore due to the speed of raw signal generation, for comparison purposes, such a limitation is ignored by \texttt{UNCALLED pafstats}. For overall runtime, we calculate CPU time and real-time using 32 threads. CPU time shows the overall amount of CPU seconds spent running a tool, while real-time shows the overall elapsed (i.e., wall clock) time. All of these tools support multi-threading, where multiple reads can be mapped simultaneously using a single thread for each read. For all of these tools, assigning a larger number of threads enables processing a larger number of reads in parallel, similar to the behavior of nanopore sequencers with hundreds to thousands of pores (i.e., channels). We note that the throughput and mapping time per read values are not affected by the thread counts as 1)~these are measured per read and 2)~single thread performs the mapping of a single read.

For accuracy, we evaluate the correctness of the mapping positions that each tool provides when compared to the ground truth mapping positions. To generate the ground truth mapping, we use a read mapping tool, minimap2~\cite{li_minimap2_2018}, to map the basecalled sequences of raw nanopore signals to their corresponding \emph{whole-genome} references. We use \texttt{UNCALLED pafstats} to compare the mapping output of a tool with the ground truth mapping to find the number of true positives or $TP$ (i.e., correct mappings), false positives or $FP$ (i.e., incorrect mappings), and false negatives or $FN$ (i.e., unmapped reads that are mapped in ground truth). Correct and incorrect mappings are identified based on the distance of the mapping positions between ground truth and the tool. To evaluate the accuracy, we calculate the precision ($P = TP/(TP+FP)$), recall ($R = TP/(TP + FN)$) and the $F_1$ ($F_{1} = 2 \times (P \times R)/(P+R)$) values.

For estimating the benefits in sequencing time and cost of each tool, we calculate the average length of sequenced bases per read when using \unc and \rh and the average number sequenced chunk of signals for \sig and \rh. We compare \rh with \sig in terms of the number of chunks because \sig does not provide the number of bases when a read is unmapped, while both tools provide the number of chunks used when a read is mapped or unmapped. These chunks include a portion of the signal produced by a nanopore within a certain time interval, which is by default set as one second of data for both \rh and \sig. The average length of bases and the number of chunks determine the estimations of how quickly each tool can make a mapping decision to activate Read Until before sequencing the remaining portion of a read, which indicates the potential savings from overall sequencing time and cost.

We evaluate \rh, \unc, and \sig for three applications 1)~read mapping, 2)~relative abundance estimation, and 3)~contamination analysis. Read mapping aims to map the raw signals to their corresponding reference genomes. Relative abundance estimation measures the abundance of each genome relative to other genomes in the same sample by mapping raw signals to a given set of reference genomes. Contamination analysis aims to identify if a sample is contaminated with a certain genome (e.g., a viral genome) by mapping raw signals to the reference genome that the sample may be contaminated with. For each tool, we use their default parameter settings in our evaluation.

To evaluate each of these applications, we use real datasets that we list in Table~\ref{rh:tab:dataset}. These datasets include both raw nanopore signals in the FAST5 format and their corresponding basecalled sequences in the FASTA format. We note that \rh can also use POD5 files. For relative abundance estimation, we create a mock community using all the read sets from datasets D1 to D5, and the reference genome is the combination of reference genomes used in these datasets. We slightly modify the reference genome we use in the relative abundance estimation such that the sequence IDs in the reference genome provide additional information about the species (e.g., taxonomy IDs) to enable calculating relative abundance in real-time. For contamination analysis, we combine the SARS-CoV-2 read sets (D1) with human read sets (D5) to identify if the combined sample is contaminated with the SARS-CoV-2 sample by mapping raw signals in the combined set to the SARS-CoV-2 reference genome. For all evaluations, we use the AMD EPYC 7742 processor at 2.26GHz to run the tools.

\begin{table}[tbh]
\centering
\caption{Details of datasets used in our evaluation.}
\resizebox{\linewidth}{!}{
\begin{tabular}{@{}clrrllr@{}}\toprule
& \textbf{Organism} & \textbf{Reads} & \textbf{Bases}              & \textbf{SRA}       & \textbf{Reference}      & \textbf{Genome}\\
& \textbf{}         & \textbf{(\#)}  & \textbf{(\#)}               & \textbf{Accession} & \textbf{Genome}         & \textbf{Size}  \\\midrule
\multicolumn{7}{c}{Read Mapping} \\\midrule
D1 & \emph{SARS-CoV-2}   & 1,382,016      & 594M                   & CADDE Centre & GCF\_009858895.2   & 29,903 \\\midrule
D2 & \emph{E. coli}      & 353,317        & 2,365M                 & ERR9127551         & GCA\_000007445.1   & 5M \\\midrule
D3 & \emph{Yeast}        & 49,989         & 380M                   & SRR8648503         & GCA\_000146045.2   & 12M\\\midrule
D4 & \emph{Green Algae}  & 29,933         & 609M                   & ERR3237140         & GCF\_000002595.2   & 111M\\\midrule
D5 & \emph{Human HG001}  & 269,507        & 1,584M                 & FAB42260 Nanopore WGS & T2T-CHM13 (v2)     & 3,117M\\\bottomrule
\multicolumn{7}{c}{Relative Abundance Estimation} \\\midrule
\multicolumn{2}{c}{D1-D5} & 2,084,762    & 5,531M                  & D1-D5              & D1-D5              & 3,246M\\\bottomrule
\multicolumn{7}{c}{Contamination Analysis} \\\midrule
\multicolumn{2}{c}{D1 and D5} & 1,651,523    & 2,178M                 & D1 and D5             & D1                 & 29,903\\\bottomrule
\multicolumn{7}{l}{Dataset numbers (e.g., D1-D5) show the combined datasets. Base counts in millions (M).}\\
\end{tabular}

}
\label{rh:tab:dataset}
\end{table}

\head{Evaluating \emph{Sequence Until}} Our goal is to avoid redundant sequencing to reduce sequencing time and cost for relative abundance estimation. We find that the Run Until mechanism can be utilized to \emph{fully} stop the sequencing run when the real-time relative abundance estimation reaches a certain confidence level to achieve accurate estimations, which we call \emph{Sequence Until}. While a similar mechanism is evaluated to enrich the coverage depth of low-abundance species~\cite{weilguny_dynamic_2023} using \ru, we evaluate the potential benefits of Run Until for low-cost relative abundance estimations. \rhcap provides a set of parameters to adjust these parameters related to \emph{Sequence Until}.

We evaluate the benefits of \emph{Sequence Until} by comparing 1)~\rh without \emph{Sequence Until} and 2)~\rh with \emph{Sequence Until} in terms of 1)~the difference in the relative abundance estimations and 2)~the estimated benefits in sequencing time and cost. To evaluate \emph{Sequence Until} in a realistic sequencing environment where reads from different species can be sequenced in a random order, we randomly shuffle the reads in the relative abundance dataset and generate a set of 50,000 reads with a random order of species so that we can simulate this random behavior. We also find that \emph{Sequence Until} can be applied to other mechanisms. To evaluate the potential benefits of \emph{Sequence Until}, we simulate the benefits when using UNCALLED with \emph{Sequence Until} and compare it with \rh.

\subsection{Performance and Peak Memory} \label{rh:subsec:perfmemory}
\begin{figure}[tbh]
  \centering
  \includegraphics[width=0.8\columnwidth]{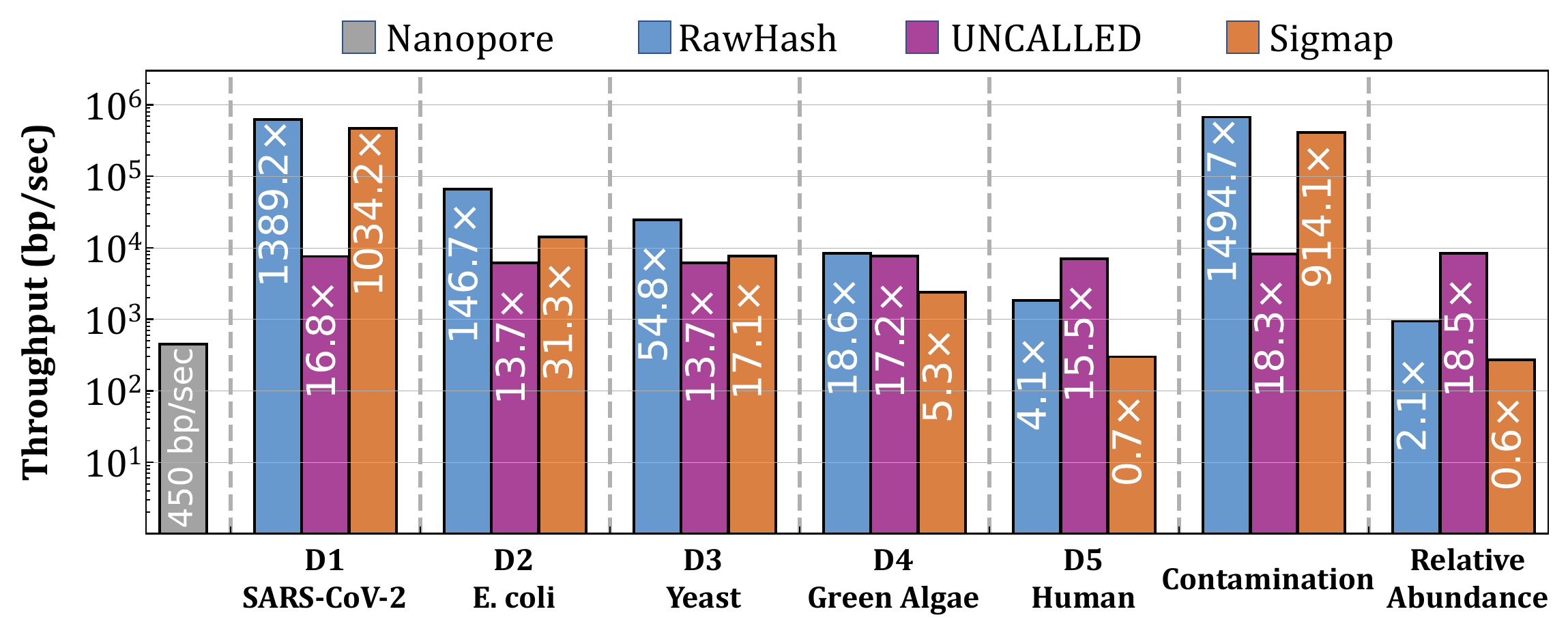}
  \caption{Throughput of each tool. Values inside the bars show the throughput ratio between each tool and a nanopore.}
  \label{rh:fig:throughput}
\end{figure}

\begin{figure}[tbh]
  \centering
  \includegraphics[width=0.6\columnwidth]{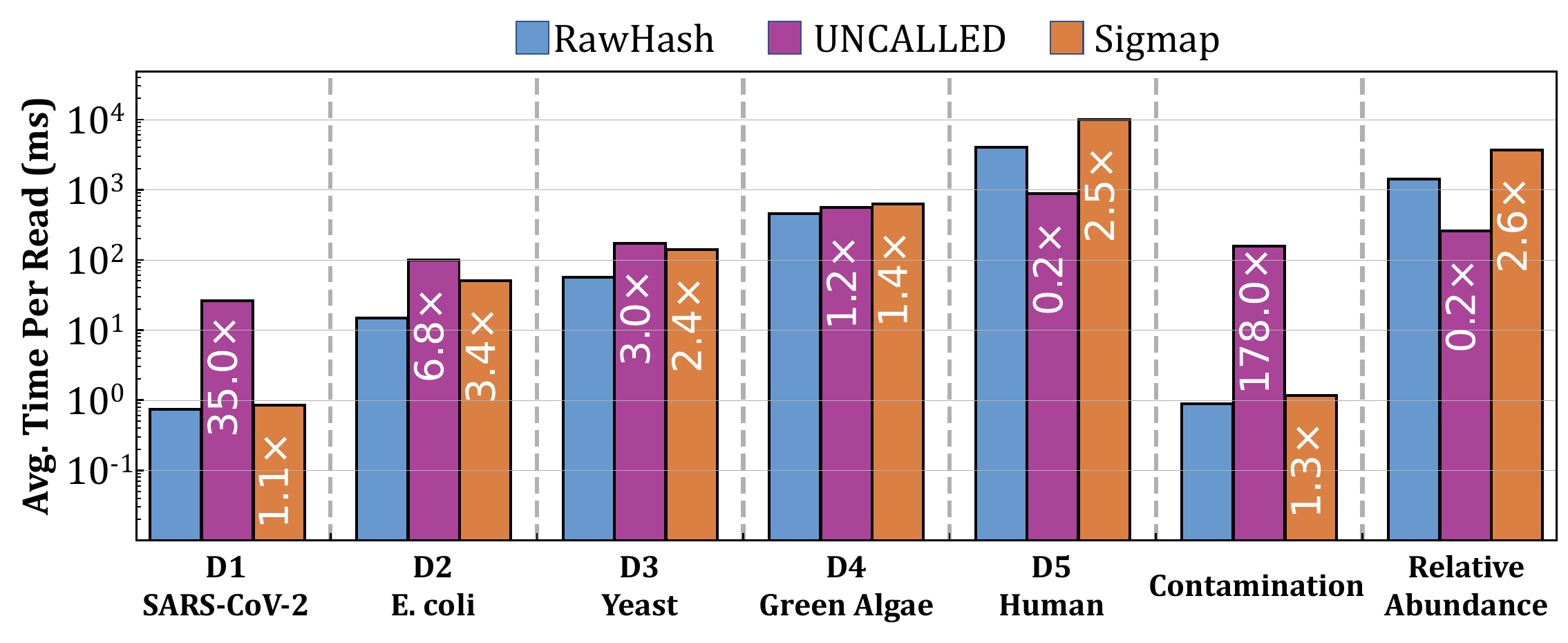}
  \caption{Average time spent per read by each tool in real-time. Values inside the bars show the speedups that \rh provides over other tools in each dataset.}
  \label{rh:fig:timeperread}
\end{figure}

Figure~\ref{rh:fig:throughput} shows the throughput of regular nanopores that we use as a baseline and the throughput of the tools when mapping raw nanopore signals to each dataset for read mapping, contamination analysis, and relative abundance estimation. Figure~\ref{rh:fig:timeperread} and Tables~\ref{rh:tab:indexing_resources} and \ref{rh:tab:mapping_resources} show the mapping time per read, and the computational resources required for indexing and mapping, respectively. We make six key observations.
First, \rh and \unc are the \emph{only} tools that can perform real-time genome analysis for large genomes, as they can provide higher throughputs than nanopores for all datasets. \sig \emph{cannot} perform real-time genome analysis for large genomes as it can provide $0.7\times$ and $0.6\times$ throughput of a nanopore for human genome mapping and relative abundance estimations, respectively. \rhcap can achieve high throughput as its seeding mechanism is based on efficiently matching hash values compared to the costly distance calculations that \sig performs for matching seeds, which shows poor scalability for larger genomes.
Second, the throughput of \unc is not affected by the genome size as it provides a near-constant throughput of around $16\times$ for all applications. This is because \unc uses FM-index~\cite{ferragina_opportunistic_2000} and a branching algorithm that provides robust scaling with respect to the reference genome size~\cite{kovaka_targeted_2021}.
Third, the throughput of \rh decreases with larger genomes as the seeding and chaining steps start taking up a larger fraction of the entire runtime of \rh.
Fourth, \rh provides an average throughput \rhavgthrU and \rhavgthrS better than \unc and \sig, while providing an average mapping speedup of \rhavgmtU and \rhavgmtS per read, respectively. Higher throughput with faster mapping times suggests that the mapping time improvements of \rh are mainly due to its computational efficiency rather than the ability to sequence shorter prefixes of reads than \unc and \sig.
Fifth, for indexing, \sig usually requires a larger amount of computational resources in terms of both runtime and peak memory usage.
Sixth, for mapping, \unc is the most efficient tool in terms of the peak memory usage as it requires at most 10GB of peak memory while 1)~\rh requires less than 12GB of memory for almost all the datasets and 2)~\sig requires significantly larger memory space than both tools. \rhcap has a larger memory footprint, $\sim52$GB, than \unc for large genomes. Although such large memory requirements for larger genomes can lead to challenges in using \rh for mobile devices with limited computational resources, such a requirement can be mitigated by using more efficient seeding techniques such as minimizers, which we leave as future work.
We conclude that \rh provides significant benefits in improving the throughput and performance for the real-time analysis of large genomes while matching the throughput of nanopores.

\begin{table}[tbh]
\centering
\caption{Computational resources required in the indexing step of each tool.}
\resizebox{0.8\linewidth}{!}{\begin{tabular}{@{}lrrrrrrr@{}}\toprule
\textbf{Tool} & \textbf{\emph{Contamination}} & \textbf{\emph{SARS-CoV-2}} & \textbf{\emph{E. coli}} & \textbf{\emph{Yeast}} & \textbf{\emph{Green Algae}} & \textbf{\emph{Human}} & \textbf{\emph{Relative Abundance}} \\\midrule
\multicolumn{8}{c}{CPU Time (sec)} \\\midrule
\unc & 8.72 & 9.00 & 11.08 & 18.62 & 285.88 & 4,148.10 & 4,382.38 \\
\sig & 0.02 & 0.04 & 8.66 & 24.57 & 449.29 & 36,765.24 & 40,926.76 \\
\rh & 0.18 & 0.13 & 2.62 & 4.48 & 34.18 & 1,184.42 & 788.88 \\\midrule
\multicolumn{8}{c}{Real time (sec)} \\\midrule
\unc & 1.01 & 1.04 & 2.67 & 7.79 & 280.27 & 4,190.00 & 4,471.82 \\
\sig & 0.13 & 0.25 & 9.31 & 25.86 & 458.46 & 37,136.61 & 41,340.16 \\
\rh & 0.14 & 0.10 & 1.70 & 2.06 & 15.82 & 278.69 & 154.68 \\\midrule
\multicolumn{8}{c}{Peak memory (GB)} \\\midrule
\unc & 0.07 & 0.07 & 0.13 & 0.31 & 11.96 & 48.44 & 47.81 \\
\sig & 0.01 & 0.01 & 0.40 & 1.04 & 8.63 & 227.77 & 238.32 \\
\rh & 0.01 & 0.01 & 0.35 & 0.76 & 5.33 & 83.09 & 152.80 \\\bottomrule
\end{tabular}
}
\label{rh:tab:indexing_resources}
\end{table}

\begin{table}[tbh]
\centering
\caption{Computational resources required in the mapping step of each tool.}
\resizebox{0.8\linewidth}{!}{\begin{tabular}{@{}lrrrrrrr@{}}\toprule
\textbf{Tool} & \textbf{\emph{Contamination}} & \textbf{\emph{SARS-CoV-2}} & \textbf{\emph{E. coli}} & \textbf{\emph{Yeast}} & \textbf{\emph{Green Algae}} & \textbf{\emph{Human}} & \textbf{\emph{Relative Abundance}} \\\midrule
\multicolumn{8}{c}{CPU Time (sec)} \\\midrule
\unc         & 265,902.26                     & 36,667.26 & 35,821.14 & 8,933.52 & 16,769.09 & 262,597.83 & 586,561.54 \\
\sig         & 4,573.18                       & 1,997.84 & 23,894.70 & 11,168.96 & 31,544.55 & 4,837,058.90 & 11,027,652.91 \\
\rh        & 3,721.62                       & 1,832.56 & 8,212.17 & 4,906.70 & 25,215.23 & 2,022,521.48 & 4,738,961.77 \\\midrule
\multicolumn{8}{c}{Real time (sec)} \\\midrule
\unc & 20,628.57 & 2,794.76 & 1,544.68 & 285.42 & 2,138.91 & 8,794.30 & 19,409.71 \\
\sig & 6,725.26 & 3,222.32 & 2,067.02 & 1,167.08 & 2,398.83 & 158,904.69 & 361,443.88 \\
\rh & 3,917.49 & 1,949.53 & 957.13 & 215.68 & 1,804.96 & 65,411.43 & 152,280.26 \\\midrule
\multicolumn{8}{c}{Peak memory (GB)} \\\midrule
\unc & 0.65 & 0.19 & 0.52 & 0.37 & 0.81 & 9.46 & 9.10 \\
\sig & 111.69 & 28.26 & 111.11 & 14.65 & 29.18 & 311.89 & 489.89 \\
\rh & 4.13 & 4.20 & 4.16 & 4.37 & 11.75 & 52.21 & 55.31 \\\bottomrule
\end{tabular}
}
\label{rh:tab:mapping_resources}
\end{table}

\subsection{Accuracy} \label{rh:subsec:accuracy}

Table~\ref{rh:tab:accuracy} shows the accuracy results of tools for each dataset and application. We make four key observations. First, \rh provides the best accuracy in terms of precision, recall, and $F_{1}$ values compared to \unc and \sig when mapping reads to large genomes (i.e., the human genome and the relative abundance estimation). \rhcap can efficiently match several events using hash values, which is specifically beneficial in reducing the number of matching regions in large genomes and increasing the specificity due to finding longer matches compared to \unc and \sig.

\begin{table}[tbh]
\centering
\caption{Mapping accuracy.}
\resizebox{0.6\linewidth}{!}{\begin{tabular}{@{}llrrr@{}}\toprule
\textbf{Dataset}   &           & \textbf{\unc}   & \textbf{\sig}   & \textbf{\rhcap}  \\\midrule
\multicolumn{5}{c}{Read Mapping} \\\midrule
D1	               & Precision & 0.9547          & \textbf{0.9929} & 0.9868          \\
\emph{SARS-CoV-2}  & Recall    & \textbf{0.9910} & 0.5540          & 0.8735 		 \\
                   & $F_1$     & \textbf{0.9725} & 0.7112          & 0.9267          \\\midrule
D2	               & Precision & 0.9816          & \textbf{0.9842} & 0.9573          \\
\emph{E. coli}     & Recall    & \textbf{0.9647} & 0.9504          & 0.9009 		 \\
                   & $F_1$     & \textbf{0.9731} & 0.9670          & 0.9282          \\\midrule
D3	               & Precision & 0.9459          & 0.9856          & \textbf{0.9862} \\
\emph{Yeast}       & Recall    & \textbf{0.9366} & 0.9123          & 0.8412 		 \\
                   & $F_1$     & 0.9412          & \textbf{0.9475} & 0.9079          \\\midrule
D4	               & Precision & 0.8836          & \textbf{0.9741} & 0.9691          \\
\emph{Green Algae} & Recall    & 0.7778          & \textbf{0.8987} & 0.7015 		 \\
                   & $F_1$     & 0.8273          & \textbf{0.9349} & 0.8139          \\\midrule
D5	               & Precision & 0.4867          & 0.4287          & \textbf{0.8959} \\
\emph{Human HG001} & Recall    & 0.2379          & 0.2641          & \textbf{0.4054} \\
                   & $F_1$     & 0.3196          & 0.3268          & \textbf{0.5582} \\\midrule
\multicolumn{5}{c}{Relative Abundance Estimation} \\\midrule
	              & Precision & 0.7683          & 0.7928           & \textbf{0.9484} \\
D1-D5             & Recall    & 0.1273          & 0.2739           & \textbf{0.3076} \\
                  & $F_1$     & 0.2184          & 0.4072           & \textbf{0.4645} \\\midrule
\multicolumn{5}{c}{Contamination Analysis} \\\midrule
                  & Precision & \textbf{0.9378} & 0.7856           & 0.8733          \\
D1, D5            & Recall    & \textbf{0.9910} & 0.5540           & 0.8735 		   \\
                  & $F_1$     & \textbf{0.9637} & 0.6498           & 0.8734          \\\bottomrule
\multicolumn{5}{l}{\footnotesize Best results are highlighted with \textbf{bold} text.} \\
\end{tabular}
}
\label{rh:tab:accuracy}
\end{table}

Second, \rh and \unc can accurately perform contamination analysis while \sig suffers from significantly lower precision and recall values. Due to the nature of a contamination analysis, it is essential to correctly eliminate the genomes other than the contaminating genome (precision) without missing the correct mappings of reads from the contaminating genome (recall). Unfortunately, \sig cannot provide high values in any of these categories, making it significantly unsafe for contamination detection.

Third, the precision of \rh does not drop with the increased length in the reference genome due to the benefits of finding long matches, which provides a higher confidence in read mapping.

Fourth, although \rh does not provide the best accuracy when mapping reads to genomes smaller than the human genome, its accuracy is on par with \unc and \sig for these genomes. \unc and \sig can achieve high recall values as their mechanisms are best optimized for accurately handling matches in relatively smaller genomes with fewer repeats and ambiguous mappings~\cite{zhang_real-time_2021, kovaka_targeted_2021}. We conclude that \rh is the \emph{only} tool that can accurately scale to performing real-time genome analysis for large genomes, especially with significantly high precision rates.

\head{Relative Abundance Estimations}
Table~\ref{rh:tab:relativeabundance} shows the relative abundance estimations that each tool makes and the Euclidean distance of their estimation to the ground truth estimation. We make two key observations.
First, we find that \rh provides the most accurate relative abundance estimations in terms of the estimation distance to the ground truth compared to \unc and \sig. This observation correlates with the accuracy results we show in Table~\ref{rh:tab:accuracy} where \rh provides the best overall accuracy for relative estimation, which results in generating the most accurate relative abundance estimations.
Second, although \sig cannot perform real-time relative abundance estimation due to its throughput being lower than a nanopore (Figure~\ref{rh:fig:throughput}), \sig provides accurate estimations that are on par with \rh. This observation shows that while \sig provides mappings with more incorrect positions due to lower precision than \rh (Table~\ref{rh:tab:accuracy}), these reads with incorrect mapping positions are mostly mapped to their correct species.
We conclude that \rh is the \emph{only} tool that can \emph{accurately} be applied to analyze relative abundance estimations while matching the throughput of nanopores at a large-scale based on the prior knowledge of the set of reference genomes to map the reads.

\begin{table}[tbh]
\centering
\caption{Relative abundance estimations.}
\begin{tabular}{@{}lrrrrrr@{}}\toprule
              & \multicolumn{6}{c}{Estimated Relative Abundance Ratios} \\\cmidrule{2-7}
\textbf{Tool} & \textbf{\emph{SARS-CoV-2}} & \textbf{\emph{E. coli}} & \textbf{\emph{Yeast}} & \textbf{\emph{Green Algae}} & \textbf{\emph{Human}} & \textbf{Distance} \\\midrule
Ground Truth  & 0.0929                     & 0.4365                  & 0.0698                & 0.1179                      & 0.2828                &  N/A              \\\midrule
\unc          & 0.0026                     & 0.5884                  & 0.0615                & 0.1313                      & 0.2161                & 0.1895            \\\midrule
\sig          & 0.0419                     & 0.4191                  & 0.1038                & 0.0962                      & 0.3390                & 0.0877            \\\midrule
\rhcap      & 0.1249                     & 0.4701                  & 0.0957                & 0.0629                      & 0.2464                & \textbf{0.0847}   \\\bottomrule
\multicolumn{7}{l}{\footnotesize Best results are highlighted with \textbf{bold} text.} \\
\end{tabular}

\label{rh:tab:relativeabundance}
\end{table}

\subsection{Sequencing Time and Cost} \label{rh:subsec:sequencingcost}
Our goal is to estimate the benefits that each tool provides in reducing the sequencing time and cost. To this end, we measure the average length of sequenced bases and the average number of sequenced chunks per read. We make two key observations.
First, \rh provides significant benefits in reducing the sequencing time and cost for large genomes (e.g., Green Algae and Human) compared to \unc, as \rh can complete the mapping process per read by using smaller prefixes of reads.
Second, \rh uses on average $1.58 \times$ more chunks compared to \sig when mapping reads, which can proportionally lead to worse sequencing time and cost for \rh compared to \sig.
We conclude that although \unc and \sig provide better advantages in reducing sequencing time and cost for smaller genomes, \rh can provide significant reductions in sequencing time and cost for larger genomes compared to \unc.

\begin{table}[tbh]
\centering
\caption{The average sequenced length of bases and the number of chunks.}
\begin{tabular}{@{}lrrrrr@{}}\toprule
\textbf{Tool}     & \textbf{\emph{SARS-CoV-2}} & \textbf{\emph{E. coli}} & \textbf{\emph{Yeast}} & \textbf{\emph{Green Algae}} & \textbf{\emph{Human}} \\\midrule
\multicolumn{6}{c}{Average sequenced base length per read} \\\midrule
\unc              & \textbf{184.51}            & \textbf{580.52}         & \textbf{1,233.20}     & 5,300.15                    & 6,060.23              \\
\rh             & 513.95                     & 1,376.14                & 2,565.09              & \textbf{4,760.59}           & \textbf{4,773.58}      \\\midrule
\multicolumn{6}{c}{Average sequenced number of chunks per read} \\\midrule
\sig              & \textbf{1.01}              & \textbf{2.11}           & \textbf{4.14}         & \textbf{5.76}               & \textbf{10.40}        \\
\rh             & 1.24                       & 3.20                    & 5.83                  & 10.72                       & 10.70                 \\\bottomrule
\multicolumn{6}{l}{\footnotesize Best results are highlighted with \textbf{bold} text.} \\
\end{tabular}

\label{rh:tab:profilingsampling}
\end{table}

\subsection{Benefits of \emph{Sequence Until}}\label{rh:subsec:sequenceuntil_benefits}

\head{Simulated \emph{Sequence Until}}
Our goal is to estimate the benefits of implementing the \emph{Sequence Until} mechanism in \unc and compare it with \rh when they both use \emph{Sequence Until} under the same conditions. To this end, we use \texttt{shuf} in Linux to randomly shuffle the mapping files that both \rh and \unc generate for relative abundance and extract a certain portion of the randomly shuffled file to identify their relative abundance estimations after $0.01\%$, $0.1\%$, $1\%$, $10\%$, and $25\%$ of the overall reads in the sample are randomly sequenced from nanopores. 

Table~\ref{rh:tab:sequenceuntil} shows the distance of relative abundance estimations after a certain portion of the read is randomly sequenced from nanopores. We make two key observations.
First, \emph{both} \rh and \unc can significantly benefit from \emph{Sequence Until} by stopping sequencing after processing a smaller portion of the entire sample since their estimations using smaller portions are close to those using the entire set of reads (Table~\ref{rh:tab:relativeabundance}) in terms of their distance to the ground truth. This suggests that many other tools can benefit from \emph{Sequence Until} as their sensitivity to relative abundance estimations may not significantly change while providing opportunities for reducing the sequencing time and cost up to a certain threshold based on the tool.

Second, \rh can provide more accurate relative abundance estimations when using only $0.1\%$ of the reads than the estimation that \unc provides using the entire set of reads (Table~\ref{rh:tab:relativeabundance}). We conclude that \emph{Sequence Until} provides significant opportunities in reducing sequencing time and cost while more accurate tools such as \rh can benefit further from \emph{Sequence Until} by using fewer portions of the entire read set than the portions that less accurate tools would need to achieve similar accuracy.

\begin{table}[tbh]
\centering
\caption{Relative abundance with simulated \emph{Sequence Until}.}
\begin{tabular}{@{}lrrrrrr@{}}\toprule
                 & \multicolumn{6}{c}{Estimated Relative Abundance Ratios} \\\cmidrule{2-7}
\textbf{Tool}    & \textbf{\emph{SARS-CoV-2}} & \textbf{\emph{E. coli}} & \textbf{\emph{Yeast}} & \textbf{\emph{Green Algae}} & \textbf{\emph{Human}} & \textbf{Distance} \\\midrule
Ground Truth     & 0.0929                     & 0.4365                  & 0.0698                & 0.1179                      & 0.2828                &  N/A              \\\midrule
\unc ($25\%$)    & 0.0026                     & 0.5890                  & 0.0613                & 0.1332                      & 0.2139                & 0.1910          \\
\rhcap ($25\%$)   & 0.0271                     & 0.4853                  & 0.0920                & 0.0786                      & 0.3170                & \textbf{0.0995} \\\midrule
\unc ($10\%$)    & 0.0026                     & 0.5906                  & 0.0611                & 0.1316                      & 0.2141                & 0.1920          \\
\rhcap ($10\%$)   & 0.0273                     & 0.4869                  & 0.0963                & 0.0772                      & 0.3124                & \textbf{0.1004} \\\midrule
\unc ($1\%$)     & 0.0026                     & 0.5750                  & 0.0616                & 0.1506                      & 0.2103                & 0.1836          \\
\rhcap ($1\%$)    & 0.0259                     & 0.4783                  & 0.0987                & 0.0882                      & 0.3088                & \textbf{0.0928} \\\midrule
\unc ($0.1\%$)   & 0.0040                     & 0.4565                  & 0.0380                & 0.1910                      & 0.3105                & 0.1242          \\
\rhcap ($0.1\%$)  & 0.0212                     & 0.5045                  & 0.1120                & 0.0810                      & 0.2814                & \textbf{0.1136} \\\midrule
\unc ($0.01\%$)  & 0.0000                     & 0.5551                  & 0.0000                & 0.0000                      & 0.4449                & 0.2602          \\
\rhcap ($0.01\%$) & 0.0906                     & 0.6122                  & 0.0000                & 0.0000                      & 0.2972                & \textbf{0.2232} \\\bottomrule
\multicolumn{7}{l}{\footnotesize Percentages show the portion of the overall reads used. Best results are highlighted with \textbf{bold} text.} \\
\end{tabular}

\label{rh:tab:sequenceuntil}
\end{table}

\head{\emph{Sequence Until} with \rh}
Our goal is to evaluate \emph{Sequence Until} when used in real-time with \rh for relative abundance estimation. Table~\ref{rh:tab:sequenceuntil_real} shows the relative abundance estimations that \rh makes with and without \emph{Sequence Until}. We note that the estimations we show for \rh in Table~\ref{rh:tab:sequenceuntil_real} are different than the estimations in Table~\ref{rh:tab:relativeabundance} since we randomly subsample the reads in the relative abundance estimation dataset, as explained in Section~\ref{rh:subsec:evaluation}. We make two key observations. First, we observe that the distance between the relative abundance estimations between these two configurations of \rh is substantially low. This indicates that our outlier detection mechanism can accurately detect the convergence to the relative abundance estimations without using a full set of reads. Second, \emph{Sequence Until} enables accurately stopping the entire sequencing after processing $7\%$ of the reads in the entire set without substantially sacrificing accuracy. We conclude that \emph{Sequence Until} has the potential to significantly reduce the sequencing time and cost by using only fewer reads from a sample while producing accurate results.

\begin{table}[tbh]
\centering
\caption{Relative abundance with \emph{Sequence Until}.}
\begin{tabular}{@{}lrrrrrr@{}}\toprule
                       & \multicolumn{6}{c}{Estimated Relative Abundance Ratios in 50,000 Random Reads} \\\cmidrule{2-7}
\textbf{Tool}          & \textbf{\emph{SARS-CoV-2}} & \textbf{\emph{E. coli}} & \textbf{\emph{Yeast}} & \textbf{\emph{Green Algae}} & \textbf{\emph{Human}} & \textbf{Distance} \\\midrule
\rhcap ($100\%$)     & 0.0270                     & 0.3636                  & 0.3062                & 0.1951                      & 0.1081                & N/A          \\\midrule
\rhcap +             & 0.0283                     & 0.3539                  & 0.3100                & 0.1946                      & 0.1133                & 0.0118          \\
\emph{Sequence Until} ($7\%$) &                            &                         &                       &                             &                       &                 \\\bottomrule
\multicolumn{7}{l}{\footnotesize Percentages show the portion of the overall reads used.} \\
\end{tabular}

\label{rh:tab:sequenceuntil_real}
\end{table}

\section{Discussion} \label{rh:sec:discussion}

We discuss the benefits we expect \rh can immediately make, the limitations of \rh, and future work. We envision that \rh can be useful mainly for two directions. First, \rh provides a low-cost solution for analyzing large genomes in real-time. Such an analysis can be significantly useful when using nanopore sequencers with limited computational resources to enable portable real-time genome analysis at a large scale.

Second, we expect that \rh can also be useful for genome analysis that does not require real-time solutions by reducing the time and energy that further steps in genome analysis may require. One of the immediate steps after generating raw nanopore signals is their translation to their corresponding DNA bases as sequences of characters with a computationally intensive step, \emph{basecalling}. Basecalling approaches are usually computationally costly and consume significant energy as they use complex deep learning models~\cite{singh_rubicon_2024, mao_genpip_2022}. Although we do not evaluate in this work, we expect that \rh can be used as a low-cost filter~\cite{cavlak_targetcall_2024} to eliminate the reads that are unlikely to be useful in downstream analysis, which can reduce the overall workload of basecallers and downstream analysis.

\head{Future work} We find three key directions for future work. First, we find that our efficient hash-based similarity identification mechanism can be used to efficiently find overlaps between signals as the reads are sequenced in real-time. Although we observe that our indexing technique is efficient in terms of the amount it requires to construct an index even for large genomes, such an overlapping technique requires substantially more optimized indexing methods and techniques that can efficiently find overlaps as more reads are sequenced and \emph{evolves} the index. Finding overlaps between signals can be beneficial in 1)~providing enriched information to basecallers to increase their accuracy and 2)~identifying \emph{redundant} signals that fully overlap with already sequenced reads in an effort to generate assemblies from signals.

Second, since \rh generates hash values for matching similar regions, it provides opportunities to use the hash-based seeding techniques that are optimized for identifying sequence similarities accurately without requiring large memory space, such as minimizers~\cite{roberts_reducing_2004, li_minimap2_2018}, spaced seeds~\cite{ma_patternhunter_2002}, syncmers~\cite{edgar_syncmers_2021}, strobemers~\cite{sahlin_effective_2021}, and fuzzy seed matching as in BLEND~\cite{firtina_blend_2023}. Although we do not evaluate in this work, we implement the minimizer seeding technique in \rh. Our initial observation motivates us that future work can exploit these seeding techniques with slight modifications in their seeding mechanisms to significantly improve the performance of certain applications without reducing the accuracy.

\revb{Third, as advancements in sequencing technologies continue at a rapid pace~\cite{wang_nanopore_2021}, raw data generated by sequencers is rapidly increasing in terms of volume and throughput, placing increasing pressure on computational pipelines and raw nanopore signal analysis. To bridge the gap between analysis and sequencing, many acceleration efforts have targeted the computational bottlenecks of raw signal analysis by using GPUs~\cite{bao_squigglenet_2021, sadasivan_accelerating_2023} and co-designing algorithm and hardware~\cite{dunn_squigglefilter_2021, shih_efficient_2023}. \rh can benefit from a hardware acceleration in two ways: 1)~GPU acceleration and 2)~acceleration with a a memory- or a storage-centric design. For a GPU acceleration, \rh can benefit from a GPU implementation as its low-cost and accurate implementation can effectively be scaled to nanopore sequencers that include thousands of nanopores such that these pores can be analyzed in parallel with an efficient GPU implementation, which we leave as future work.
However, as hardware acceleration increases and minimizes the computational bottleneck, the adverse effect of the data movement is likely to dominate the accelerated end-to-end execution latency. In such cases, to enable a substantial end-to-end acceleration, a memory-centric or a storage-centric design can 1)~eliminate the overhead of data movement, 2)~provide high parallelism capabilities, and 3)~support the vast volume of genomic data~\cite{mansouri_ghiasi_megis_2024,mansouri_ghiasi_genstore_2022}.}

\head{Limitations}
We find four limitations of \rh, which we believe can be improved with further optimizations and better solutions.
First, \rh depends on previously generated k-mer models to generate events from reference genomes. Although these k-mer models can be trained and generated~\cite{simpson_detecting_2017, oxford_nanopore_technologies_k-mer_2023}, this makes it challenging to adapt the most accurate parameters for each k-mer model based on the nanopore model used for sequencing. A more generic k-mer model that can accurately represent all nanopores is needed to easily adapt \rh to all possible nanopore models that may be released in the future.

Second, \rh starts providing lower recall values as the genome size increases, which indicates that a larger portion of correct reads cannot be mapped by \rh due to the increase in the number of false negatives. Although such an increase in false negatives does not substantially affect some applications, such as contamination analysis, where providing higher precision is more critical to correctly identify the contaminated sample, improving it is useful to provide more accurate genome analysis overall.

Third, we perform our relative abundance estimations based on a priori knowledge of reference genomes. While such an experiment can still be useful in practical scenarios, this is not the common case in metagenomic analysis, where a sample is searched against a significantly larger set of species. We expect that our mechanism can still scale to such metagenomic analyses given that many metagenomic databases are efficiently constructed by including fewer and useful information for each species~\cite{breitwieser_krakenuniq_2018}, as opposed to our analysis, where we include whole-genome references. 

Fourth, we observe that the throughput of \rh is expected to reach the throughput of a nanopore when analyzing reference genomes slightly larger than a human genome. Such a limitation can be alleviated by applying 1)~seeding techniques that provide faster and more space-efficient searches in large spaces and 2)~chaining algorithms that are optimized for hash-based seed matches without the notion of distance between seeds, unlike the chaining algorithm used in Sigmap, and \revb{3)~hardware acceleration, such as GPUs or memory-centric designs, to enable parallel processing and reduce data movement overhead, thereby ensuring \rh can efficiently handle even larger genomes.}

\section{Summary}
We propose \rh, a novel mechanism that provides a low-cost and accurate approach for real-time genome analysis for large genomes. \rhcap can efficiently and accurately perform real-time analysis of raw nanopore signals to identify similarities between the signals and a reference genome in real-time at a large scale (e.g., whole-genome analysis for human or communities with multiple samples). To efficiently and accurately identify similarities, \rh 1)~generates events from both raw signals and the reference genome, 2)~quantizes the events into values such that slightly different events that correspond to the same DNA content can have the same value, and 3)~generates hash values from multiple events to efficiently find matching regions between raw signals and a reference genome using hash values with efficient data structures such as hash tables. We compare \rh with the state-of-the-art approaches, \unc and \sig, on three important applications in terms of their performance, accuracy, and estimated benefits in reducing sequencing time and cost. Our results show that 1)~\rh is the \emph{only} tool that can be accurately applied to analyze raw nanopore signals at large scale, 2)~provides \rhavgthrU and \rhavgthrS better average throughput, and 3)~can map reads \rhavgmtU and \rhavgmtS faster than \unc and \sig, respectively.
\revb{We hope that our findings will inspire future research to integrate real-time raw signal analysis in the genome analysis pipeline, especially when analyzing larger genomes, to significantly reduce the genome analysis latency as well as sequencing time and costs.}

\clearpage
\setsuppbasednumbering
\section{Supplementary Materials}
\subsection{Configuration} \label{rh:sec:configuration}
\subsubsection{Parameters} \label{rh:subsec:parameters}

In Supplementary Table~\ref{rh:tab:parameters}, we show the parameters of each tool for each dataset. In Supplementary Table~\ref{rh:tab:presets}, we show the details of the preset values that \rh sets in Supplementary Table~\ref{rh:tab:parameters}. For \unc, \sig, and minimap2, we use the same parameter setting for all datasets. For the sake of simplicity, we only show the parameters that we explicitly set in each tool. For the descriptions of all the other parameters, we refer to the help message that each tool generates, including \rh.

We note that the parameter names shown in Supplementary Table~\ref{rh:tab:versions} are different from the parameters explained in Sections~\ref{rh:subsec:quantization} and \ref{rh:subsec:hashvalues}, although these parameters essentially perform in the same way as explained in these sections, which we describe next. First, the quantization parameter, $Q$ in Section~\ref{rh:subsec:quantization}, is set using the \texttt{-q} parameter. Second, the value of $p$ in Section~\ref{rh:subsec:quantization} (i.e., pruned bits) can be calculated as $p = Q -l -3$ where $l$ is the least significant $l$ bits of $Q$. We use $l$ instead of $q$ due to its easier programmability in our tool. This $l$ value is set using the \texttt{-l} parameter in \rh. Third, the number of events packed together, $n$ in Section~\ref{rh:subsec:hashvalues}, is set using the \texttt{-e} parameter.

We set these \texttt{-q}, \texttt{-l}, and \texttt{-e} parameters empirically for three types of datasets: 1) viral genomes, 2) small genomes (i.e., $< 50M$ bases, and 3) large genomes (i.e., $> 50M$ bases) using the preset values \texttt{-x viral}, \texttt{-x sensitive}, and \texttt{-x fast}, respectively. In our empirical analysis, we identify that accuracy and performance are significantly impacted by the values we set for \texttt{-e}, \texttt{-q} and \texttt{-l} for three reasons.
First, \texttt{e} determines the number of quantized event values packed in a single hash value. Packing a larger number of events improves the sensitivity as it becomes more challenging to find larger consecutive matches of quantized event values than finding a smaller number of consecutive matches. Finding a smaller set of matches can decrease the time spent in seeding and chaining.
Second, these values determine the level of quantization of actual event values. Smaller \texttt{-q} and \texttt{-l} values can lead to loss of information due to storing only fewer bits that cannot be useful for identifying significantly different events. Larger \texttt{-q} and \texttt{-l} values can generate different quantized values for highly similar event values that may be corresponding to the same DNA content.
Third, the number of bits that we store for each event, which are determined by \texttt{-q} and \texttt{-l}, impacts the number of events that can be packed in a single 32-bit or 64-bit value. Packing a larger number of events in a single hash value directly impacts sensitivity as discussed earlier in this paragraph (first point).

\begin{table}[tbh]
\centering
\caption{Parameters we use in our evaluation for each tool and dataset in mapping.}
\resizebox{\linewidth}{!}{\begin{tabular}{@{}lccccccc@{}}\toprule
\textbf{Tool} & \textbf{\emph{Contamination}} & \textbf{\emph{SARS-CoV-2}} & \textbf{\emph{E. coli}} & \textbf{\emph{Yeast}} & \textbf{\emph{Green Algae}} & \textbf{\emph{Human}} & \textbf{\emph{Relative Abundance}} \\\midrule
\rh         	 & -x viral -t 32             & -x viral -t 32             & -x sensitive -t 32      & -x sensitive -t 32    & -x fast -t 32               & -x fast -t 32         & -x fast -t 32\\\midrule
\unc  			 & \multicolumn{7}{c}{map -t 32} \\\midrule
\sig        	 & \multicolumn{7}{c}{-m -t 32} \\\midrule
Minimap2         & \multicolumn{7}{c}{-x map-ont -t 32} \\\bottomrule
\end{tabular}
}
\label{rh:tab:parameters}
\end{table}

\begin{table}[tbh]
\centering
\caption{Corresponding parameters of presets (-x) in RawHash.}
\resizebox{0.6\linewidth}{!}{\begin{tabular}{@{}lcc@{}}\toprule
\textbf{Preset (-x)} & \textbf{Corresponding parameters} & Usage \\\midrule
viral         	 & -e 5 -q 9 -l 3 & Viral genomes \\\midrule
sensitive  	     & -e 6 -q 9 -l 3 & Small genomes (i.e., $<50M$ bases)\\\midrule
fast        	 & -e 7 -q 9 -l 3 & Large genomes (i.e., $>50M$ bases)\\\bottomrule
\end{tabular}
}
\label{rh:tab:presets}
\end{table}

\subsubsection{Versions}\label{rh:subsec:versions}

Supplementary Table~\ref{rh:tab:versions} shows the version and the link to these corresponding versions of each tool that we use in our experiments.

\begin{table}[tbh]
\centering
\caption{Versions of each tool.}
\resizebox{\linewidth}{!}{\begin{tabular}{@{}lll@{}}\toprule
\textbf{Tool} & \textbf{Version} & \textbf{Link to the Source Code} \\\midrule
\rh & 0.9 & \url{https://github.com/CMU-SAFARI/RawHash/tree/8042b1728e352a28fcc79c2efd80c8b631fe7bac}\\\midrule
\unc  & 2.2 & \url{https://github.com/skovaka/UNCALLED/tree/74a5d4e5b5d02fb31d6e88926e8a0896dc3475cb}\\\midrule
\sig  & 0.1 & \url{https://github.com/haowenz/sigmap/tree/c9a40483264c9514587a36555b5af48d3f054f6f}\\\midrule
Minimap2 & 2.24 & \url{https://github.com/lh3/minimap2/releases/tag/v2.24}\\\bottomrule
\end{tabular}
}
\label{rh:tab:versions}
\end{table}

\setchapterbasednumbering

\chapter[\rht: Reducing Noise in Raw Nanopore Signals Better]{Better Noise Reduction for and Robust Real-Time Mapping of Raw Nanopore Signals}
\label{chap:rht}

The previous chapter develops a noise reduction mechanism for matching the hash values between raw nanopore signals. This chapter takes the next step and introduces 1)~a new noise reduction technique by building a better understanding of the characteristics of noise in raw nanopore signals, 2)~a more robust decision-making mechanism for read mapping, and 3)~more sensitive computations for chaining seed matches between raw nanopore signals.

\section{Background and Motivation}
\label{rht:sec:bg}

Nanopore technology can sequence long nucleic acid molecules up to more than two million bases at high throughput~\cite{jain_nanopore_2018}. As a molecule moves through a tiny pore, called a \emph{nanopore}, ionic current measurements are generated at a certain throughput (e.g., around 450 bases per second for DNA~\cite{kovaka_targeted_2021, zhang_real-time_2021}). These electrical measurements, known as \emph{raw signals}, can be used to 1)~identify individual bases in the molecule with computational techniques such as \emph{basecalling}~\cite{senol_cali_nanopore_2019} and 2)~analyze raw signals directly \emph{without} translating them to bases~\cite{firtina_rawhash_2023}.

Computational techniques that can analyze the raw signals while they are generated at a speed that matches the throughput of nanopore sequencing are called \emph{real-time analysis}. There are several mechanisms that can perform real-time analysis of raw nanopore signals to achieve accurate and fast genome analysis~\cite{edwards_real-time_2019, kovaka_targeted_2021, zhang_real-time_2021, payne_readfish_2021, dunn_squigglefilter_2021, bao_squigglenet_2021, shih_efficient_2023, sadasivan_rapid_2023, firtina_rawhash_2023, mikalsen_coriolis_2023, shivakumar_sigmoni_2024, lindegger_rawalign_2024}. Most of these solutions have three main limitations. First, many mechanisms offer limited scalability or support on resource-constrained devices due to their reliance on either 1)~deep neural networks (DNNs) for real-time base translation, which are usually computationally intensive and power-hungry~\cite{payne_readfish_2021, ulrich_readbouncer_2022}, or 2)~specialized hardware such as ASICs or FPGAs~\cite{bao_squigglenet_2021, dunn_squigglefilter_2021, shih_efficient_2023}. Second, while some mechanisms can directly analyze raw signals without base translation, offering an efficient alternative for real-time analysis~\cite{kovaka_targeted_2021, zhang_real-time_2021}, they often compromise accuracy or performance when applied to larger genomes. Third, methods based on machine learning often require retraining or reconfiguration~\cite{bao_squigglenet_2021, senanayake_deepselectnet_2023, sadasivan_rapid_2023}, adding a layer of complexity and reducing their flexibility for general use cases, such as read mapping to any genome.

Among the existing works, \rh~\cite{firtina_rawhash_2023} is the state-of-the-art mechanism that can accurately perform real-time mapping of raw nanopore signals for large genomes without translating them to bases with a hash-based seed-and-extend mechanism~\cite{altschul_basic_1990}. Despite its strengths in accuracy and performance, particularly for large genomes like the human genome, \rh exhibits several limitations that require further improvements.
First, \rh utilizes a simple quantization algorithm that assumes the raw signals are distributed uniformly across their normalized value range, which limits its efficiency and accuracy.
Second, \rh uses a chaining algorithm similar to that used in Sigmap~\cite{zhang_real-time_2021} without incorporating penalty scores used in minimap2~\cite{li_minimap2_2018}, which constrains its ability for more sensitive mapping.
Third, \rh performs chaining on all seed hits without filtering any of these seed hits, which substantially increases the workload of the chaining algorithm due to a large number of seed hits to chain.
Fourth, the decision-making mechanism in \rh for mapping reads to a reference genome in real-time relies on one of the mapping conditions being true (e.g., the ratio between the best and second-best chain scores), which makes it more prone to the outliers that can satisfy one of these conditions. A more robust and statistical approach that incorporates features beyond chaining scores can provide additional insights for making more sensitive and quick mapping decisions.
Fifth, while the hash-based mechanism in \rh is compatible with existing sketching techniques such as minimizers~\cite{roberts_reducing_2004, li_minimap2_2018}, strobemers~\cite{sahlin_effective_2021}, and fuzzy seed matching as in BLEND~\cite{firtina_blend_2023}, the benefits of these techniques are unknown for raw signal analysis as they are not used in \rh. Such evaluations could potentially provide additional insights on how to use the existing hash-based sketching techniques and reduce storage requirements while maintaining high accuracy.
Sixth, \rh lacks the support for recent advancements, such as the newer R10.4 flow cell version. The integration of these features can accelerate the adoption of both real-time and offline analysis.

In this work, our goal is to address the aforementioned limitations of \rh by improving its mechanism. To this end, we propose \rht to improve \rh in six directions.
First, to generate more accurate and unique hash values, we introduce a new quantization technique, \emph{adaptive quantization}. 
Second, to improve the accuracy of chaining and subsequently read mapping, we implement a more sophisticated chaining algorithm that incorporates penalty scores (as in minimap2). 
Third, to improve the performance of chaining by reducing its workload, \rht provides a filter that removes seeds frequently appearing in the reference genome, known as a \emph{frequency filter}.
Fourth, we introduce a statistical method that utilizes multiple features for making mapping decisions based on their weighted scores to eliminate the need for manual and fixed conditions to make decisions.
Fifth, we extend the hash-based mechanism to incorporate and evaluate the minimizer sketching technique, aiming to reduce storage requirements without significantly compromising accuracy.
Sixth, we integrate support for R10.4 flow cells and more recent file formats, POD5 and S/BLOW5~\cite{gamaarachchi_fast_2022}.

Compared to \rh, our extensive evaluations on five genomes of varying sizes and six different real datasets show that \rht provides higher accuracy (by \rhtavgAccRH on average and \rhtmaxAccRH at maximum) and better read mapping throughput (by \rhtavgthrR on average and \rhtmaxthrR at maximum).

\section{\rht Mechanism} \label{rht:secmethods}

\rh is a mechanism to perform mapping between raw signals by quickly matching their hash values. \rht provides substantial improvements over \rh in six key directions.
First, to generate more accurate and distinct hash values from raw signals, \rht improves the quantization mechanism with an \emph{adaptive} approach such that signal values are quantized non-uniformly based on the characteristics of a nanopore model.
Second, to provide more accurate mapping, \rht improves the chaining algorithm in \rh with more accurate penalty scores. 
Third, to reduce the workload in chaining for improved performance, we integrate a frequency filter to quickly eliminate the seed hits that occur too frequently. 
Fourth, to make more accurate and quick mapping decisions, \rht determines whether a read should be mapped at a specific point during sequencing by using a weighted sum of multiple features.
Fifth, to reduce the storage requirements of seeds, \rht incorporates and evaluates the benefits of minimizer sketching technique.
Sixth, \rht includes support for the latest features introduced by ONT, such as new file formats and flow cells.

\subsection{Adaptive Quantization} \label{rht:subsec:adaptivequantization}

To improve the accuracy and uniqueness of hash values generated from raw nanopore signals, \rht introduces a new \emph{adaptive} quantization technique that we explain in four steps.

First, to enable a more balanced and accurate assignment of normalized signal values into quantized values (i.e., buckets), \rht performs a bifurcated approach to define two different ranges: 1)~fine range and 2)~coarse range. These ranges are useful for fine-tuning the boundaries of normalized signal values, $s$ falling into a certain quantized value, $q(s)$ within the integer value range $[0, n]$, as the normalized distribution of signal values is not uniform across all ranges.
Second, within the \emph{fine range}, normalized signal values are quantized into smaller intervals to enable a high resolution, $f_{r}$, for quantization due to the larger number of normalized signal values that can be observed within this range. The boundaries of the fine range, ($f_{min}$ and $f_{max}$), are empirically defined to enable robustness and high accuracy applicable given a flow cell (e.g., R9.4) and parameters to \rht to enable flexibility.
Third, the normalized signal values outside the fine range (i.e., the \emph{coarse range}) are quantized into larger intervals with low resolution, $c_{r} = (1 - f_{r}) \times 0.5$, to enable a more balanced load of quantized values across all ranges by assigning more signal values within this range into the same quantized value.
Fourth, depending on the range that a normalized signal is in, its corresponding quantized value is assigned as shown in Equation~\ref{rht:eq:adaptivequant}. The adaptive quantization approach can enable a more balanced and accurate distribution of quantized values by better distinguishing closeby signal values with high resolution and grouping signals more efficiently in the coarser range.

\begin{equation}\label{rht:eq:adaptivequant}
q(s) = \begin{cases} 
\lfloor n \times (f_{r} \times \frac{(s - f_{min})}{f_{max} - f_{min}}) & \text{if } f_{min} \leq s \leq f_{max} \\
\lfloor n \times (f_r + c_{r} \times s) & \text{if } s < f_{min} \\
\lfloor n \times (f_r + c_{r} + c_{r} \times s) & \text{if } s > f_{max} 
\end{cases}
\end{equation}

\subsection{Chaining with Penalty Scores}
To identify the similarities between a reference genome (i.e., target sequence) and a raw signal (i.e., query sequence), the series of seed hits within close proximity in terms of their matching positions are identified using a dynamic programming (DP) algorithm, known as \emph{chaining}. Using a chaining terminology similar to that of minimap2~\cite{li_minimap2_2018}, a seed hit between a reference genome and a raw signal is usually represented by a 3-tuple $(x,y,w)$ value, known as \emph{anchor}, where $w$ represents the length of the region that a seed spans, the start and end positions of a matching interval in a reference genome and a raw signal is represented by $[x-w+1,x]$ and $[y-w+1,y]$, respectively. The chain of anchors within close proximity is identified by calculating the optimal chain score $f(i)$ of each anchor $i$, where $f(i)$ is calculated based on predecessors of anchor $i$ when anchors are sorted by their reference positions. To calculate the chain score, $f(i)$, with dynamic programming, \rh performs the following computation as used in Sigmap~\cite{zhang_real-time_2021}.

\begin{equation}\label{rht:eq:chainold}
f(i)=\max\big\{\max_{i>j\ge 1} \{ f(j)+\alpha(j,i)\},w_i\big\}
\end{equation}

where $\alpha(j,i)=\min\big\{\min\{y_i-y_j,x_i-x_j\},w_i\big\}$ is the length of the matching region between the two anchors. Although such a design is useful when identifying substantially fewer seed matches using a seeding technique based on distance calculation as used in Sigmap, \rh identifies a larger number of seed matches as it uses hash values to identify the matching region, which is usually faster than a distance calculation with the cost of reduced sensitivity.

To identify the correct mapping regions among such a large number of seed matches, \rht uses a more sensitive chaining technique as used in minimap2 by integrating the gap penalty scores such that the chain score of an anchor $i$ is calculated as shown in Equation~\ref{rht:eq:chain}:

\begin{equation}\label{rht:eq:chain}
f(i)=\max\big\{\max_{i>j\ge 1} \{ f(j)+\alpha(j,i)-\beta(j,i) \},w_i\big\}
\end{equation}

where $\beta(j,i)=\gamma_c\big((y_i-y_j)-(x_i-x_j)\big)$ is the penalty score calculated based on the gap distance, $l$, between a pair of anchors $i$ and $j$ where $\gamma_c(l) = 0.01\cdot w\cdot|l|+0.5\log_2|l|$. Based on the chain score calculation with gap costs, \rht integrates similar heuristics, mapping quality calculation, and the same complexity when calculating the chaining scores with the gap penalty as described in minimap2~\cite{li_minimap2_2018}.

\subsection{Frequency Filters}
\rht introduces a two-step frequency filtering mechanism to 1)~reduce the computational workload of the chaining process by limiting the number of anchors it processes and 2)~focus on more unique and potentially meaningful seed hits. First, to reduce the number of queries made to the hash table for identifying seed hits, \rht eliminates non-unique hash values generated from raw signals that appear more frequently than a specified threshold. Second, \rht evaluates the frequency of each seed hit within the reference genome and removes those that surpass a predefined frequency threshold, which reduces the overall workload of the chaining algorithm by providing a reduced set of more unique seed hits.

\subsection{Weighted Mapping Decision} \label{rht:subsec:weightedmapping}

\rh performs mapping while receiving chunks of signals in real-time, as provided by nanopore sequencers. It is essential to decide if a read maps to a reference genome as quickly as possible to avoid unnecessary sequencing. The decision-making process in \rh is based on a series of conditional checks involving chain scores. These checks are performed in a certain order and against fixed ratios and mean values, making the decision mainly rigid and less adaptive to variations.

To employ a more statistical approach that can generalize various variations between different data sets and genomes, \rht calculates a weighted sum of multiple features that can impact the mapping decision. To achieve this, \rht calculates normalized ratios of several metrics based on mapping quality and chain scores. These metrics are 1)~the ratio of the mapping quality to a sufficiently high mapping quality (i.e., 30), 2)~mapping quality ratio between the best chain and the mean quality of all chains, and 3)~the ratio of the chain score between the best and the mean score of all chains. These ratios are combined into a weighted sum as follows: $w_{\text{sum}} = \sum_{i=1} r_i \times w_i$, where $r_{i}$ is a ratio of a particular metric, and $w_{i}$ is the weight assigned for that particular metric. The weighted sum, $w_{sum}$, is compared against a predefined threshold value to decide if a read is considered to be mapped. \rht maps a read if the weighted sum exceeds the threshold. Such a weighted sum approach allows \rht to adaptively consider multiple aspects of the data and eliminates the potential effect of the ordering of these checks to achieve improved mapping accuracy while maintaining computational efficiency.

\subsection{Minimizer Sketching} \label{rht:subsec:minimizer}
\rh provides the opportunity to integrate the existing hash-based sketching techniques such as minimizers~\cite{roberts_reducing_2004, li_minimap2_2018} for 1)~reduced storage requirements of index in disk and memory and 2)~faster mapping due to fewer seed queries and hits.

To reduce the storage requirements of storing seeds in raw signals and due to their widespread application, \rht integrates minimizers in two steps. First, \rht generates hash values for seeds in both the reference genome and the raw signal. Second, within each window comprising $w$ hash values, the minimum hash value is selected as the minimizer. These minimizer hash values can be used to find similarities using hash tables (similar to \rh that uses hash values of all k-mers) while significantly reducing the number of hash values that need to be stored and queried during the mapping process as opposed to storing all k-mers.

\subsection{Support for New Data Formats and Flow Cells}\label{rht:subsec:fileandr10methods}
To enable better and faster adoption, \rht incorporates support for 1)~recent data formats for storing raw signals, namely POD5 and SLOW5~\cite{gamaarachchi_fast_2022} as well as the existing FAST5 format, and 2)~the latest flow cell versions due to two main reasons. First, transitioning from the FAST5 to the POD5 file format is crucial for broad adoption, as POD5 is the new standard file format introduced by Oxford Nanopore Technologies (ONT). Second, integrating the newer flow cell versions is challenging as it requires optimization of parameters involved in mapping decisions as well as segmentation. \rht enables mapping the raw signals from R10.4 flow cells by optimizing the segmentation parameters for R10.4 and adjusting the scoring parameters involved in chaining settings to enable accurate mapping for R10.4 flow cells.

\section{Results} \label{rht:secresults}
\subsection{Evaluation Methodology} \label{rht:subsec:evaluation}

We implement the improvements we propose in \rht directly on the \rh implementation. Similar to \rh, \rht provides the mapping information using a standard pairwise mapping format (PAF).

We compare \rht with the state-of-the-art works \unc~\cite{kovaka_targeted_2021}, \sig~\cite{zhang_real-time_2021}, \rh~\cite{firtina_rawhash_2023} in terms of throughput, accuracy, and the number of bases that need to be processed before stopping the sequencing of a read to estimate the benefits in sequencing time and cost. We provide the release versions of these tools in Supplementary Table~\ref{rht:tab:versions}. For throughput, we calculate the number of bases that each tool can process per second per CPU thread, which is essential to determine if a calculation in a single thread is at least as fast as the speed of sequencing from a single nanopore (i.e., single pore). In many commonly used nanopore sequencers, a nucleic acid molecule passes through a pore at around 450 bases and 400 per second with sampling rates of 4 KHz and 5 KHz for DNA in R9.4.1 and R10.4.1, respectively~\cite{kovaka_targeted_2021, sam_kovaka_uncalled4_2024}. Since each read is mapped using a single thread for all tools, the throughput calculation is not affected by the number of threads available to these tools. Rather, this throughput calculation shows how many pores a single thread can process and how many CPU threads are needed to process the entire flow cell with many pores (e.g., 512 pores in a MinION flow cell). To show these results, we calculate 1)~the number of pores that a single thread can process by dividing throughput by the number of bases sequenced per second per single pore and 2)~the number of threads needed to cover the entire flow cell.

For accuracy, we analyze three use cases: 1)~read mapping, 2)~contamination analysis, and 3)~relative abundance estimation. To identify the correct mappings, we generate the ground truth mapping output in PAF by mapping the basecalled sequences of corresponding raw signals to their reference genomes using minimap2~\cite{li_minimap2_2018}. We use \texttt{UNCALLED pafstats} to compare the mapping output from each tool with their corresponding ground truth mapping output to calculate precision ($P = TP/(TP+FP)$), recall ($R = TP/(TP + FN)$), and F1 ($F1 = 2 \times (P \times R)/(P+R)$) values, similar to \rh~\cite{firtina_rawhash_2023}. For read mapping, we compare the tools in terms of their precision, recall, and F-1 scores. For contamination analysis, the goal is to identify if a particular sample is contaminated with a certain genome (or set of genomes), which makes the precision metric more important for such a use case. For this use case, we compare the tools in terms of their precision in the main paper and show the full results (i.e., precision, recall, and F1) in Supplementary Table~\ref{rht:supptab:accuracyfull}.
For relative abundance estimation, we calculate the abundance ratio of each genome based on the ratio of reads mapped to a particular genome compared to all read mappings. We calculate the Euclidean distance of each estimation to the ground truth estimations generated based on minimap2 mappings of corresponding basecalled reads. We estimate the relative abundances based on the number of mapped reads rather than the number of mapped bases as we identify that larger genomes usually require sequencing a larger number of bases to map a read, which can lead to skewed estimations towards larger genomes.

To estimate the benefits in sequencing time and the cost per read, we identify the average sequencing length before making the mapping decision for a read. For all of our analyses, we use the default parameters of each tool as we show in Supplementary Table~\ref{rht:tab:parameters}. Table~\ref{rht:supptab:dataset} shows the details of the datasets used in our evaluation and their corresponding sequencing run settings. The \emph{Basecaller Model} column shows the details about the basecaller model and the version we use. Except for the D7 dataset, all other datasets include the basecalled sequences within their corresponding FAST5 files or the corresponding accession numbers available at NCBI. We provide the scripts to extract these basecalled sequences on the GitHub page of \rht. For the D7 dataset, we provide the necessary commands to run Dorado for basecalling on the GitHub page. Although \rht does not use the minimizer sketching technique by default to achieve the maximum accuracy, we evaluate the benefits of minimizers in \rht, which we refer to as \texttt{\rhmin}. Since the evaluated versions of \unc, \sig, and \rh do not provide the support for the R10.4 dataset, we show the corresponding results with the R10.4 dataset without comparing to these tools. When comparing \rht to other tools, we use FAST5 files containing raw signals from R9.4 flow cells on an isolated machine and SSD. We use the AMD EPYC 7742 processor at 2.26GHz to run the tools. We use 32 threads for all the tools.

\begin{table}[tbh]
\centering
\caption{Details of datasets used in our evaluation.}
\resizebox{\columnwidth}{!}{\begin{tabular}{@{}cllllllrrllr@{}}\toprule
& \textbf{Organism}     & \textbf{Device} & \textbf{Flow Cell} & \textbf{Transloc.} & \textbf{Sampling} & \textbf{Basecaller}     &\textbf{Reads} & \textbf{Bases}              & \textbf{SRA}       & \textbf{Reference}      & \textbf{Genome}\\
& \textbf{}             & \textbf{Type} & \textbf{Type}     & \textbf{Speed}  & \textbf{Frequency} &        \textbf{Model}       &\textbf{(\#)}  & \textbf{(\#)}           & \textbf{Accession} & \textbf{Genome}         & \textbf{Size}  \\\midrule
\multicolumn{12}{c}{Read Mapping} \\\midrule
D1 & \emph{SARS-CoV-2}  & MinION & R9.4.1 e8 (FLO-MIN106) & 450 & 4000 & Guppy HAC v3.2.6 & 1,382,016      & 594M               & CADDE Centre & GCF\_009858895.2   & 29,903 \\\midrule
D2 & \emph{E. coli}     & GridION & R9.4.1 e8 (FLO-MIN106) & 450 & 4000 & Guppy HAC v5.0.12  & 353,317        & 2,365M                 & ERR9127551         & GCA\_000007445.1   & 5M \\\midrule
D3 & \emph{Yeast}       & MinION & R9.4.1 e8 (FLO-MIN106) & 450 & 4000 & Albacore v2.1.7 & 49,989         & 380M                   & SRR8648503         & GCA\_000146045.2   & 12M\\\midrule
D4 & \emph{Green Algae} & PromethION & R9.4.1 e8 (FLO-PRO002) & 450 & 4000 & Albacore v2.3.1 & 29,933         & 609M                   & ERR3237140         & GCF\_000002595.2   & 111M\\\midrule
D5 & \emph{Human} & MinION & R9.4.1 e8 (FLO-MIN106) & 450 & 4000 & Guppy Flip-Flop v2.3.8 & 269,507        & 1,584M                 & FAB42260 & T2T-CHM13 (v2)     & 3,117M\\\midrule
D6 & \emph{E. coli}     & GridION & R10.4 e8.1 (FLO-MIN112) & 450 & 4000 & Guppy HAC v5.0.16 & 1,172,775        & 6,123M       & ERR9127552         & GCA\_000007445.1   & 5M \\\midrule
D7 & \emph{S. aureus}   & GridION & R10.4 e8.1 (FLO-MIN112) & 450 & 4000 & Dorado SUP v0.5.3 & 407,727        & 1,281M       & SRR21386013         & GCF\_000144955.2   & 2.8M \\\midrule
\multicolumn{12}{c}{Contamination Analysis} \\\midrule
\multicolumn{7}{c|}{D1 and D5} & 1,651,523    & 2,178M                 & D1 and D5             & D1                 & 29,903\\\midrule
\multicolumn{12}{c}{Relative Abundance Estimation} \\\midrule
\multicolumn{7}{c|}{D1-D5} & 2,084,762    & 5,531M                  & D1-D5              & D1-D5              & 3,246M\\\bottomrule
\multicolumn{12}{l}{Multiple dataset numbers in contamination analysis and relative abundance estimation show the combined datasets.}\\
\multicolumn{12}{l}{D1-D5 datasets are from R9.4, and D6 and D7 are from R10.4. Human reads are from Nanopore WGS.}\\
\multicolumn{12}{l}{Base counts in millions (M).}\\
\end{tabular}
}
\label{rht:supptab:dataset}
\end{table}

\subsection{Throughput} \label{rht:subsec:perfmemory}

\begin{figure}[tbh]
  \centering
  \includegraphics[width=\columnwidth]{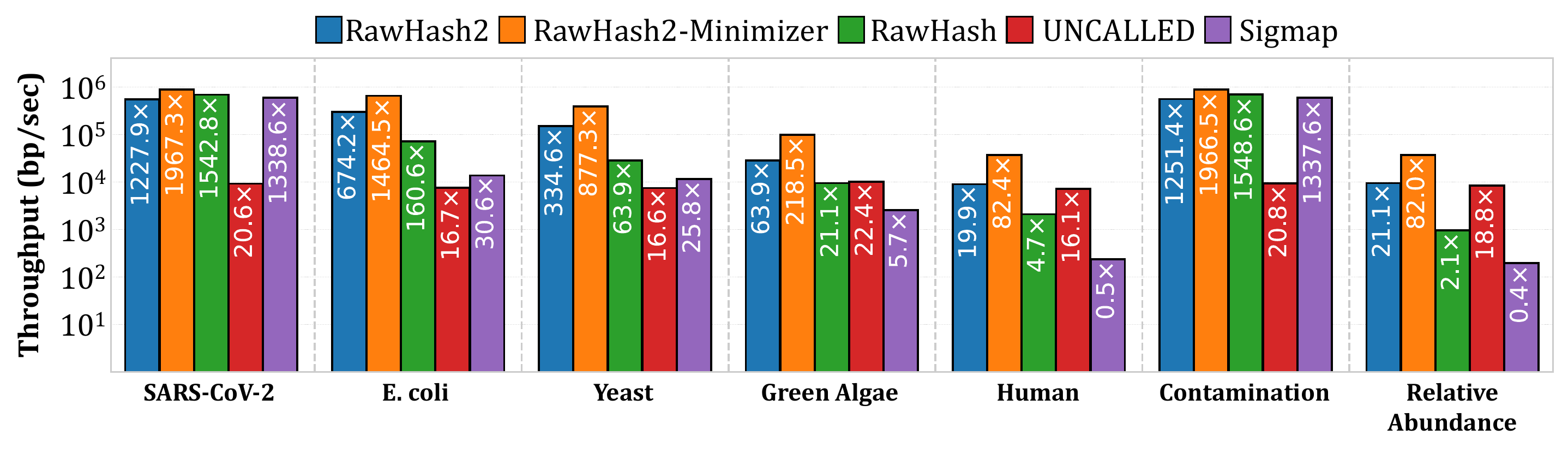}
  \caption{Throughput of each tool. Values inside the bars show how many nanopores (i.e., pores) that a single CPU thread can process.}
  \label{rht:fig:throughput}
\end{figure}

Figure~\ref{rht:fig:throughput} shows the results for 1)~throughput per single CPU thread and 2)~number of pores that a single CPU thread can analyze as annotated by the values inside the bars. We make three key observations.
First, we find that \rht provides average throughput \rhtavgthrU, \rhtavgthrS, and \rhtavgthrR better than \unc, \sig, and \rh, respectively. Such a speedup, specifically over the earlier work \rh, is achieved by reducing the workload of chaining with the unique and accurate hash values using the new quantization mechanism and the filtering technique (see the filtering ratios in Supplementary Table~\ref{rht:supptab:filtered_seed_hits}).
Second, we find that \rhmin enables reducing the computational requirements for mapping raw signals and enables improving the average throughput by $2.5\times$ compared to \rht, while the other computational resources, such as the peak memory usage and CPU time in both indexing and mapping, and the mean time spent per read are also significantly reduced as shown in Supplementary Tables~\ref{rht:supptab:performance} and Supplementary Figure~\ref{rht:suppfig:timeperread}.
Third, \rhmin requires \emph{at most} 7 threads for analyzing the entire flowcell for any evaluated dataset, while \rht requires at most 2 threads for smaller genomes and 9 to 26 threads for Green Algae and human. This shows that \rht and \rhmin can reduce computational requirements and energy consumption significantly compared to 28 threads required, on average, regardless of the genome size for \unc, which is critical for portable sequencing.
We conclude that \rht and \rhmin significantly reduce the computational overhead of mapping raw signals to reference genomes, enabling better scalability to even larger genomes.

\subsection{Accuracy} \label{rht:subsec:accuracy}

\begin{figure}[tbh]
\centering
\includegraphics[width=\columnwidth]{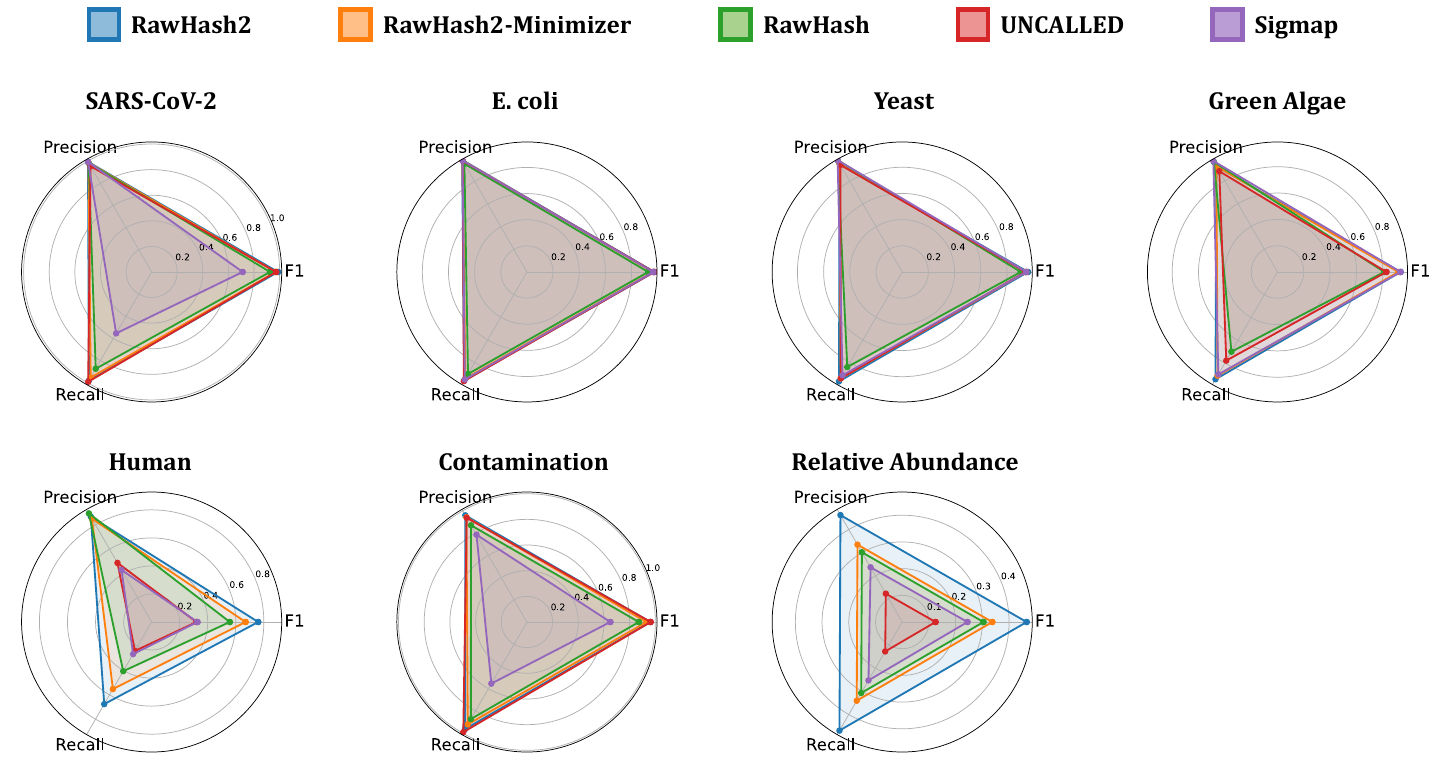}
\caption{Read mapping accuracy results in terms of F1 score, precision, and recall across different datasets. The dotted triangles show the best possible results, where each edge shows the best result for its corresponding metric.}
\label{rht:suppfig:accuracy}
\end{figure}

Table~\ref{rht:tab:accuracy} shows the accuracy results for read mapping, contamination analysis, and relative abundance estimation based on their corresponding most relevant accuracy metrics (results with all metrics are shown in Supplementary Table~\ref{rht:supptab:accuracyfull} and Figure~\ref{rht:suppfig:accuracy}). We make two key observations. First, we find that \rht provides 1)~the best accuracy in terms of the F1 score in all datasets for read mapping, 2)~the best precision for contamination analysis, and 3)~the most accurate relative abundance estimation. This is mainly achieved because 1)~the adaptive quantization enables finding more accurate mapping positions while substantially reducing the false seed hits due to less precise quantization in \rh, and 2)~the more sensitive chaining implementation with penalty scores can identify the correct mappings more accurately.

\begin{table}[tbh]
\centering
\caption{Accuracy.}
\resizebox{0.9\columnwidth}{!}{
\begin{tabular}{@{}llrrrrr@{}}\toprule
\textbf{Dataset} & \textbf{Metric} & \textbf{RH2} & \textbf{RH2-Min.} & \textbf{RH} & \textbf{UNCALLED} & \textbf{Sigmap} \\\midrule
SARS-CoV-2 & F1 & \cellcolor{bestresult}\textbf{0.9867} & 0.9691 & 0.9252 & 0.9725 & 0.7112 \\
E. coli & F1 & \cellcolor{bestresult}\textbf{0.9748} & 0.9631 & 0.9280 & 0.9731 & 0.9670 \\
Yeast & F1 & \cellcolor{bestresult}\textbf{0.9602} & 0.9472 & 0.9060 & 0.9407 & 0.9469 \\
Green Algae & F1 & \cellcolor{bestresult}\textbf{0.9351} & 0.9191 & 0.8114 & 0.8277 & 0.9350 \\
Human & F1 & \cellcolor{bestresult}\textbf{0.7599} & 0.6699 & 0.5574 & 0.3197 & 0.3269 \\
\midrule
Contamination & Precision & \cellcolor{bestresult}\textbf{0.9595} & 0.9424 & 0.8702 & 0.9378 & 0.7856 \\
\midrule
Rel. Abundance & Distance & \cellcolor{bestresult}\textbf{0.2678} & 0.4243 & 0.4385 & 0.6812 & 0.5430 \\
\bottomrule
\multicolumn{7}{l}{\footnotesize Best results are \colorbox{bestresult}{\textbf{highlighted}}.} \
\end{tabular}

}
\label{rht:tab:accuracy}
\end{table}

Second, \rhmin provides mapping accuracy similar to that of \rht with an exception for the human genome and better accuracy than \rh, providing substantially better performance results as discussed in Section~\ref{rht:subsec:perfmemory}. Such an accuracy-performance trade-off puts \rhmin in an important position when a slight drop in accuracy can be tolerated for a particular use case when a substantially better throughput is needed.
For the relatively lower accuracy that \rht and \rhmin achieve compared to minimap2, we believe the accuracy gap is due to the increased difficulty in distinguishing the chain with the correct mapping position among many chains with similar quality scores, potentially due to the false seed matches in repetitive regions. Although our in-house evaluation shows that accuracy can substantially be improved further by enabling the correct chains to be distinguished more accurately than the incorrect chains with more sensitive quantization parameters, this comes with increased performance costs due to increased seed matches and chaining calculations. Future work can focus on designing more sensitive filters to improve the accuracy for larger and repetitive genomes by eliminating seed matches from such false regions. We conclude that \rht is the most accurate tool regardless of the genome size, while the minimizer sketching technique in \rhmin can provide better accuracy than \rh and on-par accuracy to all other tools while providing the best overall performance.

\subsection{Sequencing Time and Cost} \label{rht:subsec:sequencingcost}

\begin{figure}[bth]
\centering
\includegraphics[width=\linewidth]{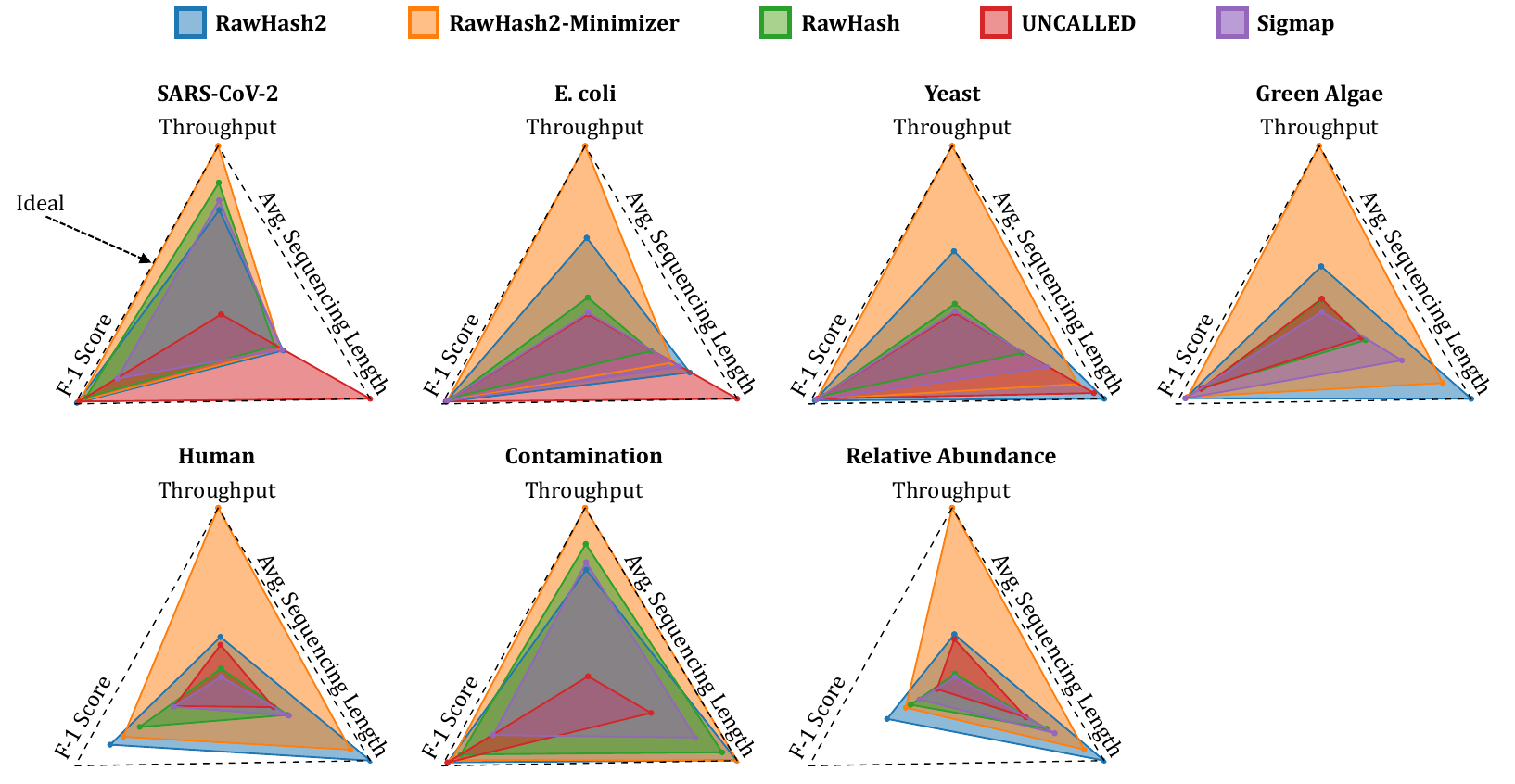}
\caption{Combined results in terms of throughput, F-1 score (i.e., accuracy), and average sequencing length across different datasets. The dotted triangles show the best possible results, where each edge shows the best result for its corresponding metric.}
\label{rht:suppfig:combined}
\end{figure}

Table~\ref{rht:tab:seq_bases} shows the average sequencing lengths in terms of bases and chunks that each tool needs to process before stopping the sequencing process of a read. Processing fewer bases can significantly help reduce the overall sequencing time and potentially the cost spent for each read by enabling better utilization of nanopores without sequencing the reads unnecessarily. Figure~\ref{rht:suppfig:combined} shows the combined results of each tool in terms of throughput, F-1 Score (i.e., accuracy), and average sequencing length for each dataset. The dotted lines in each triangle show the ideal combined result. Each edge of the triangle shows the best result for the corresponding metric, as shown in the figure. For the edge that shows the F-1 score, the best point is 1.0. All tools have F-1 scores between 0 and 1, as shown in Table~\ref{rht:tab:accuracy}. For the other two edges, which show throughput and average sequencing length, the best result is determined based on the highest result we observe for that dataset. We adjust all other results using these highest results so that the adjusted throughput and average sequencing length values are always between 0 and 1. We make three key observations.

\begin{table}[tbh]
\centering
\caption{Average length of sequencing per read.}
\begin{tabular}{@{}lrrrrr@{}}\toprule
\textbf{Dataset} & \textbf{RH2} & \textbf{RH2-Min.} & \textbf{RH} & \textbf{UNCALLED} & \textbf{Sigmap} \\\midrule
SARS-CoV-2 & 443.92  & 460.85  & 513.95  & \cellcolor{bestresult}\textbf{184.51}  & 452.38 \\
E. coli & 851.31  & 1,030.74  & 1,376.14  & \cellcolor{bestresult}\textbf{580.52}  & 950.03 \\
Yeast & \cellcolor{bestresult}\textbf{1,147.66}  & 1,395.87  & 2,565.09  & 1,233.20  & 1,862.69 \\
Green Algae & \cellcolor{bestresult}\textbf{1,385.59}  & 1,713.46  & 4,760.59  & 5,300.15  & 2,591.16 \\
Human & \cellcolor{bestresult}\textbf{2,130.59}  & 2,455.99  & 4,773.58  & 6,060.23  & 4,680.50 \\
\midrule
Contamination & 670.69  & \cellcolor{bestresult}\textbf{667.89}  & 742.56  & 1,582.63  & 927.82 \\
\midrule
Rel. Abundance & \cellcolor{bestresult}\textbf{1,024.28}  & 1,182.04  & 1,669.46  & 2,158.50  & 1,533.04 \\
\bottomrule
\multicolumn{6}{l}{\footnotesize Best results are \colorbox{bestresult}{\textbf{highlighted}}.} \
\end{tabular}

\label{rht:tab:seq_bases}
\end{table}

First, \rht reduces the average sequencing length by \rhtavgseqR compared to \rh mainly due to the improvements in mapping accuracy, which enables making quick decisions without using longer sequences.
Second, as the genome size increases, \rht provides the smallest average sequencing lengths compared to all tools.
Third, when the average length of sequencing is combined with other important metrics such as mapping accuracy in terms of F1 score and throughput, \rht provides the best trade-off in terms of all these three metrics for all datasets as shown in Figure~\ref{rht:suppfig:combined}. We conclude that \rht is the best tool for longer genomes to reduce the sequencing time and cost per read as it provides the smallest average sequencing lengths, while \unc is the best tool for shorter genomes.

\subsection{Evaluating New File Formats and R10.4}\label{rht:subsec:r104}
\head{Impact of Different File Formats on Performance} Table~\ref{rht:supptab:pod5_fast5_resources} shows the overall execution time when using different raw signal file formats: FAST5, POD5, and BLOW5~\cite{gamaarachchi_fast_2022}. To evaluate the direct impact of these formats, we run \rht (RH2) and \rht-Minimizer (RH2-Min.) 1)~using a single thread (i.e., single thread for the entire execution including \emph{both} file IO and mapping), 2)~using an isolated SSD on a PCI-e interface, 3)~using the same compression type (i.e., zstd) for all file formats, and 4)~clearing the disk cache before each execution. When using a single thread, we confirm that the underlying libraries for FAST5, POD5, and BLOW5 are not aggressively using more threads than what is allocated to them, as the thread utilization is reported as 0.99 (i.e., 99\%) by the \texttt{time -v} command for the entire execution.

We note that even if we use multiple threads when running \rht, the file IO step (i.e., reading from or writing to a file) always uses a single thread and is overlapped with the mapping step (i.e., either the read or write operation is run in parallel together with the mapping step by using one thread where the mapping step takes rest of the allocated threads). The design is due to the pipelining implementation strategy we adopt, similar to the minimap2 implementation~\cite{li_minimap2_2018}. We note that if \rht is run using a single thread, none of these steps overlap with each other, and they run sequentially using only one thread, which is our evaluation setting we show in Table~\ref{rht:supptab:pod5_fast5_resources}.

We find that POD5 and SLOW5 significantly speed up total elapsed time compared to FAST5. These results indicate that a large portion of the overhead spent for reading from a file can be mitigated with approaches that can perform faster compression and decompression, as these signal files are mostly stored in a compressed form.

\begin{table}[tbh]
\centering
\caption{Comparison of overall execution time when using different file formats in \rht in a single-threaded mode.}
\resizebox{0.3\columnwidth}{!}{
\begin{tabular}{@{}lrr@{}}\toprule
\textbf{Tool} & \textbf{\emph{E. coli}} & \textbf{\emph{Yeast}}  \\\midrule
\multicolumn{3}{c}{Elapsed Time (mm:ss)} \\\midrule
RH2-FAST5   & 19:27 & 08:35  \\
RH2-POD5    & 16:55 & 07:33  \\
RH2-BLOW5    & 17:32 & 07:38  \\\midrule
RH2-Min.-FAST5   & 12:13 & 03:56  \\
RH2-Min.-POD5    & 09:42 & 02:56  \\
RH2-Min.-BLOW5    & 10:16 & 03:02  \\\midrule
\end{tabular}

}
\label{rht:supptab:pod5_fast5_resources}
\end{table}

\head{R10.4 Accuracy and Performance} In Table~\ref{rht:supptab:r10accuracy}, we show the accuracy and performance results in terms of throughput and mean time spent per read when using R10.4 flow cells. For comparison purposes between R10.4 and R9.4, we include the results from R9.4 flow cells for \emph{E. coli}. We do not show the R9.4 results for \emph{S. aureus}, since we do not have raw signals from the same sample for this dataset.

We find that \rht can perform fast analysis with reasonable accuracy that can be useful for certain use cases (e.g., contamination analysis) when using raw signals from R10.4, although \rht achieves lower accuracy with R10.4 than using R9.4. This is likely because 1)~we use a k-mer model optimized for the R10.4.1 flow cell version rather than R10.4, and 2)~minimap2 can provide more accurate mapping due to improved accuracy of these basecalled reads. Future work can focus on generating a k-mer model specifically designed for R10.4 to generate more accurate results. We exclude the accuracy results for R10.4.1 as the number of events found for R10.4.1 is around $35\%$ larger than that of R10.4, which leads to inaccurate mapping. We suspect that our segmentation algorithm and parameters are not optimized for R10.4.1. Our future work will focus on improving these segmentation parameters and techniques to achieve higher accuracy with R10.4.1 as well as RNA sequencing data. We believe this can be achieved because \rht 1)~is highly flexible to change all the parameters corresponding to segmentation and 2)~can map accurately without requiring long sequencing lengths (Table~\ref{rht:tab:seq_bases}), which can mainly be useful for RNA read sets.
We conclude that \rht can provide accurate and fast analysis when using the recent features released by ONT.

\begin{table}[tbh]
\centering
\caption{Accuracy and performance results when using R10.4 and R9.4 datasets}
\resizebox{0.6\columnwidth}{!}{
\begin{tabular}{@{}llrr@{}}\toprule
\textbf{Flow Cell}   &         & \textbf{RH2} & \textbf{RH2-Min.}\\\midrule
\multicolumn{4}{c}{\textbf{Read Mapping Accuracy (E. coli)}} \\\midrule
                   & F1        & 0.9748          & 0.9631         \\
R9.4               & Precision & 0.9904          & 0.9865         \\
                   & Recall    & 0.9597          & 0.9408         \\\midrule
                   & F1        & 0.8960          & 0.8389         \\
R10.4              & Precision & 0.9506          & 0.9325         \\
                   & Recall    & 0.8473          & 0.7623         \\\midrule
\multicolumn{4}{c}{\textbf{Read Mapping Accuracy (S. aureus)}} \\\midrule
                   & F1        & 0.7749          & 0.6778         \\
R10.4              & Precision & 0.8649          & 0.8167         \\
                   & Recall    & 0.7018          & 0.5793         \\\midrule
\multicolumn{4}{c}{\textbf{Performance (E. coli)}} \\\midrule
R9.4               & Throughput [bp/sec]      & 303,382.45     & 659,013.57         \\
                   & Mean time per read [ms]  & 2.161          & 1.099         \\\midrule
R10.4              & Throughput [bp/sec]      & 175,351.94     & 480,471.75         \\
                   & Mean time per read [ms]  & 6.598          & 2.505         \\\midrule
\multicolumn{4}{c}{\textbf{Performance (S. aureus)}} \\\midrule
R10.4              & Throughput [bp/sec]      & 256,680.4      & 617,308.7         \\
                   & Mean time per read [ms]  & 5.478          & 2.243         \\\bottomrule
\end{tabular}

}
\label{rht:supptab:r10accuracy}
\end{table}

\section{Summary}

We introduce \rht, a tool that provides substantial improvements over the previous state-of-the-art mechanism \rh. We make five key improvements over \rh: 1)~more sensitive quantization and chaining, 2)~reduced seed hits with filtering mechanisms, 3)~more accurate mapping decisions with weighted decisions, 4)~the first minimizer sketching technique for raw signals, and 5)~integration of the recent features from ONT.
We find the \rht provides substantial improvements in throughput and accuracy over \rh. We conclude that \rht, overall, is the best tool for mapping raw signals due to its combined benefits in throughput, accuracy, and reduced sequencing time and cost per read compared to the existing mechanisms, especially for longer genomes.
\revb{We hope that, with the substantial improvements in performance and accuracy \rht provides, it leads to 1)~wider adoption of real-time and raw nanopore signal analysis and 2)~more accurate techniques for reducing the noise in signals to achieve faster and more accurate analysis.}

\clearpage
\setsuppbasednumbering
\section{Supplementary Materials}
\subsection{Accuracy} \label{rht:suppsec:accuracy}
\subsubsection{Read Mapping Accuracy} \label{suppsubsec:accuracyfull}

In Supplementary Table~\ref{rht:supptab:accuracyfull}, we show the read mapping accuracy in all metrics (i.e., F1, Precision, and Recall) for all datasets.

\begin{table}[tbh]
\centering
\caption{Read mapping accuracy in all metrics: F1, Precision, and Recall.}
\begin{tabular}{@{}llrrrrr@{}}\toprule
\textbf{Dataset} & \textbf{Metric} & \textbf{RH2} & \textbf{RH2-Min.} & \textbf{RH} & \textbf{UNCALLED} & \textbf{Sigmap} \\\midrule
\multirow{3}{*}{SARS-CoV-2} & F1 & \cellcolor{bestresult}\textbf{0.9867} & 0.9691 & 0.9252 & 0.9725 & 0.7112 \\
 & Precision & \cellcolor{bestresult}\textbf{0.9939} & 0.9868 & 0.9832 & 0.9547 & 0.9929 \\
 & Recall & 0.9796 & 0.9521 & 0.8736 & \cellcolor{bestresult}\textbf{0.9910} & 0.5540 \\
\midrule
\multirow{3}{*}{E. coli} & F1 & \cellcolor{bestresult}\textbf{0.9748} & 0.9631 & 0.9280 & 0.9731 & 0.9670 \\
 & Precision & \cellcolor{bestresult}\textbf{0.9904} & 0.9865 & 0.9563 & 0.9817 & 0.9842 \\
 & Recall & 0.9597 & 0.9408 & 0.9014 & \cellcolor{bestresult}\textbf{0.9647} & 0.9504 \\
\midrule
\multirow{3}{*}{Yeast} & F1 & \cellcolor{bestresult}\textbf{0.9602} & 0.9472 & 0.9060 & 0.9407 & 0.9469 \\
 & Precision & 0.9553 & 0.9561 & 0.9852 & 0.9442 & \cellcolor{bestresult}\textbf{0.9857} \\
 & Recall & \cellcolor{bestresult}\textbf{0.9652} & 0.9385 & 0.8387 & 0.9372 & 0.9111 \\
\midrule
\multirow{3}{*}{Green Algae} & F1 & \cellcolor{bestresult}\textbf{0.9351} & 0.9191 & 0.8114 & 0.8277 & 0.9350 \\
 & Precision & 0.9284 & 0.9280 & 0.9652 & 0.8843 & \cellcolor{bestresult}\textbf{0.9743} \\
 & Recall & \cellcolor{bestresult}\textbf{0.9418} & 0.9104 & 0.6999 & 0.7779 & 0.8987 \\
\midrule
\multirow{3}{*}{Human} & F1 & \cellcolor{bestresult}\textbf{0.7599} & 0.6699 & 0.5574 & 0.3197 & 0.3269 \\
 & Precision & 0.8675 & 0.8511 & \cellcolor{bestresult}\textbf{0.8943} & 0.4868 & 0.4288 \\
 & Recall & \cellcolor{bestresult}\textbf{0.6760} & 0.5523 & 0.4049 & 0.2380 & 0.2642 \\
\midrule
\multirow{3}{*}{Contamination} & F1 & 0.9614 & 0.9317 & 0.8718 & \cellcolor{bestresult}\textbf{0.9637} & 0.6498 \\
 & Precision & \cellcolor{bestresult}\textbf{0.9595} & 0.9424 & 0.8702 & 0.9378 & 0.7856 \\
 & Recall & 0.9632 & 0.9212 & 0.8736 & \cellcolor{bestresult}\textbf{0.9910} & 0.5540 \\
\midrule
\multirow{3}{*}{Rel. Abundance} & F1 & \cellcolor{bestresult}\textbf{0.4659} & 0.3375 & 0.3045 & 0.1249 & 0.2443 \\
 & Precision & \cellcolor{bestresult}\textbf{0.4623} & 0.3347 & 0.3018 & 0.1226 & 0.2366 \\
 & Recall & \cellcolor{bestresult}\textbf{0.4695} & 0.3404 & 0.3071 & 0.1273 & 0.2525 \\
\bottomrule
\multicolumn{7}{l}{\footnotesize Best results are \colorbox{bestresult}{\textbf{highlighted}}.} \\
\end{tabular}

\label{rht:supptab:accuracyfull}
\end{table}

\clearpage
\subsection{Performance} \label{rht:suppsec:performance}
\subsubsection{Runtime, Peak Memory Usage, and Throughput} \label{suppsubsec:cpuandmemory}
Supplementary Table~\ref{rht:supptab:performance} shows the computational resources required by each tool during the indexing and mapping steps. To measure the required computational resources, we collect CPU time and peak memory usage of each tool for all the datasets. To collect these results, we use \texttt{time -v} command in Linux. CPU time shows the total user and system time. Peak memory usage shows the maximum resident set size in the main memory that the application requires to complete its task. To measure the CPU threads needed for analyzing the entire MinION Flowcell with 512 pores, we divide 512 with the number of pores that a single thread can process (as shown with the values inside the bars in Figure~\ref{rht:fig:throughput}) and round up the values to provide the maximum number of threads needed.

\begin{table}[tbh]
\centering
\caption{Computational resources required in the indexing and mapping steps.}
\resizebox{0.75\linewidth}{!}{
\begin{tabular}{@{}lrrrrr@{}}\toprule
\textbf{Dataset} & \textbf{RH2} & \textbf{RH2-Min.} & \textbf{RH} & \textbf{UNCALLED} & \textbf{Sigmap} \\\midrule
\multicolumn{6}{c}{Indexing CPU Time (sec)} \\\midrule
SARS-CoV-2 & 0.12 & 0.06 & 0.16 & 8.40 & \cellcolor{bestresult}\textbf{0.02} \\
E. coli & 2.48 & \cellcolor{bestresult}\textbf{1.61} & 2.56 & 10.57 & 8.86 \\
Yeast & 4.56 & \cellcolor{bestresult}\textbf{3.02} & 4.44 & 16.40 & 25.29 \\
Green Algae & 27.60 & \cellcolor{bestresult}\textbf{17.73} & 24.51 & 213.13 & 420.25 \\
Human & 1,093.56 & \cellcolor{bestresult}\textbf{588.30} & 809.08 & 3,496.76 & 41,993.26 \\
Contamination & 0.13 & 0.06 & 0.15 & 8.38 & \cellcolor{bestresult}\textbf{0.03} \\
Rel. Abundance & 747.74 & \cellcolor{bestresult}\textbf{468.14} & 751.67 & 3,666.14 & 36,216.87 \\
\midrule
\multicolumn{6}{c}{Indexing Peak Memory (GB)} \\\midrule
SARS-CoV-2 & \cellcolor{bestresult}\textbf{0.01} & \cellcolor{bestresult}\textbf{0.01} & \cellcolor{bestresult}\textbf{0.01} & 0.06 & \cellcolor{bestresult}\textbf{0.01} \\
E. coli & 0.35 & 0.19 & 0.35 & \cellcolor{bestresult}\textbf{0.11} & 0.40 \\
Yeast & 0.75 & 0.39 & 0.76 & \cellcolor{bestresult}\textbf{0.30} & 1.04 \\
Green Algae & 5.11 & \cellcolor{bestresult}\textbf{2.60} & 5.33 & 11.94 & 8.63 \\
Human & 80.75 & \cellcolor{bestresult}\textbf{40.59} & 83.09 & 48.43 & 227.77 \\
Contamination & \cellcolor{bestresult}\textbf{0.01} & \cellcolor{bestresult}\textbf{0.01} & \cellcolor{bestresult}\textbf{0.01} & 0.06 & \cellcolor{bestresult}\textbf{0.01} \\
Rel. Abundance & 152.59 & 75.62 & 152.84 & \cellcolor{bestresult}\textbf{47.80} & 238.32 \\
\midrule
\multicolumn{6}{c}{Mapping CPU Time (sec)} \\\midrule
SARS-CoV-2 & 1,705.43 & \cellcolor{bestresult}\textbf{1,227.05} & 1,539.64 & 29,282.90 & 1,413.32 \\
E. coli & 1,296.34 & \cellcolor{bestresult}\textbf{787.49} & 7,453.21 & 28,767.58 & 22,923.09 \\
Yeast & 545.77 & \cellcolor{bestresult}\textbf{246.37} & 4,145.38 & 7,181.44 & 7,146.32 \\
Green Algae & 2,135.83 & \cellcolor{bestresult}\textbf{657.63} & 22,103.03 & 12,593.01 & 26,778.44 \\
Human & 100,947.58 & \cellcolor{bestresult}\textbf{21,860.05} & 1,825,061.23 & 245,128.15 & 6,101,179.89 \\
Contamination & 3,783.69 & \cellcolor{bestresult}\textbf{2,332.28} & 3,480.43 & 234,199.60 & 3,011.78 \\
Rel. Abundance & 250,076.90 & \cellcolor{bestresult}\textbf{62,477.76} & 4,551,349.79 & 569,824.13 & 15,178,633.11 \\
\midrule
\multicolumn{6}{c}{Mapping Peak Memory (GB)} \\\midrule
SARS-CoV-2 & 4.15 & 4.16 & 4.20 & \cellcolor{bestresult}\textbf{0.17} & 28.26 \\
E. coli & 4.13 & 4.03 & 4.18 & \cellcolor{bestresult}\textbf{0.50} & 111.12 \\
Yeast & 4.38 & 4.12 & 4.37 & \cellcolor{bestresult}\textbf{0.36} & 14.66 \\
Green Algae & 6.11 & 4.98 & 11.77 & \cellcolor{bestresult}\textbf{0.78} & 29.18 \\
Human & 48.75 & 25.04 & 52.43 & \cellcolor{bestresult}\textbf{10.62} & 311.94 \\
Contamination & 4.16 & 4.14 & 4.17 & \cellcolor{bestresult}\textbf{0.62} & 111.70 \\
Rel. Abundance & 49.14 & 25.82 & 54.89 & \cellcolor{bestresult}\textbf{8.99} & 486.63 \\
\midrule
\multicolumn{6}{c}{Mapping Throughput (bp/sec)} \\\midrule
SARS-CoV-2 & 552,561.25 & \cellcolor{bestresult}\textbf{885,263.48} & 694,274.92 & 9,260.31 & 602,380.96 \\
E. coli & 303,382.45 & \cellcolor{bestresult}\textbf{659,013.57} & 72,281.32 & 7,515.76 & 13,750.97 \\
Yeast & 150,547.61 & \cellcolor{bestresult}\textbf{394,766.80} & 28,757.15 & 7,471.48 & 11,624.82 \\
Green Algae & 28,742.46 & \cellcolor{bestresult}\textbf{98,323.70} & 9,488.79 & 10,069.41 & 2,569.89 \\
Human & 8,968.78 & \cellcolor{bestresult}\textbf{37,086.38} & 2,099.35 & 7,225.67 & 236.45 \\
Contamination & 563,129.81 & \cellcolor{bestresult}\textbf{884,929.30} & 696,873.20 & 9,343.95 & 601,936.49 \\
Rel. Abundance & 9,501.37 & \cellcolor{bestresult}\textbf{36,919.79} & 962.79 & 8,437.70 & 196.48 \\
\midrule
\multicolumn{6}{c}{CPU Threads Needed for the entire MinION Flowcell (512 pores)} \\\midrule
SARS-CoV-2 & \cellcolor{bestresult}\textbf{1} & \cellcolor{bestresult}\textbf{1} & \cellcolor{bestresult}\textbf{1} & 25 & \cellcolor{bestresult}\textbf{1} \\
E. coli & \cellcolor{bestresult}\textbf{1} & \cellcolor{bestresult}\textbf{1} & 4 & 31 & 17 \\
Yeast & 2 & \cellcolor{bestresult}\textbf{1} & 9 & 31 & 20 \\
Green Algae & 9 & \cellcolor{bestresult}\textbf{3} & 25 & 23 & 90 \\
Human & 26 & \cellcolor{bestresult}\textbf{7} & 110 & 32 & 975 \\
Contamination & \cellcolor{bestresult}\textbf{1} & \cellcolor{bestresult}\textbf{1} & \cellcolor{bestresult}\textbf{1} & 25 & \cellcolor{bestresult}\textbf{1} \\
Rel. Abundance & 25 & \cellcolor{bestresult}\textbf{7} & 240 & 28 & 1173 \\\bottomrule
\multicolumn{6}{l}{\footnotesize Best results are \colorbox{bestresult}{\textbf{highlighted}}.} \\
\end{tabular}

}
\label{rht:supptab:performance}
\end{table}

\clearpage

\subsubsection{Mapping Time per Read}\label{rht:subsec:mappingtimeperread}

\begin{figure}[tbh]
  \centering
  \includegraphics[width=0.8\columnwidth]{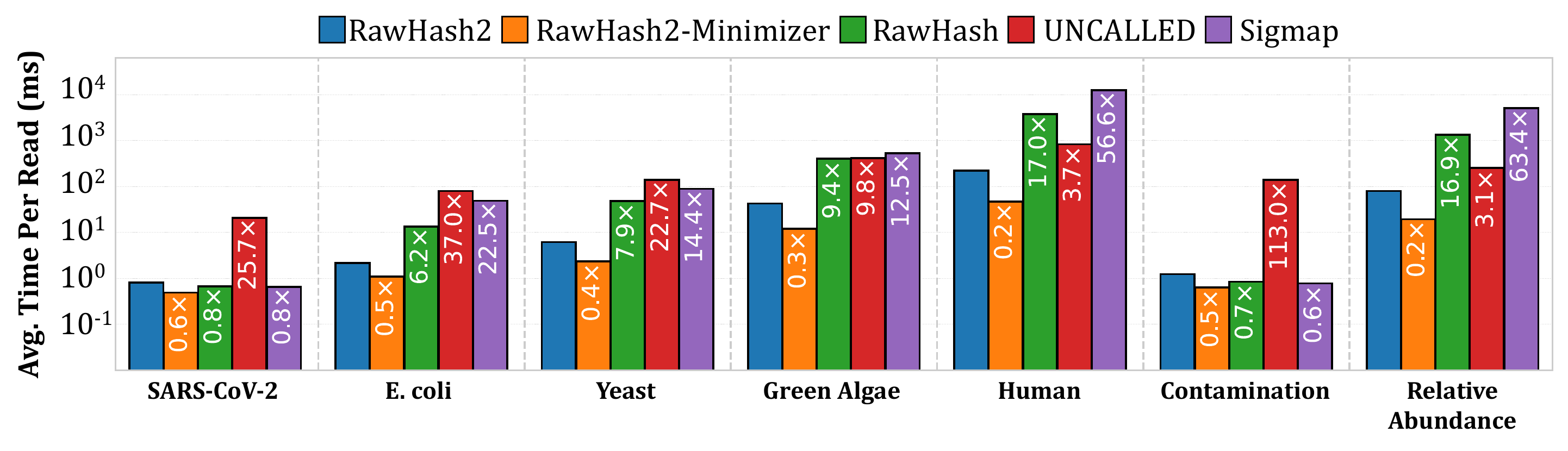}
  \caption{Average time spent per read by each tool in real-time. Values inside the bars show the speedups that \rht provides over other tools in each dataset.}
  \label{rht:suppfig:timeperread}
\end{figure}

Supplementary Figure~\ref{rht:suppfig:timeperread} shows the average mapping time that each tool spends per read for all the datasets we evaluate. The mapping times spent per read are provided by each tool as PAF output with the \texttt{mt} tag. We use these reported values to calculate the average mapping time across all reads reported in their corresponding PAF files.

\subsubsection{Ratio of Filtered Seeds from Frequency Filter}\label{suppsubsec:freqfilter}

Supplementary Table~\ref{rht:supptab:filtered_seed_hits} shows the ratio of seed hits filtered out by the frequency filtered in \rht. We calculate these ratios in three steps.
First, for each seed (i.e., a hash value that \rht constructs from raw signals), we perform a query to the hash table that is used as an index. If the hash value exists, the table returns a list of genomic regions that share the same hash value. Each region counts as a seed hit, and the list length indicates the number of seed hits.
Second, for all seeds generated from raw signals, we count 1)~the overall number of seed hits and 2)~the number of seed hits filtered out by frequency filter. We note that if the list length (i.e., number of seed hits) returned after querying a particular seed is above a certain threshold (defined by our frequency filter), all seed hits within the same list are filtered out.
Third, we calculate the ratio of filtered seed hits to the total seed hits and report these ratios in Supplementary Table~\ref{rht:supptab:filtered_seed_hits}.

\begin{table}[tbh]
\centering
\caption{Ratio of filtered seed hits from frequency filter.}
\begin{tabular}{@{}lr@{}}
\toprule
\textbf{Dataset}      & \textbf{Average Filtered Ratio} \\ \midrule
SARS-CoV-2            & 0.0627                       \\
E. coli        & 0.5505                       \\
Yeast                 & 0.5356                       \\
Green Algae           & 0.8106                       \\
Human & 0.5104                       \\
E. coli (R10.4)       & 0.6895                      \\
S. aureus (R10.4)     & 0.6003                       \\ \bottomrule
\end{tabular}

\label{rht:supptab:filtered_seed_hits}
\end{table}

\clearpage
\subsection{Configuration} \label{rht:secconfiguration}
\subsubsection{Parameters} \label{rht:subsec:parameters}

In Supplementary Table~\ref{rht:tab:parameters}, we show the parameters of each tool for each dataset. In Supplementary Table~\ref{rht:tab:presets}, we show the details of the preset values that \rht sets in Supplementary Table~\ref{rht:tab:parameters}. For \unc~\cite{kovaka_targeted_2021}, \sig~\cite{zhang_real-time_2021}, and minimap2~\cite{li_minimap2_2018}, we use the same parameter setting for all datasets. For the sake of simplicity, we only show the parameters we explicitly set in each tool. For the descriptions of all the other parameters, we refer to the help message that each tool generates, including \rht.

\begin{table}[tbh]
\centering
\caption{Parameters we use in our evaluation for each tool and dataset in mapping.}
\resizebox{\linewidth}{!}{
\begin{tabular}{@{}lccccccccc@{}}\toprule
\textbf{Tool} & \textbf{\emph{Contamination}} & \textbf{\emph{SARS-CoV-2}} & \textbf{\emph{E. coli} (R9.4)} & \textbf{\emph{Yeast}} & \textbf{\emph{Green Algae}} & \textbf{\emph{Human}} & \textbf{\emph{Rel. Abundance}} & \textbf{\emph{E. coli} (R10.4)} & \textbf{\emph{S. aureus} (R10.4)} \\\midrule
\rht    & -x viral --depletion -t 32 & -x viral -t 32 & -x sensitive -t 32 & -x sensitive -t 32 & -x sensitive -t 32 & -x fast -t 32 & -x fast -t 32 & -x sensitive --r10 -t 32 & -x sensitive --r10 -t 32\\\midrule
\rhmin   & -x viral -w3 --depletion -t 32  & -x viral -w3 -t 32  & -x sensitive -w3 -t 32 & -x sensitive -w3 -t 32 & -x sensitive -w3 -t 32 & -x fast -w3 -t 32 & -x fast -w3 -t 32 & -x sensitive --r10 -w3 -t 32 & -x sensitive --r10 -w3 -t 32\\\midrule
\rh         	 & -x viral -t 32  & -x viral -t 32  & -x sensitive -t 32 & -x sensitive -t 32 & -x fast -t 32 & -x fast -t 32 & -x fast -t 32 & NA & NA\\\midrule
\unc  			 & \multicolumn{7}{c}{map -t 32} & NA & NA\\\midrule
\sig        	 & \multicolumn{7}{c}{-m -t 32} & NA & NA\\\midrule
Minimap2         & \multicolumn{9}{c}{-x map-ont -t 32}\\\bottomrule
\end{tabular}

}
\label{rht:tab:parameters}
\end{table}

\begin{table}[tbh]
\centering
\caption{Corresponding parameters of presets (-x) in \rht.}
\resizebox{\linewidth}{!}{
\begin{tabular}{@{}lcc@{}}\toprule
\textbf{Preset} & \textbf{Corresponding parameters} & Usage \\\midrule
viral      & -e 6 -q 4 --max-chunks 5 --bw 100 --max-target-gap 500 & Viral genomes\\
& --max-target-gap 500 --min-score 10 --chain-gap-scale 1.2 --chain-skip-scale 0.3 &  \\\midrule
sensitive  & -e 8 -q 4 --fine-range 0.4 & Small genomes (i.e., $<500M$ bases)\\\midrule
fast       & -e 8 -q 4 --max-chunks 20 & Large genomes (i.e., $>500M$ bases)\\\midrule
\multicolumn{3}{c}{\textbf{Other helper parameters}}\\\midrule
depletion  & --best-chains 5 --min-mapq 10 --w-threshold 0.5 & Contamination analysis\\
& --min-anchors 2 --min-score 15 --chain-skip-scale 0 & \\\midrule
r10        & -k9 --seg-window-length1 3 --seg-window-length2 6 --seg-threshold1 6.5 & For R10.4 Flow Cells \\
& --seg-threshold2 4 --seg-peak-height 0.2 --chain-gap-scale 1.2 & \\\bottomrule
\end{tabular}

}
\label{rht:tab:presets}
\end{table}

\clearpage

\subsubsection{Versions}\label{rht:subsec:versions}

Supplementary Table~\ref{rht:tab:versions} shows the version and the link to these corresponding versions of each tool and library we use in our experiments and in \rht, respectively.

\begin{table}[tbh]
\centering
\caption{Versions of each tool and library.}
\resizebox{\linewidth}{!}{
\begin{tabular}{@{}lll@{}}\toprule
\textbf{Tool} & \textbf{Version} & \textbf{Link to the Source Code} \\\midrule
\rht & 2.1 & \url{https://github.com/CMU-SAFARI/RawHash/releases/tag/v2.1}\\\midrule
\rh & 1.0 & \url{https://github.com/CMU-SAFARI/RawHash/releases/tag/v1.0}\\\midrule
\unc  & 2.3 & \url{https://github.com/skovaka/UNCALLED/releases/tag/v2.3}\\\midrule
\sig  & 0.1 & \url{https://github.com/haowenz/sigmap/releases/tag/v0.1}\\\midrule
Minimap2 & 2.24 & \url{https://github.com/lh3/minimap2/releases/tag/v2.24}\\\midrule
\multicolumn{3}{c}{Library versions}\\\midrule
FAST5 (HDF5) & 1.10 & \url{https://github.com/HDFGroup/hdf5/tree/db30c2d} \\\midrule
POD5 & 0.2.2 & \url{https://github.com/nanoporetech/pod5-file-format/releases/tag/0.3.10}\\\midrule
S/BLOW5 & 1.2.0-beta & \url{https://github.com/hasindu2008/slow5lib/tree/e0d0d0f}\\\bottomrule
\end{tabular}

}
\label{rht:tab:versions}
\end{table}

\setchapterbasednumbering

\chapter[\rs: Overlapping and Assembling Raw Nanopore Signals]{Enabling New Applications\\by Overlapping and Assembling Raw Nanopore Signals}
\label{chap:rs}

Throughout the previous three chapters, we built a detailed understanding of the effects of noise and the limitations this noise puts on the heuristics and algorithms designed to handle noise. In this chapter, after better handling and understanding of this noise for raw nanopore signals, we enable a new application in raw nanopore signals by introducing a new technique, \rs, that can find all-vs-all overlapping of reads to build \emph{de novo} assemblies from raw signals. We discuss new directions that can be enabled by \rs.

\section{Background and Motivation}
Nanopore sequencing technology~\cite{\citenanopore} can sequence long nucleic acid molecules of up to a few million bases at high throughput~\cite{jain_nanopore_2018,pugh_current_2023,senol_cali_nanopore_2019}. As a molecule moves through a tiny pore, called a \emph{nanopore}, ionic current measurements, called \emph{raw signals}, are generated~\cite{deamer_three_2016}.
Nanopore sequencing provides three unique key benefits.

First, nanopore sequencing enables stopping the sequencing of single reads or the entire sequencing run early, known as \emph{adaptive sampling} or \emph{selective sequencing}~\cite{loose_real-time_2016},
\rev{while raw signals are generated and analyzed during sequencing, called \emph{real-time analysis}}. Adaptive sampling can substantially reduce the sequencing time and cost by avoiding unnecessary sequencing.
Second, raw nanopore signals retain more information than nucleotides, such as methylation, and can be useful for many applications~\cite{flynn_evaluation_2022}.
Third, compact nanopore sequencing devices enable on-site portable sequencing and analysis, \rev{which can be coupled with real-time analysis~\cite{bloemen_development_2023}.}

Existing works \rev{that analyze raw nanopore signals}~\cite{\citebasecallnanodnn,\citebasecallnanohmm,\citesignalanalysis} mainly utilize deep learning mechanisms~\cite{\citebasecallnanodnn} to translate these signals into nucleotides, a process called \emph{basecalling}.
\rev{Basecalling mechanisms usually 1)~are designed to use large chunks of raw signal data for accurate analysis}~\cite{zhang_real-time_2021} and 2)~have high computational requirements~\cite{shih_efficient_2023,senol_cali_nanopore_2019}. This can impose limitations \rev{to enable 1)~accurate real-time analysis~\cite{zhang_real-time_2021} and 2)~portable sequencing with constrained resources~\cite{shih_efficient_2023}.}

To fully utilize the unique benefits of nanopore sequencing, it is necessary to analyze raw signals with 1)~high accuracy and low latency for adaptive sampling and 2)~low resource usage for portability \rev{and efficiency}. To achieve this, several mechanisms focus on analyzing raw nanopore signals \rev{\emph{without}} basecalling~\cite{\citesignalanalysis}.\rev{\footnote{We use the \emph{raw signal analysis} term specifically for these \revc{mechanisms} in the remainder of the paper.}}
Although raw nanopore signal mapping to a reference genome is widely studied to achieve \rev{relatively} accurate and fast mapping of raw signals~\cite{\citesignalanalysismapped}, \rev{\emph{none}} of these works \rev{\emph{without}} a reference genome.

\rev{When a reference genome is not available, a genome can be constructed from scratch, called \emph{de novo} assembly construction\cite{fleischmann_whole-genome_1995}.} To construct a \emph{de novo} assembly, similar regions between all read pairs are identified, called \emph{all-vs-all overlapping, instead of identifying similarities between reads and a reference genome.}~\cite{li_minimap_2016, senol_cali_nanopore_2019, firtina_blend_2023}. However, an all-vs-all overlapping between raw nanopore signals remains unsolved due to several unique challenges~\cite{deamer_three_2016,bhattacharya_molecular_2012,kawano_controlling_2009,smeets_noise_2008}.

\rev{First, it is challenging to identify similarities between a pair of noisy raw signals as compared to a noisy raw signal and an accurate signal generated from a reference genome, an approach commonly used in earlier works for read mapping~\cite{kovaka_targeted_2021, zhang_real-time_2021, firtina_rawhash_2023, firtina_rawhash2_2024, lindegger_rawalign_2024}. This is because converting reference genomes to their expected raw signal values is free from certain types of noise that raw nanopore signals contain (e.g., stochastic signal fluctuations~\cite{deamer_three_2016} and variable speed of DNA molecules moving through nanopores~\cite{bhattacharya_molecular_2012,kawano_controlling_2009}), while none of the raw signals are free from such noise in all-vs-all overlapping.}

Second, existing raw signal analysis works lack the mapping strategies typically used in all-vs-all overlapping, such as reporting multiple mappings (i.e., overlaps) of a read to many reads. This is because these works are mainly designed to stop the mapping process as soon as there is \rev{an} accurate mapping for a read to minimize the unnecessary sequencing~\cite{kovaka_targeted_2021}. \rev{For all-vs-all overlapping, pairwise mappings of a read to many reads (instead of a single mapping) must be reported while avoiding certain trivial cyclic pairwise mappings to construct an assembly~\cite{li_minimap_2016}.}

Third, read overlapping can increase the space requirements for storing and using indexing. \rev{This is because a read set is likely to be sequenced such that its overall number of bases is larger than the number of bases in its corresponding genome.} Such an increased index size can raise the computational and space demands for read overlapping. \rev{It is essential to provide high-throughput and scalable computation to enable real-time downstream analysis for future work (as discussed in Section~\ref{rs:sec:discussion}).}

\rev{\textbf{Our goal} is to enable 1)~raw signal analysis \emph{without} a reference genome and 2)~new use cases with raw nanopore signal analysis, such as \emph{assembly from overlapping} raw signals, by addressing the challenges \rev{of} accurate and fast all-vs-all overlapping of raw nanopore signals.} To this end, \rev{we propose \emph{\rs}, the first mechanism} that enables fast and accurate overlap finding between raw nanopore signals. \rev{The key idea in \rs is to re-design the existing state-of-the-art hash-based seeding mechanism~\cite{firtina_rawhash_2023, firtina_rawhash2_2024} for raw signals with more effective noise reduction techniques and useful outputting strategies to find all overlapping pairs accurately, which we explain in three key steps.} 

\rev{First, to enable identifying similarities between a pair of noisy raw signals accurately,
\rs filters raw signals to select those substantially distinct from their surrounding signals. Such non-distinct and adjacent signals are usually the result of a certain error type in the analysis, known as \emph{stay errors}~\cite{shivakumar_sigmoni_2024, firtina_rawhash_2023,zhang_real-time_2021}.
Although similar filtering strategies~\cite{shivakumar_sigmoni_2024, firtina_rawhash_2023,zhang_real-time_2021} are exploited when mapping raw signals to reference genomes, \rs performs a more aggressive filtering to avoid storing the erroneous portions of signals in the index to enable accurate similarity identification from the sufficiently distinct regions of signals.}
Second, to find multiple overlaps for a read, \rs identifies highly accurate chains from seed matches based on their chaining scores and reports \emph{all} of these chains as mappings, \rev{as opposed to} choosing \rev{solely} the best mapping \rev{as} determined by weighted decisions among all such chains~\cite{firtina_rawhash2_2024}.
\rev{Third, to prevent trivial cycles between a pair of overlapping reads, \rs ensures that only one of the overlapping reads in each pair is always chosen as a query sequence,} and the other is always chosen as a target sequence based on a deterministic ordering mechanism between these reads. These steps enable \rs to find overlaps \rev{between raw signals} accurately and quickly.

\section{Methods} \label{rs:sec:methods}
\subsection{Overview} \label{rs:subsec:met_overview}
\begin{figure}[tbh]
  \centering
  \includegraphics[width=0.8\columnwidth]{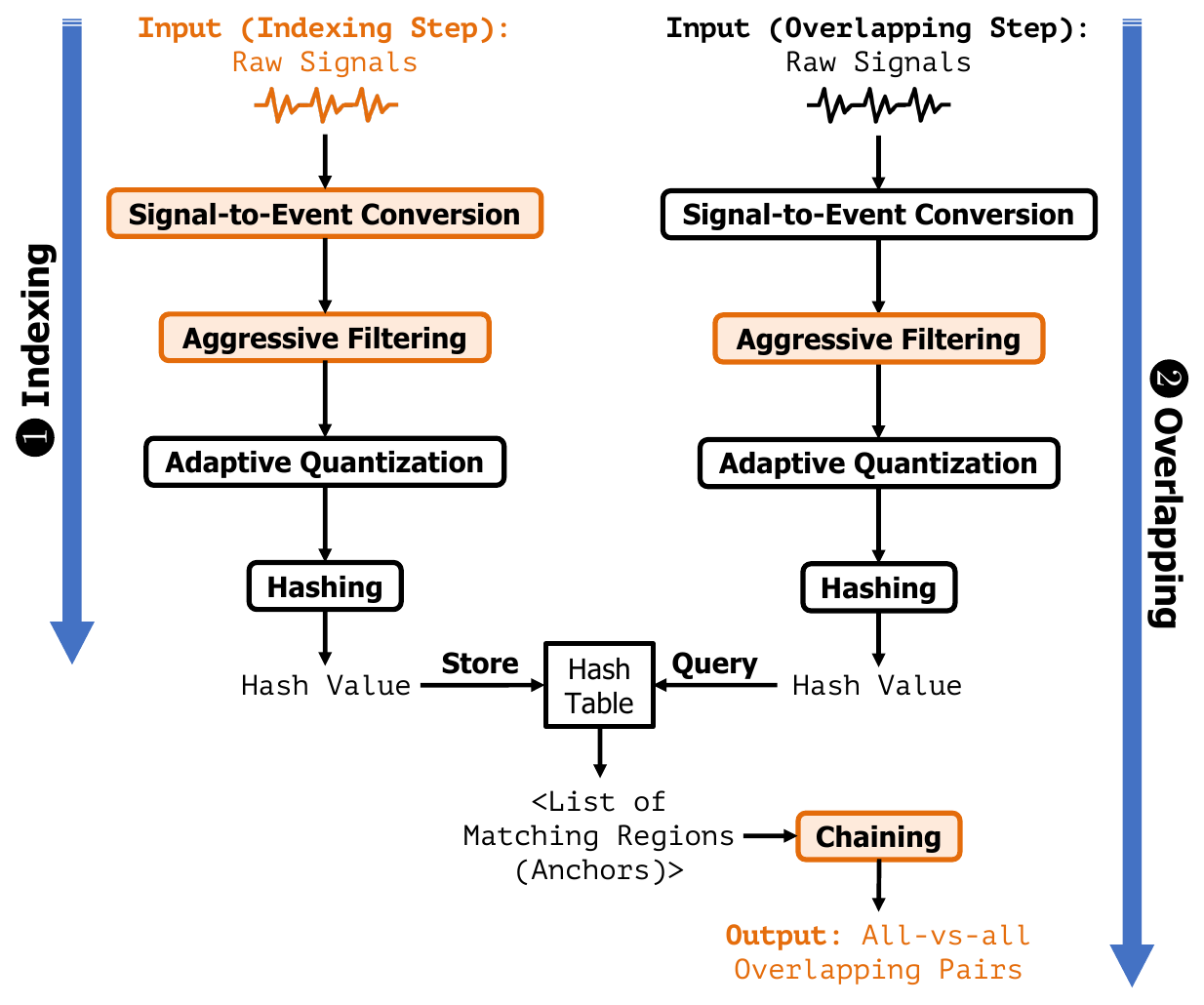}
  \caption{Overview of \rs. \revc{We use colors for the inputs, steps, and outputs to highlight the parts that \rs modifies over \rht.}}
  \label{rs:fig:overview}
\end{figure}

\rs is a mechanism to find overlapping pairs between raw nanopore signals (i.e., \emph{all-vs-all overlapping}), \rev{which can be used by existing assemblers} to construct \emph{de novo} assembly graphs without basecalling, as shown in Figure~\ref{rs:fig:overview}. To achieve this, \rs builds on the state-of-the-art raw signal mapper, \rht~\cite{firtina_rawhash_2023, firtina_rawhash2_2024}. \rs extends \rht to support all-vs-all raw signal overlapping in four key steps.
First, to enable efficient and accurate indexing from the noisy raw signals (\circled{1}), \rs aggressively filters the raw input after the signal-to-event conversion to avoid nanopore-related errors.
Second, to improve the accuracy of overlapping (\circled{2}), \rs adjusts the minimum chaining score to avoid false chains, ensuring that only high-confidence overlaps are considered.
Third, to enable finding useful and long connections from many overlaps, \rs adjusts the output strategy such that 1)~\emph{all} chains rather than only the best chain are reported and 2)~cyclic overlaps are avoided.
\rev{Fourth, \rs enables the use of existing \emph{de novo} assemblers \rev{off-the-shelf}, such as miniasm~\cite{li_minimap_2016}, by providing the overlap information in a standardized format that these assemblers use.}

\subsection{Constructing an Index from Noisy Raw Signals via Aggressive Filtering} \label{rs:subsec:met_indexing}
\begin{figure}[tbh]
  \centering
  \includegraphics[width=0.9\columnwidth]{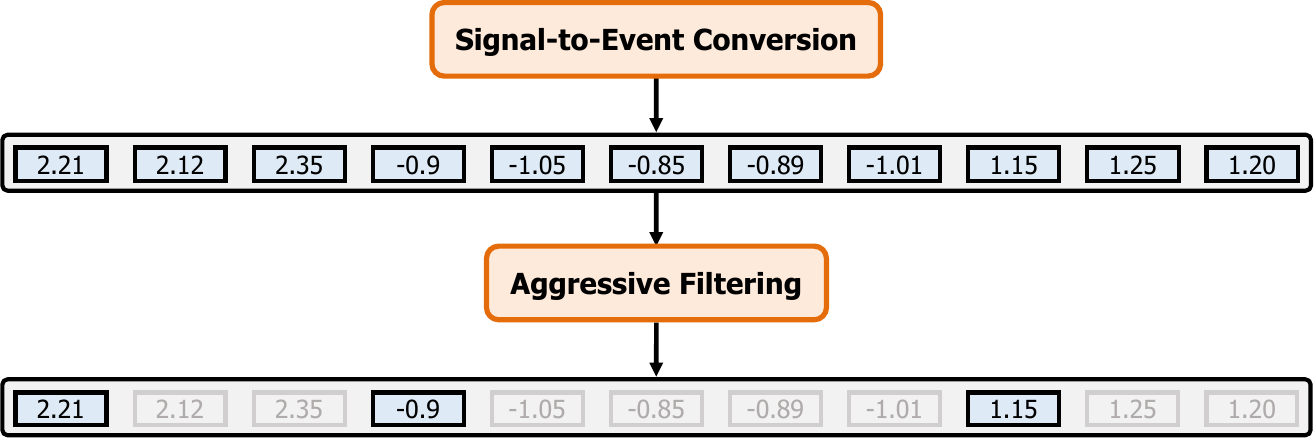}
  \caption{Filtering in \rs. Values in gray boxes show the filtered signals as their values are close to the previous signal (in a blue box) that is not filtered out.}
  \label{rs:fig:filter}
\end{figure}

\rs identifies overlapping regions between raw nanopore signals using a hash-based seeding mechanism that operates in two steps.
First, \rev{to enable quick matching between raw signals}, \rs enables constructing an index directly from raw nanopore signals instead of using an existing reference genome sequence. \rev{While converting reference genomes to their expected raw signal values to store them in an index is mainly free from certain types of noise that raw nanopore signals contain (e.g., stochastic signal fluctuations~\cite{deamer_three_2016} and variable speed of DNA molecules moving through nanopores~\cite{bhattacharya_molecular_2012,kawano_controlling_2009}), noise in raw nanopore signals can cause challenges to find accurate matches between raw signals when these signals are stored in an index.}
Second, to reduce noise stored in the hash tables and enable accurate similarity identification when both signals are noisy, \rs aggressively filters raw signals as shown in Figure~\ref{rs:fig:filter}. The filtering mechanism iteratively compares two adjacent signals, $s_i$ and $s_j$, and removes the second signal, $s_j$, if the absolute difference \revc{between the two adjacent signals} is below a certain threshold $T$. \revc{This filtering generates a list of filtered signals that aggressively aims to reduce the impact of stay errors during sequencing.} Although similar filtering approaches are used in prior works~\cite{zhang_real-time_2021, firtina_rawhash_2023, firtina_rawhash2_2024} to reduce the stay errors \rev{(e.g., by aiming to perform homopolymer compression of raw signals~\cite{shivakumar_sigmoni_2024}),} \rs employs a substantially larger threshold (i.e., aggressive) for filtering to minimize noise both in the index and during mapping.
By applying the aggressive filtering technique, \rs ensures that only high-quality, informative events are used in indexing to improve the accuracy and efficiency of the overall overlapping mechanism when both signals are noisy.

\subsection{Adjusting the Chaining Mechanism for Overlapping}
To reduce the number of \revc{chains that do not result in mapping (i.e., false chains)} and construct longer chains, \rs adjusts the chaining mechanism~\cite{li_minimap2_2018, firtina_rawhash2_2024} in two ways.
First, \rs constructs chains between seed matches (i.e., anchors) with longer gaps by adjusting the maximum gap length between anchors. This adjustment is needed because the filtering mechanism results in a sparser list with potentially long gaps between signals.
Second, to ensure that only high-confidence chains are considered as overlaps among many pairwise overlapping regions, \rs sets a higher minimum chaining score for overlapping than mapping, which effectively filters out spurious matches. 
These adjustments to the chaining mechanism enable \rs to construct long and accurate chains between noisy and sparse raw signals, improving the overall sensitivity of the overlapping process and further downstream analysis such as \emph{de novo} assembly construction.

\subsection{Adjusting the Mapping Strategy}
\revd{\rs adjusts the mapping strategy used in \rht in two ways.
First, \rs generates pairwise mappings of a read to many reads (instead of a single mapping per read). To achieve this, \rs identifies all valid chains based on their chaining scores and reports all such chains between raw signals. This enables the overlapping of a single raw signal with multiple other raw signals.
Second, \rs filters out \emph{trivial} cyclic overlaps between two reads. To do so, \rs avoids reporting overlapping signal pairs both as \emph{query} and \emph{target} in the mapping output, which can complicate the \emph{de novo} assembly construction process~\cite{li_minimap_2016, simpson_abyss_2009, i_sovic_approaches_2013}. To avoid these trivial cyclic overlaps, \rs implements a pre-defined and deterministic ordering between raw signals (e.g., based on the lexicographic ordering of read names). By comparing raw signals deterministically, \rs guarantees that only one of each pair of overlapping reads is processed as a query sequence while the other is treated as a target sequence.
Reporting all chains while avoiding cyclic overlaps between raw signals allows for a comprehensive representation of the overlapping regions, which is useful for constructing accurate and long assemblies.

\head{Output Format}
\rs provides the overlapping information between raw signals using the Pairwise mApping Format (PAF)~\cite{li_minimap_2016}.
To construct an assembly graph from the overlapping information that \rs generates, any assembler that takes PAF files as input can be used, including miniasm~\cite{li_minimap_2016}.}

\section{Results} \label{rs:sec:results}
\subsection{Evaluation Methodology} \label{rs:subsec:evaluation}
\revd{We implement the improvements we propose in \rs on the \rht implementation~\cite{firtina_rawhash2_2024}. Similar to \rht and minimap2~\cite{li_minimap2_2018}, \rs provides the mapping information in the standard Pairwise mApping Format (PAF)~\cite{li_minimap_2016}.
We basecall the signals on two different hardware setups using Dorado's 1)~high accuracy (HAC) and super high accuracy (SUP) models with an NVIDIA RTX A6000 GPU~\cite{nvidia_a6000_2024}, and 2)~fast (Fast) and high accuracy (HAC) models with an Intel Xeon Gold 6226R CPU~\cite{intel_xeon6226R_2024}. We use Dorado's 1)~Fast model on a CPU to show the best performance (i.e., speed) that a CPU-based basecalling can provide and 2)~SUP model on a GPU to show the best accuracy that basecalling provides along with its associated increased computational cost. We use POD5 files for all datasets, as suggested by Dorado~\cite{oxford_nanopore_technologies_dorado_2024} for optimal performance.
Since no prior works can overlap reads using raw signals, we compare \rs to minimap2, the current state-of-the-art read overlapper for basecalled sequences using the datasets shown in Table~\ref{rs:tab:dataset}. We use datasets both with high sequencing depth of coverage (D1 to D3) and low coverage \rev{(D4 and D5)} to evaluate the capability to construct assemblies in both scenarios. For D5, the depth of coverage is at the extreme lower levels to evaluate the assembly capabilities of both raw signal and basecalled analysis.}

\begin{table}[htb]
\centering
\caption{Details of datasets used in our evaluation.}
\resizebox{0.9\linewidth}{!}{

\begin{tabular}{@{}clllrrrr@{}}\toprule
& \textbf{Organism}     & \textbf{Device}  &\textbf{Reads} & \textbf{Bases}  & \textbf{Avg. Read} & \textbf{Estimated}           & \textbf{SRA}       \\
& \textbf{}             & \textbf{Type}    &\textbf{(\#)}  & \textbf{(\#)}   & \textbf{Length}   & \textbf{Coverage ($\times$)} & \textbf{Accession} \\\midrule
D1 & \emph{SARS-CoV-2}  & MinION          & 10,001        & 4.02M           & 402               & 135$\times$                  & CADDE Centre       \\\midrule
D2 & \emph{E. coli}     & GridION         & 353,948       & 2,332M          & 6,588             & 445$\times$                  & ERR9127551         \\\midrule
D3 & \emph{Yeast}       & MinION          & 50,023        & 385M            & 7,698             & 32$\times$                   & SRR8648503         \\\midrule
D4 & \emph{Green Algae} & PromethION      & 30,012        & 622M            & 20,731            & 5.6$\times$                  & ERR3237140         \\\midrule
D5 & \emph{Human}       & MinION          & 270,006       & 1,773M          & 6,567             & 0.6$\times$                  & FAB42260           \\\midrule
\multicolumn{8}{l}{Base counts in millions (M). Coverage is estimated using corresponding reference genomes.}\\
\multicolumn{8}{l}{All datasets are generated using \revb{ONT's} R9.4.1 e8 flow cells (450 bp/sec translocation speed with}\\
\multicolumn{8}{l}{4000 signals/sec sampling frequency).}\\
\multicolumn{8}{l}{All of the datasets are basecalled using Dorado (HAC) v3.3.}\\
\end{tabular}

}
\label{rs:tab:dataset}
\end{table}

\revd{We evaluate computational requirements in terms of overall run time with 32 CPU threads and peak memory usage. We use 32 CPU threads since the average thread utilization of CPU-based basecalling is around 32. We report the elapsed time, CPU time (when using only CPUs without GPUs), and peak memory usage of 1)~minimap2 with and without basecalling and 2)~\rs. For \rs, we additionally report throughput (average number of signals analyzed per second per single CPU thread) to provide insights about its capabilities for real-time analysis. To measure performance and peak memory usage, we use the \texttt{time -v} command in Linux when running \rs, minimap2, and Dorado. For average speedup and memory comparisons of \rs against other methods, we use the geometric mean to reduce the impact of outlier data points on the average calculation.

We evaluate \rs based on two use cases: 1)~read overlapping and 2)~\emph{de novo} assembly \rev{by constructing assembly graphs}. We generate read overlaps from 1)~raw signals using \rs and 2)~basecalled sequences (with Dorado's HAC model) of corresponding strands of raw signals using minimap2. To evaluate \rev{the accuracy} of all-vs-all overlapping, we calculate the ratio of overlapping pairs 1)~shared by both \rs and minimap2, 2)~unique to \rs, and 3)~unique to minimap2.
For \emph{de novo} assembly, we use miniasm~\cite{li_minimap_2016} to generate assembly graphs from the read overlaps that \rs and minimap2 generate. We use miniasm because it enables us to evaluate the impact of overlaps that \rs and minimap2 finds by \emph{using the same assembler} for both of them.
To identify the roofline in terms of the assembly contiguity given a dataset, we use a highly accurate assembler as \emph{gold standard} for our evaluations. To do this, we use Dorado's most accurate model (i.e., SUP) to construct highly accurate reads and use Flye~\cite{kolmogorov_assembly_2019} to construct assemblies from these reads, as suggested in the guidelines by ONT~\cite{oxford_nanopore_technologies_guidelines_2022}. This approach enables us to evaluate the gap between the assemblies that \rs generates and the golden standard assembly that can be constructed from the same datasets.

To evaluate the assemblies constructed from the overlaps that \rs and minimap2 generate, we use several metrics that are mainly related to the contiguity of the assemblies. These metrics are 1)~the total length of unitigs (Total Length), 2)~the number of nucleotides in the largest assembly graph component (Largest Comp.), 3)~unitig length at which 50\% of the assembly is covered by unitigs of equal or greater length (N50), 4)~area under the Nx curve to generate a more robust evaluation of contiguity (auN)~\cite{li_aun_2020}, 5)~longest unitig length, and 6)~the number of unitigs. We use Bandage~\cite{wick_bandage_2015} to visualize these assemblies. We note that our contiguity evaluations for \rs and minimap2 are based on the assembly graphs constructed by miniasm. \rev{We leave 1)~constructing the assembled signal sequence (i.e., spelling the assembly from the graph), 2)~designing a consensus mechanism for error correction, and 3)~evaluating the accuracy of these assembled signals as future work, which we discuss in more detail in Section~\ref{rs:sec:discussion}.}}

We provide the parameter settings and versions for each tool as well as the details of the preset parameters in \rs in Supplementary Tables~\ref{rs:tab:parameters} (parameters),~\ref{rs:tab:presets} (details of presets), and~\ref{rs:tab:versions} (versions). We provide the scripts to fully reproduce our results on the GitHub repository at \rsrelease, which contains the corresponding release version of \rs we show in Supplementary Table~\ref{rs:tab:versions}.
\rev{We provide the detailed reasoning behind the choice of Flye in order to generate the gold standard in Supplementary Material~\ref{rs:suppsec:gold_standard}.}
 
\subsection{Performance and Memory} \label{rs:subsec:perfmemory}

\head{Throughput} Table~\ref{rs:tab:throughput} shows the throughput (i.e., signals processed per second per CPU thread) that \rs reports. \revc{Our goal is to estimate the number of CPU threads needed to achieve a throughput faster than the throughput of a single sequencer. Using as few CPU threads as possible is useful to 1)~provide better scalability (i.e., analyzing a larger amount of data with the same computation capabilities) and 2)~reduce the overall computational requirements and corresponding energy consumption (i.e., analyzing the same amount of data with less computational capabilities). The latter is especially essential for resource-constrained devices (e.g., devices with external and limited batteries). The throughput of a single nanopore is around 5,000 signals per second~\cite{oxford_nanopore_technologies_dorado_2024}, and the entire sequencer is usually equipped with 512 nanopores. This means a single sequencer's throughput is around $2{,}560{,}000$ signals per second ($5{,}000 \times 512$)~\cite{magi_nanopore_2018}.}
We find that \rs provides an average throughput of \rsavgthr signals per second \revc{\textbf{per CPU thread}}. This means \rs can achieve an analysis throughput faster than a single sequencer's throughput by using an average of two CPU threads. This shows that the overlapping mechanism of \rs is fast \revc{enough} for performing real-time overlapping tasks using very few CPU threads.

\begin{table}[htb]
\centering
\caption{\rs Throughput (signals processed per second per CPU thread).}
\begin{tabular}{@{}lrrrrr@{}}\toprule
                     & D1                & D2             & D3           & D4                 & D5          \\
                     & \emph{SARS-CoV-2} & \emph{E. coli} & \emph{Yeast} & \emph{Green Algae} & \emph{Human}\\\toprule
\textbf{Throughput}  & 2,065,764         & 2,720,702      & 2,128,800    & 1,668,065          & 3,579,472    \\\bottomrule
\end{tabular}

\label{rs:tab:throughput}
\end{table}

\head{Computational resources} Table~\ref{rs:tab:resources} shows the computational resources of the different approaches. Inside the parentheses provided with the minimap2 results, we provide the ratio between the reported result and the corresponding \rs result \revc{(if higher than $1\times$, \rs is better)}.

\begin{table}[tbh]
\centering
\caption{\revc{Comparison of various tools across different organisms in terms of elapsed time, CPU time, and peak memory usage. The values in parentheses represent the ratio of the result shown in the cell compared to the corresponding result of \rs (values higher than $1\times$ indicate that \rs performs better). We highlight the cells that \rs provide a \colorbox{bestresult}{better} and  \colorbox{worseresult}{worse} results with colors.}}
\resizebox{0.9\columnwidth}{!}{
\begin{tabular}{@{}llrrr@{}}
\toprule
\textbf{Organism} & \textbf{Tool} & \textbf{Elapsed time} & \textbf{CPU time} & \textbf{Peak} \\
\                 &                & \textbf{(hh:mm:ss)}   & \textbf{(sec)}    & \textbf{Mem. (GB)} \\\toprule
D1 & \textbf{\rs} & 0:00:03 & 33 & 1.07 \\
\emph{SARS-CoV-2} & \textbf{Minimap2} & 0:00:01 \cellcolor{worseresult}{($0.33\times$)} & 19 \cellcolor{worseresult}{($0.58\times$)} & 0.16 \cellcolor{worseresult}{($0.15\times$)} \\
& \textbf{Minimap2 + Dorado CPU (Fast)} & 0:01:45 \cellcolor{bestresult}{($35.00\times$)} & 3,227 \cellcolor{bestresult}{($97.79\times$)} & 44.93 \cellcolor{bestresult}{($41.99\times$)} \\
& \textbf{Minimap2 + Dorado CPU (HAC)} & 0:05:45 \cellcolor{bestresult}{($115.00\times$)} & 5,457 \cellcolor{bestresult}{($165.36\times$)} & 57.98 \cellcolor{bestresult}{($54.19\times$)} \\
& \textbf{Minimap2 + Dorado GPU (HAC)} & 0:01:41 \cellcolor{bestresult}{($33.67\times$)} & NA & 0.8 \cellcolor{worseresult}{($0.75\times$)} \\
& \textbf{Minimap2 + Dorado GPU (SUP)} & 0:25:47 \cellcolor{bestresult}{($515.67\times$)} & NA & 1.23 \cellcolor{bestresult}{($1.15\times$)} \\
\midrule
D2 & \textbf{\rs} & 1:12:44 & 132,758 & 6.72 \\
\emph{E. coli} & \textbf{Minimap2} & 0:14:25 \cellcolor{worseresult}{($0.20\times$)} & 25,721 \cellcolor{worseresult}{($0.19\times$)} & 26.73 \cellcolor{bestresult}{($3.98\times$)} \\
& \textbf{Minimap2 + Dorado CPU (Fast)} & 7:17:05 \cellcolor{bestresult}{($6.01\times$)} & 583,358 \cellcolor{bestresult}{($4.39\times$)} & 50.43 \cellcolor{bestresult}{($7.50\times$)} \\
& \textbf{Minimap2 + Dorado CPU (HAC)} & 32:26:12 \cellcolor{bestresult}{($26.76\times$)} & 1,335,697 \cellcolor{bestresult}{($10.06\times$)} & 38.0 \cellcolor{bestresult}{($5.65\times$)} \\
& \textbf{Minimap2 + Dorado GPU (HAC)} & 0:36:14 \cellcolor{worseresult}{($0.50\times$)} & NA & 26.73 \cellcolor{bestresult}{($3.98\times$)} \\
& \textbf{Minimap2 + Dorado GPU (SUP)} & 1:30:30 \cellcolor{bestresult}{($1.24\times$)} & NA & 26.73 \cellcolor{bestresult}{($3.98\times$)} \\
\midrule
D3 & \textbf{\rs} & 0:01:18 & 2,241 & 6.39 \\
\emph{Yeast} & \textbf{Minimap2} & 0:00:21 \cellcolor{worseresult}{($0.27\times$)} & 290 \cellcolor{worseresult}{($0.13\times$)} & 5.25 \cellcolor{worseresult}{($0.82\times$)} \\
& \textbf{Minimap2 + Dorado CPU (Fast)} & 0:54:04 \cellcolor{bestresult}{($41.59\times$)} & 71,796 \cellcolor{bestresult}{($32.04\times$)} & 56.13 \cellcolor{bestresult}{($8.78\times$)} \\
& \textbf{Minimap2 + Dorado CPU (HAC)} & 3:13:56 \cellcolor{bestresult}{($149.18\times$)} & 193,640 \cellcolor{bestresult}{($86.41\times$)} & 65.43 \cellcolor{bestresult}{($10.24\times$)} \\
& \textbf{Minimap2 + Dorado GPU (HAC)} & 0:04:33 \cellcolor{bestresult}{($3.50\times$)} & NA & 5.25 \cellcolor{worseresult}{($0.82\times$)} \\
& \textbf{Minimap2 + Dorado GPU (SUP)} & 0:10:33 \cellcolor{bestresult}{($8.12\times$)} & NA & 5.92 \cellcolor{worseresult}{($0.93\times$)} \\
\midrule
D4 & \textbf{\rs} & 0:07:57 & 14,064 & 8.67 \\
\emph{Green Algae} & \textbf{Minimap2} & 0:00:47 \cellcolor{worseresult}{($0.10\times$)} & 882 \cellcolor{worseresult}{($0.06\times$)} & 8.7 \cellcolor{bestresult}{($1.00\times$)} \\
& \textbf{Minimap2 + Dorado CPU (Fast)} & 1:16:35 \cellcolor{bestresult}{($9.63\times$)} & 79,606 \cellcolor{bestresult}{($5.66\times$)} & 50.88 \cellcolor{bestresult}{($5.87\times$)} \\
& \textbf{Minimap2 + Dorado CPU (HAC)} & 4:30:07 \cellcolor{bestresult}{($33.98\times$)} & 286,362 \cellcolor{bestresult}{($20.36\times$)} & 64.07 \cellcolor{bestresult}{($7.39\times$)} \\
& \textbf{Minimap2 + Dorado GPU (HAC)} & 0:06:01 \cellcolor{worseresult}{($0.76\times$)} & NA & 8.7 \cellcolor{bestresult}{($1.00\times$)} \\
& \textbf{Minimap2 + Dorado GPU (SUP)} & 0:14:54 \cellcolor{bestresult}{($1.87\times$)} & NA & 8.7 \cellcolor{bestresult}{($1.00\times$)} \\
\midrule
D5 & \textbf{\rs} & 0:28:56 & 51,975 & 6.0 \\
\emph{Human} & \textbf{Minimap2} & 0:01:52 \cellcolor{worseresult}{($0.06\times$)} & 1,372 \cellcolor{worseresult}{($0.03\times$)} & 20.21 \cellcolor{bestresult}{($3.37\times$)} \\
& \textbf{Minimap2 + Dorado CPU (Fast)} & 6:42:24 \cellcolor{bestresult}{($13.91\times$)} & 802,983 \cellcolor{bestresult}{($15.45\times$)} & 81.98 \cellcolor{bestresult}{($13.66\times$)} \\
& \textbf{Minimap2 + Dorado CPU (HAC)} & 23:27:18 \cellcolor{bestresult}{($48.64\times$)} & 1,219,043 \cellcolor{bestresult}{($23.45\times$)} & 46.12 \cellcolor{bestresult}{($7.69\times$)} \\
& \textbf{Minimap2 + Dorado GPU (HAC)} & 0:20:24 \cellcolor{worseresult}{($0.71\times$)} & NA & 20.31 \cellcolor{bestresult}{($3.38\times$)} \\
& \textbf{Minimap2 + Dorado GPU (SUP)} & 1:05:48 \cellcolor{bestresult}{($2.27\times$)} & NA & 20.21 \cellcolor{bestresult}{($3.37\times$)} \\\bottomrule
\end{tabular}

}
\label{rs:tab:resources}
\end{table}

We make three key observations.
First, \rs provides a substantial speedup and lower peak memory usage compared to minimap2 when combined with Dorado's fast model on a CPU by, on average, \rsavgelfcpu (elapsed time), \rsavgcpufcpu (CPU time), and \rsavgpeakfcpu (peak memory usage). When using Dorado's HAC model, \rs provides even better results on average by \rsavgelhcpu (elapsed time), \rsavgcpuhcpu (CPU time), and \rsavgpeakhcpu (peak memory usage), since running the HAC model is computationally more costly than running the fast model.

Second, \rs achieves an average: 1)~speedup of \rsavgelhgpu (HAC) and \rsavgelsgpu (SUP), and 2)~reduced peak memory usage of \rsavgpeakhgpu (HAC) and \rsavgpeaksgpu (SUP), compared to GPU-based basecalling followed by minimap2. Although basecalling with GPUs followed by overlapping is usually faster in most cases when using the HAC model \revc{(\textbf{Minimap2 + Dorado GPU (HAC)} in Table~\ref{rs:tab:resources})}, \rs still achieves better average speedup due to the substantial speedup it provides for the D1 dataset. \revc{Although comparing the results between CPUs and GPUs is not ideal due to the massive parallelism that GPUs provide compared to CPUs, \rs still provides comparable and even better performance even with the limited parallelism (i.e., 32 threads) it uses on CPUs compared to the parallelism that GPUs provide (e.g., a few tens of thousands of threads). This shows that \rs can provide even faster results when accelerated with GPUs (outside the scope of this work) based on its comparison when basecalling is done using CPUs and GPUs.}

Third, minimap2, without considering the resources that basecalling requires, is more resource-efficient than \rs as it takes on average \rsavgelmmt (elapsed time), \rsavgcpummt (CPU time), and \rsavgpeakmmt (peak memory usage) of those \rs takes. This is mainly expected as analyzing raw signals requires additional steps (e.g., signal-to-event conversion), more memory due to sequencing depth (as opposed to a reference genome), and incurs noise that requires less efficient parameter settings for \rs (e.g., the window parameter in minimizers) when analyzing raw signals compared to minimap2. Future work can focus on processing these raw signals more accurately to reduce the limits for adjusting parameters that can impact performance and memory usage.

We conclude that \rs can perform read overlapping with higher throughput that can be useful when focusing on real-time analysis and better computational resources than CPU-based basecalling followed by minimap2. We find that \rs's speed is mainly dependent on the amount of bases stored in the hash table, as the speed decreases with increasing the number of locations that need to be analyzed in the index per read. \revc{To resolve this scalability issue, future work should focus on designing indexing and filtering methods that provide a limitation on the volume of signals stored in the index and processed during the overlapping step.} These results can also be useful when basecalling raw signals using GPUs to reduce the computational overhead that GPU-based basecalling requires, which we discuss in Section~\ref{rs:sec:discussion}.

\subsection{All-vs-All Overlapping Statistics} \label{rs:subsec:accuracy}
\revd{Table~\ref{rs:tab:accuracy} shows the all-vs-all overlapping statistics between \rs and minimap2. We make two key observations.
First, on average, \rsavgshared of overlap pairs generated by \rs are shared with the overlap pairs that minimap2 generates. This shows that a large fraction of overlapping pairs does not require basecalling to generate identical overlapping information identified by minimap2 after basecalling.
Second, although \rs can find a substantial amount of overlap pairs identical to the overlaps minimap2 finds, specifically for the D1 dataset, there are still overlapping pairs unique to \rs (\rsavgru on average) and minimap2 (\rsavgmmu on average), mainly due to the decreased shared overlaps as the genome size increases. These differences are likely due to differences in 1)~certain parameters (e.g., chaining scores) and 2)~increased noise inherent in raw signals compared to basecalled sequences, which can become more pronounced in genomes with greater size, complexity, or repetitiveness. \rev{Although maximizing the shared overlaps can provide insights regarding the accuracy of overlapping information that \rs finds, it is not necessary to provide near-identical shared overlap statistics for certain use cases such as constructing \emph{de novo} assemblies~\cite{vaser_time-_2021}.} Instead, contiguous assemblies can still be constructed using a smaller but useful portion of shared overlapping pairs. We conclude that \rs provides a mechanism that shares a large portion of overlaps with minimap2, while the shared portions decrease as the genome size increases.}

\begin{table}[tbh]
\centering
\caption{All-vs-all Overlapping Statistics. \revc{Percentages show the overlapping pairs that are 1)~unique to \rs, 2)~unique to minimap2, and 3)~reported in both tools (Shared Overlaps).}}

\begin{tabular}{@{}clrrr@{}}\toprule
& \textbf{Organism}      & \textbf{Unique to}     & \textbf{Unique to} & \textbf{Shared} \\
&                        & \textbf{\rs ($\%$)}  & \textbf{Minimap2 ($\%$)} & \textbf{Overlaps ($\%$)} \\\toprule
D1 & \emph{SARS-CoV-2}   & 11.55 & 15.27 & 73.18 \\\midrule
D2 & \emph{E. coli}      & 8.33 & 50.62 & 41.05 \\\midrule
D3 & \emph{Yeast}        & 24.94 & 35.17 & 39.89 \\\midrule
D4 & \emph{Green Algae}  & 3.76 & 78.64 & 17.61 \\\midrule
D5 & \emph{Human}        & 32.69 & 56.18 & 11.13 \\\bottomrule
\end{tabular}

\label{rs:tab:accuracy}
\end{table}

\subsection{Contiguity of the Assembly Graphs} \label{rs:subsec:assembly}
\revd{To evaluate the impact of overlaps that \rs and minimap2 find, we construct \emph{de novo} assembly \rev{graphs} from these overlaps. Table~\ref{rs:tab:assembly} shows the contiguity statistics of these \emph{de novo} assembly \rev{graphs}. \rev{We note that the assembly graphs constructed using miniasm provide the connection information used to assemble unitigs, as well as the connections between unitigs that make up subgraph components in the assembly graph. Since we use the miniasm assembler without any modifications, we cannot generate the assembled sequences from raw signals as miniasm supports generating assembled sequences from basecalled sequencing reads. This mainly prevents us from evaluating the accuracy of these assemblies from raw signals, which we further discuss in Section~\ref{rs:sec:discussion}.} We do not provide the assembly statistics for the D1 dataset as miniasm cannot generate an assembly using the overlaps from \rs and minimap2, potentially due to the small size of the genome. We make three observations.}

\begin{table}[tbh]
\centering
\caption{Assembly Statistics.}
\resizebox{\columnwidth}{!}{
\begin{tabular}{@{}llrrrrrr@{}}
\toprule
\textbf{Dataset}   & \textbf{Tool} & \textbf{Total}        & \textbf{Largest}     & \textbf{N50}  & \textbf{auN} & \textbf{Longest}     & \textbf{Unitig} \\
                   &               & \textbf{Length (bp)}  & \textbf{Comp. (bp)}  & \textbf{(bp)} & \textbf{(bp)} & \textbf{Unitig (bp)} & \textbf{Count}    \\\toprule
D2                 & \rs           & 14,525,505            & 4,841,669            & 1,535,079     & 1,309,738    & 2,722,499            & 31                  \\
\emph{E. coli}     & minimap2      & 10,434,542            & 5,207,206            & 5,204,754     & 5,194,738    & 5,207,206            & 4 \\
                   & Gold standard & 5,235,343             & 5,235,343            & 5,235,343     & 5,235,343    & 5,235,343            & 1 \\\midrule
D3                 & \rs           & 13,898,208            & 362,050              & 41,118        & 48,106       & 161,883              & 396                 \\
\emph{Yeast}       & minimap2      & 23,755,455            & 1,611,876            & 134,050       & 150,908      & 464,054              & 282   \\
                   & Gold standard & 11,963,521            & 11,835,059           & 640,934       & 623,210      & 1,073,346            & 68 \\\midrule
D4                 & \rs           & 3,448,899             & 448,422              & 93,111        & 108,818      & 252,038              & 50                  \\
\emph{Green Algae} & minimap2      & 2,117,190             & 198,709              & 63,310        & 88,906       & 198,709              & 55      \\
                   & Gold standard & 106,479,288           & 2,255,807            & 452,774       & 538,136      & 1,667,975            & 420 \\\midrule
D5                 & \rs           & 1,850,419             & 493,004              & 51,300        & 116,049      & 364,113              & 48                  \\
\emph{Human}       & minimap2      & 747,607               & 65,951               & 19,476        & 22,103       & 48,424               & 61    \\
                   & Gold standard & 8,365,210             & 367,305              & 19,329        & 29,697       & 150,470              & 592 \\\bottomrule
\end{tabular}

}
\label{rs:tab:assembly}
\end{table}

\revd{First, we find that we can construct long unitigs from the raw signal overlaps that \rs finds. The unitigs we can construct from these raw signal overlaps are substantially longer than the average read length of their corresponding datasets (e.g., the longest unitig length in D2 is $413.25\times$ longer than the average read length of the D2 dataset). \textbf{\rs is the first work that enables \emph{de novo} assembly construction directly from raw signal overlaps without basecalling}, which has several implications and can enable future work, as we discuss in Section~\ref{rs:sec:discussion}.}

Second, we observe that the unitigs generated from \rs overlaps are usually less contiguous compared to those generated from minimap2 overlaps, based on all the metrics we show in Table~\ref{rs:tab:assembly} and Supplementary Figures~\ref{rs:suppfig:d2_assembly}–\ref{rs:suppfig:d5_assembly}. Compared to the gold standard assemblies generated using highly accurate basecalled reads and a state-of-the-art assembler, we find that \rs can still achieve a significant portion of the assembly contiguity, especially for the D2 dataset (\emph{E.coli}). For example, \rs constructs unitigs with the longest unitig length of 2.7 million bases, which is over half the length of the \emph{E.~coli} genome, whereas the gold standard assembly produces a single unitig covering the entire genome. \textbf{This indicates that \rs can achieve substantial contiguity relative to the gold standard without basecalling.}

Third, for the larger genome (i.e., D4 and D5), we find that \rs can construct longer unitigs compared to those generated by minimap2. The coverage of this dataset is substantially lower than that of other datasets. This suggests that \rs can find more useful overlaps leading to longer paths in the assembly for the cases when coverage is relatively lower (e.g., coverage of less than or around $5\times$ in Table~\ref{rs:tab:dataset}). Basecalled sequences may generate more contiguous assemblies when the coverage is higher.

We conclude that \rs generates useful overlapping information, enabling the construction of assemblies directly from raw signals. This approach achieves substantial contiguity and, in some cases (e.g., at low coverage), provides better contiguity compared to minimap2. \revd{We discuss the potential next steps to enable evaluating the accuracy of these assemblies in Section~\ref{rs:sec:discussion}.}

\section{Discussion and Future Work} \label{rs:sec:discussion}
\head{Limitations} Our evaluation demonstrates that \rs achieves high throughput when finding overlaps between raw signal pairs, making it a viable candidate for real-time analysis. However, there are still two main challenges to fully utilize real-time \emph{de novo} assembly construction during sequencing.
First, the hash-based index should be constructed and updated dynamically in real-time while sequencing is in progress. Although \rs provides the mechanisms for storing multiple hash tables that can be constructed for each chunk of raw signals generated in real-time, it is not computationally feasible to dynamically update the index for all sequenced signals as the memory and computational burden increases with each hash table generated. Such an approach requires a decision-making mechanism to dynamically stop updating the index after sequencing a certain amount of signals. The stopping mechanism should be accurate enough to ensure that the current state of the index provides sufficient information to find the overlaps between the already sequenced signals and the new signals generated after the index construction.
Second, the assembly graph should be constructed and updated dynamically while the new overlap information is generated in real time. This is challenging as the intermediate steps for generating the unitigs (e.g., the transitive reduction step~\cite{myers_fragment_2005}) in assembly graphs may not work optimally without the full overlap information, as it is likely to remove graph connections that can be useful with new overlap information. It might be feasible to use graph structures that are more suitable for streaming data generation, such as de Bruijn graphs, for assembly construction purposes~\cite{el-metwally_lightassembler_2016, rozov_faucet_2018, scott_streaming_2022}.

\rs can find overlaps only between reads coming from the same strand, as it lacks the capability to construct the reverse complemented version of the signals to identify the matches on the other strands. This potentially leads to gaps in the assembly and construction of the unitigs from both strands. Designing a mechanism that can reverse complement raw signals without basecalling is future work.

We evaluate \rs using raw signals generated from ONT's R9.4 flowcells to show that overlapping information between raw signals can be used to construct \emph{de novo} assemblies. \revd{We use a substantially low coverage for the human dataset to evaluate the benefits of raw signal analysis compared to basecalled analysis.} We leave optimizations for newer flow cell versions (e.g., R10.4.1) \revd{and larger scale analysis with higher depth of coverage for future work.}

\revd{\head{Challenges for constructing and evaluating \emph{de novo} assemblies} Although the main focus of our work is enabling all-vs-all overlapping between raw signals, we identify \emph{de novo} assembly construction as a natural use case that can utilize the all-vs-all overlap information from raw signals that \rs finds. To show that the overlaps from \rs can lead to unitigs that are much longer than average read lengths, we construct and evaluate the \emph{de novo} the assembly graphs from the \rs overlaps, as discussed in Section~\ref{rs:subsec:assembly}. However, we identify two key challenges that limit the scope of our evaluation of these \emph{de novo} assemblies, which we leave as future work.

First, we cannot generate the sequence of assembled signals from the assembly graphs (i.e., spelling the sequence from the assembly graph). This is because we use the miniasm assembler to construct these assembly graphs from the raw signal overlaps. Although miniasm can use the overlap position information to construct the assembly graph, it cannot output the sequence of assemblies as it is mainly designed for basecalled sequences. This is challenging because each raw signal contains metadata used in basecalling mechanisms (e.g., nanopore-specific information such as the baseline current levels). Carrying the metadata information effectively from many overlapping signals is essential to enable the further analysis of these assembled signals.

Second, to evaluate the accuracy of the constructed assemblies, these assemblies need to be aligned to a ground truth assembly. This can be done either by either 1)~basecalling the assembled signal or 2)~aligning the assembled signal using dynamic time warping (DTW)~\cite{lindegger_rawalign_2024} to the ground truth assembly. These approaches require constructing the assembled signal (i.e., the first challenge mentioned above) and designing a new class of basecallers to directly basecall assembled signals, which we discuss as future work below. Although the missing accuracy evaluations of these assemblies can be a major limitation of our work, we believe the assemblies that are constructed from the \rs overlaps are likely to be close to the assembly accuracy that can be achieved by using the minimap2 overlaps due to the non-negligible portion of overlapping pairs shared by these two mechanisms.}

\head{New Directions for Future Work}
\revd{We identify several new directions for future work. First, existing basecalling techniques are oblivious to the similarity information between raw signals. We believe the overlapping information that \rs finds can be useful to improve the accuracy and performance (i.e., speed) of the basecalling techniques. To improve the accuracy, the overlap information can be identified as a useful feature that can be used when training and using the underlying deep neural models that the basecalling techniques utilize. If useful, the overlap information can make the basecallers more accurate and less complex. To improve the performance, the basecallers can be designed to avoid basecalling the overlapped regions redundantly. Single basecalling of the overlapping regions (instead of basecalling them for all the reads) can reduce the amount of computations that the basecalling performs and improve performance.}
Second, \emph{de novo} assembly construction enables identifying the reads that do not provide useful information for assembly if they are fully contained by other overlapping read pairs~\cite{li_minimap_2016}. Identifying these potentially useless reads early on provides the opportunity to avoid basecalling them, which can improve the overall execution time of basecalling by filtering out such reads.
Third, \emph{de novo} assembly construction from raw signals provides new opportunities to design new basecallers that can basecall the assembled signals. Such a basecaller has the potential to substantially improve the performance, as it enables basecalling fewer long unitigs instead of many shorter reads.
Fourth, we believe the algorithms that are specifically designed to construct assemblies from raw signals should be designed to improve the overall quality of such assemblies and to enable constructing the sequence of assembled signals.

\section{Summary}
We introduce \rs, the \emph{first} mechanism that can find overlaps between \revc{two sets of} raw nanopore signals without translating them to bases.
We find that \rs can 1)~find overlaps while meeting the real-time requirements with an average throughput
of \rsavgthr signals/sec \revc{per CPU thread}, 2)~reduce the overall time needed for finding overlaps (on average by \rsavgelfcpu \revc{ and up to by \rsmaxelfcpu}) and peak memory usage (on average by \rsavgpeakfcpu \revc{ and up to by \rsmaxpeakfcpu}) compared to the time and memory needed to run \revc{state-of-the-art read mapper (minimap2)} combined with the Dorado basecaller running on a CPU with its fastest model, 3)~share a large portion of overlapping pairs with minimap2 (\rsavgshared on average), and 4)~construct long assemblies from these useful overlaps. \revc{We find that we can construct assemblies of half the length of the entire \emph{E. coli} genome (i.e., of length 2.7 million bases) directly from raw signal overlaps without basecalling.} Finding overlapping pairs from raw signals is critical for enabling new directions that have not been explored before for raw signal analysis, such as \emph{de novo} assembly construction from overlaps that we explore in this work. We discuss many other new directions that can be enabled by finding overlaps and constructing \emph{de novo} assemblies.

\revc{We hope and believe that \rs enables future work in at least two key directions. First, we aim to fully perform end-to-end genome analysis without basecalling. This can be achieved by further improving tools such as \rs to construct \emph{de novo} assemblies directly from raw signals such that these assemblies are consistently better than those generated from basecalled sequences in all cases. Second, we should rethink how we train and use modern neural network-based basecallers by integrating additional useful information (e.g., overlaps or assemblies of signals) generated by \rs into these basecallers.}

\clearpage
\setsuppbasednumbering
\section{Supplementary Materials}
\subsection{Visualizing Assemblies} \label{rs:suppsec:assembly_visual}

\begin{figure}[tbh]
\centering
\includegraphics[width=\columnwidth]{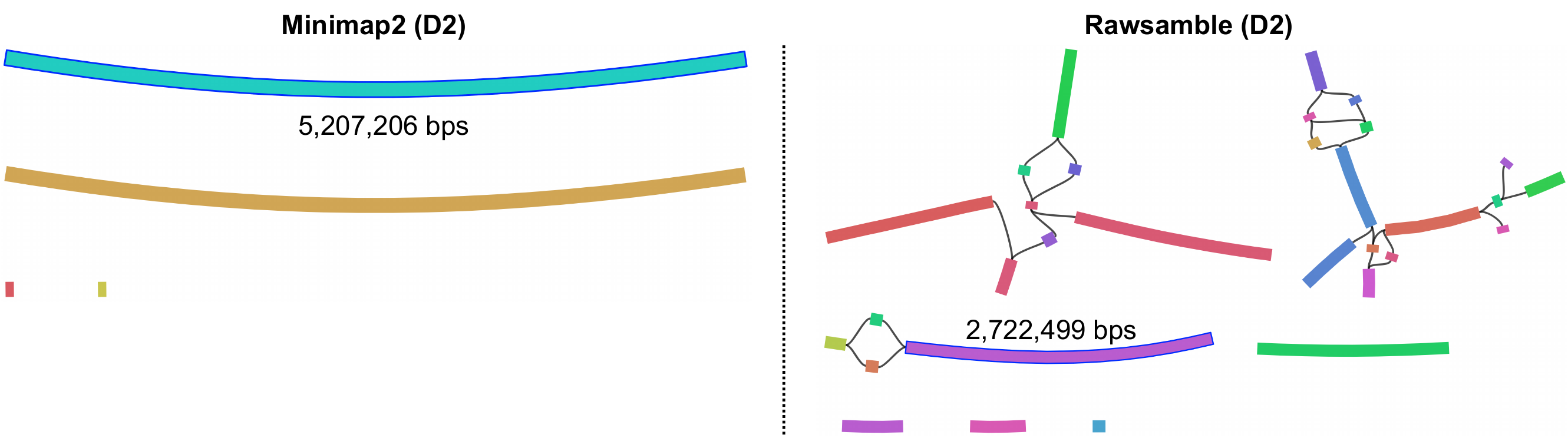}
\caption{Visualization of the assembly graphs generated using \rs and minimap2 overlaps from the D2 (\emph{E. coli}) dataset.}
\label{rs:suppfig:d2_assembly}
\end{figure}

\begin{figure}[tbh]
\centering
\includegraphics[width=\columnwidth]{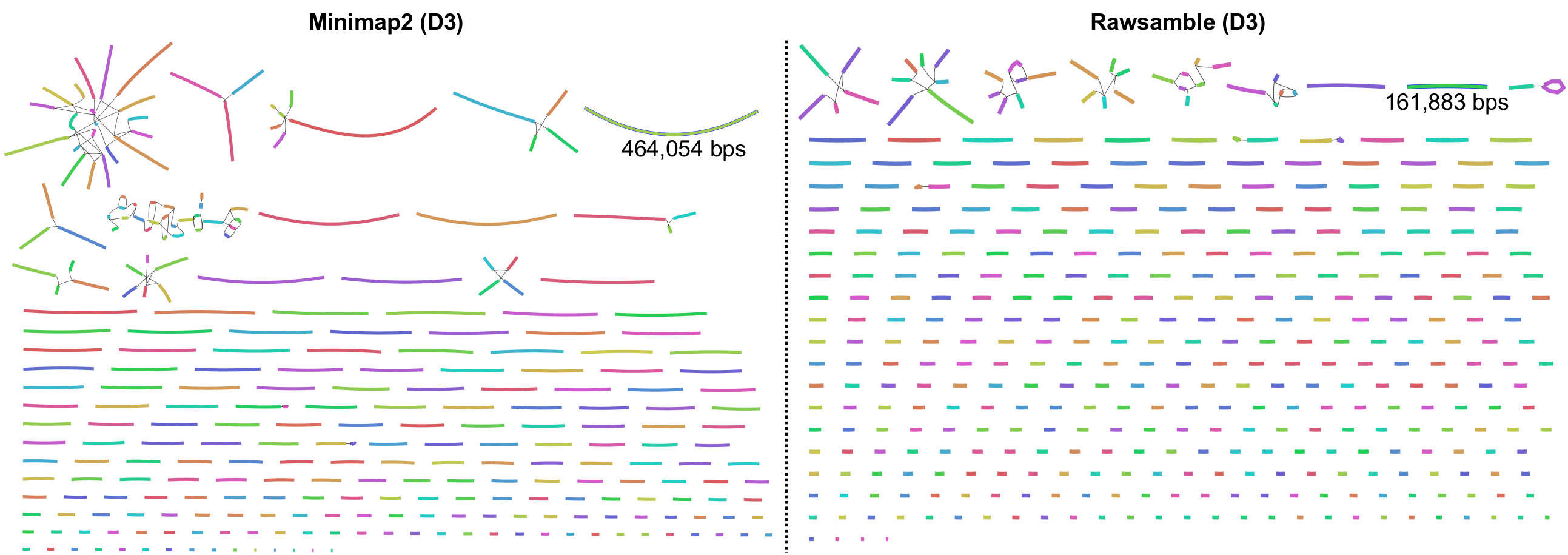}
\caption{Visualization of the assembly graphs generated using \rs and minimap2 overlaps from the D3 (\emph{Yeast}) dataset.}
\label{rs:suppfig:d3_assembly}
\end{figure}

\begin{figure}[tbh]
\centering
\includegraphics[width=\columnwidth]{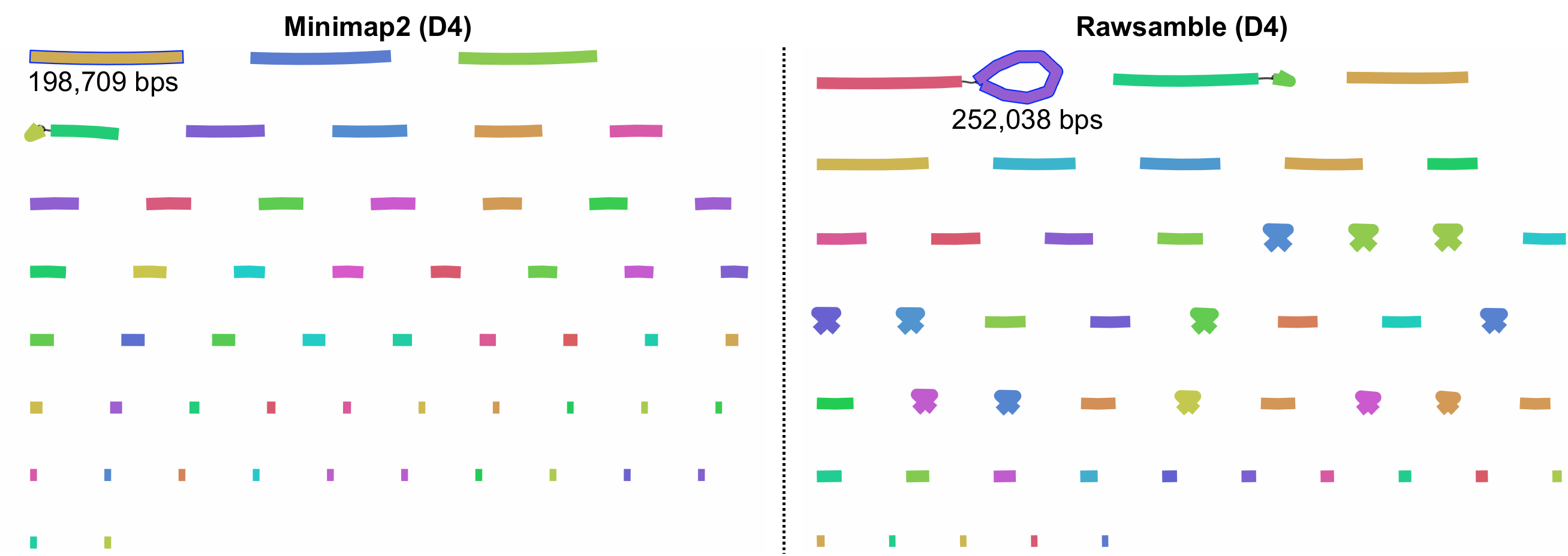}
\caption{Visualization of the assembly graphs generated using \rs and minimap2 overlaps from the D4 (\emph{Green Algae}) dataset.}
\label{rs:suppfig:d4_assembly}
\end{figure}

\begin{figure}[tbh]
\centering
\includegraphics[width=\columnwidth]{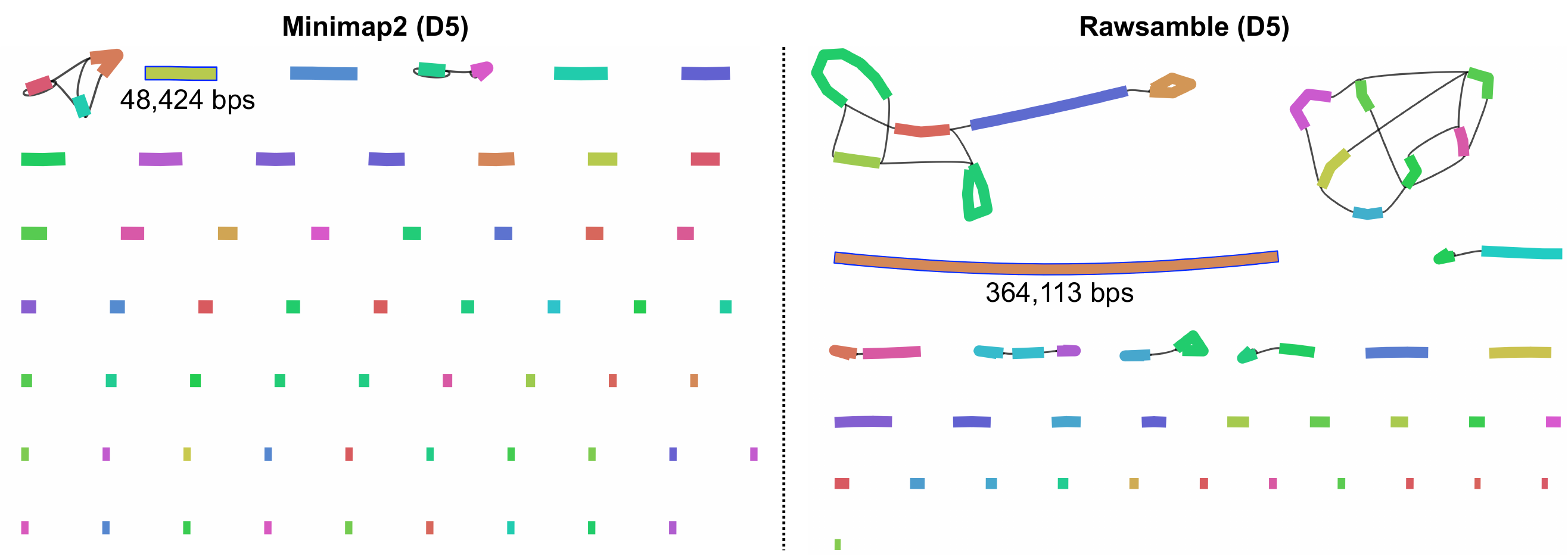}
\caption{Visualization of the assembly graphs generated using \rs and minimap2 overlaps from the D5 (\emph{Human}) dataset.}
\label{rs:suppfig:d5_assembly}
\end{figure}

Supplementary Figures~\ref{rs:suppfig:d2_assembly},~\ref{rs:suppfig:d3_assembly},~\ref{rs:suppfig:d4_assembly}, and~\ref{rs:suppfig:d5_assembly} show the assembly graphs created with miniasm in GFA formats. To construct these assembly graphs with miniasm, we provide the overlap information that \rs and minimap2 provide as a PAF file~\cite{li_minimap_2016}. We use Bandage~\cite{wick_bandage_2015} to visualize these assembly graphs. To provide the scale between unitigs generated using the same dataset, we annotate the longest contig in each assembly graph with their lengths.

\clearpage
\section{Generating a Gold Standard Assembly}
\label{rs:suppsec:gold_standard}
To generate a gold standard assembly using R9.4 Simplex reads, we explore two approaches.

First, we use the raw basecalled sequences directly with a set of state-of-the-art assemblers identified from recent benchmarks~\cite{Sun2021, Wick2021, Cosma2022, Yu2024}, namely Hifiasm~\cite{cheng_haplotype-resolved_2021}, Verkko~\cite{rautiainen_telomere--telomere_2023}, LJA~\cite{bankevich_multiplex_2022}, HiCanu~\cite{nurk_hicanu_2020}, and Flye~\cite{kolmogorov_assembly_2019}.

Second, we apply error correction to the reads using the HERRO tool~\cite{stanojevic_telomere--telomere_2024} developed by Oxford Nanopore Technologies (ONT), which is also integrated into the latest versions of Dorado as \textit{dorado correct}. HERRO corrects erroneous R9.4 and R10.4 data, enabling their use as a replacement for accurate PacBio HiFi reads required by state-of-the-art hybrid assembly approaches. Supplementary Table~\ref{rs:tab:coverage} shows the estimated sequencing depth of coverage before and after the HERRO correction for each dataset.

For the high-coverage D2 \emph{E.coli} dataset, Hifiasm outputs a fragmented assembly, while Verkko requires extensive parameter tuning (specifically in terms of coverage, \texttt{---unitig-abundance}, and \texttt{---base-k}) to achieve desirable contiguity and completeness. LJA produces an almost perfect assembly when a minimum length filter of 30kbp is applied to the corrected reads, as suggested by the HERRO paper. Flye performs comparably to LJA without the need for error correction.

For the other datasets, with or without error correction, most assemblers either fail to produce assemblies or generate suboptimal results, except for Flye. We attribute this to their sensitivity to inaccuracies in the uncorrected reads and the reduced coverage after correction. Supplementary Table~\ref{rs:tab:coverage} shows how the coverage levels change after correction for each dataset. Notably, Flye works well (and sometimes even slightly better) with uncorrected reads, highlighting its robustness to noisy ONT reads.

Based on these observations, we select Flye as the assembler to generate the gold standard assemblies in our evaluations, given its ability to handle noisy ONT reads without the need for error correction and its consistent performance across different datasets. Therefore, we use Flye to construct the gold standard assemblies from the basecalled reads in our study.

\begin{table}[tbh] 
\centering
\caption{Coverage levels before and after error correction for each dataset.}
\begin{tabular}{@{}lcc@{}}\toprule
\textbf{Dataset} & \textbf{Coverage Before Correction} & \textbf{Coverage After Correction} \\\midrule
D2 \emph{E.~coli} & 445$\times$ & 240$\times$ \\
D3 \emph{Yeast} & 32$\times$ & 12$\times$ \\
D4 \emph{Green algae} & 5.6$\times$ & 3.7$\times$ \\
D5 \emph{Human} & 0.6$\times$ & 0.002$\times$ \\\bottomrule
\end{tabular}

\label{rs:tab:coverage}
\end{table}

\clearpage
\subsection{Configuration} \label{rs:sec:configuration}
\subsubsection{Parameters} \label{rs:subsec:parameters}

In Supplementary Table~\ref{rs:tab:parameters}, we show the parameters of each tool for each dataset. In Supplementary Table~\ref{rs:tab:presets}, we show the details of the preset values that \rs sets in Supplementary Table~\ref{rs:tab:parameters}. For minimap2~\cite{li_minimap2_2018}, we use the same parameter setting for all datasets. For miniasm~\cite{li_minimap_2016}, we use the default parameter settings for all datasets.

\begin{table}[tbh]
\centering
\caption{Parameters we use in our evaluation for each tool and dataset in mapping.}
\resizebox{\linewidth}{!}{
\begin{tabular}{@{}lccccc@{}}\toprule
\textbf{Tool} & \textbf{D1 \emph{SARS-CoV-2}} & \textbf{D2 \emph{E. coli}} & \textbf{D3 \emph{Yeast}} & \textbf{D4 \emph{Green Algae}} & \textbf{D5 \emph{Human}}\\\midrule
\rs    & -x ava-viral -t 32 & -x ava -t 32 & -x ava -t 32 & -x ava -t 32 & -x ava --chain-gap-scale 0.6 -t 32\\\midrule
Minimap2          & \multicolumn{5}{c}{-x ava-ont --for-only -t 32}\\\midrule
Dorado CPU (Fast) & \multicolumn{5}{c}{basecaller -x cpu dna\_r9.4.1\_e8\_fast@v3.4}\\\midrule
Dorado CPU (HAC) & \multicolumn{5}{c}{basecaller -x cpu dna\_r9.4.1\_e8\_hac@v3.3}\\\midrule
Dorado GPU (HAC) & \multicolumn{5}{c}{basecaller dna\_r9.4.1\_e8\_hac@v3.3}\\\midrule
Dorado GPU (SUP) & \multicolumn{5}{c}{basecaller dna\_r9.4.1\_e8\_sup@v3.3}\\\midrule
Miniasm         & \multicolumn{5}{c}{}\\\bottomrule
\end{tabular}

}
\label{rs:tab:parameters}
\end{table}

\begin{table}[tbh]
\centering
\caption{Corresponding parameters of presets (-x) in \rs.}
\resizebox{\linewidth}{!}{
\begin{tabular}{@{}lcc@{}}\toprule
\textbf{Preset} & \textbf{Corresponding parameters} & Usage \\\midrule
ava-viral  & -e 6 -q 4 -w 0 --sig-diff 0.45 --fine-range 0.4 --min-score 20 --min-score2 30 --min-anchors 5 & Viral genomes\\
& --min-mapq 5 --bw 1000 --max-target-gap 2500 --max-query-gap 2500 --chain-gap-scale 1.2 --chain-skip-scale 0.3 &\\\midrule
ava  & -e 8 -q 4 -w 3 --sig-diff 0.45 --fine-range 0.4 --min-score 40 --min-score2 75 & Default case\\
& --min-anchors 5 --min-mapq 5 --bw 5000 --max-target-gap 2500 --max-query-gap 2500 & \\\bottomrule
\end{tabular}

}
\label{rs:tab:presets}
\end{table}

\subsubsection{Versions}\label{rs:subsec:versions}

Supplementary Table~\ref{rs:tab:versions} shows the version and the link to these corresponding versions of each tool.

\begin{table}[tbh]
\centering
\caption{Versions of each tool and library.}
\resizebox{0.9\linewidth}{!}{
\begin{tabular}{@{}lll@{}}\toprule
\textbf{Tool} & \textbf{Version} & \textbf{Link to the Source Code} \\\midrule
\rs & 2.1 & \url{https://github.com/CMU-SAFARI/RawHash/releases/tag/v2.1}\\\midrule
Minimap2 & 2.24 & \url{https://github.com/lh3/minimap2/releases/tag/v2.24}\\\midrule
Dorado & 0.7.3 & \url{https://github.com/nanoporetech/dorado/releases/tag/v0.7.3}\\\midrule
Miniasm & 0.3-r179 & \url{https://github.com/lh3/miniasm/releases/tag/v0.3}\\\midrule
Rawasm & main & \url{https://github.com/CMU-SAFARI/rawasm}\\\midrule
Flye & 2.9.5 & \url{https://github.com/mikolmogorov/Flye/releases/tag/2.9.5}\\\midrule
HERRO & 0.1 & \url{https://github.com/lbcb-sci/herro}\\\bottomrule
\end{tabular}

}
\label{rs:tab:versions}
\end{table}

\setchapterbasednumbering

\chapter{Conclusions and Future Directions}
\label{chap:conc}

The goal of this dissertation is twofold: 1)~understand how various sources of noise in sequencing data and its analysis impact the performance, accuracy, \revb{efficiency} and scalability of genome analysis; and 2)~develop new techniques to tolerate or reduce noise for faster, more accurate, and real-time analysis of genomic sequencing data. To this end, we propose a series of novel algorithms and techniques that improve the efficiency and effectiveness of key steps in the genome analysis pipeline.

First, we introduce \blend \revb{ (Chapter~\ref{chap:blend})}, a novel noise-tolerant hashing mechanism for fuzzy seed matching, enabling quick identification of highly similar genomic sequences. Our evaluations show that \blend significantly improves the performance and memory usage of read overlapping and mapping, providing substantial benefits for assembling and analyzing high-coverage genomic data.

Second, we develop \rh \revb{ (Chapter~\ref{chap:rh})}, the first mechanism that performs accurate and scalable real-time analysis of raw nanopore signals for large genomes. \rh uses a unique quantization technique that effectively reduces noise in raw signals, enabling \emph{the first} hash-based search mechanism for raw nanopore signals to bypass the computationally intensive basecalling step. Our evaluations demonstrate that \rh matches the throughput of nanopore sequencers while providing high accuracy and scalability for large genomic datasets.

Third, we propose \rht \revb{ (Chapter~\ref{chap:rht})}, an improved version of \rh that provides better noise reduction in raw nanopore signals and increases the robustness of mapping decisions. \rht incorporates more sensitive quantization and chaining algorithms, weighted mapping decisions, and frequency filters, which collectively improve both the accuracy and throughput of real-time mapping of raw nanopore signals.

Fourth, we introduce \rs \revb{ (Chapter~\ref{chap:rs})}, a novel mechanism that enables all-vs-all overlapping of raw nanopore signals, facilitating the construction of \emph{de novo} assemblies directly from raw signals without basecalling. \rs achieves significant speedups and reductions in memory usage compared to traditional basecalling and overlapping pipelines, paving the way for new applications in raw signal analysis.

Finally, based on the insights developed throughout this work, we emphasize the importance of integrating noise-tolerant techniques and real-time analysis mechanisms in future genome analysis pipelines. We argue that these advancements can 1)~improve the efficiency and accuracy of current methodologies and 2)~enable new applications and directions in the field of genomics.

\revc{
\section{Future Research Directions}

This dissertation opens up several promising avenues for future research. In this section, we review potential directions for extending and enhancing the work presented in this dissertation.

\revd{
\subsection{Improving the Segmentation Step in Raw Signal Analysis}
The segmentation algorithm used in our works, Welch's t-test~\cite{ruxton_unequal_2006}, can identify abrupt changes in raw signals to use them as segmentation points. These abrupt changes usually point to the sequencing of a new k-mer inside a nanopore. However, it is known that this particular segmentation algorithm is mainly prone to two main types of errors. First, \emph{stay errors} lead to incorrectly splitting a signal region corresponding to one k-mer into multiple segments, known as oversegmentation~\cite{shivakumar_sigmoni_2024}. These stay errors can happen mainly due to transient false fluctuations in the signal due to noise~\cite{schreiber_analysis_2015}. Detecting such stay errors are usually performed by applying filters similar to what we propose in \rs where consecutive events with very similar signal values are merged into one. Second, \emph{skip errors} lead to incorrectly merging signal regions corresponding to the sequencing of multiple k-mers into a single signal region, known as undersegmentation~\cite{shivakumar_sigmoni_2024}. Skip errors can happen mainly if the nucleic acid molecule moves quickly inside a nanopore, which may lead to generating insufficient raw signal samples for a particular nucleotide, which is another type of noise in nanopore raw signals. Due to generating fewer samples for a particular nucleotide, such skip errors are harder to detect.

Although simple segmentation algorithms such as the t-test~\cite{ruxton_unequal_2006} are computationally cheap to use, they are fundamentally incapable of detecting complex errors in raw signals. Our non-published internal observations show that the basic segmentation algorithms are a major barrier to further improving the overall accuracy of the raw signal analysis mechanism. For example, although \rht provides substantial accuracy improvements for the human genome compared to prior mechanisms, there is still a large room for achieving the accuracy that the basescalling techniques provide. This shows that the basecalling techniques are capable of identifying complex noise in raw signals when using complex deep neural models.

We suggest future work should focus on designing lightweight machine learning (ML) models to perform segmentation to replace the existing t-test techniques. It is critical that these segmentation algorithms are lightweight so that they do not consume the majority of the execution time in raw signal analysis.

\subsection{Designing Robust Algorithms Applicable to Future Sequencing Technologies and Chemistries}

The commonly used segmentation algorithm, Welch's t-test~\cite{ruxton_unequal_2006}, relies on nanopore chemistry-dependent parameters. These parameters define the thresholds for identifying segmentation points depending on the expected subtle changes in the signal as well as the window length that defines the expected range of signals corresponding to the sequencing of a single k-mer. These parameters change depending on the nanopore chemistry and the sequenced molecule (i.e., DNA or RNA). This requirement to adjust parameters for different nanopore chemistries makes the segmentation algorithm less robust for future sequencing technologies.

A promising future direction is to develop techniques that combine the low computational footprint of simple methods (e.g., t-test segmentation) with the adaptability of neural networks. For example, small convolutional or recurrent neural networks can efficiently learn generic signal features and handle a wide range of nanopore chemistries~\cite{zhang_nanopore_2020, cheng_raw_2024}, potentially preserving accuracy without extensive retraining. By training such models on diverse datasets representing multiple chemistries, it may be possible to build robust approaches that consistently segment signals across differing conditions. We encourage further exploration of such hybrid or lightweight neural designs to ensure that raw signal processing pipelines remain effective for emerging and future sequencing technologies.}

\subsection{Integration of Overlapping and Assembly Information into Basecalling}

Integrating overlapping and assembly information with basecalling is a potential direction for future work. By combining the overlapping and assembly information that \rs generates with existing basecalling methods, it is possible to provide additional features to the machine learning models used in basecallers. This integration can improve the accuracy and performance of basecalling, especially in noisy datasets, by leveraging the contextual information provided by overlapping and assembly of raw signals. Enhancing basecallers with this additional context may lead to more accurate sequence interpretations and better overall genomic analyses.

\subsection{Real-Time \emph{De Novo} Assembly Construction}

Building upon the capabilities of \rs, future work can explore the \revb{\emph{real-time}} construction of \emph{de novo} assemblies during sequencing. A real-time \emph{de novo} assembly construction involves dynamically updating 1)~the hash-based index and 2)~assembly graph as new raw signals are sequenced. Enabling immediate identification and assembly of genomic sequences without waiting for the completion of sequencing can 1)~reduce the overall time and computational resources required for assembly and 2)~enable new directions for adaptive sampling. However, implementing real-time assembly poses challenges, such as managing data structures efficiently in real-time and handling the continuous stream of incoming data. Addressing these challenges can significantly accelerate genomic assembly processes and benefit time-sensitive applications.

\subsection{Performing Further Steps Fully Using Raw Nanopore Signals Without Basecalling}

Developing new algorithms and techniques for raw signal analysis is crucial to include many downstream steps of genome analysis without relying on basecalling. Developing techniques to directly process and interpret raw signals for applications such as variant calling, metagenomics, and functional genomics can reduce computational overhead and improve the speed of these analyses. By bypassing the basecalling step, it is possible to leverage the rich information contained in raw signals, potentially increasing accuracy and sensitivity in detecting genomic features. This approach is particularly beneficial in time-sensitive applications where rapid results are crucial.

\subsection{Exploring Hardware Acceleration for Raw Signal Analysis}

Given the large data volumes and massive parallelism in genome analysis, exploring hardware acceleration for raw signal analysis is an important direction. Implementing the algorithms used in raw signal analysis on suitable hardware, such as GPUs~\cite{\citehwrawgpu}, FPGAs\rev{~\cite{\citehwrawcust}}, or processing-in-memory (PIM) architectures\rev{~\cite{oliveira_damov_2021,seshadri_ambit_2017,hajinazar_simdram_2021,ferreira_pluto_2021}}\mytodo{Cite more}, could significantly improve the performance of analysis while reducing the overall energy consumption. By leveraging specialized hardware, it is possible to achieve higher throughput and lower latency, which are critical for processing large volumes of genomic data efficiently, especially for portable sequencing devices where computational resources are limited.

\subsection{Integrating \blend in \rh}

Integrating the \blend mechanism in \rh is a potential avenue for improving the sensitivity and accuracy of raw nanopore signal analysis. By incorporating \blend's fuzzy seed matching capabilities, \rh can improve its ability to identify similar regions in noisy datasets. This integration may lead to better handling of the variability and noise inherent in raw signals. Additionally, it can reduce the computational cost associated with processing noisy data, making the analysis more efficient.
}
\section{Concluding Remarks}

In this dissertation, we address critical challenges in genome analysis by developing new techniques to tolerate and reduce noise in sequencing data. We build a detailed understanding of how noise affects the computational mechanisms in genome analysis and propose solutions that improve the speed, accuracy, and scalability of key steps in the genome analysis pipeline. We introduce four novel mechanisms—\blend, \rh, \rht, and \rs—that collectively enable faster, more accurate, and scalable genome analysis. We hope that the methods and insights presented in this dissertation will inspire future research and lead to the development of more efficient and innovative approaches to genome analysis.

\appendix
\cleardoublepage%
\chapter{Other Works of the Author}

During my Ph.D. studies, I was involved in various other projects across different fields, focusing mainly on genome analysis acceleration, metagenomics, basecalling, and genome assembly.

As a first and co-first author, I contributed to several fields in genome analysis and its acceleration. I was a co-first author of \emph{AirLift}~\cite{kim_airlift_2024}, a fast and comprehensive technique that significantly reduces the computational overhead of remapping genomic data between reference genomes. I also co-led the paper entitled \emph{Accelerating Genome Analysis via Algorithm-Architecture Co-Design}~\cite{mutlu_accelerating_2023}, which reviews recent advancements in genome analysis acceleration and highlights the integration of multiple steps using suitable architectures to improve performance and efficiency. In genome assembly and polishing, I was the first author of \emph{Apollo}~\cite{firtina_apollo_2020}, an assembly polishing algorithm that is both sequencing-technology-independent and scalable to large genomes. Apollo was designed to improve the accuracy of genome assembly polishing by utilizing reads from all available sequencing technologies within a single run. I was also the first author of \emph{ApHMM}~\cite{firtina_aphmm_2024}, a flexible acceleration framework that reduces computational and energy overheads associated with profile hidden Markov models (pHMMs) used in bioinformatics applications, including assembly polishing we developed in Apollo.

As a co-author, I contributed to several papers in acceleration of genome analysis with hardware and software co-design.
I was involved in \emph{GenASM}~\cite{senol_cali_genasm_2020}, a framework that accelerates approximate string matching, a key step in genome sequence analysis, providing significant performance and power benefits. I contributed to \emph{GenStore}~\cite{mansouri_ghiasi_genstore_2022}, an in-storage processing system that improves read mapping performance by reducing data movement and computational overheads, and \emph{SeGraM}~\cite{senol_cali_segram_2022}, a universal hardware accelerator designed for both sequence-to-graph and sequence-to-sequence mapping, addressing critical bottlenecks in genomic mapping pipelines. In the metagenomics field, I contributed to \emph{MegIS}~\cite{mansouri_ghiasi_megis_2024}, an in-storage processing system designed to reduce data movement overhead in metagenomic analysis. I was involved in \emph{Demeter}~\cite{shahroodi_demeter_2022}, a fast and energy-efficient food profiler using hyperdimensional computing in memory, which significantly improves throughput and reduces memory usage while maintaining accuracy.

My work in basecalling and nanopore sequencing includes contributions to 1)~\emph{RUBICON}~\cite{singh_rubicon_2024}, a framework for designing efficient deep learning-based basecallers, 2)~\emph{TargetCall}~\cite{cavlak_targetcall_2024}, a pre-basecalling filter that eliminates wasted computation by discarding off-target reads before the basecalling process, and 3)~\emph{RawAlign}~\cite{lindegger_rawalign_2024}, a tool designed for accurate, fast, and scalable raw nanopore signal mapping by combining seeding and alignment. I also contributed to \emph{Swordfish}~\cite{shahroodi_swordfish_2023}, a framework for evaluating deep neural network-based basecalling using computation-in-memory with non-ideal memristors, which explores hardware/software co-design solutions to address accuracy loss due to hardware limitations. We also proposed \emph{GenPIP}~\cite{mao_genpip_2022}, an in-memory genome analysis accelerator that tightly integrates basecalling and read mapping to reduce inefficient computation and data movement.

I worked on \emph{FastRemap}~\cite{kim_fastremap_2022}, a tool for quickly remapping reads between genome assemblies, which significantly improves the efficiency of genome assembly processes. Moreover, I contributed to \emph{Molecules to Genomic Variations}~\cite{alser_molecules_2022}, a comprehensive overview of intelligent algorithms and architectures designed to accelerate genome analysis at the population level.
In the area of computational biology tools and infrastructure, I worked on a systematic review titled \emph{Packaging and Containerization of Computational Methods}~\cite{alser_packaging_2022}, which describes and compares software packaging and containerization platforms used in omics research, highlighting the need for easier installation and greater usability of biomedical research tools.

I also worked on the acceleration of virtual-to-physical address mapping with \emph{Utopia}~\cite{kanellopoulos_utopia_2023}, which proposes a hybrid virtual-to-physical address mapping scheme that enhances address translation performance.

I am currently involved in several works that are available as preprint. I am involved in \emph{SequenceLab}~\cite{rumpf_sequencelab_2023}, a comprehensive benchmark of computational methods for comparing genomic sequences, which provides an in-depth analysis of the heterogeneity among widely-used methods, and \emph{MetaFast}~\cite{gollwitzer_metafast_2023}, which enables fast metagenomic classification through seed counting and edit distance approximation to improve accuracy-runtime tradeoffs in metagenomic analysis.

\chapter{Complete List of the Author's Contributions}
\label{appendix:complete_list}
This section lists the author's contributions to the literature in reverse chronological order under four categories:
1)~first author (Section~\ref{appendix:sec:first_author_contributions}), 
2)~co-first authored (Section~\ref{appendix:sec:co_first_authored}), 
3)~co-supervised contributions (Section~\ref{appendix:sec:co_supervised_contributions}), and
4)~other contributions (Section~\ref{appendix:sec:other_contributions}).

\section{Major Contributions Led by the Author}
\label{appendix:sec:first_author_contributions}

\begin{enumerate}
    \item Can Firtina, Maximilian Mordig, Harun Mustafa, Sayan Goswami, Nika Mansouri Ghiasi, Stefano Mercogliano, Furkan Eris, Joël Lindegger, Andre Kahles, and Onur Mutlu, \textbf{``Rawsamble: Overlapping and Assembling Raw Nanopore Signals using a Hash-based Seeding Mechanism,''} \emph{arXiv} [Under Submission], 2024. Source code available: \url{https://github.com/CMU-SAFARI/RawHash}

    \item Can Firtina, Melina Soysal, Joël Lindegger, and Onur Mutlu, \textbf{``RawHash2: Mapping Raw Nanopore Signals Using Hash-Based Seeding and Adaptive Quantization,''} \emph{Bioinformatics}, 2024. Source code available: \url{http://github.com/CMU-SAFARI/RawHash}

    \item Can Firtina, Kamlesh Pillai, Gurpreet S. Kalsi, Bharathwaj Suresh, Damla Senol Cali, Jeremie S. Kim, Taha Shahroodi, Meryem Banu Cavlak, Joël Lindegger, Mohammed Alser, Juan Gómez Luna, Sreenivas Subramoney, and Onur Mutlu, \textbf{``ApHMM: Accelerating Profile Hidden Markov Models for Fast and Energy-efficient Genome Analysis,''} \emph{ACM TACO}, 2024. Source code available: \url{http://github.com/CMU-SAFARI/ApHMM-GPU}

    \item Can Firtina, Nika Mansouri Ghiasi, Joël Lindegger, Gagandeep Singh, Meryem Banu Cavlak, Haiyu Mao, and Onur Mutlu, \textbf{``RawHash: enabling fast and accurate real-time analysis of raw nanopore signals for large genomes,''} in \emph{Proceedings of the 31st Annual Conference on Intelligent Systems for Molecular Biology and the 22nd European Conference on Computational Biology (ISMB/ECCB)}, Lyon, France, 2023. Source code available: \url{https://github.com/CMU-SAFARI/RawHash}

    \item Can Firtina, Jisung Park, Mohammed Alser, Jeremie S. Kim, Damla Senol Cali, Taha Shahroodi, Nika Mansouri Ghiasi, Gagandeep Singh, Konstantinos Kanellopoulos, Can Alkan, and Onur Mutlu, \textbf{``BLEND: a fast, memory-efficient and accurate mechanism to find fuzzy seed matches in genome analysis,''} \emph{NAR Genomics and Bioinformatics}, 2023. Source code available: \url{https://github.com/CMU-SAFARI/BLEND}

    \item Can Firtina, Jeremie S. Kim, Mohammed Alser, Damla Senol Cali, A. Ercument Cicek, Can Alkan, and Onur Mutlu, \textbf{``Apollo: a sequencing-technology-independent, scalable and accurate assembly polishing algorithm,''} \emph{Bioinformatics}, 2020. Source code available: \url{https://github.com/CMU-SAFARI/Apollo}
\end{enumerate}

\section{Major Contributions Co-Led by the Author}
\label{appendix:sec:co_first_authored}

\begin{enumerate}
    \item Jeremie S. Kim, Can Firtina, Meryem Banu Cavlak, Damla Senol Cali, Nastaran Hajinazar, Mohammed Alser, Can Alkan, and Onur Mutlu, \textbf{``AirLift: A Fast and Comprehensive Technique for Remapping Alignments between Reference Genomes,''} \emph{IEEE/ACM TCBB}, 2024. Source code available: \url{https://github.com/CMU-SAFARI/AirLift}

    \item Onur Mutlu and Can Firtina, \textbf{``Accelerating Genome Analysis via Algorithm-Architecture Co-Design,''} in \emph{DAC}, 2023.
\end{enumerate}

\section{Co-supervised Contributions}
\label{appendix:sec:co_supervised_contributions}

\begin{enumerate}
    \item Meryem Banu Cavlak, Gagandeep Singh, Mohammed Alser, Can Firtina, Joel Lindegger, Mohammad Sadrosadati, Nika Mansouri Ghiasi, Can Alkan, and Onur Mutlu, \textbf{``TargetCall: Eliminating the Wasted Computation in Basecalling via Pre-Basecalling Filtering,''} \emph{Frontiers in Genetics}, 2024. Source code available: \url{https://github.com/CMU-SAFARI/TargetCall}
    
    \item Joël Lindegger, Can Firtina, Nika Mansouri Ghiasi, Mohammad Sadrosadati, Mohammed Alser, and Onur Mutlu, \textbf{``RawAlign: Accurate, Fast, and Scalable Raw Nanopore Signal Mapping via Combining Seeding and Alignment,''} \emph{IEEE Access}, 2024. Source code available: \url{https://github.com/CMU-SAFARI/RawAlign}
\end{enumerate}

\section{Other Contributions}
\label{appendix:sec:other_contributions}

\begin{enumerate}

    \item Nika Mansouri Ghiasi, Mohammad Sadrosadati, Harun Mustafa, Arvid Gollwitzer, Can Firtina, Julien Eudine, Haiyu Mao, Joël Lindegger, Meryem Banu Cavlak, Mohammed Alser, Jisung Park, and Onur Mutlu, \textbf{``MegIS: High-Performance, Energy-Efficient, and Low-Cost Metagenomic Analysis with In-Storage Processing,''} in \emph{ISCA}, 2024.

    \item Mohammed Alser, Brendan Lawlor, Richard J. Abdill, Sharon Waymost, Ram Ayyala, Neha Rajkumar, Nathan LaPierre, Jaqueline Brito, André M. Ribeiro-dos-Santos, Nour Almadhoun, Varuni Sarwal, Can Firtina, Tomasz Osinski, Eleazar Eskin, Qiyang Hu, Derek Strong, Byoung-Do (B.D) Kim, Malak S. Abedalthagafi, Onur Mutlu, and Serghei Mangul, \textbf{``Packaging and containerization of computational methods,''} \emph{Nature Protocols}, 2024.

    \item Gagandeep Singh, Mohammed Alser, Kristof Denolf, Can Firtina, Alireza Khodamoradi, Meryem Banu Cavlak, Henk Corporaal, and Onur Mutlu, \textbf{``RUBICON: a framework for designing efficient deep learning-based genomic basecallers,''} \emph{Genome Biology}, 2024. Source code available: \url{https://github.com/CMU-SAFARI/Rubicon}

    \item Arvid E. Gollwitzer, Mohammed Alser, Joel Bergtholdt, Joel Lindegger, Maximilian-David Rumpf, Can Firtina, Serghei Mangul, and Onur Mutlu, \textbf{``MetaFast: Enabling Fast Metagenomic Classification via Seed Counting and Edit Distance Approximation,''} \emph{arXiv} [Under Submission], 2023. Source code available: \url{https://github.com/CMU-SAFARI/MetaTrinity}

    \item Maximilian-David Rumpf, Mohammed Alser, Arvid E. Gollwitzer, Joel Lindegger, Nour Almadhoun, Can Firtina, Serghei Mangul, and Onur Mutlu, \textbf{``SequenceLab: A Comprehensive Benchmark of Computational Methods for Comparing Genomic Sequences,''} \emph{arXiv} [Under Submission], 2023. Source code available: \url{https://github.com/CMU-SAFARI/SequenceLab}

    \item Taha Shahroodi, Gagandeep Singh, Mahdi Zahedi, Haiyu Mao, Joel Lindegger, Can Firtina, Stephan Wong, Onur Mutlu, and Said Hamdioui, \textbf{``Swordfish: A Framework for Evaluating Deep Neural Network-Based Basecalling Using Computation-In-Memory with Non-Ideal Memristors,''} in \emph{MICRO}, 2023.

    \item Konstantinos Kanellopoulos, Rahul Bera, Kosta Stojiljkovic, F. Nisa Bostanci, Can Firtina, Rachata Ausavarungnirun, Rakesh Kumar, Nastaran Hajinazar, Mohammad Sadrosadati, Nandita Vijaykumar, and Onur Mutlu, \textbf{``Utopia: Fast and Efficient Address Translation via Hybrid Restrictive \& Flexible Virtual-to-Physical Address Mappings,''} in \emph{MICRO}, 2023. Artifact available: \url{https://github.com/CMU-SAFARI/Utopia}

    \item Haiyu Mao, Mohammed Alser, Mohammad Sadrosadati, Can Firtina, Akanksha Baranwal, Damla Senol Cali, Aditya Manglik, Nour Almadhoun Alserr, and Onur Mutlu, \textbf{``GenPIP: In-Memory Acceleration of Genome Analysis via Tight Integration of Basecalling and Read Mapping,''} in \emph{MICRO}, 2022.

    \item Jeremie S. Kim, Can Firtina, Meryem Banu Cavlak, Damla Senol Cali, Can Alkan, and Onur Mutlu, \textbf{``FastRemap: a tool for quickly remapping reads between genome assemblies,''} \emph{Bioinformatics}, 2022. Source code available: \url{https://github.com/CMU-SAFARI/FastRemap}

    \item Mohammed Alser, Joel Lindegger, Can Firtina, Nour Almadhoun, Haiyu Mao, Gagandeep Singh, Juan Gomez-Luna, and Onur Mutlu, \textbf{``From molecules to genomic variations: Accelerating genome analysis via intelligent algorithms and architectures,''} \emph{CSBJ}, 2022.

    \item Taha Shahroodi, Mahdi Zahedi, Can Firtina, Mohammed Alser, Stephan Wong, Onur Mutlu, and Said Hamdioui, \textbf{``Demeter: A fast and energy-efficient food profiler using hyperdimensional computing in memory,''} \emph{IEEE Access}, 2022.

    \item Damla Senol Cali, Konstantinos Kanellopoulos, Joël Lindegger, Zülal Bingöl, Gurpreet S. Kalsi, Ziyi Zuo, Can Firtina, Meryem Banu Cavlak, Jeremie Kim, Nika Mansouri Ghiasi, Gagandeep Singh, Juan Gómez-Luna, Nour Almadhoun Alserr, Mohammed Alser, Sreenivas Subramoney, Can Alkan, Saugata Ghose, and Onur Mutlu, \textbf{``SeGraM: A universal hardware accelerator for genomic sequence-to-graph and sequence-to-sequence mapping,''} in \emph{ISCA}, 2022. Artifact available: \url{https://github.com/CMU-SAFARI/SeGraM}

    \item Nika Mansouri Ghiasi, Jisung Park, Harun Mustafa, Jeremie Kim, Ataberk Olgun, Arvid Gollwitzer, Damla Senol Cali, Can Firtina, Haiyu Mao, Nour Almadhoun Alserr, Rachata Ausavarungnirun, Nandita Vijaykumar, Mohammed Alser, and Onur Mutlu, \textbf{``GenStore: A high-performance in-storage processing system for genome sequence analysis,''} in \emph{ASPLOS}, 2022. Artifact available: \url{https://github.com/CMU-SAFARI/GenStore}

    \item Damla Senol Cali, Gurpreet S. Kalsi, Zülal Bingöl, Can Firtina, Lavanya Subramanian, Jeremie S. Kim, Rachata Ausavarungnirun, Mohammed Alser, Juan Gomez-Luna, Amirali Boroumand, Anant Norion, Allison Scibisz, Sreenivas Subramoneyon, Can Alkan, Saugata Ghose, and Onur Mutlu, \textbf{``GenASM: A High-Performance, Low-Power Approximate String Matching Acceleration Framework for Genome Sequence Analysis,''} in \emph{MICRO}, 2020. Artifact available: \url{https://github.com/CMU-SAFARI/GenASM}
\end{enumerate}

\cleardoublepage
\balance
\begin{singlespace}
\setstretch{0.9}
\bibliographystyle{IEEEtran}
{\small \bibliography{references}}
\end{singlespace}

\bookmarksetup{startatroot}
\end{document}